%% file: book.tex
\documentclass[graybox,envcountchap,sectrefs]{svmono}
\usepackage[pdftex]{hyperref}
\usepackage{amsmath,amssymb}
\usepackage{chapterbib}
\usepackage{geometry}
\usepackage{changepage}
\usepackage{textcomp,marvosym}
\usepackage{algorithm}
\usepackage[noend]{algpseudocode}
\usepackage{nameref,hyperref}
\usepackage{microtype}
\usepackage{lastpage,fancyhdr,graphicx}
\usepackage{longtable}
\usepackage{lscape}
\usepackage{array}
\usepackage{multirow}
\usepackage{multicol}
\usepackage{cite}
\usepackage{mathptmx}
\usepackage{helvet}
\usepackage{courier}
\usepackage{type1cm}         

\usepackage{makeidx}         % allows index generation
\usepackage{graphicx}        % standard LaTeX graphics tool
\usepackage[bottom]{footmisc}% places footnotes at page bottom

\usepackage{lineno}
\makeatletter
\newcommand\nobreakhline{%
\multispan\LT@cols
\unskip\leaders\hrule\@height\arrayrulewidth\hfill\\*}
\makeatother
\date{}
\begin{document}

\author{Elaine Fehrman,
Vincent Egan,\\
Alexander N. Gorban,
Jeremy Levesley,\\
Evgeny M. Mirkes,
Awaz K. Muhammad}
\title{Personality Traits and  Drug Consumption }
\subtitle{A Story Told by Data}
\maketitle

\frontmatter%%%%%%%%%%%%%%%%%%%%%%%%%%%%%%%%%%%%%%%%%%%%%%%%%%%%%%
\abstract{In this book a story is told about the psychological traits associated with drug consumption. The book includes:
\begin{itemize}
\item A review of published works on the psychological profiles of drug users.
\item Analysis of a new original database with information on 1885 respondents and usage of 18 drugs. (Database is available online.)
\item An introductory description of the data mining and machine learning methods used for the analysis of this dataset.
\item The demonstration that the personality traits (five factor model, impulsivity, and sensation seeking), together with simple demographic data, give the possibility of predicting the risk of consumption of individual drugs with sensitivity and specificity above 70\% for most drugs.
\item The analysis of correlations of use of different substances and the description of the groups of drugs with correlated use (correlation pleiades).
\item Proof of significant differences of personality profiles for users of different drugs. This is explicitly proved for benzodiazepines, ecstasy, and heroin.
\item Tables of personality profiles for users and non-users of 18 substances.
\end{itemize}
The book is aimed at advanced undergraduates or first-year PhD students, as well as researchers and practitioners. No previous knowledge of machine learning, advanced data mining concepts or modern psychology of personality is assumed. For more detailed introduction into statistical methods we recommend several undergraduate textbooks. Familiarity with basic statistics and some experience in the use of probabilities would be helpful as well as some basic technical understanding of psychology.
}

%\include{dedic}
%\include{foreword}
\include{preface}

\tableofcontents

\mainmatter%%%%%%%%%%%%%%%%%%%%%%%%%%%%%%%%%%%%%%%%%%%%%%%%%%%%%%%
%\include{part}
\include{chapter1}

\include{chapter2}

\include{chapter3}

\include{summary}

\backmatter%%%%%%%%%%%%%%%%%%%%%%%%%%%%%%%%%%%%%%%%%%%%%%%%%%%%%%%
\include{appendix}

\include{AuthorsBio}
%\include{glossary}
%\include{solutions}
%\printindex

%%%%%%%%%%%%%%%%%%%%%%%%%%%%%%%%%%%%%%%%%%%%%%%%%%%%%%%%%%%%%%%%%%%%%%
\end{document}

%% file: preface.tex
%%%%%%%%%%%%%%%%%%%%%%preface.tex%%%%%%%%%%%%%%%%%%%%%%%%%%%%%%%%%%%%%%%%%
% sample preface
%FINAL
% Use this file as a template for your own input.
%
%%%%%%%%%%%%%%%%%%%%%%%% Springer %%%%%%%%%%%%%%%%%%%%%%%%%%

\preface

%% Please write your preface here
Each set of data can tell us a story. Our mission is to extract this story from the data and translate it into more readily accessible human language. There are a number of tools for such a translation.
To prepare this story we have to collect data, to ask interesting questions and to apply all the possible data mining technical tools to find the answers. Then we should
verify the answers, exclude spurious (overoptimistic) correlations and patterns, and tell the story to users.

The topic of mining interesting knowledge remains very intriguing. Many researchers have approached this problem from a plethora of
different angles. One of the main ideas in these approaches has been information gain (the more information gain there is, the more interesting the result is).
Nevertheless, we need a good understanding of what makes patterns that are found “interesting” from the end-user's point of view. Here various perspectives might be involved,
from practical importance to aesthetic beauty. The extraction of deep and interesting knowledge from data was formulated as an important problem for the
5th IEEE International Conference on Data Mining (ICDM 2005) \cite{YangWu2006}. Nowadays, the fast growth of the fields of data science and machine learning
provides us with many tools for answering such questions, but the art of asking interesting questions still requires human expertise.

The practical importance of the problem of evaluating an individual's risk of consuming and/or abusing drugs cannot be underestimated \cite{WorldDrug2016}. One might well ask how this risk depends  on a multitude of possible factors \cite{Hawkins1992}?  The linking of personality traits to risk of substance use disorder  is an enduring problem \cite{Kotov2010}. Researchers return again and again to this problem following the   collection of new data, and with new questions. 

How do personality, gender, education, nationality, age, and  other attributes affect this risk? Is this dependence different for different drugs? For example, does the risk of ecstasy consumption and the risk of heroin consumption differ for different personality profiles?  Which personality traits are the most important for evaluation of the risk of consumption of a particular drug, and are these traits different for different drugs? These questions are the focus of our research.

The data set we collected contains information on the consumption of 18 central nervous system psychoactive drugs, by 2,051 respondents (after cleaning, 1,885 participants remained, male/female = 943/942). The database is available online \cite{FehrmanData2016,FehrmanDataUCI}.

The questions we pose above have been reformulated as classification problems and many well-known data mining methods have been employed to address these problems: decision trees, random forests, $k$-nearest neighbours, linear discriminant analysis, Gaussian mixtures, probability density function estimation using radial basis functions, logistic regression and na{\"i}ve Bayes. For data preprocessing, transformation and ranking we have used methods such as polychoric correlation, nonlinear CatPCA (Categorical Principal Component Analysis), sparse PCA, and original double Kaiser's feature selection.

The main results of the work are:
\begin{itemize}
\item The presentation and descriptive analysis of a database with information on 1,885 respondents and their usage of 18 drugs.
\item Demonstration that the personality traits (Five Factor Model \cite{Costa92}, impulsivity, and sensation-seeking) together with simple demographic data give the possibility of predicting the risk of consumption of individual drugs with sensitivity and specificity above 70\% for most drugs.
\item The construction of the best classifiers and most significant predictors for each individual drug in question.
\item Revelation of significantly distinct personality profiles for users of different drugs; in particular,  groups of heroin and ecstasy users are significantly different in Neuroticism (higher for heroin), Extraversion (higher for ecstasy), Agreeableness (higher for Ecstasy),   and Impulsivity (higher for heroin); groups of heroin and benzodiazepine users are significantly different in Agreeableness (higher for benzodiazepines), Impulsivity (higher for heroin), and Sensation-Seeking  (higher for heroin); groups of ecstasy and benzodiazepine users are significantly different in  Neuroticism (higher for benzodiazepines), Extraversion (higher for ecstasy), Openness to Experience (higher for ecstasy), and Sensation-Seeking  (higher for ecstasy).
\item The discovery of three correlation pleiades of drugs; these are the clusters of drugs with correlated consumption centered around heroin, ecstasy, and benzodiazepines.  The correlation pleiades should include the mini-sequences of drug involvement  found in longitudinal studies \cite{Newcomb1986}, and aim to serve as maps for analysis of  different patterns of influence.
\item The development of risk map technology for the visualization of the probability of drug consumption.
\end{itemize}

Four of the authors (ANG, JL, EMM, and AKM) are applied mathematicians and two are psychologists (EF and VE). Data were collected by EF and processed by EMM and AKM. The psychological framework for this study was developed by  EF and VE, and the analytic methodology was selected and developed by ANG and EMM. The final results were
 critically analysed and described by ANG, JL, EMM, and AKM from the data mining perspective, and EF and VE provided the psychological interpretation and conceptualization. 

For psychologists, the book gives a new understanding of the relationship between  personality traits and the usage of 18 psychoactive substances, provides a new openly available database for further study, and presents many useful methods of data analysis. For applied mathematicians and statisticians, the book details a case study in a fascinating area of application, exemplifying the use of various data mining methods in such scenarios.
 
This book is aimed at advanced undergraduates or first-year PhD students, as well as researchers and practitioners in data analysis, applied mathematics and psychology. No previous knowledge of machine learning, advanced data mining concepts or  psychology of personality is assumed. Familiarity with basic statistics and some experience of the use of probability is helpful, as well as some basic understanding of psychology. Two books \cite{CorrPersonality2009,WittenTibshiraniIntro2013} include all the necessary prerequisites (and much more). Linear Discriminant Analysis (LDA), Principal Component Analysis (PCA), and Decision Trees (DT) are systematically employed in the book. Therefore, it may be useful to refresh the knowledge of these classical methods using the textbook [10], which is concentrated more on the applications of the methods and less on the mathematical details.

A preliminary report of our work was published as an arXiv e-print in 2015 \cite{Fehrman15} and presented at the Conference of International Federation of Classification Societies 2015 (IFCS 2015) \cite{FehrmanBologna15}.

This book is not the end of the story told by the data. We will continue our work and try to extract more interesting knowledge and patterns from the data. Moreover, we are happy for you, the readers, to join us in this adventure.
We believe that every large annotated dataset is a treasure trove and that there is an abundance of interesting knowledge to discover from them. We have published our database online \cite{FehrmanData2016,FehrmanDataUCI} and invite everybody to use it for their own projects, from BSc and MSc level to PhD, or just for curiosity-driven research. We would be very happy to see the fascinating outcomes of these projects.

\begingroup
\renewcommand{\addcontentsline}[3]{}

\endgroup

\vspace{\baselineskip}
\begin{flushright}\noindent
Leicester -- Nottingham, \hfill {\it Elaine Fehrman}\\
December 2017 \hfill {\it Vincent Egan}\\
\hfill {\it Alexander N. Gorban}\\
\hfill {\it Jeremy Levesley}\\
\hfill {\it Evgeny M. Mirkes}\\
\hfill {\it Awaz K. Muhammad}\\
\end{flushright}

%% file: chapter1.tex
%%%%%%%%%%%%%%%%%%%%% chapter.tex %%%%%%%%%%%%%%%%%%%%%%%%%%%%%%%%%
%FINAL
% sample chapter
%
% Use this file as a template for your own input.
%
%%%%%%%%%%%%%%%%%%%%%%%% Springer-Verlag %%%%%%%%%%%%%%%%%%%%%%%%%%
%\motto{Use the template \emph{chapter.tex} to style the various elements of your chapter content.}

\chapter{Introduction: drug use and personality profiles}
\label{intro} % Always give a unique label
% use \chaptermark{}
% to alter or adjust the chapter heading in the running head

\abstract*{Drug use disorder is characterised by several terms: addiction, dependence, and abuse. We discuss the notion of psychoactive substance and relations between the existing definitions. The personality traits which may be important for predisposition to use of drugs are introduced: the five factor model, impulsivity, and sensation-seeking. A number of studies have illustrated that personality traits are associated with drug consumption. The previous pertinent results are reviewed. A database with information on 1885 respondents and their usage of 18 drugs is introduced. The results of our study are briefly outlined: The personality traits (five factor model, impulsivity, and sensation-seeking) together with simple demographic data make possible the prediction of the risk of consumption of individual drugs; Personality profiles for users of different drugs. In particular, groups of heroin and ecstasy users are significantly different; There exist three correlation pleiades of drugs. These are clusters of drugs with correlated consumption, centered around heroin, ecstasy, and benzodiazepines.}

\section{Definitions of drugs and drug usage}

Since Sir Karl Popper, it has become a commonplace opinion in the philosophy of science that the `value' of definitions, other than for  mathematics, is generally unhelpful. Nevertheless, for many more practical spheres of activity, from jurisprudence to health planning, definitions are necessary to impose theoretical boundaries on a subject, in spite of their incompleteness and their tendency to change with time. This applies strongly to definitions of drugs and drug use. 

Following the standard definitions \cite{Kleiman11}, 
\begin{itemize}
\item A {\em drug} is a `chemical that influences biological function (other than by providing nutrition or hydration)'. 
\item A {\em psychoactive drug} is a `drug whose influence is in a part on mental functions'.
\item An {\em abusable   psychoactive drug} is a `drug whose mental effects are sufficiently pleasant or interesting or helpful that some people choose to take it for a reason other than to relieve a specific malady'.
	\end{itemize}
	 In our study we use the term `drug' for  abusable   psychoactive drug regardless of whether it is illicit or not.
While legal substances  such as sugar, alcohol and tobacco are probably responsible for far more premature death than illegal recreational drugs \cite{Beaglehole11}, the social and personal consequences of recreational drug use can be highly problematic \cite{Bickel14}.

Use of drugs introduces risk into a life across a broad spectrum; it constitutes an important factor for increasing risk of poor health, along with earlier mortality and morbidity, and has significant consequences for society \cite{McGinnis93,Sutina13}. Drug consumption and addiction constitutes a serious problem globally. Though drug use is argued by civil libertarians to be a matter of individual choice, the effects on an individual of drug use such as greater mortality or lowered individual functioning, suggest that drug use has social and interpersonal effects on individuals who have not chosen to use drugs themselves.

Several terms are used to characterise drug use disorder: addiction, dependence, and abuse.
For a long time, `substance abuse' and `substance dependence' were considered as separate disorders. In 2013, The Diagnostic and Statistical Manual of Mental Disorders (DSM-5) joined these two diagnoses into `Substance Use Disorder' \cite{DSM52013}. This is a more inclusive term used to  identify persons with substance-related problems. More recently, abuse and dependence have been defined on a scale that measures the time and degree of substance use. Criteria are provided for substance use disorder, supplemented by criteria for intoxication, withdrawal, substance/medication-induced disorders, and unspecified substance-induced disorders, where relevant. Abuse can be considered as the early stage of substance use disorder. 

In our study we differentiate the substance users on the basis of recency of use but do not identify existence and depth of the substance dependence. 

For prevention and effective care of substance use disorder, we need to identify the risk factors and develop methods for their evaluation and control \cite{WHO04}.

\section{Personality traits}

Sir Francis Galton (1884) proposed  that a lexical approach in which one used dictionary definitions of dispositions could be a means of constructing a description of individual differences (see \cite{Goldberg1993}). He selected the personality-descriptive terms and stated the problem of their interrelations.  In 1934, Thurstone  \cite{Thurstone1934} selected 60 adjectives that are in common use for describing people and asked each of 1300 respondents  to think of a person they knew well and to select the adjectives that can best describe this person. After studying the correlation matrix he found that {\em five} factors are sufficient to describe this choice. 

There have been many versions of  the five factor model  proposed since Thurston \cite{Digman1990}, for example:
\begin{itemize}
\item Surgency,  agreeableness, dependability,  emotional  stability,  and  culture;
\item Surgency,  agreeableness,   conscientiousness,  emotional  stability,  and  culture;
\item Assertiveness,  likeability,  emotionality,  in­telligence,  and  responsibility;
\item Social  adaptability, conformity, will  to  achieve, emotional  control, and inquiring  intellect;
\item Assertiveness,   likeability,  task  interest, emotionality, and intelligence;
\item Extraversion,  friendly  compliance,  will  to  achieve, neuroticism, and intellect;
\item Power,  love,  work, affect, and intellect;
\item Interpersonal  involvement, level  of  socialization,  self-control, emotional  stability, independence.
\end{itemize}
There are also systems with different numbers of factors (three, seven, etc.). The most important three-factor system is Eysenck's model comprising  extraversion, psychoticism, and neuroticism.

Nowadays, after many years of research and development, psychologists have largely agreed that the personality traits of the modern Five Factor Model (FFM) constitutes the most comprehensive and adaptable system for understanding human individual differences \cite{Costa92}. The FFM comprises Neuroticism (N), Extraversion (E), Openness to Experience (O), Agreeableness (A), and Conscientiousness (C).

The five traits can be summarized thus:
\begin{description}
\item[N] \emph{Neuroticism}  is a long-term tendency to experience negative emotions such as nervousness, tension, anxiety and depression (associated adjectives \cite{McCrae1992}: anxious, self-pitying, tense, touchy, unstable, and worrying);
\item[E] \emph{Extraversion}  manifested in characters who are outgoing, warm, active, assertive, talkative, and cheerful; these persons are often in search of stimulation (associated adjectives: active, assertive, energetic, enthusiastic, outgoing, and talkative);
\item[O] \emph{Openness to experience}  is associated with a   general appreciation for art, unusual ideas, and imaginative, creative, unconventional, and wide interests (associated adjectives: artistic, curious, imaginative, insightful, original, and wide interest);
\item[A] \emph{Agreeableness}  is a dimension of interpersonal relations, characterized by altruism, trust, modesty, kindness, compassion and cooperativeness (associated adjectives: appreciative, forgiving, generous, kind, sympathetic, and trusting);
\item[C] \emph{Conscientiousness} is a tendency to be organized and dependable, strong-willed, persistent, reliable, and efficient (associated adjectives: efficient, organised, reliable, responsible, and thorough).
\end{description}
Individuals low on the A and C trait dimensions have less incidence of the reported attributes, so, for example, lower Agreeableness is associated with greater antisocial behaviour \cite{Egan11}.

\section{How many inputs do the predictive models have: 5, 30, 60, or 240?}

The NEO PI-R questionnaire was specifically designed to measure the FFM of personality \cite{Costa92}. It provides scores corresponding to N, E, O, A, and C (`domain scores'). The NEO PI-R consists of 240 self-report items answered on a five-point scale, with separate scales for each of the five domains. Each scale consists of six correlated sub-scales (`facets'). A list of the facets within each domain is presented in the first column of Table~\ref{tab:Facet}.

There are several versions of the FFM questionnaire: NEO PI-R with 240 questions (`items'), 30 facets, and five domains; the older NEO-FFI with 180 items, etc. A shorter version of the Revised NEO Personality Inventory (NEO-PI-R), the NEO-Five
Factor Inventory (NEO-FFI), has 60 items (12 per domain and no facet structure) selected from the original items \cite{Costa92}. This shorter questionnaire was revised \cite{McCrae04} after Egan et al. demonstrated that the robustness of the original version should be improved \cite{Egan00}. NEO-FFI was designed as a brief instrument that provides estimates of the factors for use in exploratory research.

The values of  the five factors are used as inputs in numerous statistical models for prediction, diagnosis, and risk evaluation. These models are employed in psychology, psychiatry, medicine, education, sociology, and many other areas where personality may be important. For example \cite{Poropat2009}, academic performance at primary school was found to significantly correlate with Emotional Stability (+), Agreeableness (+), Conscientiousness (+), and Openness to Experience (+) (the sign of correlations is presented in parentheses). Success in primary school is also significantly and highly correlated with intelligence (+), the Pearson correlation coefficient $r>0.5$. For higher academic levels, correlations of Academic  Performance with Emotional Stability, Agreeableness, and Openess, significantly decreases ($r\lessapprox 0.1$). Correlation with Intelligence also decreases by two or more, but correlation with Conscientiousness remains almost the same for all academic levels ($r\approx 0.21-0.28$). Correlations between Conscientiousness and Academic Performance were largely independent of Intelligence. This knowledge can be useful for educational professionals and parents.

Another example demonstrates how personality affects career success \cite{Seibert2001}.
Extraversion was related positively to salary level, promotions, and career satisfaction, and Neuroticism was related negatively to career satisfaction. Agreeableness was related negatively only to career satisfaction and Openness was related negatively to salary level. There was a significant negative relationship between Agreeableness and salary for individuals in people-oriented occupations (with direct interaction with clients, for example) but no such relationships were found in occupations without a strong `people' component. At the same time, Agreeableness is positively correlated with performance in jobs involving teamwork (interaction with co-workers) \cite{Mount1998}. These result are of interest to Human Resources departments.

\begin{figure}
\centering
\includegraphics[width=0.9\textwidth]{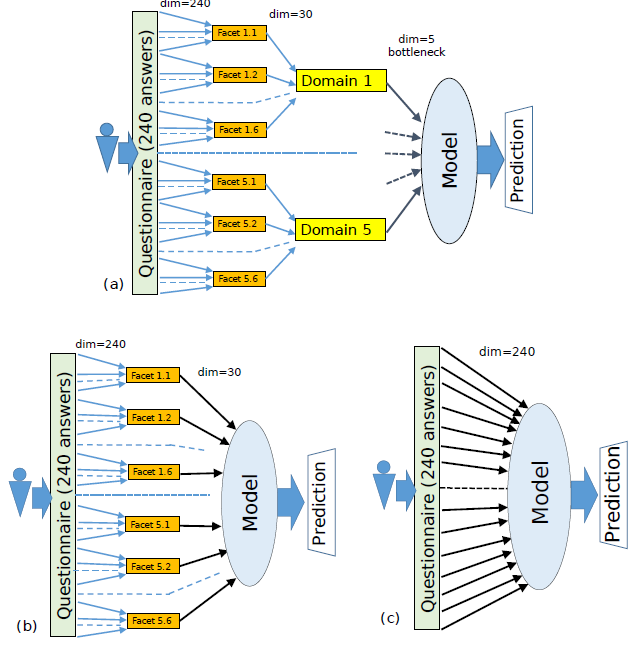}
\caption{Three types of predictive models based on the FFM NEO PI-R questionnaire: (a) statistical models which uses five FFM inputs prepared by the standard FFM procedure, (b) facet-based predictors \cite{Widiger2002, Bagby2005}, and (c) direct predictors, which avoid the step of explicit diagnosis and work with multidimensional raw input information (usually, Artificial Intelligence models like neural networks \cite{Dorrer1995, Gorban1995}).}
\label{FFM MOdels}
\end{figure}

Most of the statistical models use the values of five factors (N, E, O, A, C) as the inputs and produce assessment, diagnosis, recommendations or prognosis as the outputs (Fig.~\ref{FFM MOdels}a). For the NEO PI-R questionnaire this means that we take the 240 inputs, transform them into 30 facet values, then transform these 30 numbers into five factors and use these five numbers as the inputs for the statistical or, more broadly, data analytic model. To construct this model with five inputs and the desired outputs, one should use data with known answers and supervising learning (or,  more narrowly, various regression and classification models). The crucial question arises: is it true that for all specific diagnosis, assessment, prognosis, and recommendation problems the facets should be linearly combined with the same coefficients?

An alternative version is the facet trait model (Fig.~\ref{FFM MOdels}b), where combining of facets into the final output depends on the problem and data \cite{Bagby2005}. We can go further and consider flexible combination of the raw information, the questionnaire answers for each problem, and dataset (Fig.~\ref{FFM MOdels}c) \cite{Dorrer1995, Gorban1995}.

One of the most developed area of FFM application is psychiatry and psychology, for example, for the assessment of personality psychopathology. The facet trait model created for 10 personality disorders \cite{Widiger2002, Bagby2005} demonstrates that optimal combinations of facets into predictors is not uniform inside the domains (Table~\ref{tab:Facet}). Some facets are more important for assessment than the others and the selection of important facets depends on the specific personality disorder (see Table~\ref{tab:Facet}). Nevertheless, there are almost no internal contradictions  inside domains in Table~\ref{tab:Facet}: for almost each domain and any given disorder all significant facets have the same sign of deviation from the norm: either all have higher values ($\Uparrow$ ) or all have lower values ($\Downarrow$). The only exclusion is the contradiction between facets `Warmth' and 
`Assertiveness' from the domain `Extraversion': both are important for the diagnosis `Dependent' but for this diagnosis `Warmth'  is expected to be higher than average and `Assertiveness' is expected to be lower.

\begin{table}[!ht]
\centering
\caption{FFM facet trait predictor set for DSM-IV PD \cite{Widiger2002, Bagby2005}. }
\label{tab:Facet} % Give a unique label
\begin{tabular}{|l|cccccccccc|}\hline
FFM &PAR &SZD &SZT &ATS &BDL &HST &NAR &AVD &DEP &OBC\\ \hline
{\em Neuroticism} & & & & & & & & & & \\
\hspace{5mm}Anxiety & & & $\Uparrow$ & & $\Uparrow$ & & & $\Uparrow$ & $\Uparrow$ & \\
\hspace{5mm}Angry Hostility& $\Uparrow$ & & & $\Uparrow$ &$\Uparrow$ & &$\Uparrow$ & & & \\
\hspace{5mm}Depression & & & & & $\Uparrow$ & $\Uparrow$ & & $\Uparrow$ & & \\
\hspace{5mm}Self-consciousness&& & $\Uparrow$ & & & $\Uparrow$ & $\Uparrow$ & $\Uparrow$ & $\Uparrow$ & \\
\hspace{5mm}Impulsiveness & & & & & $\Uparrow$ & & & & & \\
\hspace{5mm}Vulnerability & & & & & $\Uparrow$ & & & $\Uparrow$ & $\Uparrow$ & \\ \hline
{\em Extraversion} & & & & & & & & & & \\
\hspace{5mm}Warmth & &$\Downarrow$ & $\Downarrow$ & & & $\Uparrow$ & & & $\Uparrow$ & \\
\hspace{5mm}Gregariousness& &$\Downarrow$ & $\Downarrow$ & & & $\Uparrow$ & & $\Downarrow$ & & \\
\hspace{5mm}Assertiveness & & & & & & & & $\Downarrow$ & $\Downarrow$ & $\Uparrow$ \\
\hspace{5mm}Activity & & & & & & & & & & \\
\hspace{5mm}Excitement seeking&& & &$\Uparrow$ & & $\Uparrow$ & & $\Downarrow$ & & \\
\hspace{5mm}Positive emotions &&$\Downarrow$ & $\Downarrow$ & & &$\Uparrow$ & & & & \\ \hline
{\em Openness to Experience} & & & & & & & & & & \\
\hspace{5mm}Fantasy & & & $\Uparrow$ & & & $\Uparrow$ & $\Uparrow$ & & & \\
\hspace{5mm}Aesthetics & & & & & & & & & & \\
\hspace{5mm}Feelings & & $\Downarrow$ & & & & $\Uparrow$ & & & & \\
\hspace{5mm}Actions & & & $\Uparrow$ & & & & & & & \\
\hspace{5mm}Ideas & & & $\Uparrow$ & & & & & & & \\
\hspace{5mm}Values & & & & & & & & & & $\Downarrow$\\ \hline
{\em Agreeableness} & & & & & & & & & & \\
\hspace{5mm}Trust & $\Downarrow$ & & $\Downarrow$ & & $\Downarrow$ & $\Uparrow$ & & & $\Uparrow$ & \\
\hspace{5mm}Straightforwardness&$\Downarrow$& & & $\Downarrow$ & & & & & & \\
\hspace{5mm}Altruism & & & & $\Downarrow$ & & & $\Downarrow$ & & $\Uparrow$ & \\
\hspace{5mm}Compliance &$\Downarrow$ & & & $\Downarrow$ & $\Downarrow$ & & & & $\Uparrow$ &$\Downarrow$ \\
\hspace{5mm}Modesty & & & & & & & $\Downarrow$ & &$\Uparrow$ & \\
\hspace{5mm}Tender mindedness& & & & $\Downarrow$ & & & $\Downarrow$ & & & \\ \hline
{\em Conscientiousness } & & & & & & & & & & \\
\hspace{5mm}Competence & & & & & $\Downarrow$ & & & & & $\Uparrow$ \\
\hspace{5mm}Order & & & & & & & & & & $\Uparrow$ \\
\hspace{5mm}Dutifulness & & & & $\Downarrow$& & & & & & $\Uparrow$ \\
\hspace{5mm}Achievement striving&& & & & & &$\Uparrow$ & & & $\Uparrow$ \\
\hspace{5mm}Self-discipline& & & & $\Downarrow$ & & & & & & \\
\hspace{5mm}Deliberation & & & & $\Downarrow$ & & & & & & \\\hline
\end{tabular}\\
$\Uparrow$=high values; $\Downarrow$=low values; Personality disorders: PAR=Paranoid;SZD=Schizoid; SZT=Schizotypal; ATS=Antisocial; BDL=Borderline; HST=Histrionic; NAR=Narcissis-Narcissistic; AVD=Avoidant; DEP=Dependent; OBC=Obsessive-compulsive.
\end{table}

In 1995, Dorrer and Gorban with co-authors \cite{Dorrer1995} employed neural network technology and  the original software library MultiNeuron \cite{Gorban1995} for direct prediction of human relations on the basis of raw questionnaire information.  A specially reduced personality questionnaire with 91  questions was prepared. The possible answers to each question were: `yes', `do not know', and `no', which were coded as $+1$, 0, and $-1$, correspondingly. The neural networks (committees of six networks of different architecture)  were prepared to predict results of sociometry of relations between university students inside an academic group. Neural networks had to predict students' answers to the sociometric  question: ``To what degree would you like to work in your future profession with this group member?'' The answer was supposed to be given as a 10-point estimate (0 - most negative attitude to a person as a would-be co-worker, 10 - maximum positive).  The status and expansivity of each group member were evaluated from the answers to these questions. Sociometric status is a measurement that reflects the degree to which someone is liked or disliked by their peers from a group. Social expansivity is the tendency of a group members to choose and highly evaluate many others. These two characteristics  were used as elements of neural network output vector for each person. The inputs were 91 answers of this person to the pesonality questionnaire. The neural networks were trained on data from several academic groups and tested on academic groups never seen before. 

The  91 questions from the questionnaire were ranked by importance for the neural networks prediction.  Cross-validation showed that reduction of the questionnaire to 46 questions (the empirically optimal number in these experiments) gave the best prediction result. Committees of networks  always gave better results than a single network. 

While such (relatively novel) systems are often more accurate they are more costly in two ways: they are hungrier in terms of data requirements and computational resources.

In this book, we focus on the classical systems with explicitly measured personality, which have bottleneck of five (or seven) factors (Fig.~\ref{FFM MOdels}a). Nevertheless, modern development of artificial intelligence and neural network systems ensures us that the computational models,  which process raw information without explicitly describing personality (Fig.~\ref{FFM MOdels}c), could be important elements of future  personality assessment.

\section{The problem of the relationship between personality traits and drug consumption}

There are numerous {\em risk factors} for addiction, which are defined as any attribute, characteristic, or event in the life of an individual that increase the probability of drug consumption. A number of such attributes are correlated with initial drug use, including genetic inheritance as well as psychological, social, individual, environmental, and economic factors \cite{Cleveland08,Ventura14,WHO04,McGue2000}.  Modest genetic and strong environmental influences on adolescent illicit substance use and abuse is impressively consistent across  multiple substances  \cite{McGue2000}. 

The important risk factors are likewise associated with a number of personality traits \cite{Dubey10,Bogg04}.

There is a well-known problem in analysing of the psychological traits associated with drug use: to distinguish  the effect of drug use from the cause of \cite{KhantzianKhantzian1984}. To solve this problem, we have to use {\em relatively constant psychological traits}. Another solution is to organise large longitudinal studies which will analyse the traits of the persons at the different stages of drug use (such an approach seems to be more or less impossible for a number of reasons).
The concepts of states, traits, and of causality, are crucial for a psychological theory of personality and drug use. They remain part of the focus of ongoing research \cite{Steyer2015}.

Stability of personality traits is one of the keystones of  personality measurement and is indicated by the importance placed on the reliability and validity of psychometric measurement. The stability problem of measured personality was approached many times within the FFM framework, from the very beginning. The  hypothesis was that in `normal' individuals the stable self-concept is crystallized in early adulthood \cite{McCrae1982}. The results were summarised in four items \cite{McCraeCostaStability1994}. In brief: (1) ``The mean levels of personality traits change with development but reach final adult level at about age 30.''  (2) ``Individual differences in personality traits, which show at least some continuity from early childhood on, are also essentially fixed by age 30.'' (3) ``Stability appears to characterize  the major domains of personality'' --  N, E, O, A, C. (4) ``Generalizations about stability apply to virtually everyone'' (to all healthy adults)  \cite{SchmittAllik2007}.
 
Leading experts examined various theories of personality change and stability, and state-of-the-art measurement issues in a special volume \cite{HeathertonWeinberger1994}. Costa and McCrae reviewed evidence suggesting that, in most cases, personality traits are indeed unchanging \cite{CostaMcGrae1994IN}. On another hand, Weinberger \cite{Weinberger1994} presented an elegant argument:  ``Psychotherapy represents the most sustained and well-documented effort to trigger psychological change.'' Therefore, we can conclude that there is a place for doubt: either personality can change or psychotherapy cannot provide personality change (i.e. it can be only superficial). However, the evidence that therapy may change persons is limited; psychological and pharmacological treatment led persons treated for depression to become more extraverted, open to experience, agreeable and conscientious, and substantially more emotional stable after treatment, though these changes were largely unrelated to initial depression severity \cite{DeFruyt2006}.
 
 Stability of personality has been tested with samples of drug users in several studies. For example, 230 opioid-dependent patients completed NEO Pl-R at admission and again, approximately 19 weeks later. Results indicated fair to good stability for all NEO PI-R factor domain scores (N, E, O, A, C) \cite{Carter2001}. Stability of the scores  was not significantly affected by drug-positive versus drug-negative status at follow-up.

A number of studies have illustrated that personality traits are associated with drug consumption. Meta-analysis of associations between personality traits  and specific depressive, anxiety, and substance use disorders in adults showed that all diagnostic groups were high on N and low on C \cite{Kotov2010}. This analysis involved 175 studies published from 1980 to 2007.

Several studies of opioid dependent samples demonstrate high N, low C, and average O \cite{Brooner1994, Carter2001, Kornor2007}. There are also some differences betweens studies and cohorts: a Norwegian group of opioid users demonstrated lower E \cite{Kornor2007}, whereas USA groups did not significantly deviate in E from the norm  \cite{Brooner1994, Carter2001}. O  and A  were observed lower in Norwegian drug dependent patients than in controls, but given the sample size (65 patients) the difference was not significant.

The combination N$\Uparrow$,  A$\Downarrow$, and C$\Downarrow$ for substance abusers was reported by \cite{McCormick1998} (here, and below arrows indicate direction of deviation from the mean level). According to \cite{McCormick1998}, the cocaine users were characterized by higher levels of E and O, whereas polysubstance users were characterized by lower levels of A  and C . 

We expected that drug usage would be associated with high N, low A, and low C ( N$\Uparrow$,  A$\Downarrow$, and C$\Downarrow$).
This combination was observed for various types of psychopathy and deviant behavior. For example, in the  analysis of  the {\em `dark triad' of personality}, Machiavellianism, Narcissism and Psychopathy were evident	 \cite{Jakobwitz06}.
\begin{itemize}
\item {\em Machiavellianism} refers to interpersonal strategies that advocate self-interest, deception and manipulation. A questionnaire MACH-IV is   the most widely used tool to measure MACH, the score of Machiavellianism \cite{McHoskey1998}. Persons high in MACH are likely to exploit others and less likely to be concerned about other people beyond their own self-interest.
\item The concept of {\em narcissism} comes from the Greek myth of Narcissus who falls in love with his own reflection. Formalised in psychodynamic theory, it describes a pathological
form of self-love. One commonly used operational definition of narcissism is based on the Narcissistic Personality Inventory, that measures persistent attention seeking,
extreme vanity, excessive self-focus, and exploitativeness in interpersonal relations. It comprises four factors: Exploitativeness/Entitlement, Leadership/Authority,
Superiority/Arrogance and Self-Absorption/Self-Admiration \cite{RaskinHall1979}.
\item Psychopathy can be measured as a detailed structured interview, or as a brief self-report.  The key constructs are operationalised in Levenson's self-report measure of psychopathy which measures two facets of psychopathy. Factor 1 reflects {\em primary psychopathy} (e.g., selfishness, callousness, lack of interpersonal affect, superficial charm and remorselessness), and Factor 2 measures antisocial lifestyle and behaviours, and is akin to {\em secondary psychopathy} (excessive risk-takers who exhibit usual amounts of stress and guilt). 
\item In clinical practice, Hare's Psychopathy Checklist Revised (PCL-R) \cite{Hare2003} is `the-state-of-the-art'. The PCL-R uses a semi-structured interview and case-history information summarised in 20 items  scored by trained professional. It also measures two correlated factors. Factor 1 (the core personality traits of psychopathy) is labelled as ``selfish, callous and remorseless use of others''. Factor 2  is labelled as ``chronically unstable, antisocial and socially deviant lifestyle". It is associated with reactive anger, criminality, and impulsive violence.
\end{itemize}

Analysis of the `dark triad' of personality showed that N was  positively associated with psychopathy and Machiavellianism (Table \ref{tab:04}).  The dark dimension of
personality can be described in terms of low A, whereas much of the antisocial behaviour in `normal' people appears underpinned by high N, low A, and low C  \cite{Jakobwitz06}.

\begin{table}[!ht]
\centering
\caption{ PCC between NEO-FFI trait scores and the  `dark triad' scores ($n$=82). PP stands for primary psychopathy, SP for secondary psycholathy,
M for Machiavellianism, and Nar for narcissism scores. }
\label{tab:04}       % Give a unique label
\begin{tabular}{|c|c|c|c|c|c|}\hline
& {N} &{E}&{O}&{A}&{C}\\\hline
PP&0.30***&	0.08 &$-0.21$* &	$-0.43$***& 	$-0.21$*\\\hline
SP& 0.47***	&$0.04$ &$-0.21$*	 &	$-0.23$** &	$-0.19$* \\\hline
M&	0.38***&	$-0.13$& $-0.17$	 &	$-0.41$***&	$-0.27$**\\\hline
Nar&	$-0.10$ &0.10&	0.10 &	$-0.43$***& $-0.24$** \\\hline
\end{tabular}\\
*$p<0.1$, **$p<0.05$, ***$p<0.01$
\end{table}

More detailed factor analysis found that Machiavellianism and psychopathy shared common variance and were primary defined by a very high loading (-0.725) for Agreeableness, whereas narcissism was primarily defined by it's high loadings for extroversion and intellect, once the common psychopathic variance associated with narcissism was rotated to the antisocial element of the dark triad (psychopathy and Machiavellianism) \cite{Egan2014}.

The so-called `\emph{negative urgency}' is the tendency to act rashly when distressed, and is also characterized by high N, low A,  and low C \cite{Settles12}.
Negative urgency predicted alcohol dependence symptoms in personality disordered women, drinking problems and smoker status in pre-adolescents, and aggression, risky sex, illegal drug use, drinking problems, and disordered behavior in college students.

Thus, the hypothesis about the personality profile  N$\Uparrow$, A$\Downarrow$ and C$\Downarrow$ for drug users has a reliable background. We
validated this hypothesis with our data and found, indeed, that for some groups of drug users it holds true. For example, for heroin and methadone users, we
found this typical combination. At the same time, we found various deviations from this profile for other drugs. For example, for groups of recent  LSD users (used less than a year ago, or used less than a month ago, or used less than a week ago),  N does not deviate significantly from the mean but O and C do: O$\Uparrow$, C$\Downarrow$.  Our findings suggest also that O is higher for many groups of drug users. Detailed profiles of all groups of users are presented in  \ref{Appendix 1}.

Roncero et al \cite{Roncero14} highlighted the importance of the relationship between high N and the presence of psychotic symptoms following cocaine-induced drug consumption. Vollrath \& Torgersen \cite{Vollrath02} observed that the personality traits of N, E, and C are highly correlated with hazardous health behaviours. A low score of C, and high score of E, or a high score of N correlate strongly with multiple risky health behaviours. Formally, this profile associated with risk can be described as  (C$\Downarrow$~AND~E$\Uparrow$)~OR~N$\Uparrow$.
 Flory et al \cite{Flory02} found alcohol use to be associated with lower A and C, and higher E. They also found that lower A and C, and higher O are associated with marijuana use. Sutin et al \cite{Sutina13} demonstrated that the relationship between low C and drug consumption is moderated by poverty; low C is a stronger risk factor for illicit drug usage among those with relatively higher socioeconomic status. They found that high N, and low A and C are associated with higher risk of drug use (including cocaine, crack, morphine, codeine, and heroin). It should be mentioned that high N is positively associated with many other addictions like internet addiction, exercise addiction, compulsive buying, and study addiction \cite{Andreassen2013}.

An individual's personality profile contributes to becoming a drug user. Terracciano et al \cite{Terracciano08} demonstrated that compared to `never smokers', current cigarette smokers were lower on C  and higher on N. They found that the profiles for cocaine and heroin users scored very high on N, and very low on C whilst marijuana users scored high on O but low on A, and C. Turiano et al \cite{Turiano12} found a positive correlation between N and O, and drug use, while increasing scores for C and A decreases risk of drug use. Previous studies demonstrated that participants who use drugs, including alcohol and nicotine, have a strong positive correlation between A and C, and a strong negative correlation for each of these factors with N \cite{Stewart00,Haider02}. Three high-order personality traits are proposed as endophenotypes for substance use disorders: Positive Emotionality, Negative Emotionality, and Constraint \cite{Belcher2016}.

The problem of risk evaluation for individuals is much more complex. This was explored very recently by Yasnitskiy et al \cite{Yasnitskiy15}, Valeroa et al \cite{Valeroa14} and Bulut \& Bucak \cite{Bulut14}. Both individual and environmental factors predict substance use, and different patterns of interaction among these factors may have different implications \cite{Rioux2016}. Age is a very important attribute for diagnosis and prognosis of substance use disorders. In particular, early adolescent onset of substance use is a robust predictor of future substance use disorders \cite{Weissman2015}.

Valeroa et al \cite{Valeroa14} evaluated the individual risk of drug consumption for alcohol, cocaine, opiates, cannabis, ecstasy, and amphetamines. Input data were collected using a Spanish version of the Zuckerman-Kuhlman Personality Questionnaire (ZKPQ). Two samples were used in this study. The first one consisted of 336 drug dependent psychiatric patients of one hospital. The second sample included 486 control individuals. The authors used a decision tree as a tool to identify the most informative attributes. The sensitivity (proportion of correctly identified positives) of 40\% and specificity (proportion of correctly identified negatives) of 94\% were achieved for the training set. The main purpose of this research was to test if predicting drug consumption was possible and to identify the most informative attributes using data mining methods. Decision tree methods were applied to explore the differential role of personality profiles in drug consumer and control individuals. The two personality factors, Neuroticism and anxiety and the ZKPQ's Impulsivity, were found to be most relevant for drug consumption prediction. The low sensitivity (40\%) score means that such a decision tree cannot be applied to real life situations.

Without focussing on specific addictions, Bulut \& Bucak \cite{Bulut14} estimated the proportion of teenagers who exhibit a high risk of addiction. The attributes were collected by an original questionnaire, which included 25 questions. The form was filled in by 671 students. The first 20 questions asked about the teenagers' financial situation, temperament type, family and social relations, and cultural preferences. The last five questions were completed by their teachers and concerned the grade point average of the student for the previous semester according to a five-point grading system, whether the student had been given any disciplinary punishment so far, if the student had alcohol problems, if the student smoked cigarettes or used tobacco products, and whether the student misused substances.
In Bulut et al's study there were five risk classes as outputs. The authors diagnosed teenagers' risk of being a drug abuser using seven types of classification algorithms: $k$-nearest neighbor, ID3 and C4.5 decision tree based algorithms, na{\"i}ve Bayes classifier, na{\"i}ve Bayes/decision trees hybrid approach, one-attribute-rule, and projective adaptive resonance theory. The classification accuracy of the best classifier was reported as 98\%.

Yasnitskiy et al \cite{Yasnitskiy15}, attempted to evaluate the individual's risk of illicit drug consumption and to recommend the most efficient changes in the individual's social environment to reduce this risk. The input and output features were collected by an original questionnaire. The attributes consisted of: level of education, having friends who use drugs, temperament type, number of children in the family, financial situation,  levels of alcohol and cigarette smoking consumption, family relations (cases of physical, emotional and psychological abuse, level of trust and happiness in the family). There were 72 participants. A neural network model was used to evaluate the importance of attributes to diagnose the tendency towards drug addiction. A series of virtual experiments was performed  for several test patients (drug users) to evaluate how possible it is to control the propensity for drug addiction.  The most effective change of social environment features was predicted for each person. The recommended changes depended on the personal profile, and significantly varied for different participants. This approach produced individual bespoke advice to affect decreasing drug dependence.

Profiles of drug users have some similarity for all drugs (for example, N$\Uparrow$ and C$\Downarrow$), but substance abuse populations differ in details and severity of these deviations form the normal profile. This important observation was done by Donovan et al. \cite{Donovan1998} in 1998. They used MMPI for personality description and analysed groups of alcoholics, heroin, cocaine, and polydrug addicts by Discriminant Analysis. They identified three functions in the MMPI data,  which distinguished between the groups along important dimensions. These functions are: Level of Disturbance, Mania versus Alienated Depression, and Odd Thinking and Introversion versus Psychopathic Acting Out. 

Two additional characteristics of personality are proven to be important for analysis of substance use, Impulsivity (Imp) and Sensation-Seeking (SS).
\begin{description}
\item[Imp] \emph{Impulsivity}  is defined as a tendency to act without adequate forethought;
\item[SS{ }] \emph{Sensation-Seeking}   is  defined by the search for experiences and feelings, that are varied, novel, complex and intense, and by the readiness to take risks for the sake of such experiences. 
\end{description}
It was shown that high SS is associated with 
increased risk of substance use \cite{Zuckerman1972,Liraud2000,Kopstein01}.

Imp has been operationalised in many different ways \cite{Evanden1999}. It was demonstrated that substance use disorders are strongly associated with high personality trait  Imp scores on various measures \cite{Verdejo-Garcia2008,Loree2015}. Moreover, 	Imp score has significant impact on the treatment of substance use disorders: higher Imp implies lower success rate \cite{Loree2015}. It is possible that psychosocial and pharamacological treatments that may decrease Imp will improve substance use treatment outcomes \cite{Loree2015}. 

Impulsivity has been shown to predict aggression and heavy drinking \cite{McMurranEgan2002}.
Poor social problem solving has been identified as a potential mediating variable between impulsivity and aggression.
It is likely that the cognitive and behavioural features of impulsivity militate
against the acquisition of good social problem-solving skills early in life and that these deficits
persist into adulthood, increasing the likelihood of interpersonal problems.

A model was proposed, which attributes substance use/misuse to four distinct personality factors: SS, Imp, anxiety sensitivity (AS), and introversion/hopelessness (I/H). These four factors form a so-called Substance Use Risk Profile Scale \cite{Conrod2000}. The model was tested on groups of cannabis users \cite{Conrod2000, Woicik2009, Hecimovic2014}. It was demonstrated that SS was positively associated with expansion motives,  Imp was associated with drug availability motives, AS was associated with conformity motives and I/H was associated with coping motives for cannabis use \cite{Hecimovic2014}. Therefore, the authors of this model concluded that four personality risk factors in the model  are associated with distinct cannabis use motives.

The personality trait Imp and  laboratory tests of neurobehavioral impulsivity measured different aspects of general impulsivity phenomenon. Relationships between these two aspects are different in groups of heroin users and amphetamines users (even the sign of correlations is different) \cite{Vassileva2014}. Very recently,  demographic, personality (Imp, trait psychopathy, aggression, SS), psychiatric (attention deficit hyperactivity disorder, conduct disorder, antisocial personality disorder, psychopathy, anxiety, depression), and neurocognitive impulsivity measures (Delay Discounting, Go/No-Go, Stop Signal, Immediate Memory, Balloon Analogue Risk, Cambridge Gambling, and Iowa Gambling tasks) are used as predictors in a machine-learning algorithm to separate 39 amphetamine mono-dependent, 44 heroin mono-dependent, 58 polysubstance dependent, and 81 non-substance dependent individuals \cite{AhnVassileva2016}.

Two integrative personality dimensions capture important risk factors for substance use diorder \cite{Krueger2002}:
\begin{itemize}
\item {\em Internalizing} relates to generalized psychological distress, refers to insufficient amounts of behavior and is sensitive to a wide range of problems in living (associated with overcontrol of emotion, social withdrawal, phobias, symptoms of depression, anxiety, somatic disorder, traumatic distress, suicide). 
\item {\em Externalizing} refers to acting-out problems that involve
excess behavior and is often more directly associated with behaviors that cause distress for others, and to self as a consequence (associated with  undercontrol of emotion,  oppositional defiance, negativism, aggression, symptoms of attention deficit, hyperactivity, conduct, and other impulse control disorders).
\end{itemize}
Empirical results suggested co-occurrence of internalizing and externalizing problems among substance users \cite{Chan2008}. Nevertheless,  the externalizing dimension differentiated heroin users from alcohol, marijuana, and cocaine users \cite{Hopwood2008}. Internalizing and externalizing symptoms can be evaluated in FFM \cite{Derefinko2007}.

Correlations between drug consumption and gender, age, family income and geographical location  was studied in USA in a series of epidemiologic surveys (see \cite{Haberstick2014}). For example, it was demonstrated that rates of alcohol and cannabis abuse and dependence were greater among men than women. Nevertheless, the gender differences reported in large-scale epidemiological 
studies, were not pronounced in some adolescent samples \cite{Young2002}.

In our study \cite{Fehrman15,FehrmanBologna15} we tested associations with personality traits and biographical data (age, gender, and education) for 18 different types of drugs separately, using the Revised NEO Five-Factor Inventory (NEO-FFI-R) \cite{McCrae04}, the Barratt Impulsiveness Scale Version 11 (BIS-11) \cite{Stanford09}, and the Impulsivity Sensation-Seeking Scale (ImpSS) \cite{Zuckerman94} to assess Imp and SS respectively. For this analysis, we employed various methods of statistics, data analysis and machine learning.

\section{New dataset open for use}

The database was collected by an anonymous online survey methodology by Elaine Fehrman, yielding 2051 respondents. In January 2011, the research proposal was approved by the University of Leicester Forensic Psychology Ethical Advisory Group, and subsequently received strong support from the University of Leicester School of Psychology Research Ethics Committee (PREC).

The database is available online \cite{FehrmanData2016,FehrmanDataUCI}.  
An online survey tool from Survey Gizmo \cite{Surveygizmo,Bhaskaran2010} was employed to gather data which maximised anonymity; this was particularly relevant to canvassing respondents  views, given the sensitive nature of drug use. All participants were required to declare themselves at least 18 years of age prior to giving informed consent.

The study recruited 2051 participants over a 12-month recruitment period. Of these persons, 166 did not respond correctly to a validity check built into the middle of the scale, so were presumed to be inattentive to the questions being asked. Nine of these were also found to have endorsed the use of a fictitious drug, which was included precisely to identify respondents who overclaim, as have other studies of this kind \cite{Hoare10}. This led a useable sample of 1885 participants (male/female = 943/942).  It was found to be biased when compared with the general population, the comparison (see Chapter~\ref{Chapter:Results}, Fig.~\ref{MeanTscorefig:2}) being based on the data published by Egan et al. \cite{Egan00} and Costa \& McCrae \cite{McCrae04}. Such a bias is usual for clinical cohorts \cite{Gurrera00,Terracciano08} and `problematic' or `pathological' groups..

The sample recruited was highly educated, with just under two-thirds (59.5\%) educated to, at least, degree or professional certificate level: 14.4\% (271) reported holding a professional certificate or diploma, 25.5\% (481) an undergraduate degree, 15\% (284) a master's degree, and 4.7\% (89) a doctorate. Approximately 26.8\% (506) of the sample had received some college or university tuition although they did not hold any certificates; lastly, 13.6\% (257) had left school at the age of 18 or younger.

Twelve  attributes are known  for each respondent: personality measurements which include N, E, O, A, and C scores from NEO-FFI-R,  Impulsivity (Imp) from the BIS-11, Sensation Seeking (SS) from the
ImpSS, level of education (Edu.), age, gender, country of residence, and ethnicity. The data set contains information on the consumption of {\em 18 central nervous system psychoactive drugs including alcohol, amphetamines, amyl nitrite, benzodiazepines, cannabis, chocolate, cocaine, caffeine, crack, ecstasy, heroin, ketamine, legal highs, LSD, methadone, magic mushrooms (MMushrooms), nicotine, and Volatile Substance Abuse (VSA)} i.e. glues, gases, and aerosols. One fictitious drug (Semeron) was introduced to identify overclaimers.  For each drug, participants selected either: they never used this drug, used it over a decade ago, or in the last decade, year, month, week, or day. 

Participants were asked about various substances, which were classified as either central nervous system depressants, stimulants, or hallucinogens. The depressant drugs comprised alcohol, amyl nitrite, benzodiazepines, tranquilizers, solvents and inhalants, and opiates such as heroin and methadone/prescribed opiates. The stimulants consisted of amphetamines, nicotine, cocaine powder, crack cocaine, caffeine, and chocolate. Although chocolate contains caffeine, data for chocolate was measured separately, given that it may induce parallel psychopharmacological and behavioural effects in individuals congruent to other addictive substances \cite{Bruinsma99}. The hallucinogens included cannabis, ecstasy, ketamine, LSD, and magic mushrooms. Legal highs such as mephedrone, salvia, and various legal smoking mixtures were also measured.

The objective of the study was to assess the potential effect of the FFM personality traits, Imp, SS, and demographic data on drug consumption for different drugs, groups of drugs and for different definitions of drug users. The study had two purposes: (i) to identify the association of personality profiles (i.e.  FFM+Imp+SS) with drug consumption and (ii) to predict the risk of drug consumption for each individual according to their personality profile. 

Participants were asked to indicate which ethnic category was broadly representative of their cultural background. An overwhelming majority (91.2\%; 1720) reported being White, (1.8\%; 33) stated they were Black, and (1.4\%; 26) Asian. The remainder of the sample (5.6\%; 106) described themselves as `Other' or `Mixed' categories. This small number of persons belonging to specific non-white ethnicities precluded any analyses involving racial categories.

In order to assess the personality traits of the sample, the Revised NEO Five-Factor Inventory (NEO-FFI-R) questionnaire was employed \cite{Costa92}. The NEO-FFI-R is a highly reliable measure of basic personality domains; internal consistencies are 0.84 (N); 0.78 (E); 0.78 (O); 0.77 (A), and 0.75 (C) \cite{Egan11}. The scale is a 60-item inventory comprised of five personality domains or factors. The NEO-FFI-R is a shortened version of the Revised NEO-Personality Inventory (NEO-PI-R) \cite{Costa92}. The five factors are: N, E, O, A, and C with 12 items per domain.

All of these domains are hierarchically defined by specific facets \cite{McCrae91}. Egan et al. \cite{Egan00} observed that the score for the O and E domains of the NEO-FFI instrument are less reliable than for N, A, and C. The personality traits are not independent. They are correlated, with higher N being associated with lower E, lower A and lower C, and higher  E being associated with higher C (see Table \ref{tab:0} for more details).

\begin{table}[!ht]
\centering
\caption{Pearson's correlation coefficients (PCC) between NEO-FFI trait scores for a large British sample, $n=1025$ \cite{Egan00}; the $p$-value is the probability of observing by chance the same or greater correlation coefficient if the data are uncorrelated}

\smallskip

\label{tab:0}       % Give a unique label
\begin{tabular}{|c|c|c|c|c|c|}\hline
& {N} &{E}&{O}&{A}&{C}\\\hline
N&	&	$-0.40$** &	0.07*&	$-0.22$**& $-0.36$**\\\hline
E&	$-0.40$** &	 &	0.16** &	0.22**&	0.30** \\\hline
O&	0.07*&	0.16**&	 &	0.08*&	$-0.15$** \\\hline
A&	$-0.22$** &0.22**&	0.08*&	 &	0.13** \\\hline
C&	$-0.36$**&	0.30* &	$-0.15$**&	0.13** &	 \\\hline
\end{tabular}\\
$^*p<0.02$, **$p<0.001$
\end{table}

In our study, participants were asked to read the 60 NEO-FFI-R statements and indicate on a five-point Likert--type scale how much a given item applied to them (i.e. 0 = `Strongly Disagree', 1 = `Disagree', 2 = `Neutral', 3 = `Agree', to 4 = `Strongly Agree').

The second measure used was the Barratt Impulsiveness Scale (BIS-11) \cite{Stanford09}. BIS-11 is a 30-item self-report questionnaire, which measures the behavioural construct of impulsiveness, and comprises three subscales: motor impulsiveness, attentional impulsiveness, and non-planning. The `motor' aspect reflects acting without thinking, the `attentional' component, poor concentration and thought intrusions, and the `non-planning', a lack of consideration for consequences \cite{Snowden11}. The scale's items are scored on a four-point Likert-type scale. This study modified the response range to make it compatible with previous related studies \cite{GMontes09}.
A score of five usually connotes the most impulsive response although some items are reverse-scored to prevent response bias. Items are aggregated, and the higher the BIS-11 scores are, the higher the impulsivity level \cite{Fossati01} is. BIS-11 is regarded a reliable psychometric instrument with good test-retest reliability (Spearman's rho is equal to 0.83) and internal consistency (Cronbach's alpha is equal to 0.83 \cite{Stanford09,Snowden11}).

The third measurement tool employed was the Impulsiveness Sensation-Seeking (ImpSS). Although the ImpSS combines the traits of impulsivity and sensation-seeking, it is regarded as a measure of a general sensation-seeking trait \cite{Zuckerman94}. The scale consists of 19 statements in true-false format, comprising eight items measuring impulsivity  ({Imp}), and 11 items gauging Sensation-Seeking  ({SS}). The ImpSS is considered a valid and reliable measure of high risk behavioural correlates such as substance misuse \cite{McDaniel08}.

It was recognised at the outset of this study that drug use research regularly (and spuriously) dichotomises individuals as users or non-users, without due regard to their frequency or duration/desistance of drug use \cite{Ragan10}. In this study, finer distinctions concerning the measurement of drug use have been deployed, due to the potential for the existence of qualitative differences amongst individuals with varying usage levels. In relation to each drug, respondents were asked to indicate if they never used this drug, used it over a decade ago, or in the last decade, year, month, week, or day. This format captured the breadth of a drug-using career, and the specific recency of use. 

\begin{figure}
    \centering
    \includegraphics[width=0.99\textwidth]{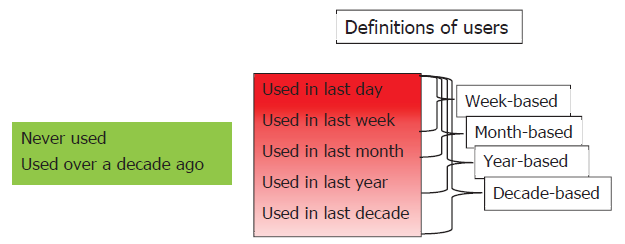}
    \caption {{Categories of drug users.}  Categories with green background  always correspond to drug non-users. Four different definitions of drug users are presented.}
    \label{Categoriesfig:1}
\end{figure}

For decade based separation, we merged two isolated categories (`Never used' and `Used over a decade ago') into the class of non-users, and all other categories were merged to form the class of users. For year-based classification we  additionally merged the category `Used over a decade ago' into the  group of non-users and placed four other categories (`Used in last year-month-week-day') into the group of users. We continued separating into users and non-users depending on the timescale  in this nested ``Russian doll" style. We also considered `month-based' and `week-based'  user/non-user separations. Different categories of drug users are depicted in Fig~\ref{Categoriesfig:1}.

Analysis of the classes of drug users shows that part of the classes are nested: participants which belong to the category `Used in last day' also belong to the categories `Used in last week', `Used in last month', `Used in last year' and `Used in last decade'. There are two special categories: `Never used' and `Used over a decade ago' (see Fig~\ref{Categoriesfig:1}). The data does not contain a definition of the users and non-users groups. Formally only a participant in the class `Never used' can be called a non-user, but   a participant who used a drug more than decade ago also cannot be considered a drug user for most applications. There are several possible ways to discriminate participants into groups of users and non-users for binary classification:

\begin{enumerate}
  \item Two isolated categories (`Never used' and `Used over a decade ago') are placed into the class of non-users with a green background in Fig~\ref{Categoriesfig:1}, and all other categories into the class `users' as the simplest version of binary classification. This classification problem is called `\emph{decade-based}' user/non-user separation.

  \item The categories `Used in last decade', `Used over a decade ago' and `Never used' are merged to form a group of non-users and all other categories are placed into the  group of users. This classification problem is called `\emph{year-based}'.

  \item The categories `Used in last year', `Used in last decade', `Used over a decade ago' and `Never used' are combined to form a group of non-users and all three other categories are placed into the group of users. This classification problem is called `\emph{month-based}'.

  \item The categories `Used in last week' and `Used in last month' are merged to form a  group of users and all other categories are placed into the group of non-users. This classification problem is called `\emph{week-based}'.
\end{enumerate}

We began our analysis from the decade-based user/non-user separation because it is a relatively well-balanced classification problem;  that is, there are sufficiently many users in the united group `Used in last decade-year-month-week' for all drugs in the database. If the problem is not directly specified then it is the decade-based classification problem. We also performed analysis for the year-, month-, and week-based user/non-user separation. It is useful to group drugs with highly correlated usage for this purpose (see Section~\ref{Pleiad of drug users}).

The proportion of drug users among all participants is different for different drugs and for different classification problems. The data set comprises 1,885 individuals without any missing data. Table~\ref{tab:1a} shows the percentage of drug users for each drug and for each problem in the database. It is necessary to mention that the sample is intentionally  biased to a higher proportion of drug users. This means that for the general population the proportion of illegal drug users is expected to be significantly lower \cite{HomeOffice14}. A common problem of online surveys is the unintentional biasing of samples \cite{Wright2005}. We return to this problem in Chapter~\ref{Chapter:Results}.

\begin{table}[!ht]
\centering
\caption{The number and fraction of drug users}
\label{tab:1a}
\begin{tabular}{|l|c|c|c|c|}\hline
{  Drug} 	& \multicolumn{4}{c|}{ {   User definition based on}}                 \\\cline{2-5}
         	& {  Decade } &	{  Year }&	{   Month }&	{  Week } \\\hline
Alcohol     &	1817; 96.39\% &	1749; 92.79\%&	1551; 82.28\%&  1264; 67.06\%\\\hline
Amphetamines&	679;  36.02\% &	436; 23.13\%&	238; 12.63\%&	163; 8.65\%\\\hline
Amyl nitrite&	370; 19.63\%  &	133; 7.06\%&	41; 2.18\%&	    17; 0.90\%\\\hline
Benzodiazepines&769; 40.80\%  &	535; 28.38\%&	299; 15.86\%&	179; 9.50\%\\\hline
Cannabis   &   1265; 67.11\%  &	999; 53.00\%&	788; 41.80\%&	648; 34.38\%\\\hline
Chocolate  & 1850; 98.14\%    &	1840; 97.61\%&	1786; 94.75\%&	1490; 79.05\%\\\hline
Cocaine    &687; 36.45\%      &	417; 22.12\%&	159; 8.44\%&	60; 3.18\%\\\hline
Caffeine   &	1848; 98.04\% &	1824; 96.76\%&	1764; 93.58\%&	1658; 87.96\%\\\hline
Crack      & 191; 10.13\%     &	79; 4.19\%  &	20; 1.06\%&	    11; 0.58\%\\\hline
Ecstasy    &751; 39.84\%      &	517; 27.43\%&	240; 12.73\%&	84; 4.46\%\\\hline
Heroin     &	212; 11.25\%  &	118; 6.26\%&	53; 2.81\%&	    29; 1.54\%\\\hline
Ketamine   &	350; 18.57\%  &	208; 11.03\%&	79; 4.19\%&	    37; 1.96\%\\\hline
Legal highs&	762; 40.42\%  &	564; 29.92\%&	241; 12.79\%&	131; 6.95\%\\\hline
LSD        &	557; 29.55\%  &	380; 20.16\%&	166; 8.81\%&	69; 3.66\%\\\hline
Methadone  &417; 22.12\%      &	320; 16.98\%&	171; 9.07\%&	121; 6.42\%\\\hline
MMushrooms&694; 36.82\%       &	434; 23.02\%&	159; 8.44\%&	44; 2.33\%\\\hline
Nicotine  &1264; 67.06\%      &	1060; 56.23\%&	875; 46.42\%&	767; 40.69\%\\\hline
VSA       &230; 12.20\%       &	95; 5.04\%  &	34; 1.80\%&	    21; 1.11\%\\\hline
\end{tabular}
\end{table}

\section{First results in brief}

The first result is the production of the database, cleaned and published in open access \cite{FehrmanData2016,FehrmanDataUCI}, and described in this book. We are sure that our book does not exhaust all the important information which can be extracted from this database, and will be happy if the database is used further in a variety of research and educational projects.

The second result is   the personality profiles calculated for users and non-users of 18 substances (and for four definitions of users based on the recency of use). These profiles can be used in many research projects; for example, for more detailed analysis of relations between the use of different drugs. The  profiles, with 95\% confidence intervals and $p$-values for differences between users and non-users (probabilities to observe such or greater difference by chance), are presented in the Appendix (Tables~\ref{table4}, \ref{table4b}). 

In addition, the profile of users of all illicit drugs is calculated and presented in these tables.  We have conventionally called the following substances `illicit': amphetamines, amyl nitrite, benzodiazepines, cannabis,  cocaine, caffeine, crack, ecstasy, heroin, ketamine, legal highs, LSD, methadone, magic mushrooms,  and VSA. The term `illicit' is not fully precise because in some regions the recreational use of some of these substances is decriminalized and in some countries alcohol, for example, is illicit.  We use the term `illicit drugs'  for brevity, while the exact definition of this group is just the list above.  

Our study reveals that the personality profiles are strongly associated with group membership to the users or  non-users of the 18 drugs.  For the  analysis, we used the following  subdivision of the sample T-$score$: the interval 44-49 indicated a moderately low score, $(-)$, the interval 49-51 indicated  a neutral score $(0)$, and the interval 51-56 indicated  a moderately high $(+)$ score.
 We found that the N and O scores of drug users of all 18 drugs were moderately high $(+)$ or neutral $(0)$, except for crack usage for the week-based classification, for which the O score was moderately low $(-)$. The A and C scores were moderately low $(-)$  or neutral $(0)$ for all groups of drug users and all user/non-user separations based on recency of use.
 
For most groups of  illicit drug users the A and C scores  were moderately low $(-)$ with the exception of two groups:  the A score was neutral $(0)$  in the year-based  classification (`used in last year') for LSD users  and  in the week-based classification (`used in last week') for LSD and magic mushrooms users.

The  A and C scores for groups of legal drugs users (i.e. alcohol, chocolate, caffeine, and nicotine) were neutral $(0)$,  apart from nicotine users, whose  C score was moderately low $(-)$  for all categories of user/non-user separation.

The impact of the E score was drug specific. For example, for the week-based user/non-user separation the E scores were:
\begin{itemize}
\item moderately low $(-)$ for amphetamines, amyl nitrite,  benzodiazepines, heroin,  ketamine, legal highs,  methadone, and crack;
\item moderately high $(+)$ for  cocaine, ecstasy, LSD, magic mushrooms, and VSA;
\item neutral $(0)$ for  alcohol, caffeine, chocolate, cannabis, and nicotine.
\end{itemize}
For more details see Section \ref{Sec:Use/NonUse}.

Usage of some drugs were correlated significantly. The structure of these correlations is analysed in Section \ref{Sec:Corr}. Two correlation measures were utilised: the Pearson's Correlation Coefficient (PCC) and the Relative Information Gain (RIG).
 We found three groups of drugs with highly correlated use (Section \ref{Pleiad of drug users}).  The central element was clearly identified for each group. These centres are: {\em heroin, ecstasy, and benzodiazepines}. This means that drug consumption has a `modular structure', which is made clear in the correlation graph. The idea of merging correlated attributes into `modules' referred to as {\em correlation pleiades}  is popular in biology \cite{Terentjev31,Berg60,Mitteroecker07}.
 
The concept of correlation pleiades was introduced in biostatistics in 1931 \cite{Terentjev31}.  They  were used for identification of a modular structure in evolutionary physiology \cite{Terentjev31,Mitteroecker07,Berg60,Armbruster99}. According to Berg \cite{Berg60}, correlation pleiades are clusters of correlated traits. In our approach, we distinguished the core and the peripheral elements of correlation pleiades and allowed different pleiades to have small intersections in their periphery. `Soft' clustering algorithms relax the restriction that each data object is assigned to only one cluster (like probabilistic \cite{Krishnapuram1993} or fuzzy \cite{Bezdek1981} clustering). See the book of R. Xu \& D. Wunsch \cite{XuWunsch2009} for a modern review of hard and soft clustering. We refer to \cite{Omote2006} for a discussion of clustering in graphs with intersections.

The three groups of correlated drugs centered around heroin, ecstasy, and benzodiazepines were defined for the decade-, year-, month-, and week-based classifications:
\begin{itemize}
\item  The heroin pleiad  includes crack, cocaine, methadone, and heroin;
\item The ecstasy pleiad  consists of amphetamines, cannabis, cocaine, ketamine, LSD, magic mushrooms, legal highs, and ecstasy;
\item The benzodiazepines pleiad contains methadone, amphetamines,  cocaine, and benzodiazepines.
\end{itemize}

The topology of the correlation graph can help in analysis of mini-sequences of drug involvement. These mini-sequences can be found in longitudinal studies \cite{Newcomb1986}: cigarettes are a gateway to cannabis, and then to hard drugs; and there is a synergistic effect of increasing involvement. Of course, correlation does not imply a causal relationship, but use of the substances from one mini-sequence of involvement should be correlated. Therefore, the correlation graph is a proper map for the study of the involvement routes, and synergistic and reciprocal effects.

Analysis of the intersections between correlation pleiades of drugs leads to important questions and hypotheses:
\begin{itemize}
\item Why is cocaine  a peripheral member of all pleiades?
\item Why does methadone belong to the  periphery of both the heroin and benzodiazepines pleiades?
\item Do these intersections reflect the structure of individual drug consumption and sequences of involvement, or are they related to the structure of the groups of drug consumers?
\end{itemize}

In this study, we evaluated the individual drug consumption risk separately, for each drug and pleiad of drugs. We also analysed interrelations between the individual drug consumption risks for different drugs. We applied several data mining approaches: decision tree, random forest, $k$-nearest neighbours, linear discriminant analysis, Gaussian mixture, probability density function estimation, logistic regression and na{\"i}ve Bayes. The quality of classification was surprisingly high (Section \ref{Sec:BestClass}). We tested all of the classifiers by  {\em Leave-One-Out Cross Validation}. The best results, with sensitivity and specificity greater than 75\%, were achieved for cannabis, crack, ecstasy, legal highs, LSD, and VSA. Sensitivity and specificity greater than 70\%  were achieved for the following drugs: amphetamines, amyl nitrite, benzodiazepines, chocolate, caffeine, heroin, ketamine, methadone and nicotine. An exhaustive search was performed to select the most effective subset of input features and data mining methods for each drug. The results of this analysis provide an answer to an important question about the {\em predictability} of drug consumption risk on the basis of FFM+Imp+SS profile and demographic data. 

Users for each correlation pleiad of drugs are defined as users of any of the drugs from the pleiad.
We solved the classification problem for drug pleiades for the decade-, year-, month-, and week-based user/non-user separations. The quality of classification is also high. 

Simple Fisher's linear discriminant classifiers have slightly worse performance than more advanced machine learning methods but are very robust, and give additional information. Coefficients of the discriminants are presented in the Appendix for all user/non-user separation problems for every substance, group of illicit substances, and three correlation pleiades. Stability of the discriminant models was tested in cross-validation. Even a brief look at the linear discriminants reveals differences between drugs. For example, for month-based user definition (`used in last month'), the linear discriminants for heroin and cocaine  users for seven variables FFM+Imp+SS are (Table~\ref{TabDiscr7}):
\begin{description}
\item[heroin users{ }] $\, {\rm-30.020+0.302N-0.409E+0.232O-0.499A-0.107C+0.345Imp+0.555SS>0}$;
\item[cocaine users] ${\rm -57.737+0.366N+0.148E+0.082O-0.370A-0.014C-0.001Imp +0.837 SS>0}$,
\end{description}
where the variables are represented by their T-scores (calculated for the database).
These data, together with the correlation tables (Tables \ref{tab:14}, \ref{tab:14a}), give  information about the importance of different variables for user/non-user separation for different drugs.

We also studied the problem of separation of users of different drugs. For this purpose, we selected three drugs:  heroin, ecstasy, and benzodiazepines (the centres of the pleiades). The profiles of the users of these drugs are different (see Fig.~\ref{BEHGraphs}) (the confidence level below is 95\%):
\begin{itemize}
\item The mean values in the groups of benzodiazepines and ecstasy users are statistically significantly different for N (higher for benzodiazepines) and E (higher for ecstasy) for all definitions of users, for O (higher for ecstasy) for all definitions of users excluding the year-based, and for SS (higher for ecstasy)  for all definitions excluding the month-based. 
\item The groups of benzodiazepines and heroin users are statistically significantly different  for A  (higher for benzodiazepines) for all definitions of users,  for Imp (higher for heroin) and SS (higher for heroin) for all definitions of users excluding the week-based. 
\item Heroin and ecstasy are statistically significantly different for N (higher for heroin), E (higher for ecstasy) and A (higher for ecstasy) for all definitions of users and for Imp (higher for heroin) for all definitions excluding the week-based. 
\end{itemize}

The creation of classifiers has provided the ability to evaluate the risk of drug consumption in relation to individuals. The risk map is a useful tool for data visualisation and for the generation of hypotheses for further study (see Chapter `Risk evaluation for the decade-based user/non-user separation').

In the next chapter, we briefly introduce the main methods used in the book, in Chapter \ref{Chapter:Results} the results are described in more detail, then after a brief  Discussion and Summary, the main tables are presented in the Appendix.

%% file: chapter2.tex
%%%%%%%%%%%%%%%%%%%%% chapter.tex %%%%%%%%%%%%%%%%%%%%%%%%%%%%%%%%%
%FINAL
% sample chapter
%
% Use this file as a template for your own input.
%
%%%%%%%%%%%%%%%%%%%%%%%% Springer-Verlag %%%%%%%%%%%%%%%%%%%%%%%%%%
%\motto{Use the template \emph{chapter.tex} to style the various elements of your chapter content.}
\chapter{Methods of Data Analysis}
\label{Methods} % Always give a unique label
% use \chaptermark{}
% to alter or adjust the chapter heading in the running head

\abstract*{In this chapter, we give a brief outline of the methods of data analysis used, from elementary T-scores to non-linear Principal Component analysis (PCA), including
data normalization, quantification of categorical attributes, CatPCA (Categorical Principal Component Analysis), sparse PCA, the method of principal variables, the original `double' Kaiser selection rule, $k$ Nearest Neighbours for various distances, decision tree with various split criteria (information gain, Gini gain or DKM gain), linear discriminant analysis, Gaussian mixture, probability density function estimation by radial-basis functions, logistic regression, na{\"i}ve Bayes approach, random forest, and data visualisation on the non-linear PCA canvas.}

This chapter presents a brief outline of the data analysic methods used in our work. We have tried to present a comprehensive summary but the details of the algorithms, as well as many computational and statistical estimates, remain beyond the current scope of this work. What lies herein cannot replace textbooks \cite{WittenTibshiraniIntro2013,Howell2012}, and detailed handbooks like  \cite{Hastie09}.

\section{Preprocessing}

Data should be preprocessed before application of any analytic techniques in order to solve any classification, regression or other standard data analytic problem. Preprocessing is an important and time consuming step. It includes dataset inspection (analysis of types of data, completeness, consistency, accuracy, believability, and other crucial properties of a dataset), feature extraction from raw data, cleaning (managing of missed values, noise and inconsistencies),  feature selection and transformation (selection of necessary features, without loss and redundancies, transformation to the most convenient scale and form), and dimensionality reduction. The steps of feature selection and dimensionality reduction depend on the problem we aim to solve.

In our book we deal with data of the following types: numeric (quantitative) and categorical (with a finite numbers of values). The categorical data  belong to  three subtypes: categorical ordinal (education level, for example), categorical nominal  (nation and race), and Boolean (for example, user/non-user of a drug).

\subsection{Descriptive statistics: mean, variance, covariance, correlation, information gain}

 Given a sample of size $N$, consider $N$   observations $x_i$ of a quantity $X$.  The sample mean is defined to be
 $$\bar{x}=\frac{1}{N}\sum_{i=1}^N x_i.$$
The sample variance for the same sample is
$$s^2_x= \frac{1}{N}\sum_{i=1}^N (x_i-\bar{x})^2.$$
For a vector variable the definitions of the sample mean and the sample variance are the same (the square of the vector means the square of its Euclidean length).

If a sample point has two numerical attributes, $x_i$ and $y_i$ ($i=1,\ldots, N$), then the sample covariance, $s_{xy}$, and Pearson Correlation Coefficient (PCC), $r_{xy}$, are defined:
\begin{equation}\label{PCC}
s_{xy}=\frac{1}{N}\sum_{i=1}^{N}(x_i-\bar{x})(y_i-\bar{y}); \;\;
r_{xy}=\frac{s_{xy}}{s_x s_y}.
\end{equation}
The PCC measures how much of a linear connection between $x$ and $y$ exists. The linear approximation of $y$ by a linear function of $x$, which minimises the mean square error (simple linear regression), is 
\begin{equation}\label{linreg}
\hat{y}=\bar{y}+\frac{s_y}{s_x} r_{xy}(x-\bar{x}).
\end{equation}
 For several attributes, the sample covariance matrix $S$, with the diagonal $s_{ii}=s_i^2$, and the sample PCC matrix with unit diagonal are used. 

High correlations between the inputs and outputs are beneficial and allow researchers to create a good predictive model. But the situation is different when the  {\em inputs} for a regression or classification model are highly correlated. This {\em multicollinearity}  can cause some problems.  Just as a trivial example we can take two inputs, $x$ and $y$, $x\approx y$ ($r\approx 1$). If the the coefficients of $x$ and $y$ in a regression model are $a$ and $b$ respectively, then $a+M$ and $b-M$ are also approximate solutions for the regression problem, for possibly large values of $M$. Thus, depending on the particular solution method, we might get large and oscillating solutions to the regression problem. In this case we need to apply some procedure to constrain the sorts of solutions we wish to allow. Such processes are often called {\sl regularisation processes}. In the multidimensional case, tests of multicollinearity are based on the analysis of efficient and stable invertibility of the correlation matrix.  Therefore, multidimensional multicollinearity is measured not by the values of PCCs but by the {\em condition number} of the PCC matrix between input attributes,  that is the ratio $\kappa=\lambda_{\max}/\lambda_{\min}$, where $\lambda_{\max}$ and $\lambda_{\min}$ are the maximal and the minimal eigenvalues of this matrix  \cite{Belsley2005}.   (We recall that the correlation matrix is symmetric and non-negative and its eigenvalues  are real and non-negative numbers.)  Collinearity with $\kappa<10$ is considered as `modest' collinearity; most methods work in this situation and only a few methods were reported as having failed (like support vector machine, which we do not use) \cite{Dormann 2013}. For higher $\kappa$, dimensionality reduction seems to be unavoidable (the classical principal component analysis is the first choice, see below in this chapter).

{\em  Relative Information Gain} (RIG) is widely used in data mining to measure the dependence between categorical attributes \cite{Mitchell97}. The greater the value of RIG is, the stronger the indicated correlation is.  RIG is zero for independent attributes, is not symmetric, and is a measure of mutual information. To construct RIG, we start from the {\em Shannon entropy}. Consider a categorical attribute $X$ with $m$ values $v_1, \ldots, v_m$. Given sample size $N$ assume that the value $X=v_i$ appears in the sample $n_i$ times. The normalised frequency is $f_i=n_i/N$. The sample Shannon entropy is
$$H(X)=-\sum_{i=1}^m f_i \log_2 f_i.$$
In this formula it is assumed that $0\log 0=0$ (by continuity). If any data point includes values of two categorical attributes, $X$ and $Y$, then for each value $X=v_i$ we calculate the conditional entropy $H(Y|X=v_i)$. For this purpose, we select a subsample, where  $X=v_i$, and calculate the entropy of $Y$ in this subsample (if it is not empty). The sample conditional entropy   $H(Y|X)$ is
$$ H(Y|X)=\sum_{i=1}^m f_i H(Y|X=v_i).$$

{\em Information gain} $IG(Y|X)=H(Y)-H(Y|X)$  measures how much information about $Y$ will be gained from knowledge of $X$ (on average, under given  normalised  frequencies of values $X$ and $Y$). 
Finally, the relative information gain is $RIG(Y|X)=IG(Y|X)/H(Y)$. This measures the fraction of information about $Y$ which can be extracted from $X$ (on average). It may be convenient to represent the formula for RIG more directly. Consider two categorical attributes, $X$ with values $v_1,\ldots, v_m$, and $Y$ with values $w_1, \ldots, w_k$. Let the sample have $n_{ij}$ cases with $X=v_i$ and $Y=w_j$. The total number of cases is $N=\sum_{i,j} n_{ij}$.  The  normalised  frequencies are $f_i=\sum_j n_{ij}/N$, $g_j=\sum_i n_{ij}/N$. The sample RIG is
\begin{equation}\label{RIG}
 RIG(Y|X)=\frac{\sum_j g_j \log_2 g_j-\sum_i \left(f_i \sum_j n_{ij}\log_2 n_{ij}\right)}{\sum_j g_j \log_2 g_j}.
\end{equation}

Systematic comparison of the linear correlations provided by PCC, and the non-linear mutual information approach discussed above was performed on genomic data \cite{Steuer2002}.

\subsection{Input feature normalization}
Data centralization and normalization are convenient and, for some of the analysis methods, unavoidable. The most popular normalization and centralization is given by the $z$-score. By definition, the
$z$-score of an observable attribute  $X$ is $z=(x-\mu)/\sigma$, where $\mu$ it the expectation, and $\sigma$ is the standard deviation of $X$. However, the expectation and standard are calculated from a sample $x$, giving estimates $\bar{x} \approx \mu$ and $s \approx \sigma$. Then we have 
$$z=\frac{x-\bar{x}}{s}.$$
 
The notion of $T\textrm{-} score$ has two meanings, one in general statistics, the other in psychometrics. In our book we use  the psychometric definition:
\begin{equation}\label{eq:1}
T\textrm{-} score=10z+50=10\frac{x-\mu}{\sigma}+50,
\end{equation}
with an additional condition: $T\textrm{-} score=0$ if the right hand side of (\ref{eq:1}) is negative, and $T\textrm{-}score=100$, if the right hand side of (\ref{eq:1}) exceeds 100. $T\textrm{-}score $ returns results between 0 to 100 (most scores   fall between 20 and 80 without any additional condition).  The $T\textrm{-}score$ of the expectation $\mu$ is 50, and  the interval $\mu\pm \sigma$ transforms  into $[40,60]$.

We meet problems in defining the population mean (the norm) $\mu$ and the standard deviation $\sigma$, which are discussed in Chapter \ref{Chapter:Results}. For instance, what is "normal" can depend on age, social group and other factors. Thus the sense of an average becomes less clear. Additionally (see Table~\ref{tab:5}), the sample in our study deviates from the population norm in the same direction as drug users deviate from the sample mean. However, the deviations of the mean of users groups are different for different drugs. Therefore, the definition of $T\textrm{-}score$ we use in what follows is based on sample mean and standard deviation:
\begin{equation}
\label{eq:2}
T\textrm{-}score_{sample} = 10\frac{x-\bar{x}}{s}+50.
\end{equation}

Usually, the $T\textrm{-} score$ is categorized into five categories to summarise an individual's personality score concerning each factor: the interval 20-35 indicates very low scores; 35-45 low scores; 45-55 average scores; 55-65 high scores, and 65-80 indicates very high scores. 

We study average $T\textrm{-}score_{sample}$ for  groups of users and non-users. The standard deviation of these mean values decreases with the number $m$ of cases in a group as $1/\sqrt{m}$ (due to the central limit theorem). To discuss these mean profiles, we use  a refined  subdivision of the $T\textrm{-}score_{sample}$: the interval 44-49 indicates moderately low $(-)$ scores; 49-51 neutral $(0)$ scores, and 51-56 indicates moderately high $(+)$ scores.

The unification of the mean and standard deviation of the T-$score_{sample}$ for all factors simplifies comparisons of  users and non-users groups for each drug.  A $t$-test is employed to estimate the significance of the differences between the mean T-$score_{sample}$ for groups of users and non-users for each FFM (NEO-FFI-R)+Imp+SS factor and each drug (Tables \ref{table4}, \ref{table4b}). In this $t$-test, a  ${p}$-value is the probability of observing, by chance, the same or a greater difference of mean. The analysis was performed using SAS 9.4.

\section{Input feature transformation}
\label{Input feature transformation}

\subsection{Principal Component Analysis -- PCA}
\label{Sec:PCA}

\begin{figure}
\centering
\includegraphics[width=0.6\textwidth]{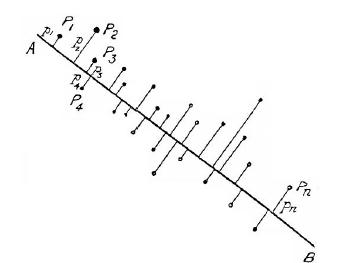}
\caption {Pearson's illustration of PCA definition. $P_i$ are data points, $p_i$ are their distances to the approximating line. The best approximation problem is UV$=\sum_i p_i^2 \to min$}
\label{Fig:PCA}
\end{figure}

Karl Pearson \cite{Pearson01} invented PCA as a method for approximation of data sets by straight lines and lower-dimensional planes (Fig.~\ref{Fig:PCA}).  The {\em Unexplained Variance} (UV) is the quadratic error of PCA, that is the sum of the squared distances from the data points to the approximating low-dimensional plane or line.   The {\em Explained  Variance} is the variance of the orthogonal projections of the data on the approximating line or plane. The {\em Fraction of Variance Unexplained (FVU)} is the ratio of the unexplained sample variance to the total sample variance. The {\em Fraction of Variance Explained}, otherwise known as the {\em Coefficient of Determination} $R^2$ is $1-$FVU.

 PCA is equivalent to spectral decomposition of the sample covariance matrix (or to the singular value decomposition (SVD) of the data matrix). The eigenvalues $\lambda_i$ of the sample covariance matrix are real and non-negative. Let us order them by size in descending order: $\lambda_1 \geq \lambda_2 \geq \ldots \ge \lambda_n$, ($n$ is the dimension of the data space) and choose a corresponding orthonormal basis of eigenvectors: $w_1, w_2, \ldots, w_n$. These eigenvectors are called the {\em Principal Components} (PC for short), and the best approximation of data by a $k$-dimensional plane is given by the first $k$ PCs via the formula
\begin{equation}\label{PCAapp}
x-\bar{x}\approx \sum_{i=1}^k w_i (w_i,x-\bar{x}),
\end{equation} 
where $(w_i,x)$ is the inner product. For this approximation, FVU$=\sum_{j>k}\lambda_j$.
Usually, PCA is applied after normalization and centralization to $z$-scores. The sample covariance matrix for $z-scores$ is the PCC (correlation) matrix.

For applications, there exists a crucial question without a definite theoretical answer: how do we decide $k$? That is, how many PCs should be used in our approximation? There are several popular heuristic rules for component retention \cite{Cangelosi2007}. The most celebrated of them is Kaiser's rule: retain PCs with $\lambda_i$ above the average value of $\lambda$ \cite{Guttman54,Kaiser60}. The trace of a matrix $A$ is the sum of its diagonal elements, and we denote this by ${\rm tr} \, A$. We note that the average $\lambda$ is ${1 \over n} {\rm tr} \, Q = \frac{1}{n}\sum_i s_i^2$, where $Q$ is the sample covariance matrix, and $s_i^2$ is the sample variance of the $i$th attribute. For the PCC matrix, the average $\lambda$ is 1 because $r_{ii}=1$, and Kaiser's rule is: retain PCs with $\lambda>1$.

Various approaches to PCA were discussed in \cite{Jolliffe1986,GorbanZin2009}. There are several directions for generalisation of PCA: nonlinear PCA \cite{Hastie1989,GorbanZin2009,Gorban08}, branching PCA \cite{GorbanZin2005,Gorban10}, nonlinear PCA with categorical variables \cite{Linting12}. Different norms for approximation errors have been employed instead of quadratic FVU \cite{Jolliffeetal2003,Kwak2008,Brooks2013,GorbanMIr2016}.

\subsection{Quantification of categorical input variables}
\label{Quantification}

It is convenient to code categorical variables by using real numbers. Such a coding allows us to use methods developed for continuous data, to calculate PCCs, etc. There exists a universal method for quantification of categorical variables by multidimensional vectors,  the so-called `dummy coding' . For a Boolean variable, (with values `true' or `false') the real number 1 for true, and 0 for false, is returned. A  categoric variable $V$ with values $v_1,\ldots, v_m$ can be represented in dummy coding by a battery of $m$ indicator Boolean variables $(B_1,\ldots , B_m)$, where $B_i=1$ if and only if $X=v_i$.  This is an $m$-dimensional vector with components 0 or 1, which are far from being independent (if one of them is 1 then all others are 0). Further dimensionality reduction is desirable, by either linear  or non-linear means. Such coding is systematically used in ANOVA and ANCOVA models \cite{Howell2012,Gujarati03}.

In some cases, it is possible to  quantify a categorical variable by using one numeric variable. Firstly, it might be that ordinal features are `almost numeric', and often have conventional numeric values assigned. Secondly, if the set of numerical variables is sufficiently rich then it is possible to represent categories (even nominal ones) via their average projections in the space of numerical variables (see Section~\ref{Nominal feature quantification} below). 

\subsubsection{Ordinal feature quantification}
\label{Ordinal feature quantification}

Values of many categorical attributes are ordered linearly, by their nature: severity of a disease, quality of a product, level of education, etc. For these attributes, we can assume that their values are obtained by binning of a `latent' numeric variable with a nice probability distribution. Usually, for this role of hypothetical latent variable, the standard normal distribution is selected  with zero mean and unit variance. Other `nice' distributions might be selected for quantification, for example, the uniform distribution on $[0,1]$. The normal distribution has density $f$ and cumulative distribution function CDF) $\phi$:
$$f(x)=\frac{1}{\sqrt{2\pi}}e^{-\frac{x^2}{2}}; \;\; \phi(x)=\int_{-\infty}^{x}f(x) dx.$$

\begin{figure}
\centering
\includegraphics[width=0.6\textwidth]{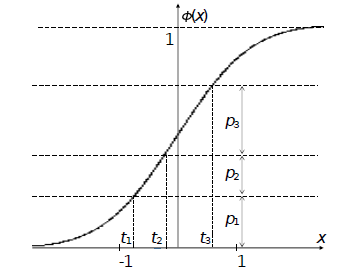}
\caption {{Quantification of an ordinal feature. Three thresholds, $t_1$, $t_2$, and $t_3$ are found for three categories with  normalised  frequencies $p_1$, $p_2$, and $p_3$}}
\label{Quantificationfig:1}
\end{figure}

Let  $V$ be a categoric ordinal variable with linearly ordered values $v_1<v_2\ldots < v_m$. Assume that there are  $ n_{i} $ cases of category $v_{i}  $ in the sample, and the corresponding frequency is $p_i=n_i/N$ ($i=1, \ldots, m$). To quantify $V$, it is necessary to assign  an interval $[t_{i-1},t_i]$ to each category $v_i$ ($-\infty=t_0<t_1<\ldots <t_{m}=\infty$). A threshold $t_i$ separates $v_i$ from $v_{i+1}$ ($i=1,\ldots, m-1$). A simple calculation gives (see Fig.~\ref{Quantificationfig:1})
\begin{equation}
t_{i}=\phi^{-1}\left(\sum_{j=1}^{i}p_{j}\right),\label{qqq1}
\end{equation}
where $\phi^{-1}$ is the inverse function of  $\phi$.

After the thresholds $t_i$ are found, the numerical value should be selected for each category $v_i$ from the interval $[t_{i-1}, t_i]$. A popular simple estimate  is given by the value $x_i$ with average probability:
\begin{equation}
x_{i}=\phi^{-1}\left(\sum_{j=1}^{i-1}p_{j}+\frac{p_{i}}{2}\right). \label{qqq2}
\end{equation}
This value is the numeric substitution for the ordinal categoric value $v_i$. 
This simple quantification method allows a number of improvements, for instance, the values $x_i$ can be
selected by the implicit maximum likelihood method.  The univariate determination of thresholds (\ref{qqq1}) can be generalised to the multivariate selection of the thresholds under the assumption that the multidimensional latent variables are normally distributed. These methods are part of a set of widely used techniques for dealing with latent continuous variables, traditionally called  `polychoric correlation' \cite{Lee95,Martinson71}. 

The matrix of polychoric coefficients can be  used  further  to calculate PCs, etc. The technique of polychoric correlation is based on the assumption that values of ordinal features result from the discretization of continuous random values with fixed thresholds. A clear description of polychoric correlation coefficients, including methods for their estimation is given in \cite{Olsson1979,  Drasgow1988}. The drawback with the multivariate selection of the thresholds is in its dependence on pairs of correlated ordinal variables. If we have to analyse several dozens of such variables then a univariate choice similar to (\ref{qqq1}) may be preferable.  A critical inquiry into polychoric and canonical correlation was published in \cite{Nishisato2002}. In our work, we have used the univariate choice of thresholds (\ref{qqq1}) and simple  average probability values (\ref{qqq1}) for quantification of ordinal variables. Following this, we have applied classical methods, and also machine learning algorithms, for data analysis.

We should ask the question: are FFM, Imp, and SS scores numerical or ordinal variables? Traditionally, these scores are thought of as numerical but it is clear that this should be the case, especially if they were obtained by short questionnaires and take a small number of values. We have taken an intermediate approach: in the analysis of average personality profiles in groups of drug users and non-users, we have handled the scores traditionally, as numeric variables, have evaluated $p$-values using the standard tests, calculated the PCCs with estimation of their $p$-values and presented the results in Tables \ref{table4}, \ref{table4b}, \ref{tab:14}, \ref{tab:14a}. For application of machine learning algorithms, we used the aforementioned procedure of quantification of ordinal features in data preprocessing, including scores, and also of the nominal features using dummy coding or CatPCA from Subsection \ref{Nominal feature quantification}. The raw database is published online in \cite{FehrmanData2016} and the database with quantified categorical variables is published  online in \cite{FehrmanDataUCI}. 

\subsubsection{Nominal feature quantification}
\label{Nominal feature quantification}
We cannot use the techniques described above to quantify nominal features such as gender, country of location, and ethnicity, because the categories of these features are unordered. To quantify nominal features we have used two methods and compare the results: universal  dummy coding which tremendously increases dimension of data, and  nonlinear Categorical Principal Component Analysis (CatPCA) \cite{Linting12}. 

CatPCA can be applied if there exists a rich family of numerical inputs. This method combines classical PCA with the idea of representation of the categories of a nominal feature by the centroids of examples from these categories in the space of numerical features. 
CatPCA assumes that the latent variable behind the nominal feature is correlated with the numerical input variables.

This procedure includes four steps (Algorithm \ref{alg1}).

\begin{figure}
\centering
\includegraphics[scale=0.9]{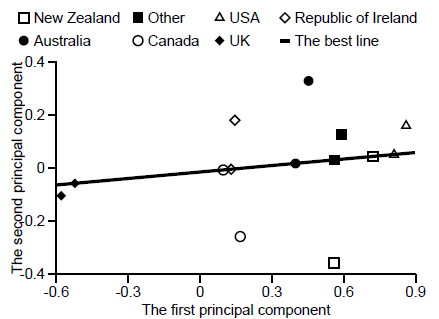}
\caption {{Quantification of the nominal feature `Country'.  Centroids of groups from different countries projected on the plane of the first two PCs of numerical input data. The line corresponds to the first PC of the set of centroids.  It is worth mentioning that this line is originally situated in the 9-dimensional space of the selected PCs of the numeric variables. In that 9-dimensional space, the projections of the centroids on this line are orthogonal but this orthogonality may be lost in the two-dimensional projection.}}
\label{Quantificationfig:3}
\end{figure}

\begin{algorithm}                        % enter the algorithm environment
\caption{Nominal feature quantification} % give the algorithm a caption
\label{alg1}                             % and a label for \ref{} commands later in the document
\begin{algorithmic}[1]                      % enter the algorithmic environment
    \State Exclude nominal features from the set of input features and calculate the PCs  in the space of  the numeric input features which have been retained. To select informative PCs we use Kaiser's rule \cite{Guttman54,Kaiser60}.
    \State Calculate the projection $C_i$ of the centroid of data points from each category $v_i$ on the selected PCs.
    \State Consider these projections of centroids as a new dataset. Calculate the first PC of them, $w$.
    \State The numerical value for each category is the projection of its centroid on this component: $v_i$ takes the numerical value of the inner product $(w,C_i)$.
\end{algorithmic}
\end{algorithm}
The process of nominal feature quantification for the feature `country' is depicted in Fig~\ref{Quantificationfig:3}.  In this figure, the points corresponding to the UK category are located very far from all other points.

As an alternative variant of nominal feature quantification we have used dummy coding of nominal variables: `country' is transformed into seven binary features with values 1 (if `true') or 0 (if `false'): UK, Canada, USA, Other (country), Australia, Republic of Ireland and New Zealand; Ethnicity is transformed into seven binary features: Mixed-White/Asian, White, Other (ethnicity), Mixed-White/Black, Asian, Black and Mixed-Black/Asian.

\subsection{Input feature ranking}
\label{Input feature ranking}
In this study, we have used three different techniques for input feature ranking. The first technique was \textit{principal variables} \cite{McCabe84}. The idea is very simple. For each attribute $x_i$ calculate the linear regression line (\ref{linreg}) for predicting all other attributes. Evaluate UV, that is the sum of square errors of these predictions (for all data points and all attributes). Select the attribute which gives the smallest FVU. This is the first principal variable. Subtract the values predicted using the first principal variable from the data table. The new dataset has one feature less (the values of the first principal variable become zero). Iterate the above approach. We can calculate the fraction of variance explained by $k$th principal variable, and the cumulative fraction of variance explained by the first $k$ principal variables. 

The second technique used was \textit{double Kaiser's selection}. Calculate PCs and select informative PCs, $w_j$, by Kaiser's rule \cite{Guttman54,Kaiser60}.  Next apply the same idea to attribute selection.  PCs are normalized vectors so that the average of the square of their coordinates is $1/n$. 
For each attribute find the maximal absolute value of the corresponding coordinates in the informative PCs: $\gamma_i=\max_j  |w_{ji}|$.
The $i$th  attribute is called trivial if $\gamma_i \le 1/\sqrt{n}$.   If there are trivial attributes then the most trivial attribute is the attribute with minimal $\gamma_i$. We removed the most trivial attribute and repeated the procedure. This procedure stops if there are no trivial attributes. This algorithm ranks attributes from the most trivial to the best.

The third technique utilised was \textit{sparse PCA} \cite{Naikal11}.  In this study, we employed the simplest thresholding sparse PCA. This method returns us a list of trivial (non-important) coordinates and the sparse PCs in the space of important coordinates.   The formal description is given in Algorithm \ref{alg2}.

\begin{algorithm}                        % enter the algorithm environment
\caption{Search of sparse PC} % give the algorithm a caption
\label{alg2}                            % and a label for \ref{} commands later in the document
\begin{algorithmic}[1]                      % enter the algorithmic environment
    \State Set the number of attributes, $n$, the sample variance of the dataset, $s^2$, and Kaiser's threshold for the variance explained by a component, $h=s^2/n$. 
    \State Search for the usual first PC with the maximal variance $\lambda$ explained by this PC. (The iterative algorithm gives the PCs in descending order of $\lambda$.)
    \If {$\lambda<h$} \State All the informative PCs have been found. Remove the last PC and go to Step 10.
    \EndIf
    \State Find the PC coordinate with the minimal absolute value, $|w_{\min}|$.
    \If {$|w_{\min}|>1/\sqrt{n}$} \State  Go to Step 9. (All the  attributes are non-trivial in this PC.) \EndIf 
    \State Set the value of the coordinate with minimal size, found at Step 5, to zero. Calculate the first PC in the reduced space of coordinates. Go to Step 5.
    \State Subtract the projection onto the PC from the above PC reduction loop from the data (the dataset is redefined), and go to Step 2.
    \State Search the attributes, which have zero coefficients in each of the PCs found by the previous selection process. These attributes are trivial.
    \If {there are trivial attributes} \State Remove trivial attributes from the set of attributes and go to Step 1. ($n$ and $s^2$ should be redefined.)\Else \State Stop \EndIf
\end{algorithmic}
\end{algorithm}

\section{Classification and risk evaluation}
\label{Risk evaluation methods}

The main classification problem studied in this book is user/non-user classification for various substances and recency of use. The inputs include age, gender, education, N, E, O, A, C, Imp., and SS quantified, by the methods described in Sec.~\ref{Quantification}.

Several families of classification methods are applied to the solution of this problem. Most of the  methods are supplemented by risk evaluation tools.

\subsection{Single attribute predictors}

Solutions of a classification problem on the basis of one attribute are very attractive. It is nice to have one measurement and a red threshold line: one class is situated below this line, another class is above. Of course, such a simple rule is rarely reliable. Nevertheless, it is always a good idea to start from one-attribute classifiers and to evaluate their capabilities. It will give use a good `bottom line' for evaluation of more sophisticated classifiers.

Creation of one-attribute classifiers is very simple: create histograms of this attribute for both classes. Traditionally, one class is called positive (users) and another is negative (non-users).  For each threshold (positive to one side, negative to another) several numbers can be evaluated easily: number of  True Positive (TP), True Negative (TN), False Positive (FP), and False Negative (FN) samples.  
Two important numbers help us to evaluate the quality of a classifier:
Sensitivity (or the true positive rate) Sn=TP/(TP+FN) and Specificity (or the true negative rate) Sp=TN/(TN+FP). Sn is the fraction of correctly recognised samples of the positive class and Sp is the fraction of correctly recognised samples of the negative class. 

\begin{figure}
\centering
\includegraphics[width=0.5\textwidth]{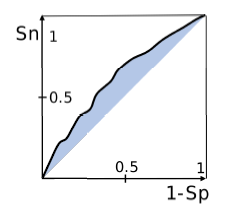}
\caption {ROC curve: dependence of True Positive Rate (Sn) on False Positive Rate (1-Sp). The AUC is $\frac{1}{2}$ plus the area of the shadowed region between the ROC curve and the diagonal.}
\label{fig:ROC}
\end{figure}

When we move threshold, Sn and Sp change in opposite directions. To represent the possible balance between Sn and Sp, the Receiver Operating Characteristic (ROC) curve is used (Fig.~\ref{fig:ROC}), with the False Positive Rate (1-Sp) on the abscissa, and the  True Positive Rate (Sn) on the ordinate. The diagonal line indicates what the balance is if we are classifying at random. Above the diagonal lines indicates a good classifier, and below the line a bad classifier. Suppose we randomly select one positive case and one negative case, that classification has positive cases above the threshold, and that the classification scores for these cases are $x_{\rm positive}$ and $x_{\rm negative}$ respectively. Then, the Area Under the Curve (AUC) estimates the probability $x_{\rm positive}>x_{\rm negative}$). The area between the ROC curve and the diagonal (shadowed on Fig.~\ref{fig:ROC}) is proportional to the so-called {\em Gini coefficient}: Gini=2AUC-1. AUC is widely used in classification \cite{Hanley1983} despite the fact that it is not a stable measure for small samples. Various alternatives have been proposed \cite{Hand2009}.

A dimensionless variable $z$ is convenient for representing the difference between two groups $S_1$ and $S_2$, with the  sample means of the attribute in the groups, $\bar{x}_1,\bar{x}_2$, and the sample standard deviations, $s_1,s_2$ respectively:
\begin{equation}\label{score}
z=\frac{|\bar{x}_1-\bar{x}_2|}{s_1+s_2}.
\end{equation}

This score measures how the variability between group means relates to the variability of the observations within the groups. It is a useful measure of the separation of the two  distributions \cite{Becker1968}. It  is calculated in the tables in Chapter~\ref{Chapter:Results} for each of seven psychological attributes and for two classes, users and non-users of illicit drugs. The higher the score is, the better  users are separated from non-users by the values of the attribute. 

A simple motivation for this measure is that for the normal distributions inside groups the optimally balanced separation of groups with given $z$ has equal specificity and sensitivity $P$, where $P=\phi(z)$, recalling that $\phi$ is the cumulative distribution function of the standard normal distribution (with $\mu=0$, $\sigma=1$). Tables~\ref{tab:2}, \ref{tab:802}, \ref{tab:803}, \ref{tab:804} demonstrate that $P$ gives realistic estimate of the empirical Sn and Sp for the data analysed.

 The score $z$, its reciprocal $z^{-1}$, and 
their multidimensional generalizations \cite{Becker1968}, are also widely  used in cluster analysis for the construction of criteria for the validity of clusters \cite{XuWunsch2009}.

If we are not happy with the performance of one-attribute classifiers then it can be improved using many approaches. Two of them are: 
\begin{itemize}
\item Classification by combination of the existing one-attribute classifiers (with weights defined by their individual performance);
\item Combination of attributes to form a new multi-input discriminant function.
\end{itemize}
It is crucially important to note that when we extend the space of possible classification rules, then we add new potential for overfitting and overoptimism. Special testing and validation procedures should be employed to avoid potentially inaccurate and misleading results. (The critical review \cite{Concato1993} is still relevant, especially with respect to the new fashionable multidimensional machine learning methods).

\subsection{Criterion for selecting the best method}\label{CriterionOfTheBestMethod}

A number of different criteria exist for the selection of the best classifier. If the cost of different misclassifications is specified then  minimisation of the expected cost can be a reasonable criterion for selection. If the costs of different misclassifications do not vary significantly then this approach leads to the minimisation of the number of errors. In situations with significant class imbalance this minimization can lead to paradoxical decisions, for example, to the selection of a trivial rule: everything is the dominating class. The class balance in the database we use is strongly biased with respect to the general population. Therefore, we decided to work with relative criteria, which provide balanced classification rules. 

The criterion we used was to pick the method such that the minimum between sensitivity and specificity was maximised: $\min\{{\rm Sn,Sp}\}\to \max$.
If $\min\{{\rm Sn,Sp}\}$ is the same for several classifiers, then we select from these the classifier which maximises Sn+Sp. Classifiers with Sn or Sp less than 50\% were not considered. There are several approaches to test the quality of a classifier: usage of an isolated test set, $n$-fold cross validation and Leave-One-Out Cross Validation (LOOCV) \cite{Arlot10}. LOOCV is used for all tests in this study; we exclude a sample from the classifier preparation, learn the classifier, and then test the result on the excluded example. Specific problems with the estimation of classifier quality for techniques like decision tree and random forest were  considered in detail by Hastie et al. \cite{Hastie09}.

\subsection{Linear Discriminant Analysis (LDA)}

The first and the most famous tool of discriminant analysis is Fisher's linear discriminant \cite{Fisher36}, where a new attribute is constructed as a linear functional of the given attributes, with the best classification ability. It is possible to calculate the score (\ref{score}) for values of any function.  The linear function with the highest score is a version of the Fisher's linear discriminant. 

We used Fisher's linear discriminant for the binary version of the problem, to separate users from non-users of each drug. We calculate the mean of the points in the $i$th class, $\bar{x}_{i}$, and empirical covariance matrix of the $i$th class, $S_{i}$, for both classes ($i=1,2$). Then we calculate the discriminating direction as 
\begin{equation}\label{FisherLD}
\omega=\left(S_{1}+S_{2}\right)^{-1}\left(\bar{x}_{1}-\bar{x}_{2}\right).
\end{equation}
Each point is projected onto the discriminating direction by calculating the dot product $(\omega,x_{i})$. This projection gives a new (combined) attribute.  The optimal threshold to separate two classes is calculated. 

In our study we have prepared linear discriminants for all possible selections of the subset of input attributes, and selected the best set of inputs for each classification problem.

\subsection{Logistic Regression (LR)}
In the sequel we have implemented the weighted version of logistic regression \cite{Hosmer04}. This method can be used for binary problems only, and is based on the following model assumption:
\begin{equation}
\frac{\mbox{probability of the first class}}{\mbox{probability of the second class}}
=\exp(\omega,x),
\end{equation}
where $\omega$ is the vector of regression coefficients and $x$ is a data vector.

The maximum  log likelihood estimate of the regression coefficients is used. This approach assumes that the outcomes of different observations are independent and maximises the weighted sum of logarithms of their probabilities. In order to prevent class  imbalance difficulties, the weights of categories are defined. The most common weight for the $i$th category is the inverse of the fraction of the $i$th category cases among all cases. Logistic regression gives only one result because there is no option to customize the method except by choice of the set of input features. We performed an exhaustive search for the best set of inputs for each classification problem.

\subsection{$k$ Nearest Neighbours ($k$NN)}
The basic concept of $k$NN is that the class of an object is the class of the majority of its $k$ nearest neighbours \cite{Clarkson05}. This algorithm is very sensitive to the definition of distance. There are several commonly used variants of distance for $k$NN: Euclidean distance; Minkovsky distance; and distances calculated after some transformation of the input space. In this study, we have used three distances: the Euclidean distance, the Fisher's transformed distance \cite{Fisher36}, and the adaptive distance \cite{Hastie96}. Moreover, we have used a weighted voting procedure with weighting of neighbours by one of the standard kernel functions \cite{Li07}.

The $k$NN algorithm is well-known \cite{Clarkson05}. The adaptive distance transformation algorithm is described in \cite{Hastie96}. $k$NN with Fisher's transformed distance is less well-known. The following parameters are used: $k$ is the number of nearest neighbours, $K$ is the kernel function, and $k_f \ge k$ is the number of neighbours which are used for the distance transformation. To define the risk of drug consumption we have to perform the steps described in Algorithm \ref{alg3}.
\begin{algorithm}                        % enter the algorithm environment
\caption{$k$NN with Fisher's transformed distance} % give the algorithm a caption
\label{alg3}                             % and a label for \ref{} commands later in the document
\begin{algorithmic}[1]                      % enter the algorithmic environment
    \State Find the $k_f$ nearest neighbours of the test point $x$.
    \State Calculate the empirical covariance matrix of $k_f$ neighbours and Fisher's discriminant direction.
    \State Find the $k$ nearest neighbours of the $x$ using the distance along Fisher's discriminant direction among the $k_f$ neighbours found earlier.
    \State Define the maximal distance $d$ from $x$ to the $k$ neighbours.
    \State Calculate the {\sl membership} for each class $C$ as a sum of the weights for $y \in C$. The weight of $y$ is the ratio $K(\|x-y\|/d)$.
    \State The drug consumption risk is defined as the ratio of the membership of the assigned class to the sum of memberships of all classes.
\end{algorithmic}
\end{algorithm}

The adaptive distance version implements the same algorithm but uses another transformation in Step 2, and another distance in Step 3 \cite{Hastie96}. The Euclidean distance version simply defines $k_f=k$ and omits Steps 2 and 3 of the algorithm.
We have tested various versions of the $k$NN models for each drug, which differ by:
\begin{itemize}
\item The number of nearest neighbours, which varies between 1 and 30;
\item The set of input features;
\item One of the three distances: Euclidean distance, adaptive distance, and Fisher's distance;
\item The kernel function for adaptive distance transformation;
\item The kernel functions for voting.
\item The weight of class `users' is varied between 0.01 and 5.0.
\end{itemize}

\subsection{Decision Tree (DT)}

\begin{figure}
\centering
\includegraphics[width=0.5\textwidth]{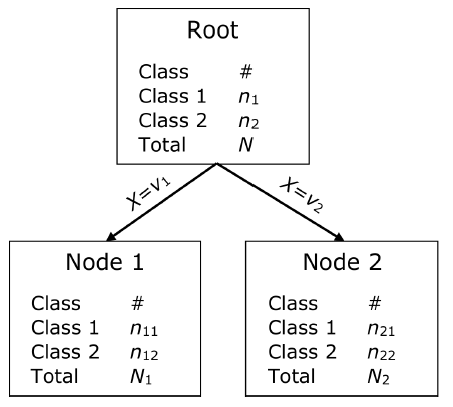}
\caption {Elementary branching in a decision tree}
\label{Fig:Branching}
\end{figure}

The decision tree is one of the most popular methods of data analysis \cite{WuEtAl2008}. It was invented before the computer era for clarification of complex diagnosis and decision making situations, in the form of a tree of simple questions and decisions. We aim to solve the classification problem (user/non-user classification). For this purpose, we consider a decision tree as a tool for combination of various classifiers. Each elementary classifier can be thought of as a categorical variable $X$ (symptom) with several nominal values $v_1, \ldots, v_k$. An elementary branching divides the dataset (the root) into two subsets (the nodes, see Fig.~\ref{Fig:Branching}). Perfect branching creates 0-1 frequencies, for example $n_{11}/N_1=n_{22}/N_2=1$ and $n_{12}=n_{21}=0$. Such a perfect situation (errorless diagnosis by one feature) is not to be expected. However, we can find the elementary classifier, which results in the closest to the perfect classification. Then we can then iterate, i.e., approach each node as a dataset and try all possible elementary classifiers, etc., until we find a perfect solution, the solution cannot be improved, or the number of examples in a node becomes too small (which will lead to overfitting).

There are many methods for developing a decision tree \cite{Breiman84,Quinlan87,Rokach10,Sofeikov14,Gelfand91,Dietterich96,Kearns99}, which differ depending on the method for selection of the best elementary classifier for branching (Fig.~\ref{Fig:Branching}), and by the stopping criteria. We have evaluated the best attributes for branching using one of the following `gain' functions: RIG, Gini gain, and DKM gain. They are defined in a similar way below. 

Consider one node (the root, Fig.~\ref{Fig:Branching}) with $N$ cases and the binary classification problem. If the attribute $X$ has $c$ possible categorical values $v_1, \ldots , v_c$ then we consider branching into $c$ nodes.   We use the following notation: $N_i$ is the number of cases with $X=v_i$ ($i=1, \ldots , c$), $n_{ij}$ ($i=1, \ldots , c$, $j=1,2$) is a number of cases of class $j$ with $X=v_i$. 

For a vector of normalised frequencies $f=(f_1, \ldots ,f_k)$ three concave (`Base')  functions are defined:
\begin{itemize}
\item ${Entropy}(f)=-\sum_{i=1}^{k}f_i \log_{2}f_i$;
\item ${Gini}(f)=1-\sum_{i=1}^{k}f_i^{2}$;
\item ${DKM}(f)=2\sqrt{f_1 f_2}$ (for $k=2$).
\end{itemize}

The corresponding gain functions for branching are defined as
$$Gain=Base\left(\frac{n_1}{N},\frac{n_2}{N}\right)-\sum_{i=1}^c \frac{N_i}{N}Base\left(\frac{n_{i1}}{N_i},\frac{n_{i2}}{N_i}\right),$$
where $Base(f)$ is one of the functions {\em Entropy, Gini, or DKM}.

All of these {\em Gain} functions qualitatively measure  the same thing: how far are distributions of nodes from the initial distribution of the root, and how close they are to the perfect situation (when each node is strongly biased to one of the classes). 

For a variety of reasons one might want to weight classes differently, for instance, to reduce the impact of classes with many outliers. We need to multiply class frequencies by weights and then to normalise by dividing by the sum of weights.

The branching with maximal {\em Gain} is considered as the best for a given criterion function.

The set of elementary classifiers (attributes) may be large, and include all the one-attribute classifiers with different thresholds, all linear discriminants, various non-linear discriminant functions, or other classifiers like $k$NN, etc.

The specified minimal number of instances in a tree's leaf is used as a criterion to stop node splitting; no leaf of the tree can contain fewer than a specified number of  instances.

We tested decision tree models, which differ by:
\begin{itemize}
\item The three split criterion (information gain, Gini gain or DKM gain);
\item The use of the real-valued features in the splitting criteria separately, or in linear combination by Fisher's discriminant;
\item The set of  input features;
\item The minimal number of instances in the leaf, which varied between 3 and 30;
\item Weight of class `users' that is varied.
\end{itemize}

Direct exhaustive search of the best decision tree can lead to the overfitiing and `overoptimism'. Special validation procedures are necessary. The standard LOOCV meets well-known problems \cite{Hastie09} because the topology of the tree may change in this procedure. Special notions of stability and structural stability are needed.  We return to this problem in more detail, after presentation of the classification results (Section~\ref{Overoptimism}). 

\subsection{Random Forest (RF)}
Random forests were proposed by Breiman \cite{Breiman01} for building a predictor ensemble with a set of decision trees that grow in randomly selected subspaces of the data \cite{Biau12}.  The random forests classification procedure consists of a collection of tree structured classifiers ${h(x,\Theta_k ),k=1,...}$, where the ${\Theta_k}$ are independent identically distributed random vectors and each tree casts a unit vote for the most popular class at input $x$” \cite{Breiman01}.

In a random forest, each tree is constructed using a different bootstrap sample from the original data \cite{Hastie09}. In standard trees, each node is split using the best split among all variables. In a random forest, each node is split using the best among a subset of predictors randomly chosen at that node \cite{Liaw02}.

Random forests try to improve on bagging by `de-correlating' the trees. Each tree has the same expectation \cite{Hastie09}. 
The forest error rate depends on two things \cite{Breiman01}. The first is the correlation between any two trees in the forest. Increasing the correlation increases the forest error rate. The second is the strength of each individual tree in the forest. A tree with a low error rate is a strong classifier. Increasing the strength of the individual trees decreases the forest error rate. The random forest algorithm builds hundreds of decision trees and combines them into a single model \cite{Williams11}. 

\subsection{Gaussian Mixture (GM)}
Gaussian mixture is a method of estimating probability under the assumption that each category of a target feature has a multivariate normal distribution \cite{Dinov08}. In each category we should estimate the empirical covariance matrix and invert it. The primary probability of belonging to the $i$th category is:

$$p_{i}(x)=p_i^0(2\pi)^{-\frac{k}{2}}|S_{i}|^{-\frac{1}{2}} \exp \left[{-\frac{1}{2}}(x-\bar{x}_{i})^{\prime}S_i^{-1}(x-\bar{x}_{i})\right]$$

where $p_i^0$ is a prior probability of the $i$th category, $k$ is the dimension of the input space, $\bar{x}_i$ is the mean point of the $i$th category, $x$ is the tested point, $S_i$ is the empirical covariance matrix of the $i$th category and $|S_i |$ is its determinant. The final probability of belonging to the $i$th  category is calculated as $$p^f_{i}(x)=p_{i}(x)/\sum_{j}p_{j}(x).$$

The prior probabilities are estimated as the proportion of cases in the $i$th category. We also used s varied multiplier to correct priors for the binary problem.

In the study, we tested Gaussian mixture models, which differ by the set of input features and corrections applied to the prior probabilities.

\subsection{Probability Density Function Estimation (PDFE)}
We have implemented the radial basis function method \cite{Buhmann03} for probability density function estimation \cite{Scott92}. The number of probability densities to estimate is equal to the number of categories of the target feature. Each probability density function is estimated separately by using nonparametric techniques. The prior probabilities are estimated from the database: $p_i=n_i/N$ where $n_i$ is the number of cases in category $i$ of the target feature, and $N$ is the total number of cases in the database.

We also use the database to define the $k$ nearest neighbours of each data point. These $k$ points are used to estimate the radius of the neighbourhood of each point as a maximum of the distance from the data point to each of its $k$ nearest neighbours. The centre of one of the kernel functions is placed at the data point \cite{Li07}. The integral of any kernel function over the whole space is equal to one. The total probability of the $i$th category is proportional to the integral of the sum of the kernel functions, which is equal to $n_i$. The total probability of each category has to be equal to the prior probability $p_i$. Thus,  the sum of the kernel functions has to be divided by $n_i$ and multiplied by $p_i$. This gives the probability density estimation for each category.

We have tested a number PDFE models, which differ by:
\begin{itemize}
\item The number of nearest neighbours (varied between 5 and 30);
\item The set of the input features;
\item The kernel function which was placed at each data point.
\end{itemize}

\subsection{Na{\"i}ve Bayes (NB)}

The NB approach is based on the simple assumption: attributes are independent. Under this assumption, we can evaluate the distribution of the separate attributes and then produce a joint distribution function just as a product of attributes' distributions. Surprisingly, this approach performs satisfactorily in many real life problems despite the obvious oversimplification.
We have used the standard version of NB \cite{Russell95}. All attributes which contain $\leq$20 different values were interpreted as categorical and the standard contingency tables were calculated for such attributes. The contingency tables we calculated are used to estimate conditional probabilities. Attributes which contain more than 20 different values were interpreted as continuous.  The mean and the variance were calculated  for continuous attributes instead of the contingency tables. We calculated the isolated mean and variance for each value of the output attribute. The conditional probability of a specified outcome $o$ and a specified value of the attribute $x$ were evaluated as the value of the probability density function for a normal distribution at point $x$ with matched mean and variance, which were calculated for the outcome $o$. This method has no customization options and was tested on different sets of input features. In the study we tested  2,048 NB models per drug.

\section{Visualisation on the non-linear PC canvas: Elastic maps}

Elastic maps \cite{GorbanZin2005} provide a tool for nonlinear dimensionality reduction. By construction, they are a system of elastic springs embedded in the data
space. This  system approximates a low-dimensional manifold, and is a model of the {\em principal manifold} \cite{Hastie1989,Gorban08}. The elastic coefficients of this system allow the switch from completely unstructured $k$-means clustering (zero elasticity) to estimators approximating linear PCs (for high bending and low stretching modules). With certain intermediate values of the elasticity coefficients the system effectively approximates non-linear principal manifolds. In this section we follow \cite{GorbanZin2009}.

Let the data set be a set of vectors  $S$ in a finite-dimensional Euclidean space. The `elastic map' is represented by a set of nodes $W_j$ (the 'elastic net') in the same space, but not necessarily a subset of $S$. Each data point $s \in S$ has a `host node', namely the node $W_j$ closest to $s$  (if there are several closest nodes then one takes the node with the smallest index). The data set  is divided in to classes $$K_j=\{s \ | \ W_j \mbox{ is the host of } s\}.$$

The `approximation energy' $D$ is the distortion
$$D=\frac{1}{2}\sum_{j=1}^k \sum_{s \in K_j}\|s-W_j\|^2,$$
which is the energy of springs with unit elasticity connecting each data point with its host node.

On the set of nodes an additional structure is defined. Some pairs of nodes, $(W_i,W_j)$, are connected by `elastic edges'. Denote this set of pairs $E$. Some triplets of nodes, $(W_i,W_j,W_k)$, form `bending ribs'. Denote this set of triplets $G$. 

The stretching energy $U_{E}$ and the bending energy $U_G$ are 
$$U_{E}=\frac{1}{2}\lambda \sum_{(W_i,W_j) \in E} \|W_i -W_j\|^2 , \; \; U_G=\frac{1}{2}\mu \sum_{(W_i,W_j,W_l) \in G} \|W_i -2W_j+W_l\|^2,$$
where $\lambda$ and $\mu$ are the stretching and bending moduli respectively. The stretching energy is sometimes referred to as the `membrane' term, while the bending energy is referred to as the `thin plate' term.

For example, on the 2D rectangular grid the elastic edges are just vertical and horizontal edges (pairs of closest vertices) and the bending ribs are the vertical or horizontal triplets of consecutive (closest) vertices. The total energy of the elastic map is thus $U=D+U_E+U_G$.
The position of the nodes $\{W_j\}$ is determined by the mechanical equilibrium of the elastic map, i.e. its location is such that it minimizes the total energy $U$.

For a given splitting of the dataset  $S$ in to classes $K_j$, minimization of the quadratic functional $U$ is a linear problem with the sparse matrix of coefficients. Therefore, similarly to PCA or $k$-means, a splitting method is used:
\begin{itemize}
\item For given $\{W_j\}$ find $\{K_j\}$;
\item For given $\{K_j\}$ minimize $U$ to find new $\{W_j\}$;
\item If no change, terminate.
\end{itemize}

This expectation-maximization algorithm  guarantees a local minimum of $U$. To improve the approximation various additional methods might be proposed, for example, the ''softening'' strategy. This strategy
starts with a rigid grid (small length, small $\lambda$ and large $\mu$) and finishes with soft grids (small $\lambda$ and $\mu$). The training goes in several epochs, each epoch with its own grid rigidness. Another adaptive strategy is the `growing net': one starts from a small number of nodes and gradually adds new nodes. Each epoch goes with its own number of nodes.

The elastic map is a continuous manifold. It is constructed from the elastic net using some interpolation procedure between nodes. For example, the simplest piecewise linear
elastic map is build by triangulation and a piecewise linear map. Data points are projected into the closest points of the elastic map \cite{GorbanZin2005}.

\begin{figure}
\centering{\includegraphics[width=0.9\textwidth]{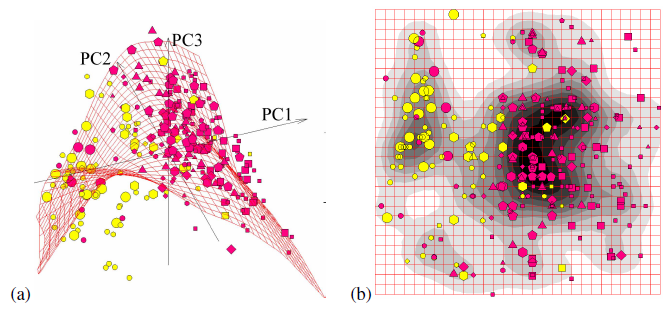}} 
\caption{Visualization \cite{Gorban08} of breast cancer dataset  \cite{Wang2005}
using elastic maps: (a)
configuration of nodes in the space of first three PCs; (b) the distribution of points in the internal
non-linear manifold coordinates together with
an estimation of the two-dimensional density of points.  (A public domain illustration)
\label{DatasetI}}
\end{figure}

The elastic map algorithm is extremely fast in the optimisation step due to the very simple form of the
smoothness penalty. In this book we have employed the original software libraries ViDaExpert freely available online \cite{Gorban15}.
This software allows the creation of an appropriate elastic manifold embedded in the dataspace (Fig.~\ref{DatasetI}a), and to color this map to visualise data density and distribution of all the attributes  (Fig.~\ref{DatasetI}). An example of such work \cite{Gorban08} is presented in Fig.~\ref{DatasetI} for breast cancer microarrays \cite{Wang2005}.  New open access software for calculation of elastic graphs is available online in MatLab \cite{githubElgraphMatLab} and R \cite{githubElgraphR}. 

Elastic maps and software have been applied in various areas, from bioinformatics \cite{Chac2007,Gorban08} to political sciences \cite{Zinovyev2011}, financial analysis \cite{Resta2016} and multiphase flows \cite{Shaban2014}. Examples of elastic maps for the drug consumption database are presented in the next chapter.

%% file: chapter3.tex
%%%%%%%%%%%%%%%%%%%%% chapter.tex %%%%%%%%%%%%%%%%%%%%%%%%%%%%%%%%%
%FINAL
% sample chapter
%
% Use this file as a template for your own input.
%
%%%%%%%%%%%%%%%%%%%%%%%% Springer-Verlag %%%%%%%%%%%%%%%%%%%%%%%%%%
%\motto{Use the template \emph{chapter.tex} to style the various elements of your chapter content.}
\chapter{Results of data analysis}
\label{Chapter:Results} % Always give a unique label
% use \chaptermark{}
% to alter or adjust the chapter heading in the running head

\abstract*{This chapter includes results of data analysis. The relationship between personality profiles and drug consumption is described and the individual drug consumption risks for different drugs is evaluated. Significant differences between groups of drug users and non-users are identified. Machine learning algorithms solve the user/non-user classification problem for many drugs with impressive sensitivity and specificity.
Analysis of correlations between use of different drugs reveals existence of clusters of substances with highly correlated use, which we term correlation pleiades. It is proven that the mean profiles of users of different drugs are significantly different (for benzodiazepines, ecstasy, and heroin). Visualisation of risk by risk maps is presented. The difference between users of different drugs is analysed and three distinct types of users are identified for benzodiazepines, ecstasy, and heroin.}

\section{Descriptive statistics and psychological profile of illicit drug users}

The data set contains seven categories of drug users: `Never used', `Used over a decade ago', `Used in last decade', `Used in last year', `Used in last month', `Used in last week', and `Used in last day'. A respondent selected their category for every drug from the list. We formed four classification  problems based on the following classes (see Section `Drug use' ): the decade-, year-, month-, and week-based user/non-user separations.

We have identified the relationship between personality profiles (NEO-FFI-R) and drug consumption for the decade-, year-, month-, and week-based classification problems. We have evaluated the risk of drug consumption for each individual according to their personality profile. This evaluation was performed separately for each drug for the decade-based user/non-user separation. We have also analysed the interrelations between the individual drug consumption risks for different drugs.  Part of these results has been presented in \cite{FehrmanBologna15} (and in more detail in the 2015 technical report~\cite{Fehrman15}).
In addition, in Section~\ref{Pleiad of drug users}  we employ the notion of {\em correlation pleiades} of drugs. We define three pleiades: heroin pleiad, ecstasy pleiad, and benzodiazepines pleiad, with respect to the decade-, year-, month-, and week-based user/non-user separations. It is also important to understand how the group of users of illicit drugs differs from the group of non-users.

The descriptive statistics for seven traits (FFM, Imp, and SS) are presented in Table~\ref{tab:2}: means, standard deviations, and 95\% confidence intervals for means for NEO-FFI-R for the full sample and for two subsamples:  non-users of illicit drugs and users of illicit drugs. We have conventionally called the following substances `illicit':  amphetamines, amyl nitrite, benzodiazepines, cannabis,  cocaine,  crack, ecstasy, heroin, ketamine, legal highs, LSD, methadone, magic mushrooms (MMushrooms),  and Volatile Substance Abuse (VSA). 

A dimensionless variable $z$ (see (\ref{score})), which measures how the variability between group relates to the variability within the groups, is convenient for representing the difference between classes. For the normal distributions inside groups the optimally balanced separation of groups with given $z$ has equal specificity and sensitivity $P$, where $P=\phi(z)$, and $\phi$ is the cumulative distribution function for the standard normal distribution.

\begin{table}[!ht]
\centering
\caption{Descriptive statistics: Means, 95\% CIs for means, and  standard deviations for the whole sample, for non-users of illicit drugs and for users of illicit drugs (decade-based user definition). The dimensionless score $z$ (\ref{score}) for separation of users from non-users of illicit drugs is presented as well as the sensitivity and specificity $P$ of the best separation of normal distributions with this $z$. Sensitivity (Sn) and Specificity (Sp) are calculated for all one-feature classifiers. $\Theta$ is the threshold for class separation: one class is given by the inequality score$\leq \Theta$ and another class by score$> \Theta$.}
\label{tab:2}
\begin{tabular}{|c|c|c|c|c|c|c|c|c|c|c|c|c|c|c|}
\cline{1-15}
Factors &	\multicolumn{3}{c|}{Total sample}&	\multicolumn{3}{c|}{Non-users of illicit drugs} & \multicolumn{3}{c|}{Users of illicit drugs} & \multicolumn{5}{c|}{One feature classifier}\\ \cline{2-15}
	&Mean &	95\% CI	&SD	&Mean	& 95\% CI	& SD	&Mean	&95\% CI	&SD &$z$ & $P$(\%) & $\Theta$ & Sn(\%) & Sp(\%)  \\ \hline
N   &23.92& 23.51, 24.33& 9.14& 21.00& 20.29, 21.71& 7.85& 24.88& 24.40, 25.37& 9.32 & 0.226 & 59 & 22 & 60 & 58 \\\hline
E   &27.58&	27.27, 27.88& 6.77&	28.52& 27.99, 29.04& 5.73& 27.27& 26.90, 27.63& 7.05 & 0.098 & 54 & 28 & 51 & 55 \\\hline
O   &33.76&	33.47, 34.06& 6.58& 30.22& 29.67, 30.77& 6.06& 34.93& 34.60, 35.26& 6.32 & 0.381 & 65 & 32 & 63 & 67 \\\hline
A   &30.87& 30.58, 31.16& 6.44& 32.87& 32.35, 33.38& 5.65& 30.21& 29.87, 30.55& 6.54 & 0.218 & 59 & 31 & 61 & 56 \\\hline
C   &29.44& 29.12, 29.75& 6.97& 32.89& 32.39, 33.40& 5.55& 28.30& 27.93, 28.66& 7.01 & 0.366 & 64 & 31 & 65 & 65 \\\hline
Imp &3.80 &	3.70, 3.90&	  2.12&	2.74 & 2.58, 2.90  & 1.74& 4.15	& 4.04, 4.26  & 2.12 & 0.364 & 64 & 3 & 71 & 56 \\\hline
SS	&5.56 &	5.44, 5.68&	  2.70& 3.77 & 3.56, 3.98  & 2.32& 6.15 & 6.02, 6.28  & 2.55 & 0.490 & 69 & 4 & 63 & 74 \\\hline
\end{tabular}
\end{table}

This is the first result: users of illicit drugs differ from non-users across all seven scales.  The 95\% CI for means in these groups do not intersect. The most significant difference 
was found for SS, then for O, for Imp and C, and for  N. The smallest difference was found for E. Later we will demonstrate that E for users of different drugs may deviate from E for non-users in a number of different ways. Moreover, the 95\% CIs for means of all three groups, total sample, users of illicit drugs and non-users of illicit drugs do not intersect for N, O, A, C, Imp and SS. Table \ref{tab:2} allows us to claim that the profile of users of illicit drugs has a characteristic form: 
\begin{equation}\label{maindrugprofile}
{\rm N}\Uparrow,{\rm O}\Uparrow, {\rm A}\Downarrow, {\rm C}\Downarrow, {\rm Imp}\Uparrow, {\rm SS}\Uparrow.
\end{equation}
The $P$ column in Table \ref{tab:2} gives a simple estimate of the separability of users from non-users of illicit drugs by a single trait. 
The best separation is given by the value of SS: estimated sensitivity and specificity are 69\%. {\em According to this estimate, SS for 69\% of illicit drug users is higher then SS of 69\% of non-users}. Of course, for more precise estimation methods the numbers will differ. We might also expect that the use of several attributes and more sophisticated classification approaches would give better sensitivity and specificity. Nevertheless, the $P$ column gives us a good indication of the possible  performance of classification.

\begin{figure}[h!]
\centering
\includegraphics[width=0.5\textwidth]{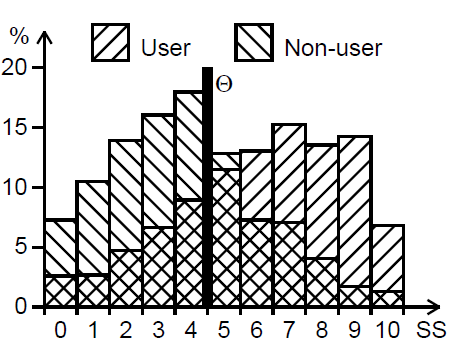}
\caption {The distributions of SS for users and non-users of illicit drugs (normalized to 100\% in each group) for the decade-based user/non-user separation. The optimal threshold is $\Theta=4$}
\label{HistSSseparation}
\end{figure}

Separation of classes ``users of illicit drugs'' and ``non-users of illicit drugs'' are presented in Tables \ref{tab:2}--\ref{tab:804}, where $\Theta$ is the {\em threshold} of the separation: one class is given by the inequality score$\leq \Theta$ and another class by score$> \Theta$. This convention, where to use strong inequality, is important because the values of scores are integer. The histogram for separation by values of SS for the decade-based definition of users is presented in Fig. \ref{HistSSseparation}.

Table \ref{tab:2_00} gives $p$ values, i.e. the probabilities of finding the same or larger difference between mean values of the traits between users of illicit drugs, non-users of illicit drugs, and the total sample. This table completely supports the profile from  (\ref{maindrugprofile}).
\begin{table}[!ht]
\centering
\caption{Significance of differences of means for total sample, users and nonusers  of illicit drugs ($p$-values) for the decade-based user definition.}
\label{tab:2_00}
\begin{tabular}{|c|c|c|c|}
\cline{1-4} 
Factors	&Total/User&	Total/Non-user &	User/Non-user\\ \hline
N &	0.003&$	<0.001$	&$<0.001$\\ \hline
E	&0.204&	0.002	&$<0.001$\\ \hline
O	&$<0.001$	&$<0.001$	&$<0.001$\\ \hline
A	&0.004	&$<0.001$	&$<0.001$\\ \hline
C	&$<0.001$	&$<0.001$	&$<0.001$\\ \hline
Imp	&$<0.001$	&$<0.001$&	$<0.001$\\ \hline
SS&	$<0.001$	&$<0.001$&	$<0.001$\\ \hline
\end{tabular}
\end{table}

There are both similarity and an important qualitative difference from the {\em `dark triad' of personality}, Machiavellianism, Narcissism and Psychopathy \cite{Jakobwitz06}.
According to Table \ref{tab:04}, the dark triad is associated with N$\Uparrow$, A$\Downarrow$ and C$\Downarrow$ and {\em low} (or neutral) O, whereas the profile (\ref{maindrugprofile}) of the users of illicit drugs has the same N$\Uparrow$, A$\Downarrow$ and C$\Downarrow$ but  {\em high} O$\Uparrow$.

The descriptive tables for the year, month and week-based user definitions are similar (Tables \ref{tab:802}--\ref{tab:8041}).

\begin{table}[!ht]
\centering
\caption{Descriptive statistics: Means, SDs, 95\% CIs for means, for the sample and two groups: non-users of illicit drugs and users of illicit drugs. Year-based classification. }
\label{tab:802}
\begin{tabular}{|c|c|c|c|c|c|c|c|c|c|c|c|c|c|c|}
\cline{1-15}
Factors &	\multicolumn{3}{c|}{Total sample}&	\multicolumn{3}{c|}{Non-users of illicit drugs} & \multicolumn{3}{c|}{Users of illicit drugs} & \multicolumn{5}{c|}{One feature classifier}\\ \cline{2-15}
	&Mean &	95\% CI	&SD	&Mean	& 95\% CI	& SD	&Mean	&95\% CI	&SD &$z$ & $P$(\%) & $\Theta$ & Sn(\%) & Sp(\%)  \\ \hline
N &23.92&23.51, 24.33&9.14&22.26&21.66, 22.87&8.15&25.18&24.64, 25.73&9.51 & 0.165 & 57 & 35 & 60 & 56 \\\hline
E&27.58&27.27, 27.88&6.77&28.21&27.76, 28.65&6.01&27.17&26.76, 27.58&7.15 & 0.078 & 53 & 40 & 51 & 55 \\\hline
O&33.76&33.47, 34.06&6.58&30.89&30.44, 31.34&6.11&35.63&35.28, 35.98&6.14 & 0.387 & 65 & 45 & 63 & 67 \\\hline
A&30.87&30.58, 31.16&6.44&32.15&31.73, 32.57&5.71&29.96&29.58, 30.33&6.58 & 0.178 & 57 & 43 & 61 & 58 \\\hline
C&29.44&29.12, 29.75&6.97&31.96&31.51, 32.42&6.16&27.77&27.37, 28.18&7.03 & 0.318 & 62 & 42 & 65 & 63 \\\hline
Imp&3.80&3.70, 3.90&2.12&2.97&2.84, 3.11&1.84&4.32&4.20, 4.44&2.11 & 0.342 & 63 & 3 & 71 & 59 \\\hline
SS&5.56&5.44, 5.68&2.70&4.00&3.82, 4.19&2.46&6.50&6.36, 6.63&2.40 & 0.514 & 70 & 5 & 63 & 69 \\\hline
\end{tabular}
\end{table}

\begin{table}[!ht]
\centering
\caption{$p$-values of significance of differences of means for total sample, non-users and users of illicit drugs. Year-based classification}
\label{tab:8021}
\begin{tabular}{|l|c|c|c|}\hline
Factors & Total/Users & Total/Non-users & Users/Non-users\\\hline
N &$<0.001$&$<0.001$&$<0.001$\\\hline
E&$0.122$&$0.049$&$0.003$\\\hline
O&$<0.001$&$<0.001$&$<0.001$\\\hline
A&$<0.001$&$<0.001$&$<0.001$\\\hline
C&$<0.001$&$<0.001$&$<0.001$\\\hline
Imp&$<0.001$&$<0.001$&$<0.001$\\\hline
SS&$<0.001$&$<0.001$&$<0.001$\\\hline
\end{tabular}
\end{table}

\begin{table}[!ht]
\centering
\caption{Descriptive statistics: Means, SDs, 95\% CIs for means, for the sample and two groups: non-users of illicit drugs and users of illicit drugs. Month-based classification}
\label{tab:803}
\begin{tabular}{|c|c|c|c|c|c|c|c|c|c|c|c|c|c|c|}
\cline{1-15}
Factors &	\multicolumn{3}{c|}{Total sample}&	\multicolumn{3}{c|}{Non-users of illicit drugs} & \multicolumn{3}{c|}{Users of illicit drugs} & \multicolumn{5}{c|}{One feature classifier}\\ \cline{2-15}
	&Mean &	95\% CI	&SD	&Mean	& 95\% CI	& SD	&Mean	&95\% CI	&SD &$z$ & $P$(\%) & $\Theta$ & Sn(\%) & Sp(\%)  \\ \hline
N &23.92&23.51, 24.33&9.14&22.60&22.06, 23.14&8.22&25.13&24.53, 25.73&9.69 & 0.141 & 56 & 35 & 56 & 60 \\\hline
E&27.58&27.27, 27.88&6.77&28.27&27.89, 28.65&5.78&27.19&26.74, 27.64&7.27 & 0.083 & 53 & 39 & 56 & 51 \\\hline
O&33.76&33.47, 34.06&6.58&31.01&30.60, 31.43&6.24&36.04&35.67, 36.41&6.00 & 0.411 & 66 & 45 & 63 & 63 \\\hline
A&30.87&30.58, 31.16&6.44&31.82&31.43, 32.22&6.02&29.81&29.40, 30.22&6.65 & 0.159 & 56 & 43 & 56 & 61 \\\hline
C&29.44&29.12, 29.75&6.97&31.78&31.38, 32.17&5.97&27.50&27.06, 27.95&7.15 & 0.326 & 63 & 42 & 61 & 65 \\\hline
Imp&3.80&3.70, 3.90&2.12&3.04&2.91, 3.16&1.92&4.43&4.30, 4.56&2.08 & 0.348 & 64 & 3 & 64 & 71 \\\hline
SS&5.56&5.44, 5.68&2.70&4.15&3.99, 4.32&2.51&6.68&6.54, 6.83&2.35 & 0.521 & 70 & 5 & 67 & 63 \\\hline
\end{tabular}
\end{table}

\begin{table}[!ht]
\centering
\caption{$p$-values of significance of differences of means for total sample,non-users and users of illicit drugs. Month-based classification}
\label{tab:8031}
\begin{tabular}{|l|c|c|c|}\hline
Factors & Total/Users & Total/Non-users & Users/Non-users\\\hline
N &$0.001$&$0.002$&$<0.001$\\\hline
E&$0.166$&$0.024$&$0.002$\\\hline
O&$<0.001$&$<0.001$&$<0.001$\\\hline
A&$<0.001$&$0.003$&$<0.001$\\\hline
C&$<0.001$&$<0.001$&$<0.001$\\\hline
Imp&$<0.001$&$<0.001$&$<0.001$\\\hline
SS&$<0.001$&$<0.001$&$<0.001$\\\hline
\end{tabular}
\end{table}

\begin{table}[!ht]
\centering
\caption{Descriptive statistics: Means, SDs, 95\% CIs for means, for the sample and two groups: non-users of illicit drugs and users of illicit drugs. Week-based classification}
\label{tab:804}
\begin{tabular}{|c|c|c|c|c|c|c|c|c|c|c|c|c|c|c|}
\cline{1-15}
Factors &	\multicolumn{3}{c|}{Total sample}&	\multicolumn{3}{c|}{Non-users of illicit drugs} & \multicolumn{3}{c|}{Users of illicit drugs} & \multicolumn{5}{c|}{One feature classifier}\\ \cline{2-15}
	&Mean &	95\% CI	&SD	&Mean	& 95\% CI	& SD	&Mean	&95\% CI	&SD &$z$ & $P$(\%) & $\Theta$ & Sn(\%) & Sp(\%)  \\ \hline
N &23.92&23.51, 24.33&9.14&22.79&22.27, 23.31&8.47&25.21&24.56, 25.86&9.72 & 0.133 & 55 & 35 & 55 & 60 \\\hline
E&27.58&27.27, 27.88&6.77&28.34&27.97, 28.71&5.96&27.09&26.60, 27.58&7.33 & 0.094 & 54 & 39 & 56 & 51 \\\hline
O&33.76&33.47, 34.06&6.58&31.18&30.78, 31.57&6.37&36.17&35.77, 36.57&5.99 & 0.404 & 66 & 46 & 66 & 63 \\\hline
A&30.87&30.58, 31.16&6.44&31.63&31.25, 32.01&6.18&29.80&29.35, 30.24&6.67 & 0.143 & 56 & 42 & 60 & 61 \\\hline
C&29.44&29.12, 29.75&6.97&31.57&31.20, 31.95&6.10&27.41&26.94, 27.89&7.10 & 0.315 & 62 & 41 & 64 & 65 \\\hline
Imp&3.80&3.70, 3.90&2.12&3.10&2.99, 3.22&1.94&4.45&4.31, 4.59&2.08 & 0.335 & 63 & 3 & 61 & 71 \\\hline
SS&5.56&5.44, 5.68&2.70&4.27&4.12, 4.43&2.52&6.73&6.58, 6.89&2.35 & 0.505 & 69 & 5 & 62 & 63 \\\hline
\end{tabular}
\end{table}

\begin{table}[!ht]
\centering
\caption{$P$-values of significance of differences of means for total sample, non-users and users of illicit drugs. Week-based classification}
\label{tab:8041}
\begin{tabular}{|l|c|c|c|}\hline
Factors & Total/Users & Total/Non-users & Users/Non-users\\\hline
N &$0.001$&$0.011$&$<0.001$\\\hline
E&$0.100$&$0.016$&$0.001$\\\hline
O&$<0.001$&$<0.001$&$<0.001$\\\hline
A&$<0.001$&$0.018$&$<0.001$\\\hline
C&$<0.001$&$<0.001$&$<0.001$\\\hline
Imp&$<0.001$&$<0.001$&$<0.001$\\\hline
SS&$<0.001$&$<0.001$&$<0.001$\\\hline
\end{tabular}
\end{table}

Pearson's correlation coefficient (PCC or $r$) is employed as a measure of the strength of a linear association between two factors. PCC for all pairs of factors are presented in Table~\ref{tab:3}. Two pairs of factors do not have significant correlation: (1) N and O ($r$=0.017, $p$=0.471); (2) A and O ($r$=0.033, $p$=0.155). However, all other pairs of personality factors are significantly correlated in the sample  (compare to Table \ref{tab:0}).

\begin{table}[!ht]
\centering
\caption{PCC for NEO-FFI-R for raw data}
\label{tab:3}
\begin{tabular}{|c|c|c|c|c|c|}\hline
{  Factors}& {  N} &{  E}&{  O}&{  A} &{  C} \\ \hline
N&&	-0.432\textsuperscript{*}&	0.017&-0.215\textsuperscript{*}& -0.398\textsuperscript{*} \\ \hline
E&-0.432\textsuperscript{*}&&0.236\textsuperscript{*}&0.159\textsuperscript{*}&0.318\textsuperscript{*}\\\hline
O&0.017&0.236\textsuperscript{*}&&0.033&-0.060\textsuperscript{**} \\\hline
A&-0.215\textsuperscript{*}&0.159\textsuperscript{*}&	0.033&&  0.249\textsuperscript{*} \\ \hline
C&-0.398\textsuperscript{*}&0.318\textsuperscript{*}&	-0.060\textsuperscript{**}&0.249\textsuperscript{*}&\\\hline
\end{tabular}\\
*$p<0.001$; **$p<0.01$.
\end{table}

Strictly speaking, the scores should be considered as ordinal features. Therefore, the polychoric correlation coefficients (PoCC) should be used. Table \ref{tab:01} presents values of PoCCs between all of the psychological traits we measured (compare to Tables \ref{tab:0} and \ref{tab:3}). The corresponding $p$-values are presented in Table \ref{tab:2}. 

\begin{table}[!ht]
\centering
\caption{Polychoric correlation coefficients (PoCC) of measured psychological traits  ($n$=1,885).}
\label{tab:01}       % Give a unique label
\begin{tabular}{|c|c|c|c|c|c|c|c|}\hline
& {N} &{E}&{O}&{A}&{C}& Imp & SS\\\hline
N&	&	$-0.431$* &	0.010 & $	-0.217$* &	$-0.391$* &0.174*&	0.080**\\\hline
E&	$-0.431$* &  & 0.245*&	0.157*	&0.308*&	0.114*&	0.210*\\\hline
O&	0.010	& 0.245*&	&0.039&	$-0.057$***&	0.278*	&0.422* \\\hline
A&	$-0.217$*&	0.157* & 0.039	&	&0.247*&	$-0.230$*&	$-0.208$* \\\hline
C&	$-0.391$*	&0.308*&$	-0.057$***&	0.247*& &	$-0.335$*&	$-0.229$*	 \\\hline
Imp& 0.174*&	0.114*&	0.278*&$-0.230$*&	$-0.335$*& &	0.623*	 \\\hline
SS&	0.080**&	0.210*&	0.422*&$	-0.208$*&$	-0.229$*&	0.623*&	 \\\hline
\end{tabular}\\
*$p<0.0001$, **$p<0.001$, ***$p<0.02$.
\end{table}

\section{Distribution of number of drugs used}
\begin{figure}
\centering
\includegraphics[width=0.84\textwidth]{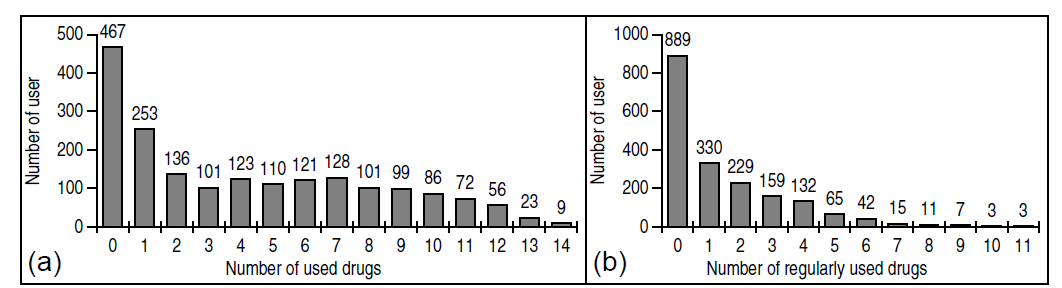}
\caption {{The histograms of the number of users of illicit drugs (the number of users versus the number of illicit drugs used):} (a)  for the decade-based user/non-user separation, and (b)  for the month-based user/non-user separation}
\label{NumberOfUserfig:3a}
\end{figure}

The diagrams in Fig.~\ref{NumberOfUserfig:3a} show the graph of the number of users versus the number of illicit drugs used for the decade-based (a) and month-based (b) user/non-user separations. In Fig.~\ref{NumberOfUserfig:3a}~a  we can see that the distribution of the number of users is bimodal with maxima at zero and seven drugs used. In Fig.~\ref{NumberOfUserfig:3a}~b the distribution of the number of users of various numbers of illegal drugs during the last months looks like the exponential distribution.

\begin{figure}
\centering
\includegraphics[width=0.84\textwidth]{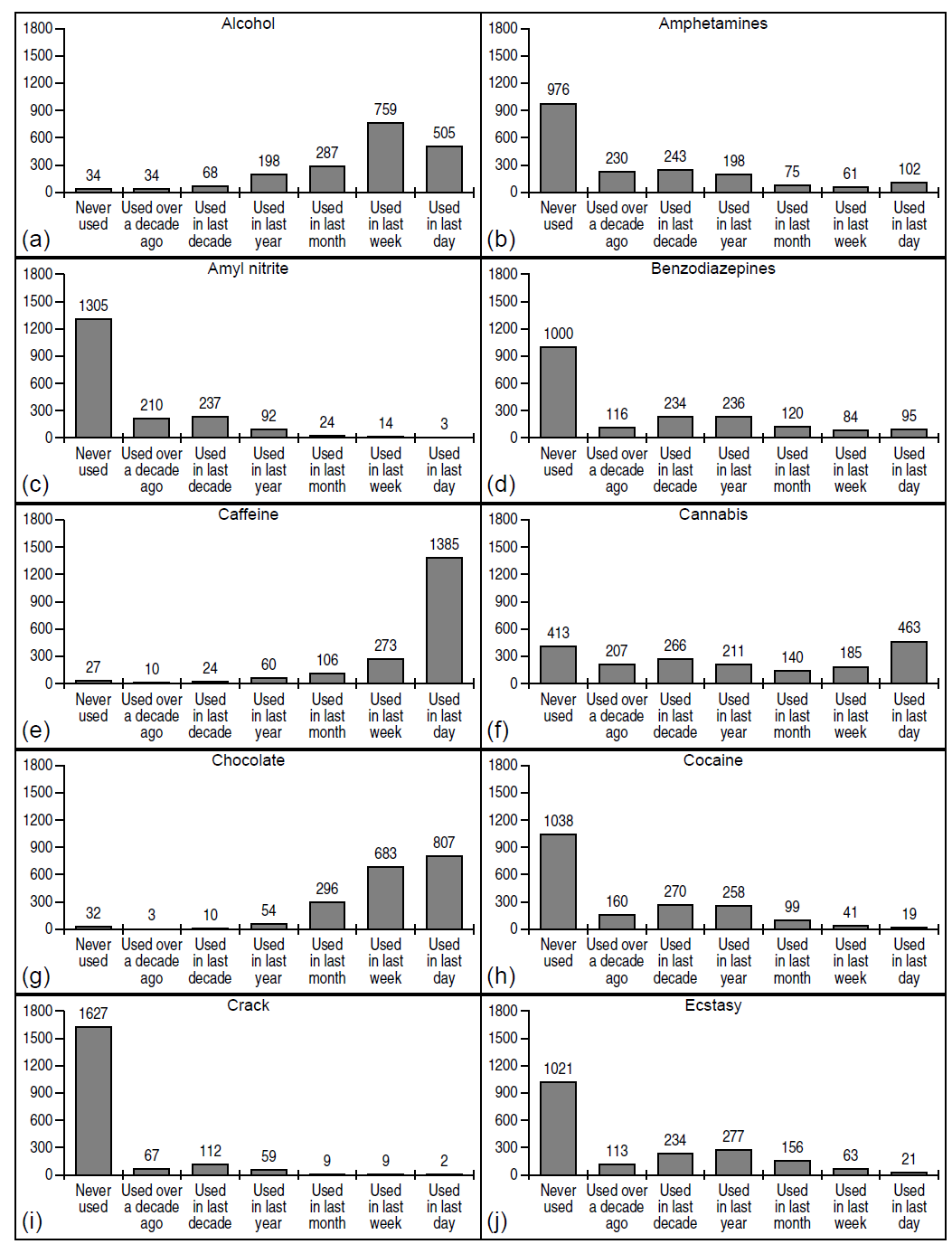}
\caption {{   Distribution of drug usage:} (a) Alcohol, (b) Amphetamines, (c) Amyl nitrite, (d) Benzodiazepines, (e) Cannabis, (f) Chocolate, (g) Cocaine, (h) Caffeine, (i) Crack, and (j) Ecstasy}
\label{DrugUserDistrib1fig:3b}
\end{figure}

\begin{figure}
\centering
\includegraphics[width=0.84\textwidth]{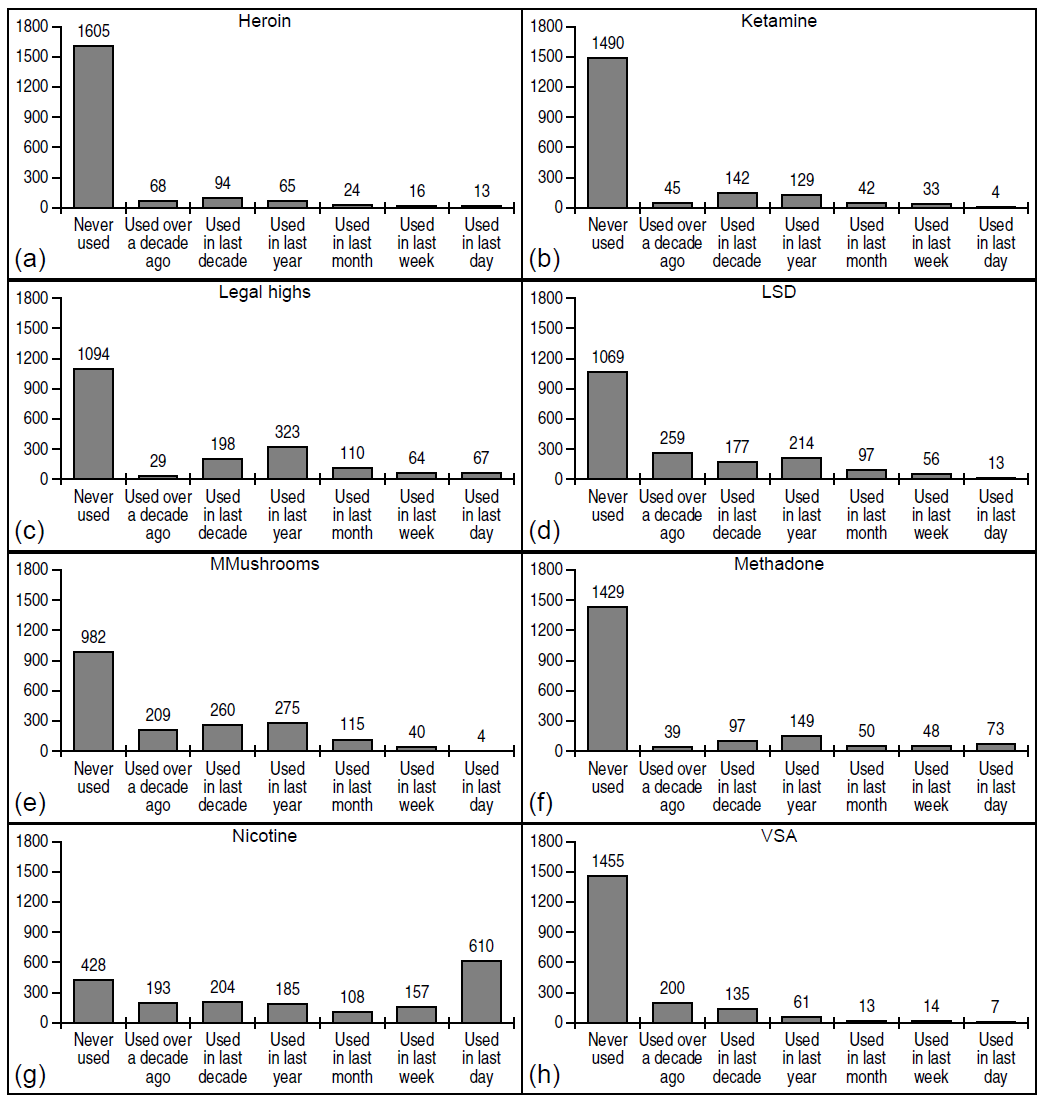}
\caption {{   Distribution of drug usage:} (a) Heroin, (b) Ketamine, (c) Legal highs, (d) LSD, (e) Methadone, (f) MMushrooms, (g) Nicotine, and (h) VSA}
\label{DrugUserDistrib2fig:3c}
\end{figure}

The distributions of the number of users for each drug are presented in Fig.~\ref{DrugUserDistrib1fig:3b} and Fig.~\ref{DrugUserDistrib2fig:3c}.   Most of the distributions have an exponential-like shape, but several have bimodal distributions. The distributions of the number of users for the three legal drugs have maximum at `Used in last day' or `Used in last week' (see Fig.~\ref{DrugUserDistrib1fig:3b}~a, e, and g). The distribution of the number of nicotine users (smokers) has three maxima: `Used in last day' for smokers, `Used in last decade' for smokers who have quit smoking, and `Never used' (see Fig.~\ref{DrugUserDistrib2fig:3c}~g). All distributions for illegal drug users have maximum in the category `Never used'. However, the distribution of cannabis users has two maxima. The main maximum is in  the category `Used in last day', and the second is in  the category `Never used' (see Fig.~\ref{DrugUserDistrib1fig:3b}f).

\section{Sample mean and population norm}
\label{popnorm}

It may seem to be a good idea to use the T-scores with respect to the population norm. Caution is needed however, since the mean values may depend on age and social group, so that the notion of `population norm' is a complex hierarchical construct rather than a simple set of means.

Following \cite{McCrae04} we include data about two groups. The first consists of high school students 
($n$=1959)  \cite{McCrae2002}.  The age range is from 14 to 18 (M=16.5, S.D.=1.0 years); approximately two-thirds were girls. In Table \ref{tab:Norm}, we denote this group with the abbreviation HS.

The second sample consists of 
adults from the Baltimore Longitudinal Study of Aging (BLSA) \cite{Shock1984}. BLSA participants are generally healthy and well-educated men and women who have volunteered to return
to the Gerontology Research Center for periodic medical and psychosocial testing. The data was collected between 1991
and 2002, $n$=1,492  (695 men and 797 women) aged 19--93 (M=56.2, S.D.=17.0 years); 65.1\% of the sample was White, 27.6\%
Black, and 7.3\% Other Race. In Table \ref{tab:Norm} we refer to this group as BLSA.

The third sample includes 1,025 participants (802 males and 221 females; two individuals did not provide information on their gender) and combines data from several studies published between 1996--2000 \cite{Egan00}. This cohort had a good range of skills, mental ability, and psychopathology, and aimed to be a representative cross-section of British society. In Table \ref{tab:Norm}, we use Brit to label this group.

For HS and BLSA data both  NEO-FFI and NEO-FFI-R profiles are available. For Brit only the NEO-FFI scores are available. For ease of comparison we include the Five Factor scores for our sample and for the subsample of illicit drug users (Table \ref{tab:2}) in Table \ref{tab:Norm}.

\begin{table} 
\centering
\caption{Mean values of Five Factors for the three `normal' samples and for our data.` N-u, Illicit' stands for non-users of illicit drugs with the decade-based definition of users (they either never used illicit drugs or used them more than a decade ago), Samp stands for the total sample, `U, Illicit'  stands for users of illicit drugs for the decade-based definition of users; compare with Table \ref{tab:2}}
\label{tab:Norm}       % Give a unique label
\begin{tabular}{|l|c|c|c|c|c|c|c|c|c|c|c|}\cline{3-12}
 \multicolumn{2}{c|}{} & \multicolumn{2}{c|}{N} &\multicolumn{2}{c|}{E}&\multicolumn{2}{c|}{O}&\multicolumn{2}{c|}{A}&\multicolumn{2}{c|}{C}\\\hline 
Group&Version& Mean & SD & Mean & SD & Mean & SD & Mean & SD& Mean & SD\\\hline
 \multirow{2}{*}{BLSA}& NEO-FFI	&15.77& 7.47&	28.50 & 6.26&29.32& 6.11&33.39 & 4.98&	 33.48 & 6.36 \\ \cline{2-12}
       &	NEO-FFI-R&  16.83& 7.36& 29.29 & 6.46& 31.29& 	6.12&32.41 & 5.42&  33.26  & 6.30 \\\hline
\multirow{2}{*}{HS}&	NEO-FFI&     24.65 & 8.07&  30.58 & 6.67 &28.40 & 6.57&28.31 &6.34&	 27.45 &7.30\\ \cline{2-12}
    & NEO-FFI-R& 25.08& 7.95& 31.80 & 6.94& 31.18& 6.96& 28.09 &6.93& 27.00 & 7.40\\\hline
Brit&NEO-FFI &19.5 &8.6 &27.1 &5.9 &  26.5 &6.5 & 29.7& 5.9 &32.1 &6.6 \\\hline
N-u, Illicit &NEO-FFI-R&21.00& 7.85& 28.52& 5.73& 30.22& 6.06& 32.87& 5.65& 32.89& 5.55\\\hline
Samp & NEO-FFI-R&23.92&9.14&27.58& 6.77&33.76&	6.58	&30.87	&6.44	&29.44	&6.97\\\hline
U, Illicit & NEO-FFI-R& 24.88 & 9.32 & 27.27&7.05&34.93& 	6.32 & 30.21&6.54  &28.30&7.01\\\hline
\end{tabular}
\end{table}

\begin{figure}
\centering
\includegraphics[width=0.5\textwidth]{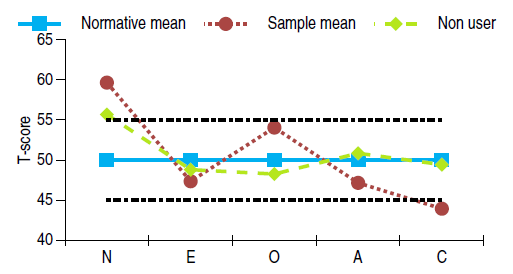}
\caption{{Mean T-score NEO-FFI-R for the total sample and for non-users of illicit drugs with respect to the BLSA mean as a norm}}
\label{MeanTscorefig:2}
\end{figure}

The means of the NEO-FFI-R T-scores based on normative data are depicted in Fig.~\ref{MeanTscorefig:2}. For this example, the `norm' is taken from the BLSA group (NEO-FFI). It is obvious from this figure that Samp is significantly biased when compared
 to the population (represented by the BLSA group). Such a bias is usual for clinical cohorts, for example, the `problematic' or `pathological' groups \cite{Fridberg11}, and the drug users \cite{Terracciano08,Flory02}.  For the group of non-users of illicit drugs the bias is much smaller.

It is important to observe that the mean values of scores for four factors in our sample, N,  E, A, and C, are {\em between} the mean scores for BLSA and HS. The mean O score in our sample is significantly {\em higher}. 

The special role of ``Openness to Experience'' (O) for drug consumption has been observed by many researchers. For example, in  
the paper ``Undergraduate marijuana and drug use as related to openness to experience'' \cite{Grossman1974} we read:  ``Marijuana  use, in the present  sample  of college students, was  associated with  personality  characteristics  which  many  would  tend  to  value 
(e.g.,  creativity).  Open  mindedness  or  `openness  to  experience' may account  for  our  findings.  People  who  are  open  to  new  experience  become  creative,  try  marijuana,  and,  in  general,  experience more  than  people  who  have  a  less  open  life  style.  This  accounts  for  our  finding  that  the  more  a  person  uses  marijuana,  the  more 
likely they  are  to  try  one  or  more  other  drugs.'' Non-users were characterised as ``the 
typical  non-creative, high  authoritarian  individuals.''

The high variability of means in the `normal' groups has  encouraged us to analyse the T-scores with respect to the sample mean and to study the differences between users and non-users rather than deviation from the norm. This analysis is presented in the next two sections.

\section{Deviation of the groups of drug users from the sample mean}
\label{Sec:Use/Mean}

Tables \ref{table4} and \ref{table4b}  demonstrate the mean T-$score_{sample}$ of five NEO-FFI-R factors, supplemented by Imp and SSs for users and non-users, for each drug with respect to the decade-, year-, month-, and week-based classification problems respectively (see Appendix \ref{Appendix 1}). Significant differences in personality factor scores are observed between these groups. The hypothesis about the universal relationship between personality profile and the risk of drug consumption can generally be described as in (\ref{maindrugprofile}): an increase in scores of N, O, Imp, and SS suggest an increase in the risk of use, whereas an increase in the scores of A and C results in a decrease in the risk of use. Thus for each drug, drug users scored higher on N and O, and lower on A and C, when compared to non-users of drugs. The influence of the score of E is drug specific (non-universal).

We now analyse the sign of T-$score_{sample}$ for various drugs and the definitions of users (decade-, year-, month- and week-based). 	We used a + sign for moderately high and high T-$score_{sample}$ (T-$score_{sample}> 55$, and a -- sign for moderately low and low T-$score_{sample}$ (T-$score_{sample}<45$, and 0 for a  score close to the mean value $55\geq$T-$score_{sample}\geq 45$.  These signs reflect the distance from the group of users to the sample mean. 

In the next section we analyse significance of deviations. There is a standard and well-known problem with reporting of the values and significance of deviations: `significant' does not mean `large' and, conversely, apparently large deviations could be insignificant. Everything is defined by interplay between the number of elements in classes  and the deviation value. Practically, significant but small deviations may be unstable: after small change of conditions they may even change their sign. We consider   the variability of the `normal' scores in Section \ref{popnorm}. Insignificant but apparently large deviation may become significant for larger samples or may vanish.  Therefore, it is necessary to answer two questions: how large is the deviation (this section) and how significant  is it (next section)?

For a significant deviation of users from non-users we use the signs $\Uparrow$ and $\Downarrow$.

The inclusion of moderate subcategories of T-$score_{sample}$ as suggested above enables us to separate the drugs into five groups for the decade-based user/non-user separation. These are presented in Table~\ref{tab:5}. Each group can be coded using the (N, E, O, A, C, Imp, SS) profile:
\begin{itemize}
\item The group with the profile $(0, 0, 0, 0, 0, 0, 0)$ includes the users of three legal drugs: alcohol, chocolate and caffeine;
\item The group of drugs with the profile $(0, 0, +, -, -, +, +)$ includes the users of amyl nitrite, LSD, and magic mushrooms;
\item Nicotine users form their own group with the profile $(0, 0, +, 0, -, +, +)$; 
\item The largest group of drugs with the profile $(+, 0, +, -, -, +, +)$ includes the users of amphetamines, benzodiazepines, cannabis, cocaine, ecstasy, ketamine, and legal highs;
\item Finally, the group with the profile $(+, -, +, -, -, +, +)$ includes the users of crack, heroin, VSA and methadone.
\end{itemize}

\begin{table}[!ht]
\centering
\caption{Deviation of T-$score_{sample}$ from the sample mean for various groups of users for the decade-based user/non-user separation}
\label{tab:5}
\begin{tabular}{|l|c|c|c|c|c|c|c|}\hline
{  Drug}&{  N}&{  E}&{  O}&{  A}&{  C }&{  Imp }&{  SS }\\\hline
Alcohol, Chocolate, Caffeine &   $0$   &  $0 $    &  $0$     &  $0$     &  $0$ &  $0$ &  $0$  \\\hline
Amyl nitrite, LSD, and MMushrooms& $0$&	$0$&	$+$&	$-$&	$-$&	$+$&	$+$\\\hline
Nicotine                              &$0$&	$0$&	$+$&	$0$&	$-$&	$+$&	$+$\\\hline
\begin{tabular}{@{}l@{}}Amphetamines, Benzodiazepines, Cannabis,\\
Cocaine, Ecstasy, Ketamine, and Legal highs \end{tabular}&$+$&	$0$&	$+$&$-$&$-$&	$+$&	$+$\\\hline
Crack, Heroin, VSA, and Methadone&$+$&	$-$&	$+$&	$-$&	$-$&	$+$&	$+$\\\hline
Ecstasy pleiad&$0$&	$0$&	$+$&	$-$&	$-$&	$+$&	$+$\\\hline
Heroin pleiad, Benzodiazepines pleiad&$+$&	$0$&	$+$&	$-$&	$-$&	$+$&	$+$\\\hline
Illicit drugs&$+$&	$0$&	$+$&	$-$&	$-$&	$+$&	$+$\\\hline
\end{tabular}
\end{table}

For the year-based user/non-user classification, drugs are separated into eight groups as presented in Table~\ref{tab:5a}. Each group can be coded using the (N, E, O, A, C, Imp, SS) profile:
\begin{itemize}
\item The group with the profile $(0, 0, 0, 0, 0,0,0)$ includes the users of three legal drugs alcohol, chocolate and caffeine;  
\item The group of drugs with the profile $(0, 0, +, -, -,+, +)$ contains just the users of magic mushrooms;
\item The LSD users also form their own group with the profile $(0, 0, +, 0, -, +,+)$;
\item The  group with the profile $(+, 0, +, -, -,+,+)$ includes the users of amphetamines, amyl nitrite, cannabis, cocaine, crack, legal highs and VSA;
\item The group of drugs with the profile $(+, -, +, -, -)$ includes the users of benzodiazepines, heroin, and methadone;
\item The ecstasy users form their own group with the profile $(0, +, +, -, -,+,+)$;
\item The ketamine users form their own group with the profile $(+, +, +, -, -,+,+)$;
\item The nicotine users form their own group with the profile $(+, 0, +, 0, -,+,+)$.
\end{itemize}

Similarly, the deviations of T-$score_{sample}$ from the sample mean for the month-based user/non-user classification and for the week-based user/non-user classification are described in Table~\ref{tab:5b} and Table~\ref{tab:5c} respectively.
 \begin{table}[!ht]
\centering
\caption{Deviation of T-$score_{sample}$ from the sample mean for various groups of users for the year-based user/non-user separation}
\label{tab:5a}
\begin{tabular}{|l|c|c|c|c|c|c|c|}\hline
{  Drug}&{  N}&{  E}&{  O}&{  A}&{  C }&{  Imp }&{  SS }\\\hline
Alcohol, Chocolate, Caffeine &   $0$   &  $0$    &  $0$     &  $0$     &  $0$ &  $0$     &  $0$ \\\hline
MMushrooms &   $0$    &   $0$    &   $+$    &   $-$    &  $-$  &   $+$ &   $+$\\\hline
LSD &    $0$   &   $0$    &$+$   &  $0$     &  $-$    &   $+$ &   $+$\\\hline
\begin{tabular}{@{}l@{}}Amphetamines, Amyl nitrite, Cannabis, \\
Cocaine, Crack, Legal highs, and VSA \end{tabular} &   $+$  & $0$      &  $+$     &  $-$     &  $-$  &   $+$ &   $+$\\\hline
Benzodiazepines, Heroin, and Methadone  &   $+$    &  $-$   &  $+$     &  $-$     & $-$   &  $+$ &   $+$ \\\hline
Ecstasy &   $0$   &    $+$   &   $+$  &  $-$     &  $-$  &   $+$ &   $+$ \\\hline
Ketamine&   $+$  & $+$  & $+$  &  $-$  &$-$  &   $+$ &   $+$\\\hline
Nicotine&   $+$    &   $0$    &   $+$    &  $0$    &  $-$  &   $+$ &   $+$\\\hline
Heroin pleiad, Ecstasy pleiad, Benzodiazepines pleiad&   $+$    &   $0$    &   $+$    &  $-$    &  $-$  &   $+$ &   $+$\\\hline
Illicit drugs&$+$&	$0$&	$+$&	$-$&	$-$&	$+$&	$+$\\\hline
\end{tabular}
\end{table}

\begin{table} 
\centering
\caption{Deviation of T-$score_{sample}$ from the sample mean for various groups of users for the month-based user/non-user separation}
\label{tab:5b}
\begin{tabular}{|l|c|c|c|c|c|c|c|}\hline
{  Drug}&{  N}&{  E}&{  O}&{  A}&{  C }&{  Imp }&{  SS }\\\hline
Alcohol, Chocolate, Caffeine &   $0$   &  $0 $    &  $0$     &  $0$     &  $0$  &  $0$     &  $0$\\\hline
Cannabis and MMushrooms &   $0$    &   $0$    &   $+$    &   $-$    &  $-$  &   $+$ &   $+$\\\hline
Nicotine &    $+$   &   $0$    &$+$       &  $0$     &  $-$  &   $+$ &   $+$\\\hline
Amphetamines, Ketamine, and Legal highs &   $+$  & $0$      &  $+$     &  $-$     & $-$   &   $+$ &   $+$\\\hline
Benzodiazepines, Heroin, and Methadone  &   $+$    &  $-$   &  $+$     &  $-$     & $-$   &   $+$ &   $+$\\\hline
Ecstasy and  LSD&   $0$   &    $+$   &   $+$  &  $-$     &  $-$  &   $+$ &   $+$\\\hline
Cocaine and VSA&   $+$  & $+$  & $+$  &  $-$  &$-$ &   $+$ &   $+$\\\hline
Amyl nitrite &$0$ &$0$ &$0$ &$-$ &$-$ &   $+$ &   $+$\\\hline
Crack&   $+$    &   $-$    &   $0$    &  $-$    &  $-$  &   $+$ &   $+$\\\hline
Ecstasy pleiad&   $0$    &   $0$    &   $+$    &  $-$    &  $-$  &   $+$ &   $+$\\\hline
Heroin pleiad, Benzodiazepines pleiad&   $+$    &   $-$    &   $+$    &  $-$    &  $-$  &   $+$ &   $+$\\\hline
Illicit drugs&$+$&	$0$&	$+$&	$-$&	$-$&	$+$&	$+$\\\hline
\end{tabular}
\end{table}

\begin{table}[!ht]
\centering
\caption{Deviation of T-$score_{sample}$ from the sample mean for various groups of users for the week-based user/non-user separation}
\label{tab:5c}
\begin{tabular}{|l|c|c|c|c|c|c|c|}\hline
{  Drug}&{  N}&{  E}&{  O}&{  A}&{  C }&{  Imp }&{  SS }\\\hline
Alcohol, Chocolate, Caffeine &   $0$   &  $0 $    &  $0$     &  $0$     &  $0$ &  $0$     &  $0$ \\\hline
Cannabis &   $0$    &   $0$    &   $+$    &   $-$    &  $-$  &   $+$ &   $+$\\\hline
LSD and MMushrooms&    $0$   &   $+$    &$+$       &  $0$     &  $-$  &   $+$ &   $+$\\\hline
Ketamine &   $0$  & $-$      &  $+$     &  $-$     & $-$   &   $+$ &   $+$\\\hline
\begin{tabular}{@{}l@{}}Amphetamines, Benzodiazepines, \\ Heroin, Legal highs, and Methadone\end{tabular} & $+$ & $-$ & $+$ & $-$ & $-$  &   $+$ &   $+$ \\\hline
Ecstasy&   $0$   &    $+$   &   $+$  &  $-$     &  $-$  &   $+$ &   $+$\\\hline
VSA&   $0$   &    $+$   &   $+$  &  $-$     &  $0$  &   $+$ &   $+$\\\hline
Cocaine &   $+$  & $+$  & $+$  &  $-$  &$-$ &   $+$ &   $+$\\\hline
Nicotine &    $+$   &   $0$    &$+$       &  $0$     &  $-$  &   $+$ &   $+$\\\hline
Amyl nitrite &$0$ &$-$ &$0$ &$-$ &$-$ &   $+$ &   $+$\\\hline
Crack&   $+$    &   $-$    &   $-$    &  $-$    &  $-$  &   $+$ &   $+$\\\hline
Heroin pleiad, Benzodiazepines pleiad&   $+$    &   $-$    &   $+$    &  $-$    &  $-$  &   $+$ &   $+$\\\hline
Ecstasy pleiad&   $0$    &   $0$    &   $+$    &  $-$    &  $-$  &   $+$ &   $+$\\\hline
Illicit drugs&$+$&	$0$&	$+$&	$-$&	$-$&	$+$&	$+$\\\hline
\end{tabular}
\end{table}

The personality profiles are strongly associated with membership of groups of the users and non-users of the 18 drugs. We found that the N and O score of drug users of all 18 drugs are moderately high $(+)$ or neutral $(0)$, and that the A and C scores of drug users are moderately low $(-)$ or neutral $(0)$.  Detailed  results  can be seen in Tables  \ref{tab:5a}, \ref{tab:5b}, and \ref{tab:5c}.

The effect of the E score is drug specific. Drugs are divided into three groups with respect to the E score of users (in the year-, month-, and week-based classification problems;  see Tables~\ref{tab:5}, \ref{tab:5a}, \ref{tab:5b} and \ref{tab:5c}. For example, for the week-based user/non-user separation the E score is:
\begin{itemize}
\item Moderately low  $(-)$ in groups of users of amphetamines, amyl nitrite,  benzodiazepines, heroin,  ketamine, legal highs,  methadone, and crack;
\item Moderately high $(+)$ in groups of users of cocaine, ecstasy, LSD, magic mushrooms, and VSA;
\item Neutral $(0)$  in groups of users of  alcohol, caffeine, chocolate, cannabis, and nicotine.
\end{itemize}

\section{Significant differences between groups of drug users and non-users}
\label{Sec:Use/NonUse}

Tables~\ref{tab:6}, \ref{tab:6A}, \ref{tab:6B}, and  \ref{tab:6C} (``user minus non-user profiles'') show where there are significant differences between the means of the personality traits for the groups of users and non-users for the decade-, year-, month-, and week-based classification problems respectively. Three significance level are used:
\begin{itemize}
\item 99\% significance level ($p$-value is less than 0.01). Symbol  `$\Downarrow$ ' corresponds to 99\%  significant difference where the mean in users group is less than mean in non-users group and symbol `$\Uparrow$' corresponds to 99\%  significant difference where the mean in users group is greater than the mean in non-users group.
\item 98\% significance level ($p$-value is less than 0.02). Symbol `$\downarrow$ ' corresponds to 98\%  significant difference where the mean in users group is less than mean in non-users group and symbol `$\uparrow$' corresponds to 98\%  significant difference where the mean in users group is greater than the mean in non-users group.
\item 95\% significance level ($p$-value is less than 0.05). Symbol `$\downharpoonleft$ ' corresponds to 95\%  significant difference where the mean in users group is less than mean in non-users group and symbol `$\upharpoonleft$' corresponds to 95\%  significant difference where the mean in users group is greater than the mean in non-users group.
\end{itemize}
Empty cells in the tables below correspond to insignificant differences.

For example for the decade-based user/non-user separation  (see Table~\ref{tab:6}) chocolate does not have a significant difference between users and non-users for any of the factors. Alcohol users and non-users only have a 99\% significant difference in the C , Imp, and SS scores, and 95\% significant difference in the A score. According to Table \ref{tab:5} all these deviations are small.

LSD and magic mushrooms for the decade-based user/non-user separation  (see Table~\ref{tab:6}) have 99\% significant difference between users and non-users in the O, A, C, Imp, and SS scores and 95\% significant difference in the N score. According to  Table \ref{tab:5} both for LSD and magic mushrooms the deviation in the N score is small and all the deviations in  the O, A, C, Imp, and SS scores are not small.

In Table~\ref{tab:6}, benzodiazepines and methadone  have 99\% significant differences between users and non-users in all seven scores. For methadone, all of these differences are not small (Table \ref{tab:5}) and for benzodiazepines the difference in the E score is small inspite of having 99\% significance.

The significance of the differences of the means for groups of users and non-users for the year, month, and week-based user definition is presented in  Tables~\ref{tab:6A}--\ref{tab:6C}. We hope that the previous descriptions of where the significant differences lie are enough for the reader to interpret this table. It is useful to consider significance of differences together with their value (Tables  \ref{tab:5a}-- \ref{tab:5c}). Additional information about these differences could be extracted from the detailed Tables \ref{table4}, \ref{table4b} in the Appendix.

\begin{table}[!ht]
\centering
\caption{Significant differences of means for groups of users and non-users for the decade-based user/non-user separation.} \label{tab:6} \begin{tabular}{|l|c|c|c|c|c|c|c|}\hline
{  Drug} & {  N} & {  E } & {  O} & {  A} & {  C}& {  Imp}& {  SS}\\\hline
Chocolate &  &  &  &  &  &  & \\\hline
Alcohol &  &  &  & $\downharpoonleft$ & $\Downarrow$ & $\Uparrow$ & $\Uparrow$\\\hline
Amyl nitrite &  &  & $\Uparrow$ & $\Downarrow$ & $\Downarrow$ & $\Uparrow$ & $\Uparrow$\\\hline
Caffeine  &  & $\upharpoonleft$ & $\Uparrow$ &  & $\downarrow$ & $\uparrow$ & $\Uparrow$\\\hline
LSD, MMushrooms & $\upharpoonleft$ &  & $\Uparrow$ & $\Downarrow$ & $\Downarrow$ & $\Uparrow$ & $\Uparrow$\\\hline
Amphetamine, Cocaine, Crack, Ecstasy, Ketamine, Legal highs, Nicotine, VSA & $\Uparrow$ &  & $\Uparrow$ & $\Downarrow$ & $\Downarrow$ & $\Uparrow$ & $\Uparrow$\\\hline
Cannabis, Heroin & $\Uparrow$ & $\downharpoonleft$ & $\Uparrow$ & $\Downarrow$ & $\Downarrow$ & $\Uparrow$ & $\Uparrow$\\\hline
Benzodiazepines, Methadone & $\Uparrow$ & $\Downarrow$ & $\Uparrow$ & $\Downarrow$ & $\Downarrow$ & $\Uparrow$ & $\Uparrow$\\\hline
Benzodiazepines pleiad, Heroin pleiad & $\Uparrow$ &  & $\Uparrow$ & $\Downarrow$ & $\Downarrow$ & $\Uparrow$ & $\Uparrow$\\\hline
Ecstasy pleiad & $\Uparrow$ & $\downharpoonleft$ & $\Uparrow$ & $\Downarrow$ & $\Downarrow$ & $\Uparrow$ & $\Uparrow$\\\hline
Illicit drugs & $\Uparrow$ & $\Downarrow$ & $\Uparrow$ & $\Downarrow$ & $\Downarrow$ & $\Uparrow$ & $\Uparrow$\\\hline
\end{tabular}
\end{table}

\begin{table}[!ht]
\centering
\caption{Significant differences of means for groups of users and non-users for the year-based user/non-user separation.} \label{tab:6A} \begin{tabular}{|l|c|c|c|c|c|c|c|}\hline
{  Drug} & {  N} & {  E } & {  O} & {  A} & {  C}& {  Imp}& {  SS}\\\hline
Chocolate &  &  &  &  &  &  & \\\hline
Alcohol &  &  &  &  &  &  & $\Uparrow$\\\hline
Amyl nitrite &  &  & $\uparrow$ & $\Downarrow$ & $\Downarrow$ & $\Uparrow$ & $\Uparrow$\\\hline
LSD &  &  & $\Uparrow$ & $\downharpoonleft$ & $\Downarrow$ & $\Uparrow$ & $\Uparrow$\\\hline
MMushrooms &  &  & $\Uparrow$ & $\Downarrow$ & $\Downarrow$ & $\Uparrow$ & $\Uparrow$\\\hline
Caffeine &  & $\uparrow$ & $\upharpoonleft$ &  & $\downharpoonleft$ & $\Uparrow$ & $\Uparrow$\\\hline
Ecstasy &  & $\Uparrow$ & $\Uparrow$ & $\Downarrow$ & $\Downarrow$ & $\Uparrow$ & $\Uparrow$\\\hline
Ketamine & $\uparrow$ &  & $\Uparrow$ & $\Downarrow$ & $\Downarrow$ & $\Uparrow$ & $\Uparrow$\\\hline
VSA & $\Uparrow$ &  & $\Uparrow$ & $\downarrow$ & $\Downarrow$ & $\Uparrow$ & $\Uparrow$\\\hline
Amphetamine, Cannabis, Crack, Legal highs, Nicotine & $\Uparrow$ &  & $\Uparrow$ & $\Downarrow$ & $\Downarrow$ & $\Uparrow$ & $\Uparrow$\\\hline
Cocaine & $\Uparrow$ & $\upharpoonleft$ & $\Uparrow$ & $\Downarrow$ & $\Downarrow$ & $\Uparrow$ & $\Uparrow$\\\hline
Heroin & $\Uparrow$ & $\downarrow$ & $\Uparrow$ & $\Downarrow$ & $\Downarrow$ & $\Uparrow$ & $\Uparrow$\\\hline
Benzodiazepines, Methadone & $\Uparrow$ & $\Downarrow$ & $\Uparrow$ & $\Downarrow$ & $\Downarrow$ & $\Uparrow$ & $\Uparrow$\\\hline
Heroin pleiad& $\Uparrow$ &  & $\Uparrow$ & $\Downarrow$ & $\Downarrow$ & $\Uparrow$ & $\Uparrow$\\\hline
Benzodiazepines pleiad, Ecstasy pleiad & $\Uparrow$ & $\downharpoonleft$ & $\Uparrow$ & $\Downarrow$ & $\Downarrow$ & $\Uparrow$ & $\Uparrow$\\\hline
Illicit drugs & $\Uparrow$ & $\Downarrow$ & $\Uparrow$ & $\Downarrow$ & $\Downarrow$ & $\Uparrow$ & $\Uparrow$\\\hline
\end{tabular}
\end{table}

\begin{table}[!ht]
\centering
\caption{Significant differences of means for groups of users and non-users for the month-based user/non-user separation.} \label{tab:6B} \begin{tabular}{|l|c|c|c|c|c|c|c|}\hline
{  Drug} & {  N} & {  E } & {  O} & {  A} & {  C}& {  Imp}& {  SS}\\\hline
Chocolate &  &  &  &  &  &  & \\\hline
Amyl nitrite &  &  &  & $\downarrow$ & $\downharpoonleft$ & $\uparrow$ & $\Uparrow$\\\hline
LSD, MMushrooms &  &  & $\Uparrow$ &  & $\Downarrow$ & $\Uparrow$ & $\Uparrow$\\\hline
VSA &  &  & $\Uparrow$ & $\downharpoonleft$ &  & $\Uparrow$ & $\Uparrow$\\\hline
Ketamine &  &  & $\Uparrow$ & $\downharpoonleft$ & $\Downarrow$ & $\Uparrow$ & $\Uparrow$\\\hline
Caffeine &  & $\Uparrow$ &  &  &  & $\Uparrow$ & $\Uparrow$\\\hline
Alcohol &  & $\Uparrow$ & $\upharpoonleft$ &  &  &  & $\Uparrow$\\\hline
Ecstasy &  & $\Uparrow$ & $\Uparrow$ & $\Downarrow$ & $\Downarrow$ & $\Uparrow$ & $\Uparrow$\\\hline
Crack & $\Uparrow$ &  &  & $\downharpoonleft$ & $\downharpoonleft$ & $\uparrow$ & $\Uparrow$\\\hline
Amphetamine, Cannabis, Cocaine, Legal highs, Nicotine & $\Uparrow$ &  & $\Uparrow$ & $\Downarrow$ & $\Downarrow$ & $\Uparrow$ & $\Uparrow$\\\hline
Heroin & $\Uparrow$ & $\downarrow$ &  & $\Downarrow$ & $\Downarrow$ & $\Uparrow$ & $\Uparrow$\\\hline
Benzodiazepines, Methadone & $\Uparrow$ & $\Downarrow$ & $\Uparrow$ & $\Downarrow$ & $\Downarrow$ & $\Uparrow$ & $\Uparrow$\\\hline
Ecstasy pleiad, Heroin pleiad & $\Uparrow$ &  & $\Uparrow$ & $\Downarrow$ & $\Downarrow$ & $\Uparrow$ & $\Uparrow$\\\hline
Benzodiazepines pleiad & $\Uparrow$ & $\Downarrow$ & $\Uparrow$ & $\Downarrow$ & $\Downarrow$ & $\Uparrow$ & $\Uparrow$\\\hline
Illicit drugs & $\Uparrow$ & $\Downarrow$ & $\Uparrow$ & $\Downarrow$ & $\Downarrow$ & $\Uparrow$ & $\Uparrow$\\\hline
\end{tabular}
\end{table}

\begin{figure}
\centering
\includegraphics[width=0.9\textwidth]{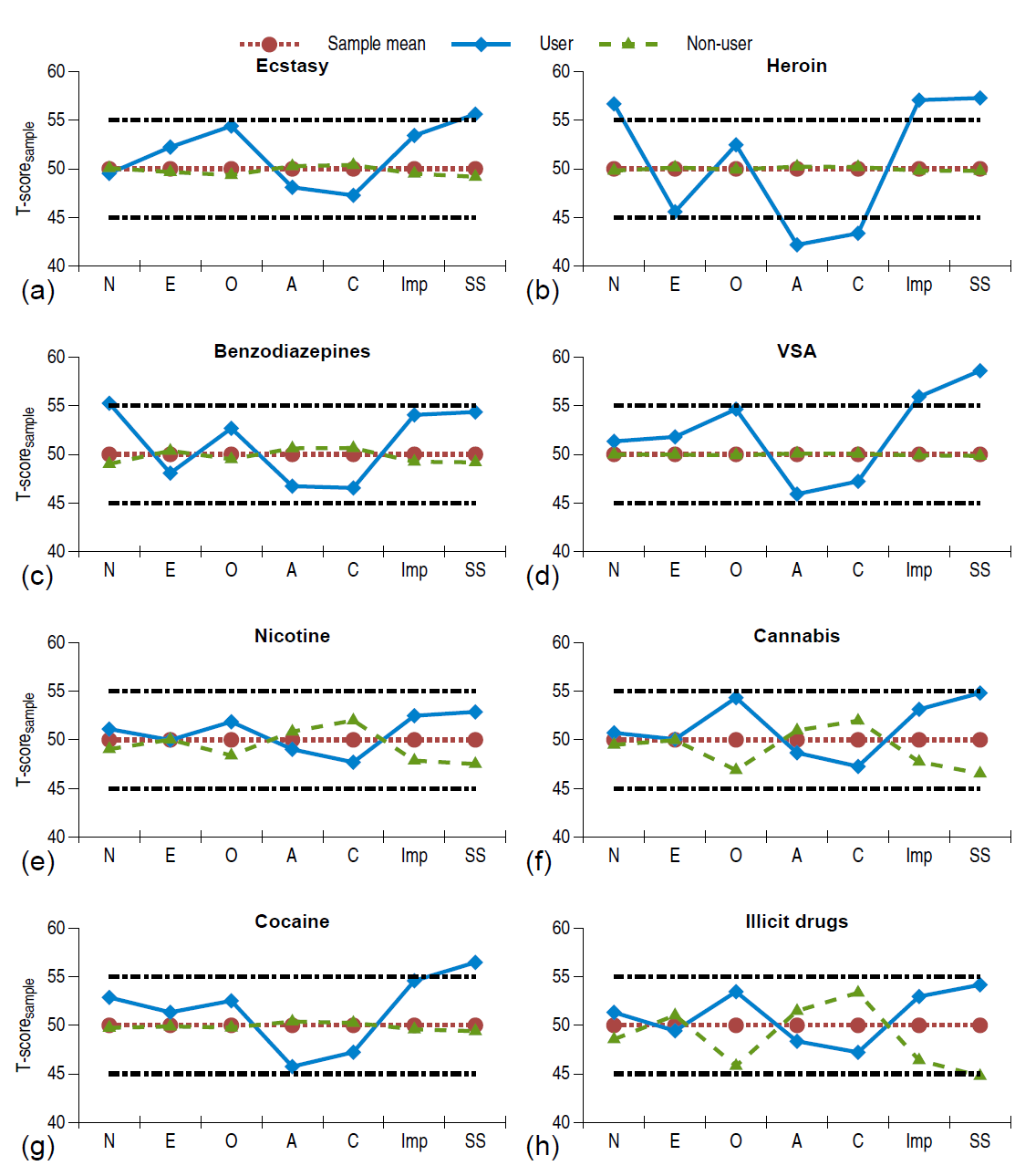}
\caption {{  Average personality profiles for the month-based user/non-user separation.} T-scores with respect to the population norm mean for Ecstasy, Heroin, Benzodiazepines, VSA, Nicotine, Cannabis, Cocaine,  and the whole group of Illicit drugs}
 \label{Averagepersonalityfig:4}
\end{figure}

\begin{table}[!ht]
\centering
\caption{Significant differences of means for groups of users and non-users for the week-based user/non-user separation. } \label{tab:6C} \begin{tabular}{|l|c|c|c|c|c|c|c|}\hline
{  Drug} & {  N} & {  E } & {  O} & {  A} & {  C}& {  Imp}& {  SS}\\\hline
Chocolate &  &  &  &  &  &  & $\downarrow$\\\hline
Crack &  &  &  & $\downharpoonleft$ &  & $\upharpoonleft$ & $\Uparrow$\\\hline
Amyl nitrite &  &  &  & $\downharpoonleft$ & $\downharpoonleft$ & $\upharpoonleft$ & $\Uparrow$\\\hline
Ketamine &  &  & $\uparrow$ & $\downarrow$ & $\downarrow$ & $\uparrow$ & $\Uparrow$\\\hline
VSA &  &  & $\Uparrow$ &  &  & $\Uparrow$ & $\Uparrow$\\\hline
LSD &  &  & $\Uparrow$ &  & $\Downarrow$ & $\upharpoonleft$ & $\Uparrow$\\\hline
Cannabis &  &  & $\Uparrow$ & $\Downarrow$ & $\Downarrow$ & $\Uparrow$ & $\Uparrow$\\\hline
MMushrooms &  & $\upharpoonleft$ & $\Uparrow$ &  &  & $\Uparrow$ & $\Uparrow$\\\hline
Caffeine &  & $\uparrow$ &  &  &  & $\upharpoonleft$ & \\\hline
Ecstasy &  & $\uparrow$ & $\Uparrow$ &  & $\Downarrow$ & $\Uparrow$ & $\Uparrow$\\\hline
Alcohol &  & $\Uparrow$ &  &  &  &  & $\Uparrow$\\\hline
Cocaine & $\uparrow$ &  &  & $\Downarrow$ & $\Downarrow$ & $\Uparrow$ & $\Uparrow$\\\hline
Amphetamine, Nicotine & $\Uparrow$ &  & $\Uparrow$ & $\Downarrow$ & $\Downarrow$ & $\Uparrow$ & $\Uparrow$\\\hline
Heroin & $\Uparrow$ & $\downharpoonleft$ &  & $\Downarrow$ & $\Downarrow$ & $\Uparrow$ & $\Uparrow$\\\hline
Methadone & $\Uparrow$ & $\Downarrow$ & $\upharpoonleft$ & $\Downarrow$ & $\Downarrow$ & $\Uparrow$ & $\Uparrow$\\\hline
Benzodiazepines, Legal highs & $\Uparrow$ & $\Downarrow$ & $\Uparrow$ & $\Downarrow$ & $\Downarrow$ & $\Uparrow$ & $\Uparrow$\\\hline
Ecstasy pleiad & $\Uparrow$ &  & $\Uparrow$ & $\Downarrow$ & $\Downarrow$ & $\Uparrow$ & $\Uparrow$\\\hline
Benzodiazepines pleiad, Heroin pleiad& $\Uparrow$ & $\Downarrow$ & $\Uparrow$ & $\Downarrow$ & $\Downarrow$ & $\Uparrow$ & $\Uparrow$\\\hline
Illicit drugs & $\Uparrow$ & $\Downarrow$ & $\Uparrow$ & $\Downarrow$ & $\Downarrow$ & $\Uparrow$ & $\Uparrow$\\\hline
\end{tabular}
\end{table}
Mean values of Five Factor scores for groups of drug users and non-users for the decade-based user/non-user separation  are depicted in Fig.~\ref{Averagepersonalityfig:4}. Some similarities and differences between users of different drugs are obvious from this figure: compare, for example, profiles of ecstasy users and heroin users, which are very different to profiles for nicotine and cannabis users, which are qualitatively similar (but with larger deviations of cannabis users profiles from the sample mean). Mean values of all seven factor scores for groups of drug users and non-users for all definitions of  users are presented in Tables~\ref{table4} and \ref{table4b}.

\section{Correlation between usage of different drugs \label{Sec:Corr}}

Usage of each drug is a binary variable (users or non-users) for all versions of user definition. Tables~\ref{tab:14} and \ref{tab:14a} contain PCCs, which are computed for each pair of the 153 (=18 times 17 divided by 2) potential drug usages for the decade- and year-based user/non-user separations respectively (see Appendix \ref{Appendix 2}). The majority of the PCCs are significant, since the sample size is 1,885.

The correlation in 124 pairs of drug usages from a totality of 153 pairs have, for the decade-based  classification problem, \emph{p}-values less than 0.01 (\emph{p}-value is the probability of observing by chance the same or greater correlation coefficient for uncorrelated variables). It is necessary to employ a {\em multi-testing} approach when testing 153 pairs of drug usages in order to estimate the significance of the correlation \cite{Benjamini95}. We apply the most conservative technique, the Bonferroni correction, and  used the Benjamini-Hochberg (BH) step-up procedure \cite{Benjamini95} to control the False Discovery Rate (FDR) in order to estimate the genuine significance of these correlations. There are 115 significant  correlation coefficients with Bonferroni corrected \emph{p}-value 0.001. The BH step-up procedure with threshold of FDR equal to 0.01 defines 127 significant correlation coefficients.

\begin{figure}
   \centering
   \includegraphics[width=0.9\textwidth]{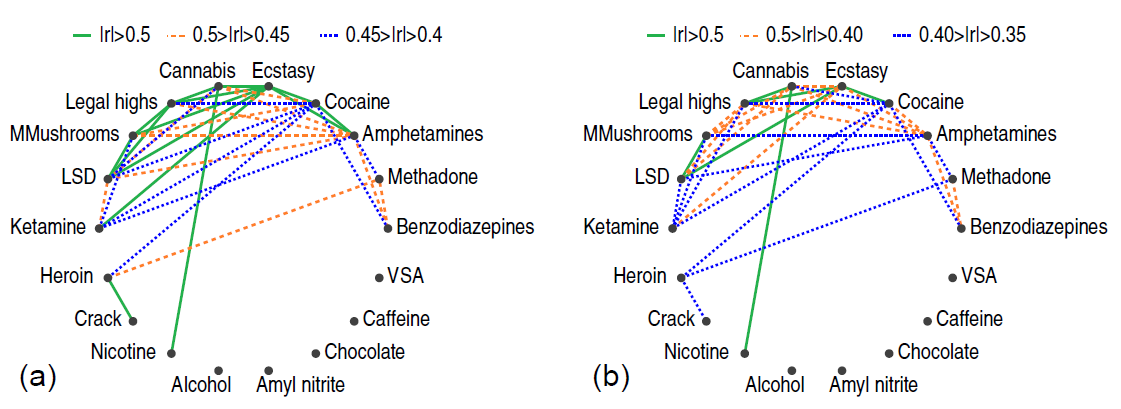}
   \caption {{Strong drug usage correlations:}(a) for the decade-based classification  problem,  and (b)  for the year-based classification problem}
   \label{Strongdrugusfig:5}
\end{figure}

However, a significant correlation does not necessarily imply a strong association or causality. For example, the correlation coefficient for alcohol usage and amyl nitrate usage is significant (i.e. the $p$-value is equal to 0.0013) but the value of this coefficient is equal to 0.074, and thus cannot be considered as an important association. We consider correlations with absolute values of PCC  $|r|\ge 0.4$. Fig.~\ref{Strongdrugusfig:5} sets out all significant identified correlations greater than 0.4. In this study, for the decade-based classification problem, we consider the correlation as weak if $|r|<0.4$, medium if $0.45>|r|\ge 0.4$, strong if $0.5>|r|\ge 0.45$, and very strong if $|r|\ge 0.5$.

The correlation coefficient is high for each pair  from the group: amphetamines, cannabis, cocaine, ecstasy, ketamine, legal highs, LSD, and magic mushrooms, excluding correlations between cannabis and ketamine usage (r=0.302) and between legal highs and ketamine usage (r=0.393) (Fig.~\ref{Strongdrugusfig:5}a).
Crack, benzodiazepines, heroin, methadone, and nicotine usages are correlated with one, two, or three other drugs usage (see Fig.~\ref{Strongdrugusfig:5}a). Amyl nitrite, chocolate, caffeine and VSA usage are uncorrelated or weakly correlated with usage of any other drug.

The structure of correlations of the year-based user/non-user separation is approximately the same as for the decade-based classification problem (see Fig.~\ref{Strongdrugusfig:5}).  We consider correlations with absolute values of PCC  $|r|\ge 0.35$ for the year-based classification. Fig.~\ref{Strongdrugusfig:5}B  sets out all identified significant correlations with $|r|>0.35$. The correlation can be interpreted as weak if $|r|<0.35$; medium if   $0.40>|r|\ge 0.35$; strong if $0.5>|r|\ge 0.40$; and very strong if $|r|\ge 0.5$. On base of this similarity of correlation structures we define pleiades for three central drugs: heroin, ecstasy, and benzodiazepines (as described in the Section~\ref{Pleiad of drug users}).

\begin{figure}
 \centering
\includegraphics[width=0.9\textwidth]{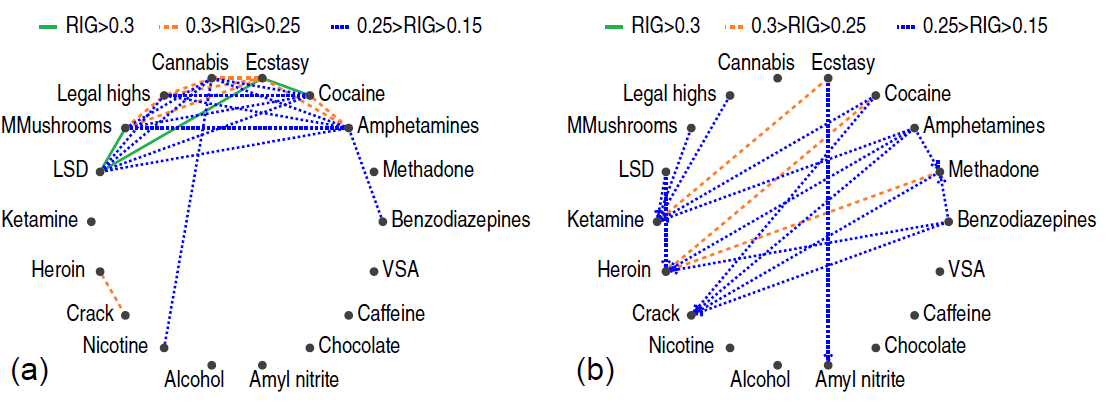}
\caption {{   Pairs of drug usages with high RIG}: ((a) approximately symmetric RIG and (b) significantly asymmetric RIG. In Figure (b), the arrow from cocaine usage to heroin usage, for example, means that knowledge of cocaine usage can decrease uncertainty in knowledge about heroin usage}
\label{RIGfig:6}
\end{figure}

{\em Relative Information Gain} (RIG) is widely used in data mining to measure the dependence between categorical attributes (\ref{RIG}). For example, the value of RIG  for Drug 1 usage from usage of Drug 2 is equal to a fraction of uncertainty (entropy) in drug 1 usage, which can be removed if the value of drug 2 usage is known. The significance of RIG for binary random variables is the same as for PCC. The majority of RIGs are statistically significant, but have small values. Fig.~\ref{RIGfig:6} presents all pairs with RIG \textgreater 0.15.

Fig.~\ref{RIGfig:6}a shows `approximately symmetric' RIGs. Here, we call $\mathrm{RIG}(X|Y)$ approximately symmetric if
$$ \frac{| \mathrm{RIG}(X|Y)-\mathrm{RIG}(Y|X)|}{\min \left(\mathrm{RIG}(X|Y), \mathrm{RIG}(Y|X)\right)}<0.2.$$  RIG is approximately symmetric  for each pair from the following group: amphetamines, cannabis, cocaine, ecstasy, legal highs, LSD and magic mushrooms. This group is the same as in Fig.~\ref{Strongdrugusfig:5} (except  ketamine).  Fig.~\ref{RIGfig:6}b shows asymmetric RIGs. Asymmetric RIGs illustrate a markedly different pattern to that of Fig.~\ref{Strongdrugusfig:5}.

\section{Ranking of input features}

It should be stressed that the five factors (FFM), Imp, and SS are all correlated.
To identify the most informative features we applied the methods which are described in  Section \ref{Input feature ranking}. The results of the principal variables calculation are given in Table~\ref{tab:8} for CatPCA quantification, and in Table~\ref{tab:9} for the dummy coding of nominal features. Tables \ref{tab:8} and \ref{tab:9} contain lists of attributes in order from best to worst. The results of the Double Kaiser ranking are shown in the same tables.

\begin{table}[!ht]
\centering
\caption{The results of feature ranking. Data include country of residence and ethnicity quantified by CatPCA. FVE is the fraction of explained variance. Gndr stays for gender. CFVE is the cumulative FVE. The least informative features are located towards the bottom of the table.}
\label{tab:8}
\begin{tabular}{|l|c|c||l|}\hline
\multicolumn{3}{|c||}{{  Principal variable ranking}}& {   Double Kaiser's ranking}\\\cline{1-3}
\multicolumn{1}{|c|} {  Attribute}&	{  FVE}&	{  CFVE} &\\\hline
SS&	0.192&	0.192&	E\\\hline
N&	    0.153&	0.345&	C \\\hline
A&  	0.106&	0.451&	SS\\\hline
Edu&	        0.104&	0.555&	N \\\hline
O&	        0.092&	0.647&	Imp \\\hline
C&	0.088&	0.735&	O \\\hline
E&       0.076&	0.811&	A  \\\hline
Age&	            0.073&	0.884&	Age\\\hline
Imp &	    0.055&	0.939&	Edu \\\hline
Country&	        0.037	&0.976	&Country\\\hline
Gndr	&           0.021&	0.997&	Gndr\\\hline
Ethnicity&	        0.003&	1.000&	Ethnicity\\\hline
\end{tabular}
\end{table}

\begin{table}[!ht]
\centering
\caption{The results of feature ranking. Data include dummy coded country of residence and ethnicity. FVE is the fraction of explained variance. CFVE is the cumulative FVE. The least informative features are lower located. }
\label{tab:9}
\begin{tabular}{|l|c|c||l|}\hline
\multicolumn{3}{|c||}{ {   Principal variable ranking}}&{   Double Kaiser's ranking}\\\cline{1-3}
\multicolumn{1}{|c|}{  Attribute}&	{  FVE}&	{  CFVE}& \\\hline
SS&	0.186&	0.186&	E \\\hline
N&	    0.149&	0.335&	C \\\hline
A &	    0.103&	0.438&	SS\\\hline
Edu &	        0.101&	0.539&	N \\\hline
O &	        0.089&	0.627&	Imp \\\hline
C &	0.086&	0.714&	O \\\hline
E &	    0.074&	0.787&	A \\\hline
Age&	            0.071&	0.858&	Age\\\hline
Imp &	    0.053&	0.911&	Edu \\\hline
UK&	                0.027&	0.938&	UK\\\hline
Gndr&         	0.020&	0.959&	USA\\\hline
USA	&               0.013&	0.972&	Gndr\\\hline
White&           	0.010&	0.982&	Other (country)\\\hline
Other (country)&	0.005&	0.988&	White\\\hline
Canada&	            0.004&	0.991&	Other (ethnicity)\\\hline
Other (ethnicity)&	0.003&	0.994&	Canada\\\hline
Black&	            0.002&	0.995&	Asian\\\hline
Australia&	        0.002&	0.997&	Mixed-White/Black\\\hline
Asian&          	0.001&	0.998&	Australia\\\hline
Mixed-White\/Black&	0.001&	0.999&	Black\\\hline
Republic of Ireland&0.000&	1.000&	Mixed-White/Asian\\\hline
Mixed-White\/Asian&	0.000&	1.000&	Republic of Ireland\\\hline
New Zealand&	    0.000&	1.000&	New Zealand\\\hline
Mixed-Black\/Asian&	0.000&	1.000&	Mixed-Black/Asian\\\hline
\end{tabular}
\end{table}

The results of application of sparse PCA are shown in Tables~\ref{tab:10} and \ref{tab:11}. As a result of feature selection we can exclude ethnicity from further consideration. There is a more intriguing effect regarding country of location. Only two countries are informative (in our sample): the UK and the USA. Furthermore, inclusion of country in the personality measures does not add much to the  prediction of drug usage. To understand the reasons for these two countries' importance in the prediction of drug consumption we compare the statistics for the subsamples: UK - non-UK and USA - non-USA. We calculated the $p$-value for coincidence of distribution of personality measurements in each subsample.  We obtained the same results for both divisions into subsamples: all input features have significantly different distributions with a 99.9\% confidence level for UK and non-UK subsamples and likewise for USA -- non-USA subsamples. This means that the UK and non-UK samples are biased, and similarly for the USA and non-USA samples.

Our goal is to predict the risk of drug consumption for an individual. This means that we have to consider individual specific factors. Occupation within a specific country can be thought of as an important risk factor, but we do not have enough data for  countries other than the UK and the USA because of the  composition of the dataset: participants from the UK (1,044; 55.4\%), the USA (557; 29.5\%), Canada (87; 4.6\%), Australia (54; 2.9\%), New Zealand (5; 0.3\%) and Ireland (20; 1.1\%). A total of 118 (6.3\%) came from a diversity of other countries, none of whom individually formed as much as 1\% of the sample, or did not declare the country of location. Thus we exclude the `country' feature from further study. As a result, we continue with the 10 input features: Age, Edu, N, E, O, A, C, Imp, SS, and Gndr.

\begin{table}[!ht]
\centering
\caption{The result of sparse PCA feature ranking. Data include country of residence and ethnicity quantified by CatPCA. }
\label{tab:10}
\begin{tabular}{|c|c|l|}\hline
{   Step}&	{  \# of components}&	{   Removed attributes}\\\hline
1&	5&	Gndr and Ethnicity\\\hline
2&	4&	\begin{tabular}{@{}c@{}}No removed attributes. The retained set of attributes: Age, Edu, \\
N, E, O, A, C, Imp, SS, and country \end{tabular}\\\hline
\end{tabular}
\end{table}

\begin{table}[!ht]
\centering
\caption{ The result of sparse PCA feature ranking. Data include dummy coded country of residence and ethnicity.}
\label{tab:11}
\begin{tabular}{|c|c|l|}\hline
{   Step}&	{   \# of components}& {   Removed attributes}\\\hline
1&	8&	\begin{tabular}{@{}c@{}} Canada, Other (country), Australia, Republic of Ireland, \\
New Zealand, Mixed-White/Asian, White, Other (ethnicity), \\ Mixed-White/Black, Asian, 
Black and Mixed-Black/Asian \end{tabular} \\\hline
2&	5&Gndrr, UK and USA\\\hline
3&	4&\begin{tabular}{@{}c@{}}No removed attributes. The retained set of attributes: Age,\\ Edu, N, E, O,
A, C, Imp, and SS \end{tabular}\\\hline
\end{tabular}
\end{table}

\section{Selection of the best classifiers for the decade-based classification problem \label{Sec:BestClass}}

The first step for risk evaluation is the construction of classifiers. We have tested the eight methods described in the `Risk evaluation methods' Section and selected the best one. The results of classifier selection are presented in Table~\ref{tab:12}. This table shows that for all drugs except alcohol, cocaine and magic mushrooms, the sensitivity and specificity are greater than 70\%, which  is an unexpectedly high accuracy.

Recall that we have 10 input features: Age, Edu, N, E, O, A, C, Imp, SS, and Gndr; each of which is an important predictor for at least five drugs. However, there is no single most effective classifier which uses all input features. The maximal number of attributes used is 6 out of 10 and the minimal number is 2. In Section~\ref{CriterionOfTheBestMethod} the best method is defined as 	the method which maximises the value of the minimum of sensitivity and specificity. If the minimum of sensitivity and specificity is the same for several classifiers then the classifier with the maximal sum of the sensitivity  and specificity is selected from these.
Table~\ref{tab:12} shows the different sets of attributes which are used in the best user/non-user classifier for each different drug.

%\begin{landscape}
\begin{table}
\centering
\caption {The best results of the drug users classifiers (decade-based definition of users). Symbol `X' means the used input feature. Results are calculated by LOOCV.}
\label{tab:12}
\begin{tabular}{|l|c|c|c|c|c|c|c|c|c|c|c|c|c|c|} \hline
{   Target feature} &{    Classifier} &{   Age}&{   Edu}&{   N} &{   E} &{   O} &{   A} &{    C} &{   Imp} &{   SS }&{   Gndr}&{    Sn} &{   Sp}&{   Sum}  \\
               &    &   & &  &  & & & & &  &&{   (\%)}& {   (\%)}&{   (\%)} \\   \hline
Alcohol        &LDA	& X&X &X & & & & &  &  X   & X &  75.34  & 63.24 &138.58 \\\hline
Amphetamines   &DT  & X&  &	X& &X& &X & X & X&   & 	81.30  &71.48  &152.77  \\\hline
Amyl nitrite   &DT	&	& &X &	& X& &X  & &X  &  &  73.51 &	87.86 &	161.37 \\\hline
Benzodiazepines& DT&X&	&X&	X&	&	&	&	X&	X&	X&	70.87&	71.51&	142.38\\ \hline
Cannabis       &DT&X&	X&	&&	X&	X&	X&	X&&	&79.29&	80.00&	159.29\\ \hline
Chocolate      &$k$NN&	X&	&	&	X&	&	&	X&&	&X&	72.43&	71.43&	143.86 \\ \hline
Cocaine	       &DT&X&	&&	&X&	X&	&	X&	X&	&	68.27&	83.06&	151.32\\ \hline
Caffeine       &$k$NN&X & X&	&&	X&	X&	&	X&	&	&	70.51&	72.97&	143.48\\ \hline
Crack	       &DT&	&&	&	X&	&	&	X&	&	&	&	80.63&	78.57&	159.20 \\ \hline
Ecstasy        &DT&	X&	&&	&&	&	&	&	X&	X&	76.17&	77.16&	153.33\\ \hline
Heroin	       &DT&X&	&	&&	&&	&	X&	&	X&	82.55&	72.98&	155.53 \\ \hline
Ketamine       &DT&	X&	&&	X&	&	X&	&	X&	X&	&	72.29&	80.98&	153.26 \\ \hline
Legal highs    &DT&	X&	&&	&	X&	X&	X&	&	X&	X&	79.53&	82.37&	161.90 \\ \hline
LSD	           &DT&X&	&X	&X	&X	&	&	&X	&	&X	&85.46&	77.56&	163.02 \\ \hline
Methadone      &DT&	X&	X&	&	X&	X&	&&	&&	X&	79.14&	72.48&	151.62\\ \hline
MMushrooms     &DT	&	&	&	&X	&&	&&	&	&X	&65.56&	94.79&	160.36 \\ \hline
Nicotine       &DT	&	&&X	&X	&	&	&X	&	&	&X	&71.28&	79.07&	150.35 \\ \hline
VSA            &DT&X&	X&	&	X&	&	X&	X&	&	X&	&	83.48&	77.64&	161.12 \\
\hline
\end{tabular}
\end{table}
%\end{landscape}

The use of a feature in the best classifier can be interpreted as `ranking by fact'. We note that this ranking by fact is very different from the other rankings presented in Tables~\ref{tab:8} and \ref{tab:10}. For example, in Tables~\ref{tab:8} and \ref{tab:10} we see that Age is not one  of the most informative measures, but it is used in the best classifiers for 14 of the drugs. The second most used input feature is Gndr, which is regarded as non-informative by Sparse PCA (Table~\ref{tab:10}) and as one of the least informative by other methods (Table~\ref{tab:8}).
This means that consumption of these 10 drugs is Gndr dependent. 

\begin{figure}
\centering
\includegraphics[width=0.72\textwidth]{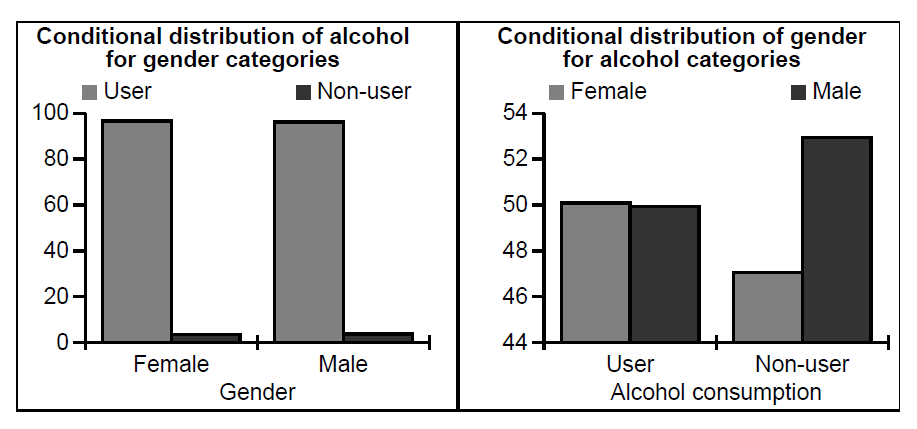}
\caption {{   Conditional distribution for Gndr and alcohol}}
\label{ConditionaldisAlcoholgenderfig:9}
\end{figure}

\begin{figure}
\centering
\includegraphics[width=0.72\textwidth]{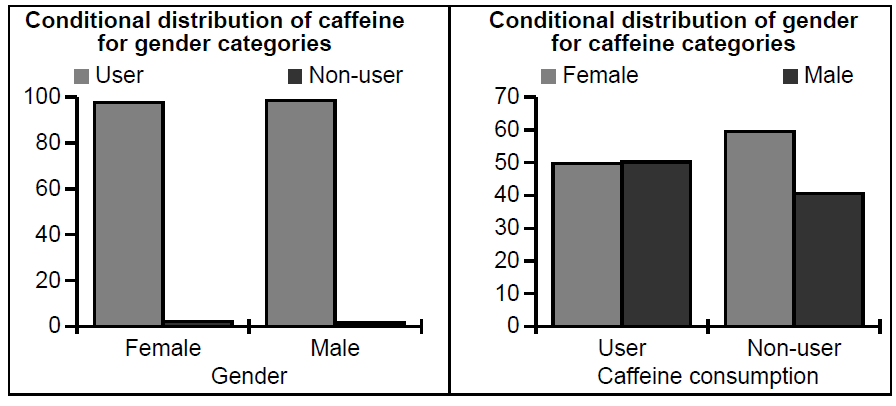}
\caption {{   Conditional distribution for Gndr and caffeine}}
\label{ConditionaldisCaffaingenderfig:10}
\end{figure}

\begin{figure}
\centering
\includegraphics[width=0.72\textwidth]{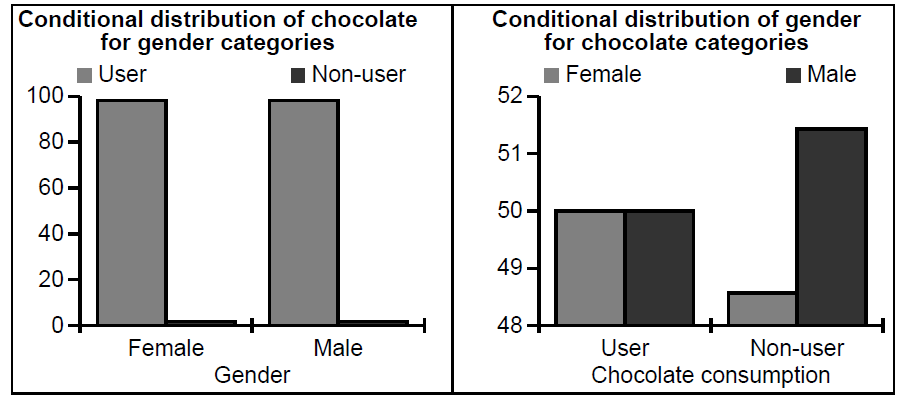}
\caption {{   Conditional distribution for Gndr and chocolate}}
\label{ConditionaldisChocolategenderfig:11}
\end{figure}

\begin{figure}
\centering
\includegraphics[width=0.72\textwidth]{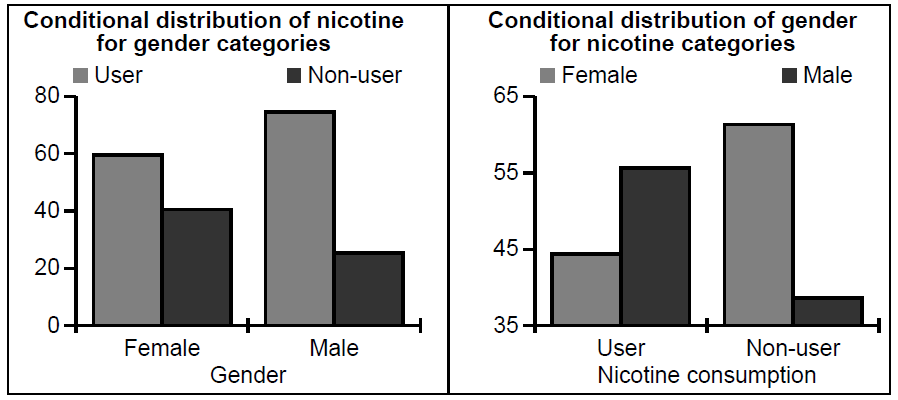}
\caption {{   Conditional distribution for Gndr and nicotine}}
\label{ConditionaldisNicotinegenderfig:19}
\end{figure}

We have found some unexpected outcomes. For example, in the dataset the fraction of females who are alcohol users is greater than that fraction of males (Fig.~\ref{ConditionaldisAlcoholgenderfig:9}) but a greater proportion of males consume caffeine drinks (for example, coffee)  (Fig.~\ref{ConditionaldisCaffaingenderfig:10}). The fraction of males who do not eat chocolate is greater than for females (Fig.~\ref{ConditionaldisChocolategenderfig:11}). The conditional distributions for nicotine show the fraction of males who smoke is higher (Fig.~\ref{ConditionaldisNicotinegenderfig:19}).

%%%%%%%%%%%%%%%%%%%%%%%%%%%%%

The next most informative input features are E and SS which are used in the best classifiers for nine drugs. Features O, C, and Imp are used in the best classifiers for eight drugs. Features N and A are used in the best classifiers for six drugs. Thus, personality factors are associated with drug use and each one impacts on specific drugs. Finally, Edu  is used in the best classifiers for five drugs (see Table~\ref{tab:12}).

To predict the usage of most drugs DT is the best classifier (see Table~\ref{tab:12}). LDA is the best classifier for alcohol use with five input features, and has sensitivity 75.34\% and specificity 63.24\%. $k$NN is the best classifier for chocolate and caffeine users. These $k$NN classifiers use four features for chocolate and five features for caffeine.

The drugs can be separated into disjoint groups by the number of attributes used for the best classifiers:
\begin{itemize}
  \item The group of classifiers with two input features contains classifiers for two drugs: crack and magic mushrooms. Both classifiers for this group use the E score.
  \item The group of classifiers with three input features includes classifiers for two drugs: ecstasy and heroin. Both classifiers in this group use Age  and Gndr and do not use any NEO-FFI factors.
  \item The group of classifiers with four input features includes classifiers for three drugs: amyl nitrite, chocolate, and nicotine. All classifiers for this group use the C score.
  \item The group of classifiers with five input features includes classifiers for five drugs: alcohol, cocaine, caffeine, ketamine, and methadone. All classifiers for this group use Age.
  \item The group of classifiers with six input features includes classifiers for six drug users: amphetamines, benzodiazepines, cannabis, legal highs, LSD, and VSA. All classifiers for this group use Age.
\end{itemize}

It is important to stress that the {\em attributes which are not used in the best classifiers are not non-informative}. For example, for ecstasy consumption the best classifier is based on Age, SS, and Gndr and has sensitivity 76.17\% and specificity 77.16\%. There exists a DT for usage of the same drug based on Age, Edu, O, C, and SS, with sensitivity 77.23\% and specificity 75.22\%; a DT based on Age, Edu, E, O, and A, with sensitivity 73.24\% and specificity 78.22\%; a LR classifier based on Age, Edu, O, C, Imp, SS, and Gndr, with sensitivity 74.83\% and specificity 74.52\%;  a $k$NN classifier based on Age, Edu, N, E, O, C, Imp, SS, and Gndr, with sensitivity 75.63\% and specificity 75.75\%. This means that for the  risk evaluation of ecstasy usage all input attributes are informative but the required information can be extracted from a smaller subset of the attributes.

\begin{figure}
\centering
\includegraphics [width=0.53\textwidth]{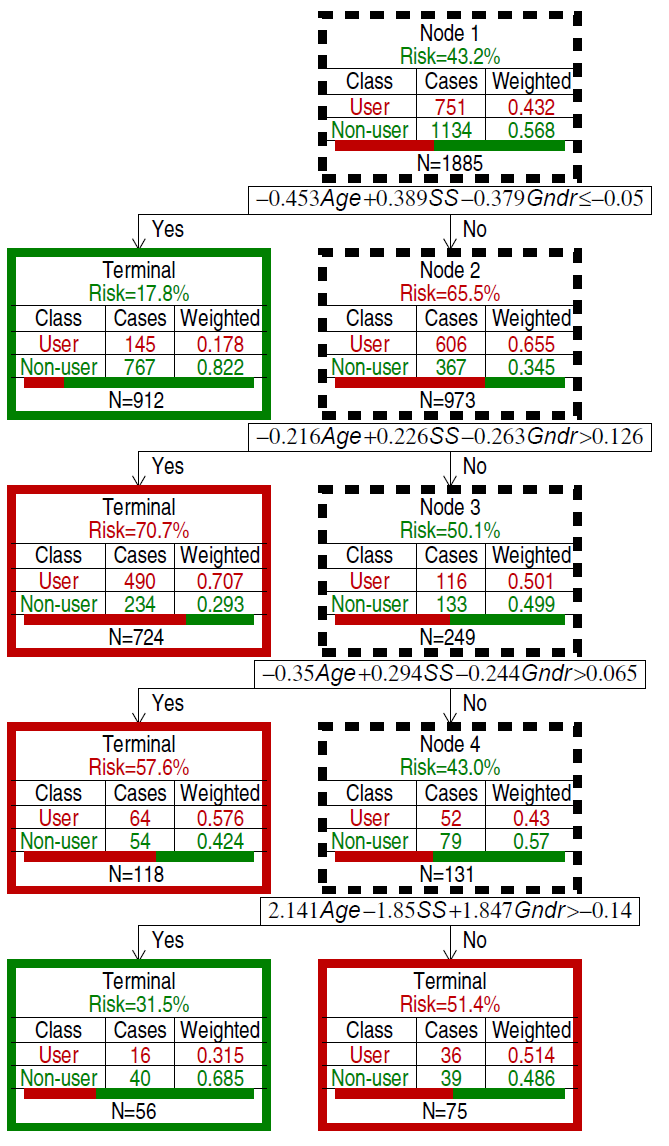}
\caption {{   Decision tree for ecstasy}. Input features are: Age, SS, and Gndr. Non-terminal nodes are depicted with dashed border. Values of Age, SS, and Gndr are calculated by quantification procedures described in Section~\ref{Quantification}. The weight of each case of users class is 1.15 and of non-users class is 1. In column `Weighted' normalizing weights are presented (the weight of each class is divided by sum of weights)}
\label{Treefig7}
\end{figure}

The results presented in Table~\ref{tab:12} were calculated by LOOCV. It should be stressed that different methods of testing give rise to different values for sensitivity and specificity. Common methods include calculation of test set errors (the holdout method), $k$-fold cross-validation, testing on the entire sample (if it is sufficiently large, the so-called `na{\"i}ve' method), random sampling, and many others.
For example, a DT formed for the entire sample can have a sensitivity and specificity different from LOOCV \cite{Hastie09}. For illustration, consider the DT for ecstasy, depicted in  Fig.~\ref{Treefig7}. It has sensitivity 78.56\% and specificity 71.16\%, calculated using the whole sample. The results of LOOCV for a tree with the same choices are given in Table~\ref{tab:12}: sensitivity 76.17\% and specificity 77.16\%.

The role of SS is very important for most of the party drugs. In particular, the risk of ecstasy consumption can be evaluated with high accuracy on the basis of Age, Gndr and SS (see Table~\ref{tab:12}, Fig.~\ref{Treefig7}, and \ref{RiskMapsfig:8}), and does not need the personality traits from the FFM.

\section{Correlation pleiades of drugs}
\label{Pleiad of drug users}

Consider correlations between drug usage for  the year- and decade-based definitions  (Fig.~\ref{Strongdrugusfig:5}). It can be seen from Fig.~\ref{Strongdrugusfig:5} that the structure of these correlations for the year- and decade-based definitions of drug users is approximately the same.
We have found three groups of strongly correlated drugs, each containing several drugs which are pairwise strongly correlated. This means that drug consumption has a `modular structure', and we have identified three modules:
\begin{itemize}
  \item  Crack, cocaine, methadone, and heroin;
  \item Amphetamines, cannabis, cocaine, ketamine, LSD, magic mushrooms, legal highs, and ecstasy;
  \item Methadone, amphetamines, cocaine and benzodiazepines.
\end{itemize}
This modular structure has a clear representation in the correlation graph, Fig.~\ref{Strongdrugusfig:5}.

The idea of merging correlated attributes into `modules' is popular in biology These modules are called  {\it correlation pleiades} \cite{Terentjev31,Berg60,Mitteroecker07}.
This concept was introduced in biostatistics  in 1931 \cite{Terentjev31}.  Correlation pleiades were used  in evolutionary physiology for the identification of modular structures in a variety of contexts~\cite{Terentjev31,Mitteroecker07,Berg60,Armbruster99}.  Berg \cite{Berg60} presented correlation data from three unspecialized and three specialized pollination species, and proposed that correlation pleiades are clusters of correlated traits.  This means that, in the standard approach to clustering, the pleiades do not intersect. The classical clustering methods are referred to as `hard' or `crisp' clustering, meaning that each data object is assigned to only one cluster.
This restriction is relaxed for fuzzy \cite{Bezdek1981} and probabilistic clustering \cite{Krishnapuram1993}. Such approaches are useful when the clusters are not well separated.

In our study, correlation pleiades are appropriate since the drugs can be grouped in clusters with highly correlated use (see Fig.~\ref{Strongdrugusfig:5}a and b):
\begin{itemize}
\item The \emph{Heroin pleiad (heroinPl)} includes crack, cocaine, methadone, and heroin;
\item The \emph{Ecstasy pleiad (ecstasyPl)} includes amphetamines, cannabis, cocaine, ketamine, LSD, magic mushrooms, legal highs, and ecstasy;
\item The \emph{Benzodiazepines pleiad (benzoPl)} includes methadone, amphetamines, cocaine, and benzodiazepines.
\end{itemize}

\begin{figure}
\centering
\includegraphics[width=0.6\textwidth]{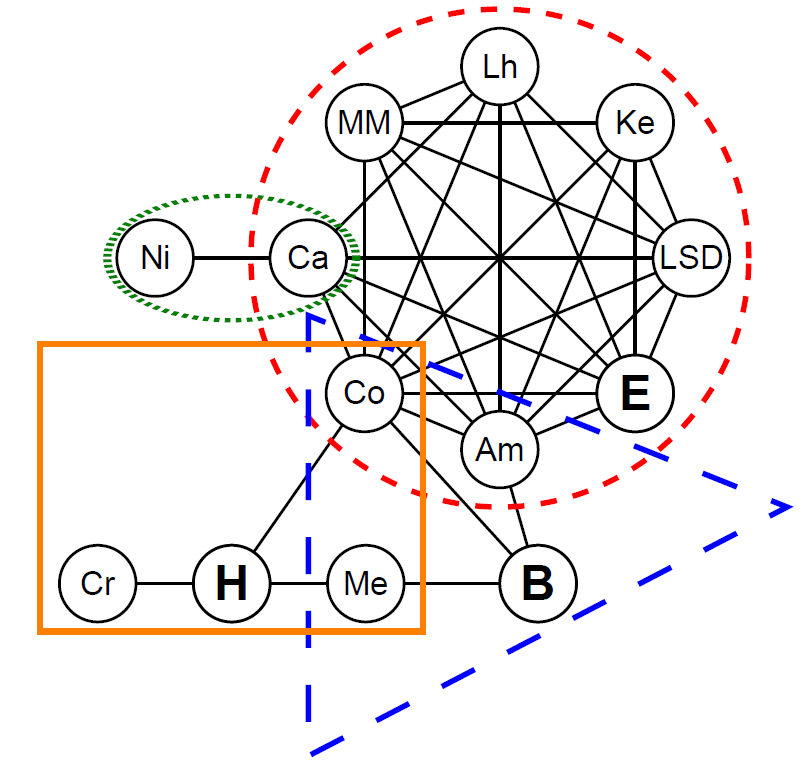}
\caption{{ Correlation pleiades for drug use (in a circle, in a triangle and in a rectangle). Additionally, a highly correlated `smoking couple', cannabis and nicotine, is separated by an ellipse. 
E stands for ecstasy, H for heroin, B for benzodiazepines, and MM for magic mushrooms. Other drugs are denoted by the first two letters of their names. Edges represent correlations}}
\label{Fig:Pleiades}
\end{figure}

These correlation pleiades include 12 drugs (Fig. \ref{Fig:Pleiades}). Additionally, we can consider the `smoking couple', the highly correlated pair cannabis--nicotine. Other drugs do not have strong symmetric correlations. There exists an asymmetric correlation link from ecstasy to amyl nitrite (Fig. \ref{RIGfig:6}). Therefore,  amyl nitrite can be considered as a peripheral element of the ecstasy pleiad.

Fuzzy and probabilistic clustering may help to reveal more sophisticated relationships between objects and clusters. For example, analysis of the intersections between correlation pleiades of drugs can generate important questions and hypotheses:
\begin{itemize}
\item Which patterns of behaviour are reflected by the existence of pleiades? (For example,  is the ecstasyPl just the group of party drugs united by habits of use?)
\item Why is cocaine  a peripheral member of all pleiades?
\item Why does methadone belong  to the  periphery of  both the heroin and benzodiazepines pleiades?
\item Why do amphetamines belong  to the  periphery of both  the ecstasy and benzodiazepines pleiades?
\item Do these intersections reflect the structure of individual drug consumption or the structure of the groups of drug consumers?
\end{itemize}

We have defined groups of users and non-users for each pleiad.
 {\em A group of users for a pleiad includes the users of any individual drug from the pleiad} (see Table~\ref{tab:13a}). A group of non-users contains all participants which are not included in the group of users. Table~\ref{tab:13a} shows the total number of users and their percentages in the database for three pleiades and for each user definition of (the decade-, year-, month-, or week-based user/non-user separation).

The class imbalance problem is well known \cite{Hastie09}. Users form a  small fraction of the dataset (significantly less than half) for most drugs (see Table~\ref{tab:1a}). The classes of users and non-users are more balanced  for pleiades of drugs  than for individual drugs (compare Tables~\ref{tab:13a} and \ref{tab:1a}).
Table~\ref{tab:13a} shows that the number of drug users in the database for all three pleiades  are more balanced (closer to 50\%) than the number of users of the corresponding individual drug  (Table~\ref{tab:1a}). For example, for the year-based classification problem the number of benzoPl users is  830 (44.03\%), while the number of benzodiazepine users is 535 (28.38\%) and the number of heroinPl users is 585 (31.03\%), while the number of heroin users is 118 (6.26\%).

\begin{table}[!ht]
\centering
\caption{ Number of drug users for pleiades in the database}
\label{tab:13a}
\begin{tabular}{|l|c|c|c|c|}\hline
{  Pleiad} 	& \multicolumn{4}{c|}{ {   User definition based on}}\\\cline{2-5}
	           & {  Decade} &	{  Year}&	{   Month}&	{  Week} \\\hline
HeroinPl&	832 (44.14\%) &	585 (31.03\%) &	309 (16.39\%)&	184 (9.76\%)\\\hline
EcstasyPl& 	1317 (69.87\%)&	1089 (57.77\%)&	921 (48.86\%)&	792 (42.02\%)\\\hline
BenzoPl&	1089 (57.77\%)&	830 (44.03\%) &	528 (28.01\%)&	363 (19.26\%)\\\hline
\end{tabular}
\end{table}

%%%%% Significant differences of means for groups of users and non-users for the decade, year month week%%
\begin{table}[!ht]
\centering
\caption{Statistically significant differences of means for groups of users and non-users for each 
pleiad for decade- year-, month-, and week-based classification problem. The symbol  `$\Downarrow$ 'corresponds to a significant 
difference where the mean in the users group is less than the mean in the non-users group, and the symbol  `$\Uparrow$' corresponds to a significant difference where the mean in the users group is greater than the mean in the non-users group. Empty cells corresponds to insignificant differences. The difference is considered to be significant if the p-value is less than 0.01)} \label{tab:13ab}

\begin{tabular}{|l|c|c|c|c|c|}\hline
{  Pleiades od drugs} & {  N} & {  E } & {  O} & {  A} & {  C}\\\hline
\multicolumn{6}{|c|}{ {The decade-based user/non-user separation}}\\\hline
HeroinPl, EcstasyPl, BenzoPl & $\Uparrow$ &  & $\Uparrow$ & $\Downarrow$ & $\Downarrow$\\\hline
\multicolumn{6}{|c|}{ {The year-based user/non-user separation}}\\\hline
HeroinPl, EcstasyPl, BenzoPl & $\Uparrow$ &  & $\Uparrow$ & $\Downarrow$ & $\Downarrow$\\\hline
\multicolumn{6}{|c|}{ {The month-based user/non-user separation}}\\\hline
HeroinPl, EcstasyPl & $\Uparrow$ &  & $\Uparrow$ & $\Downarrow$ & $\Downarrow$\\\hline
BenzoPl             & $\Uparrow$ & $\Downarrow$ & $\Uparrow$ & $\Downarrow$ & $\Downarrow$\\\hline
\multicolumn{6}{|c|}{ {The week-based user/non-user separation}}\\\hline
HeroinPl, BenzoPl             & $\Uparrow$ & $\Downarrow$ & $\Uparrow$ & $\Downarrow$ & $\Downarrow$\\\hline
EcstasyPl                     & $\Uparrow$ &  & $\Uparrow$ & $\Downarrow$ & $\Downarrow$\\\hline
\end{tabular}
\end{table}

%%%%% Significant differences of means for groups of users and non-users for the decade, year month week%%
The introduction of moderate subcategories of T-$score_{sample}$ for pleiades of drugs enables us to  separate the pleiades of drugs into two groups for the decade-, month-, and week-based user/non-user separation. For year-based user/non-user separation there is only one group with profile $(+, 0, +, -, -)$, and includes the users of heroinPl, ecstasyPl and benzoPl.

For the decade-based classification problem, the group with the profile $(+, 0, +, -, -)$  includes the users of heroinPl and benzoPl. The group with the profile $(0, 0, +, -, -)$  includes the users of EcstasyPl. 

For the month- and week-based classification problem, the group with the profile $(+, -, +, -, -)$  includes the users of heroinPl and benzoPl. The group with the profile $(0, 0, +, -, -)$  includes the users of EcstasyPl.

\begin{figure}
\centering
\includegraphics[width=0.84\textwidth]{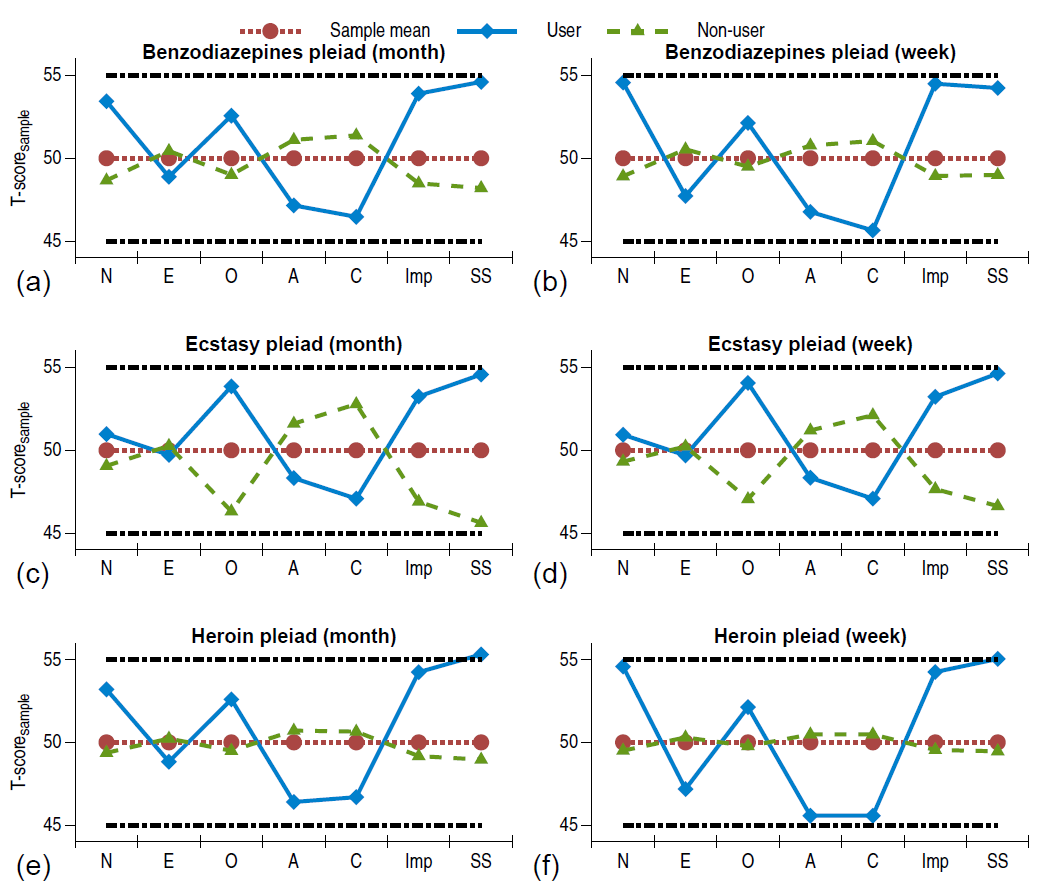}
\caption {{  Average personality profiles for drug pleiades for the month- and week-based user/non-user separation} (T-score$_{sample}$  with respect to the sample means)}
 \label{PersProfPleiades}
\end{figure}

The personality profiles for pleiades of drugs are qualitatively similar but some differences should be mentioned: the N level for EcstasyPl users is lower than for HeroinPl users, whereas levels of E and A 
are higher  for EcstasyPl users (see Fig.~\ref{PersProfPleiades}).

%%%%%%%%%%%%%%%%%%%%%%%%%%%%%%%%%%%%%%%%%%%%%

We have applied the eight methods described in Section `Risk evaluation methods' and selected the best one for each  pleiad for the decade-, year-, month-, and week-based classification problems. The results of   classifier selection are depicted in Table~\ref{tab:12a}. The quality of classification is high.

The classification results are excellent for each pleiad for the decade-, year-, month, and week-based problems. We can compare the classifiers for one pleiad and for different problems (see Table~\ref{tab:12a}). For example:
\begin{itemize}
\item The best classifier for ecstasyPl  for the year-based user/non-user separation is a DT with seven attributes and has sensitivity 80.65\% and specificity 80.72\%;
\item The best classifier for heroinPl  for the month-based user/non-user separation is a DT with five attributes and has sensitivity 74.18\% and specificity 74.11\%;
\item The best classifier for benzoPl for the week-based user/non-user separation is a DT with five attributes and has sensitivity 75.10\% and specificity 75.76\%.
\end{itemize}

%\begin{landscape}
\begin{table}
\centering
\caption {The best results for classifiers of the pleiad users. Symbol `X' means input feature used in the best classifier. Sensitivity (Sn) and Specificity (Sp) were calculated by LOOCV.}
\label{tab:12a}
\begin{tabular}{|l|c|c|c|c|c|c|c|c|c|c|c|c|c|c|c|} \hline
{   Pleiades } &{    Classifier} &{   Age}&{   Edu}&{   N} &{   E} &{   O} &{   A} &{    C} &{   Imp} &{   SS }&{   Gndr}&\#&{    Sn} &{   Sp}&{   Sum}  \\
{  of drugs} &&&&&&&&&&&&&{   (\%)}& {   (\%)}&{   (\%)} \\   \hline

\multicolumn{16}{|c|}{ {The decade-based user/non-user separation}}\\\hline
HeroinPl         &	DT&	 &	 	& &	X&	X&	 &	X&	 &	 &	X&	4&	71.23&	78.85&	150.07\\\hline
EcstasyPl        &  DT&	X&	 &	X&	X&	X&	 &	 &	 &	X&	X&	6&	80.63&	79.80&	160.44\\\hline
BenzoPl          &	DT&	 &	X&	 &	X&	X&	X&	 &	 &	X&	 &	5&	73.37&	72.45&	145.82\\\hline

\multicolumn{16}{|c|}{ {The year-based user/non user-separation}}\\\hline
HeroinPl           &DT&	 &	X&	X&	 &	 &	X&	 &	 &	 &	X&	4&	73.69&	71.80&	145.49 \\\hline
EcstasyPl          &DT&	X&	X&	 &	 &	X&	 &	X&	X&	X&	X&	7&	80.65&	80.72&	161.37 \\\hline
BenzoPl	           &DT&	 &	X&	X&	 &	 &	 &	X&	 &	X&	 &	4&	73.93&	73.98&	147.91 \\\hline									
\multicolumn{16}{|c|}{ {The month-based user/non user-separation}}\\\hline
HeroinPl&	DT  &	X&	 &	 &	X&	X&	 &	X&	 &	X&	 &	5&	74.18&	74.11&	148.29\\\hline
EcstasyPl&	PDFE&	X&	X&	 &	X&	X&	 &	X&	X&	X&	X&	8&	79.34&	79.50&	158.83\\\hline
BenzoPl	&	DT	&   X&	X&	 &	 &	 &	X&	X&	 &	 &	 &	4&	73.18&	73.11&	146.28\\\hline												
\multicolumn{16}{|c|}{ {The week-based user/non user-separation}}\\\hline
HeroinPl&	DT&	X&	X&	X&	X&	X&	 &	 &	X&	X&	X&	8&	75.84&	73.91&	149.75\\\hline
EcstasyPl&	LR&	X&	X&	 &	X&	X&	X&	X&	 &	X&	X&	8&	77.68&	77.78&	155.45\\\hline
BenzoPl&	DT&	 &	X&	 &	 &	 &	X&	X&	 &	X&	X&	5&	75.10&	75.76&	150.86\\\hline
\end{tabular}
\end{table}
%\end{landscape}

Comparison of Tables~\ref{tab:12} and~\ref{tab:12a} shows that the best classifiers for the ecstasy and benzodiazepines pleiades are more accurate than the best classifiers for the consumption of the `central' drugs of the pleiades, ecstasy and benzodiazepines respectively, even for the decade-based user definition.  Classifiers for heroinPl may have slightly worse accuracy  but these classifiers are more robust because they solve classification problems which have more balanced classes. All other classifiers for pleiades of drugs are more robust too for the same reasons, for all pleiades and definitions of users.

Tables~\ref{tab:12} and~\ref{tab:12a} for the decade-based user definition show that most of the classifiers for pleiades use more input features than the classifiers for individual drugs.
We can see from these tables that the accuracies of the classifiers for pleiades and for individual drugs do not differ drastically,  but the use of a greater number of input features suggests more robust classifiers.

It is important to remark that  pleiades are usually assumed to be disjoint. We have considered pleiades which are named by the central drug and the peripheral drugs can be shared. For example, the heroin and ecstasy pleiades have cocaine as an intersection. This approach corresponds to the concept of `soft clustering'.

\section{Overoptimism problem}
\label{Overoptimism}

The best machine learning methods  selected  give impressive solutions for the user/non-user classification problems. Nevertheless, the procedure of selection has used the same data as the training process: we have tested each method by LOOCV. Such an approach could produce  so-called {\em overoptimism}: the cross-validation errors of the best method on the same set, which was used for the method selection, may be underestimated. 

To demonstrate that the data of Tables~\ref{tab:12} and~\ref{tab:12a} are valid for generalisation errors and samples we have never seen before, we may need additional validation on a large hold out sample, which was not compromised by its usage in the method selection. We do not have an  additional large sample and splitting the existing sample into a training set (for training and cross-validation in method selection) and a validation set (for validation of the best method) will decrease the {\em statistic power} of analysis \cite{Ellis2010}.

Following \cite{Hoadley2001}, high performance on the test sample does not
guarantee high performance on future samples, things
do change and there is always a chance that a variable
and its relationships will be different in  future samples. Selection of the best models and best sets of `dominant variables' can damage robustness of the model  to   future variations. 

The idea of stability of the model can significantly help in the testiong of classifiers  \cite{Blockeel2002}. In the process of cross-validation we can test additionally stability of the model and answer the questions:
\begin{itemize}
\item How many examples change their class in cross-validation (we can calculate the number for each transition between classes: class A$\to$ class B, etc.). This is {\em classification stability}.
\item How many qualitatively different models (for example, decision trees with different structure) were generated in cross-validation. This is {\em structural stability}.
\end{itemize}
We can also extract the set of examples with unstable classification and study this set separately.

Hand \cite{Hand2006} clearly demonstrated that  `simple methods typically
yield performance almost as good as more sophisticated methods, to the
extent that the difference in performance may be swamped by other sources
of uncertainty that generally are not considered in the classical supervised
classification paradigm.' 

Therefore, let us consider the results of the best methods (Tables~\ref{tab:12} and~\ref{tab:12a}) as an upper border of the possible classifier performance and apply the simple linear discriminant method. This method is robust and leaves no space for overoptimism if the samples are sufficiently large and there is no multicollinearity. We have also analysed the classification stability of the linear discriminant in cross-validation.

Multicollinearity means strong linear dependence between input attributes. This makes the model very sensitive to fluctuations in data is an important source of instability of classifiers. The standard measure of multicollinearity  is the condition number of the correlation matrix, that is the ratio $\kappa=\lambda_{\max}/\lambda_{\min}$, where $\lambda_{\max}$ and $\lambda_{\min}$ are the maximal and the minimal eigenvalues of this matrix. Collinearity with $\kappa<10$ is considerated as `modest' collinearity  \cite{Dormann 2013}.

Eigenvalues of the correlation matrix between the seven psychological traits are:
 $$\lambda_i = 2.267, 1.809, 0.887, 0.678, 0.548, 0.468, 0.342; \; \kappa=\lambda_{\max}/\lambda_{\min}=6.628.$$
  Eigenvalues of the correlation matrix between the ten attributes (quantified) 
including the seven psychological traits, Age, Edu and Gndr are:
 $$\lambda_i =  2.595, 1.867, 1.111, 0.980, 0.814, 0.757, 0.599, 0.524, 0.427, 0.327; \; \kappa=\lambda_{\max}/\lambda_{\min}= 7.945.$$
 
 We can see that there is no strong multicollinearity despite the existence of significant and not small correlations between psychological traits (see Tables \ref{tab:3} and \ref{tab:01}). Three correlation coefficients, between the N and E scores, between the O and SS scores and between the Imp and SS scores exceed 0.4 in absolute value. Absolute values of correlation coefficients above 0.4 are sometimes interpreted as indicating a multicollinearity problem. This heuristic rule is not rigorous but existence of such correlations rises a suspicion of multicollinearity and it is necessary to apply a stronger test. We have calculated the   condition number and it is sufficiently low to exclude strong multicollinearity. In the next section we have demonstrated that Fisher's linear discriminant is sufficiently robust and works stably for these values of $\kappa$ in our database.

\section{User/non-user classification by linear discriminant for  ecstasy and heroin \label{Sec:Ecs/Her}}

Fisher \cite{Fisher1936} defined the linear discriminant (LD) as a linear function of attributes, for which the ratio of the difference between classes to the standard deviation within classes is maximal (see, for example, the score \ref{score}). We have used the classical formula for the LD direction (\ref{FisherLD}). The selection threshold (intercept) $\Theta$ is defined by the balance condition Sn=Sp.

Linear discriminants separate users from non-users  by linear inequalities: 
\begin{equation}\label{LDu/nu}
D(z)=\Theta+\sum c_i z_i > 0
\end{equation}
for users and $\leq 0$ for non-users, where $\Theta$ are the thresholds (intercepts), $z_i$ are the attributes, and $c_i$ are the coefficients.

Tables \ref{TabDiscr1}-\ref{TabDiscr4} in the Appendix contain the coefficients $c_i$ of the linear discriminants for user/nonuser separation in 10-dimensional space (7 psychological attributes, Age, Edu, and Gndr). The attributes in these tables are quantified and transformed to $z$-scores with zero mean and unit  variance (positive values of the Gndr $z$-score corresponds to female). The last rows of the tables include the standard deviation of the coefficients in LOOCV. For the 7-dimensional space of psychological attributes (T-scores), the coefficients of linear discriminants are presented in Tables \ref{TabDiscr5}-\ref{TabDiscr8}.

Performance of linear discriminants in user/non-user separation is evaluated by several methods (Tables \ref{Tab:DiscrPerf1}--\ref{Tab:DiscrPerf4} for the 10-dimensional data space and Tables \ref{Tab:DiscrPerf5}--\ref{Tab:DiscrPerf8} for the 7-dimensional space of T-scores of psychological attributes). First of all, we have calculated the linear discriminant using the whole sample (see Tables \ref{TabDiscr1}--\ref{TabDiscr4}) and find all of their errors. For each solution of the classification problem we have several numbers, $P$ (positive), the number of samples recognised as positive, and  $N$ (negative), the number of samples recognised as negative. $P+N$ is the total number of samples. $P$=TP+FP (True Positive plus False Positive) and $N$=TN+FN (True Negative plus False Negative). Sensitivity is Sn=TP/(TP+FN)$\times$100\% and Specificity is Sp=TN/(TN+FP)$\times$100\%. Accuracy is Acc=(TP+TN)/($P$+$N$)$\times$100\%.  

We have calculated these performance indicators for the total sample and for the LOOCV procedure. In LOOCV the linear discriminant is calculated for the set of all samples excluding the example left out for testing. The test was performed for all samples with the corresponding redefining of Sn, Sp, and Acc. In LOOCV the linear discriminants are calculated for each test example. Each of these discriminants is a separate classification model. Stability of classification can be measured by the number of examples which change their class at least once. We have taken the basis model for the total sample and found how many TP examples of this model became  FN examples of a LOOCV model at least once. This number measured in \% of  TP+FP of the basic model is TP$\to$FN. Analogously, we have defined FP$\to$TN,  TN$\to$FP, and FN$\to$TP. The last two numbers are measured in \% of  TN+FN of the basic model.

In this section we have analysed performance of linear discriminants for two drugs, ecstasy and heroin.
They are typical (and central) elements of two pleiades, groups of drugs with correlated drug usage. The differences between them might tell us a story about different types of drug users. 

\begin{table}[!ht]\centering
\caption{Coefficients of the linear discriminant for the ecstasy user/non-user separation  and decade-, year-, month-, and week-based definition of users (10 attributes) \label{TabDiscrEcstasy}}
\begin{tabular}{|l|c|c|c|c|c|c|c|c|c|c|c|}\hline
Period&$\Theta$&Age&Edu&N&E&O&A&C&Imp&SS&Gndr\\\hline
Decade&-0.171&-0.631&-0.188&0.053&0.018&0.351&-0.065&-0.210&-0.088&0.559&-0.265\\\hline
Year&-0.464&-0.782&-0.101&-0.015&0.099&0.238&-0.025&-0.173&-0.004&0.453&-0.275\\ \hline
Month&-0.633&-0.820&0.047&-0.139&0.093&0.284&-0.123&-0.165&-0.028&0.328&-0.257\\\hline
Week&-0.779&-0.697&0.077&-0.115&0.161&0.545&-0.093&-0.252&0.217&0.230&-0.022\\\hline \hline
SD&$\leq 0.018$&$\leq 0.002$&$\leq 0.003$&$\leq 0.004$&$\leq 0.004$&$\leq 0.003$&$\leq 0.003$&$\leq 0.003$&$\leq 0.004$&$\leq 0.005$&$\leq 0.003$\\\hline
\end{tabular}
\end{table}

\begin{table}[!ht]\centering
\caption{Performance and stability of the linear discriminant for the  ecstasy user/non-user separation  and  decade-, year-, month-, and week-based definition of users  (10 attributes). All indicators are in \%. \label{Tab:EcstPerf1}}
\begin{tabular}{|l|c|c|c|c|c|c|c|c|c|c|}\hline
&\multicolumn{3}{c|}{Total sample}&\multicolumn{3}{c|}{LOOCV}&\multicolumn{4}{c|}{Stability indicators}\\\cline{2-11}
Period& Sn &Sp&Acc&Sn &Sp& Acc &TP$\to$FN&FN$\to$TP&FP$\to$TN&TN$\to$FP\\ \hline \hline
Decade& 74.4&74.7&74.6&74.2&74.3&74.2&0.3&0.7&0.5&0.4\\\hline
Year& 75.4&75.7&75.6&75.0&75.6&75.4&0.2&0.6&0.8&0.3\\\hline
Month& 72.5&72.4&72.4&71.3&72.3&72.1&1.7&1.3&1.2&1.2\\\hline
Week&72.6&71.5&71.5&67.9&71.4&71.2&3.6&8.3&2.4&5.3\\\hline
\end{tabular}
\end{table}

\begin{table}[!ht]\centering
\caption{Coefficients of the linear discriminant for the heroin user/non-user separation  and decade-, year-, month-, and week-based definition of users (10 attributes) \label{TabDiscrHeroin}}
\begin{tabular}{|l|c|c|c|c|c|c|c|c|c|c|c|}\hline
Period&$\Theta$&Age&Edu&N&E&O&A&C&Imp&SS&Gndr\\\hline
Decade&-0.615&-0.210&-0.370&0.413&-0.211&0.477&-0.265&-0.029&0.222&0.381&-0.332\\\hline
Year&-0.849&-0.584&-0.168&0.352&-0.252&0.222&-0.275&0.014&0.216&0.359&-0.378\\ \hline
Month&-1.037&-0.560&-0.371&0.181&-0.350&0.159&-0.397&0.016&0.368&0.154&-0.226\\\hline
Week&-1.096&-0.386&-0.077&0.467&-0.255&0.184&-0.412&0.013&0.437&-0.077&-0.400\\\hline \hline
SD&$\leq 0.005$&$\leq 0.004$&$\leq 0.004$&$\leq 0.004$&$\leq 0.004$&$\leq 0.004$&$\leq 0.004$&$\leq 0.004$&$\leq 0.005$&$\leq 0.006$&$\leq 0.003$\\\hline
\end{tabular}
\end{table}

\begin{table}[!ht]\centering
\caption{Performance and stability of the linear discriminant for the heroin user/non-user separation  and decade-, year-, month-, and week-based definition of users  (10 attributes). All indicators are in \%. \label{Tab:HerPerf1}}
\begin{tabular}{|l|c|c|c|c|c|c|c|c|c|c|}\hline
&\multicolumn{3}{c|}{Total sample}&\multicolumn{3}{c|}{LOOCV}&\multicolumn{4}{c|}{Stability indicators}\\\cline{2-11}
Period& Sn &Sp&Acc&Sn &Sp& Acc &TP$\to$FN&FN$\to$TP&FP$\to$TN&TN$\to$FP\\ \hline \hline
Decade&70.8&69.9&70.0&68.4&69.8&69.7&2.8&6.1&1.3&2.2\\\hline
Year&73.7&73.7&73.7&70.3&73.6&73.4&2.5&2.5&2.1&2.3\\\hline 
Month&79.2&77.6&77.7&69.8&77.5&77.2&9.4&7.5&3.2&3.8\\\hline
Week&79.3&80.1&80.1&65.5&80.0&79.8&6.9&6.9&4.6&4.8\\\hline
\end{tabular}
\end{table}

Coefficients of LD can be used for indication of how a change in an attribute value will affect the value of $D(z)$ under the condition that the values of all other attributes do not change. It is possible to change one attribute without changing other attributes because there is no strong multicollinearity. The most interesting in this ranking for the Tables \ref{TabDiscrEcstasy} and \ref{TabDiscrHeroin} is the essential difference between ranking for ecstasy and for heroin user/non-user discriminants.

Comparison of Tables \ref{TabDiscrEcstasy} and \ref{TabDiscrHeroin} immediately gives a  result. The coefficients of LD for ecstasy and heroin have significant differences: for ecstasy, the effect of Imp is less than for heroin (and can even have different sign), whereas the effect of SS is bigger for ecstasy. For ecstasy, the coefficients of A have smaller values than the coefficients of C. For heroin the situation is opposite: C has much smaller coefficient than A (and can even have the opposite sign). Also, coefficients of N for ecstasy are smaller than for heroin and can have different sign. Edu has negative coefficients  for heroin with bigger values (it `prevents' usage of heroin), whereas for ecstasy the influence of Edu is smaller and can have opposite signs (for week- and month-based definition of usage). Coefficients of E are positive for the ecstasy user/non-user separation and negative for the heroin user/non-user separation. Age has large negative coefficients both for ecstasy and heroin but for ecstasy they are 1.5--2 times bigger. For example, for the month-based definition of users the ranking of variables is:

\begin{itemize}
\item For ecstasy: Age$\Downarrow$ , SS$\Uparrow$, O$\Uparrow$, Gndr$\Downarrow$ , C$\Downarrow$, N$\Downarrow$, A$\Downarrow$,  E$\Uparrow$, Edu$\Uparrow$, Imp$\Downarrow$;
\item For heroin:  Age$\Downarrow$, A$\Downarrow$, Edu$\Downarrow$, Imp$\Uparrow$, E$\Downarrow$, Gndr$\Downarrow$ , N$\Uparrow$, O$\Uparrow$, SS$\Uparrow$, C$\Uparrow$.
\end{itemize}
The arrows $\Uparrow$  $\Downarrow$ here indicates the sign of the effect of the attribute in the user/non-user separation (and not the shift of the mean as it was in previous sections): for positive coefficients it is $\Uparrow$ and for negative coefficients it is  $\Downarrow$. The difference between heroin and ecstasy discriminants is impressive. The most important five attributes for ecstasy user/non-user discrimination have only one attribute in common with the top five attributes of heroin user/non-user discrimination (Age). We invite the reader to pay attention to the different signs of coefficients of some of the  attributes for the ecstasy and heroin linear   user/non-user discrimination in this case: C, N, E, and Imp. For most of these attributes, the traditional expectation is well-known: C$\Downarrow$, N$\Uparrow$, Imp$\Uparrow$ (at least, for illegal drugs). The values of the coefficients with unexpected signs are relatively small, the attributes with these values are ranked as less important, but the expectation is not met in any case: we can state that it is a wrong assumption that high N and Imp are predictors for ecstasy use, and it is also wrong that low C is a predictor for heroin use.

If we do not use Age, Edu, and Gndr, then the difference of the predictors persists (Tables \ref{TabEcst7} and \ref{TabHer7}): for ecstasy LD the most important attribute becomes SS, then O and C. The attributes Imp, A, N, and E have smaller coefficients, and for Imp, E, and N the sign of the coefficients depends on recency of usage.  For heroin C seems to be less important than other attributes.

The ranking of these seven psychological attributes for the same month-based user/non-user discrimination looks similar:
\begin{itemize}
\item For ecstasy: SS$\Uparrow$, O$\Uparrow$, C$\Downarrow$, N$\Downarrow$, A$\Downarrow$, Imp$\Downarrow$,  E$\Uparrow$;
\item For heroin:  SS$\Uparrow$, A$\Downarrow$ , E$\Downarrow$ ,  Imp$\Uparrow$,  N$\Uparrow$, O$\Uparrow$,  C$\Downarrow$ .
\end{itemize}
The reader should note the opposite signs of N, E, and Imp for the ecstasy and heroin user/non-user discriminants in this case. The only big jump from the 10-attribute ranking is the change of rank of SS for heroin user/non-user discrimination. We believe that this is because of large negative correlations between SS and Age, which is important for classification but not available in the seven-attribute model. Again, the upper four  variables for ecstasy user/non-user discrimination have only one attribute in common with the top four attributes of heroin user/non-user discrimination (SS).

\begin{table}[!ht]\centering
\caption{Coefficients of the linear discriminant for   ecstasy user/non-user separation,  for the decade-, year-, month-, and week-based definition of users (7 attributes). \label{TabEcst7}}
\begin{tabular}{|l|c|c|c|c|c|c|c|c|}\hline
Period&$\Theta$&N&E&O&A&C&Imp&SS\\\hline
Decade&-35.896&0.045&-0.017&0.407&-0.085&-0.342&-0.156&0.827\\\hline
Year&-42.579&-0.019&0.078&0.373&-0.045&-0.356&-0.096&0.846\\\hline
Month&-27.572&-0.169&0.101&0.462&-0.191&-0.354&-0.139&0.752\\\hline
Week&-60.127&-0.095&0.168&0.695&-0.128&-0.355&0.242&0.528\\\hline \hline
SD&$\leq 0.565$&$\leq 0.004$&$\leq 0.005$&$\leq 0.004$&$\leq 0.004$&$\leq 0.004$&$\leq 0.005$&$\leq 0.006$\\\hline
\end{tabular}
\end{table}

\begin{table}[!ht]\centering
\caption{Coefficients of the linear discriminant for   heroin user/non-user separation,  for the decade-, year-, month-, and week-based definition of users (7 attributes). \label{TabHer7}}
\begin{tabular}{|l|c|c|c|c|c|c|c|c|}\hline
Period&$\Theta$&N&E&O&A&C&Imp&SS\\\hline
Decade&-55.851&0.381&-0.283&0.526&-0.322&-0.124&0.249&0.563\\\hline
Year&-44.733&0.361&-0.327&0.368&-0.397&-0.108&0.218&0.641\\\hline
Month&-30.020&0.302&-0.409&0.232&-0.499&-0.107&0.345&0.555\\\hline
Week&-37.099&0.548&-0.325&0.265&-0.547&-0.069&0.389&0.261\\\hline \hline
SD&$\leq 0.523$&$\leq 0.004$&$\leq 0.004$&$\leq 0.005$&$\leq 0.004$&$\leq 0.005$&$\leq 0.006$&$\leq 0.007$\\\hline
\end{tabular}
\end{table}

Employment of simple LDA demonstrates that users of ecstasy and heroin differ significantly, and users of different groups of drugs should be studied separately. The hypothesis that drug usage is associated with N$\Uparrow$,   A$\Downarrow$, and  C$\Downarrow$ seems plausible at first glance, but appears to be an oversimplification. Such an analysis, and an even more detailed consideration using all definitions of drug use, is possible for every pair of drugs (and for four groups of drugs)  on the basis of the linear discriminant coefficients presented in Tables \ref{TabDiscr1}-\ref{TabDiscr4} and \ref{TabDiscr5}-\ref{TabDiscr8}.

\begin{figure}
\centering
\includegraphics[width=0.84\textwidth]{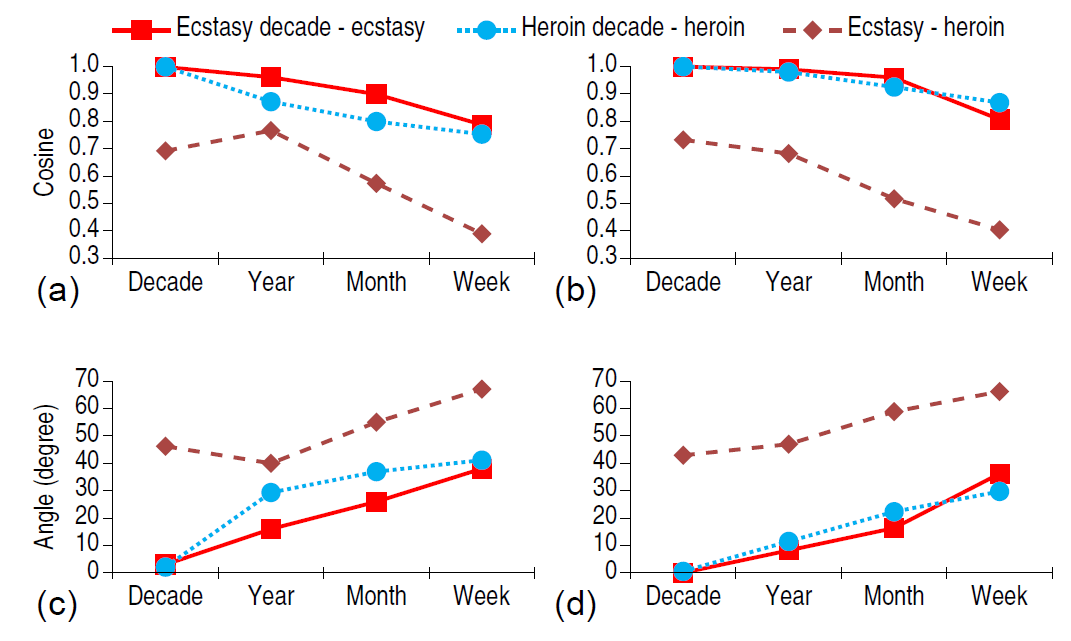}
\caption{Angles between directions of linear discriminants for user/non-user classification for ecstasy and heroin. Two types of angles are presented: between the discriminant directions for all periods and the discriminant vector for the decade-based definitions of users for both drugs, and angles between directions of linear discriminants for ecstasy and heroin (and the same periods). For convenience, both cosines of angles (a, b) and angles in degrees (c, d) are presented}
 \label{Fig:HerEcsLDAngle}
\end{figure}

The difference between linear discriminant directions can be evaluated by measuring the angles between them. For each drug, we calculate the angles between linear discriminant directions for all definition of users and the direction for the decade-based definition of users. These angles are significantly smaller than the angles between the linear discriminant directions for the same recency of use and different drugs (Fig. \ref{Fig:HerEcsLDAngle}). 

\section{Separation of heroin users from ecstasy users: what is the difference? \label{Sec:Ecs/Her2}}

\begin{figure}
\centering
\includegraphics[width=0.95\textwidth]{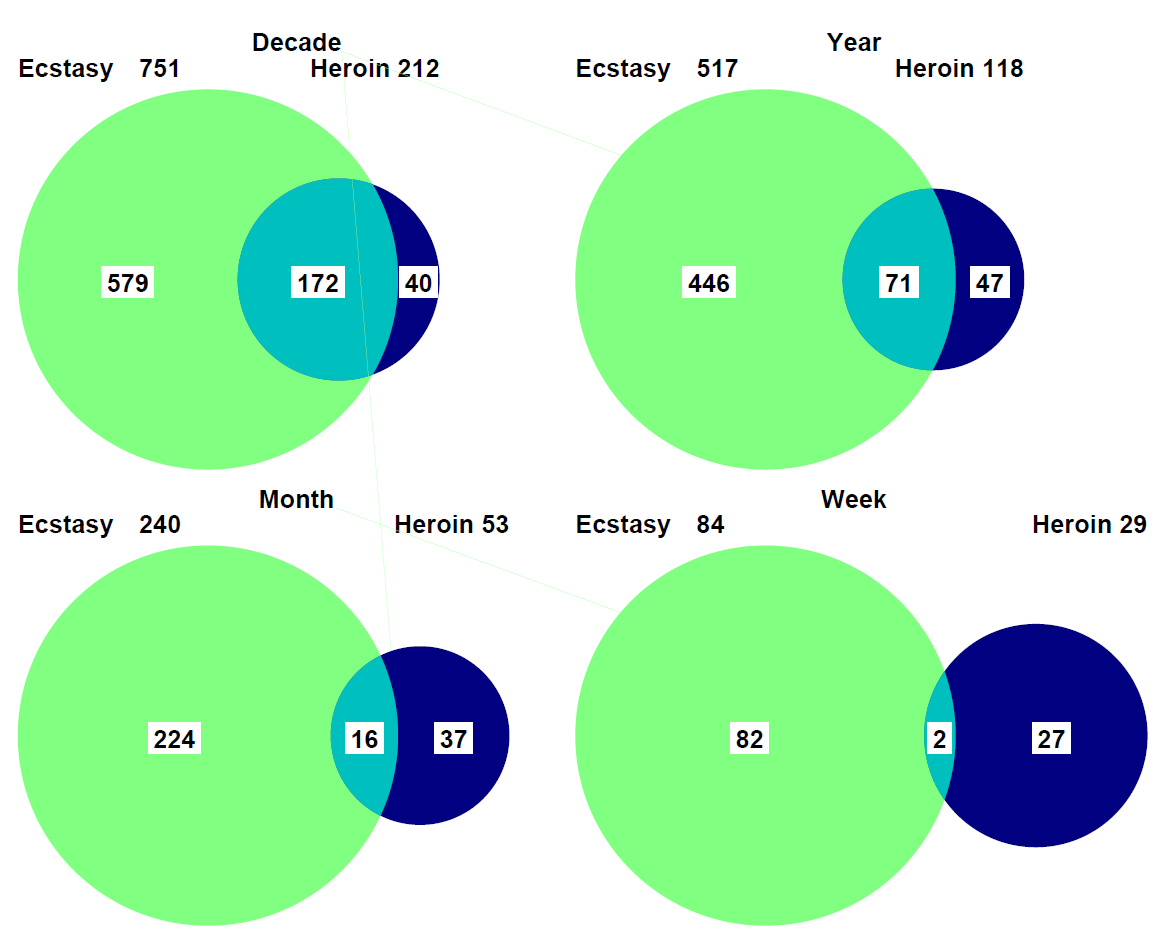}
\caption {Venn diagrams of relations between ecstasy and heroin use for decade-, year-, month-, and week-based definitions of users}
 \label{Fig:VennHE}
\end{figure}

In this section we continue the story about the differences between heroin and ecstasy use. The first simple question is: what is the intersections between the sets of users of these drugs? The answer 
is  illustrated by Fig. \ref{Fig:VennHE}. It is obvious from the figure that for recent users the proportion of simultaneous use of heroin and ecstasy is smaller. We hypothesize that people who used drugs more than a month or year ago, but not recently, just tried various drugs, whereas recent users prefer more specific drugs. Fig.  \ref{Fig:VennHE} is an argument in favour of this hypothesis.

For each recency of use, there are six important sets: users of ecstasy, users of heroin, users of ecstasy OR heroin (the union), users of ecstasy AND heroin (the intersection), users of ecstasy NOT heroin (users of ecstasy only, the difference: users of ecstasy $\setminus$ users of heroin),  and users of heroin NOT ecstasy (users of heroin only, the difference: users of heroin $\setminus$ users of ecstasy).

The intersection of heroin and ecstasy users is sometimes large, so the discrimination task is a non-standard classification. At least, the standard TPR (Sn) and TNR (Sp) do not make much sense. Let us consider a binary classification rule which separates all users of ecstasy OR heroin into two classes: E and H. We will consider an example from the set of users of heroin OR ecstasy to be FE (`false ecstasy') if it is a non-user of ecstasy classified as a user of ecstasy (or, which is the same, a user of heroin but NOT ecstasy classified as a user of ecstasy). Analogously, an example is considered as FH (`false heroin') if it is non-user of heroin classified as  a user of heroin.
In the Tables \ref{EHDecade}  we use an unusual measures of classification accuracy: True Ecstasy Rate (TER) (correctly recognised fraction of users of ecstasy NOT heroin) and True Heroin Rate (THR) (correctly recognised fraction  of users of heroin NOT ecstasy). We do not identify the case when a user of both drugs is recognised as a user of one them as an error:
$$\rm TER=\frac{\mbox{\# correctly recognised users of ecstasy NOT heroin}}{\mbox{\# users of ecstasy NOT heroin}};$$ $$\rm THR=\frac{\mbox{\# correctly recognised users of heroin NOT ecstasy}}{\mbox{\# users of heroin NOT ecstasy}}.$$

\begin{figure}
\centering
\includegraphics[width=0.84\textwidth]{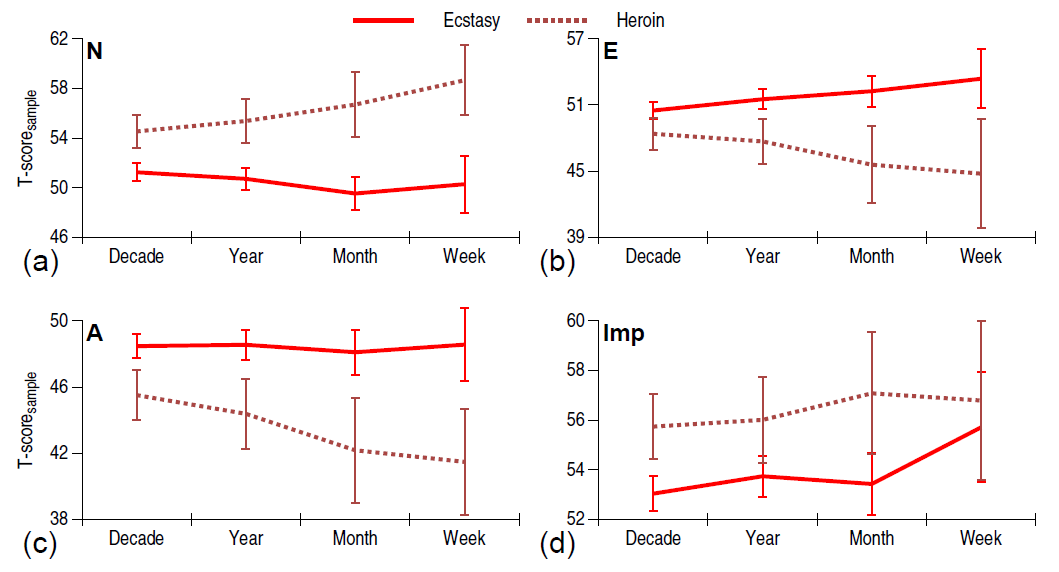}
\caption {Mean values with their 95\% confidence intervals for significantly different psychological traits  of ecstasy and heroin users: (a) N, (b) E, (c) A, (d) Imp}
\label{EHGraphs}
\end{figure}

The descriptive statistics for seven traits (FFM, Imp, and SS) are presented in Table~\ref{EHDecade} (compare to Table \ref{tab:2}). 
We can see that for ecstasy -- heroin discrimination the best classifier with one attribute is N. Moreover, differences between means of ecstasy users and heroin users are statistically significant with confidence level 99\% for N and A, for all definitions of users, for Imp for decade-based and month-based definitions, and for E for all definitions excluding the decade-based one. This difference is obvious in the graphs of mean values of N, E, A, and Imp  for ecstasy and heroin users shown in Fig.~\ref{EHGraphs}.

\begin{figure}
\centering
\includegraphics[width=0.8\textwidth]{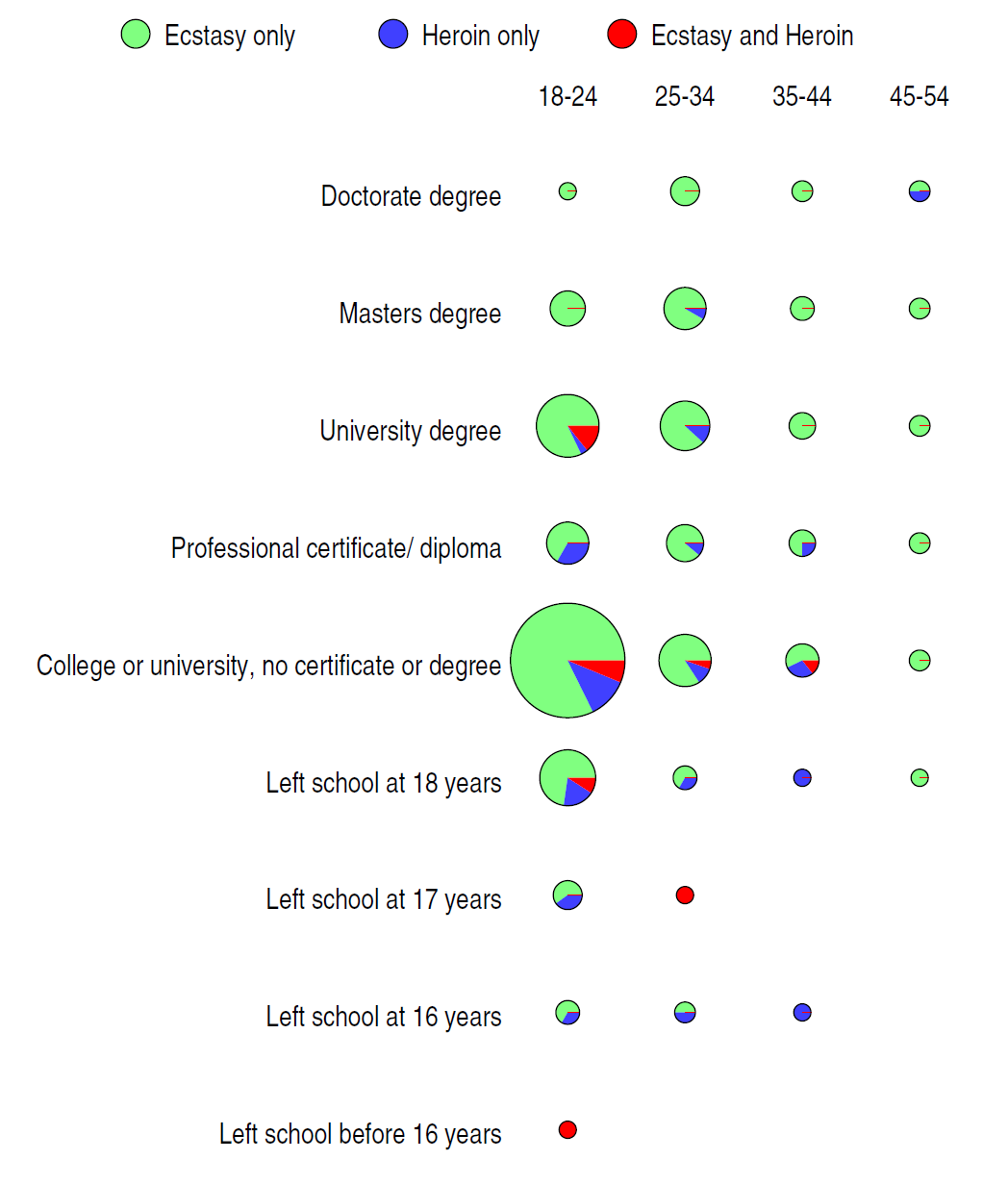}
\caption {Distribution of ecstasy (NOT heroin) users, heroin (NOT ecstasy) users, and heroin AND ecstasy users in various age and education groups}
 \label{Fig:AgeEduHE}
\end{figure}

\begin{table} 
\centering
\caption{Means and  standard deviations for users of ecstasy,  users of heroin, and users of ecstasy OR heroin (E OR H users). The dimensionless $ z$-score (\ref{score}) for separation of ecstasy users from heroin users is given as well as $P=\phi(z)$ (\%), TER (\%),  and THR (\%). The H-E $p$-value  is the probability of observing by chance the same or greater difference.}
\label{EHDecade}
\begin{tabular}{|c|c|c|c|c|c|c|c|c|c|c|c|c|c|c|}
\cline{1-13}
Factors &	\multicolumn{2}{c|}{Ecstasy users}&	\multicolumn{2}{c|}{Heroin users} & \multicolumn{2}{c|}{E OR H users} & H-E &\multicolumn{5}{c|}{One feature classifier}\\ \cline{2-7} \cline{9-13}
	&Mean &	SD	&Mean	& SD & Mean & SD & \emph{p}-value&$z$ & $P$ & $\Theta$ & TER & THR \\ \hline
	\multicolumn{13}{|c|}{Decade-based user definition}\\\hline
N  & 25.06 & 9.20 & 28.06 & 8.89 & 25.26 & 9.23 & $<0.001$ & 0.154 & 56 & 27 & 65 & 61 \\\hline
E & 27.91 & 7.12 & 26.48 & 7.37 & 27.76 & 7.14 & 0.012 & 0.088 & 54 & 27 & 57 & 60 \\\hline
O & 36.14 & 6.02 & 36.55 & 5.86 & 36.07 & 6.02 & 0.366 & 0.041 & 52 & 35 & 55 & 55 \\\hline
A & 29.88 & 6.61 & 27.98 & 7.28 & 29.77 & 6.67 & 0.001 & 0.128 & 55 & 28 & 62 & 58 \\\hline
C & 27.50 & 6.87 & 26.52 & 7.00 & 27.45 & 6.87 & 0.071 & 0.067 & 53 & 27 & 53 & 55 \\\hline
Imp & 4.45 & 2.06 & 5.02 & 2.06 & 4.44 & 2.06 & $<0.001$ & 0.139 & 56 & 4 & 48 & 58 \\\hline
SS & 6.90 & 2.23 & 7.18 & 2.34 & 6.86 & 2.26 & 0.117 & 0.070 & 53 & 6 & 59 & 50 \\\hline
\multicolumn{13}{|c|}{Year-based user definition}\\\hline
N  & 24.57 & 9.46 & 28.83 & 8.76 & 25.02 & 9.47 & $<0.001$ & 0.209 & 58 & 27 & 64 & 62 \\\hline
E & 28.60 & 7.12 & 26.01 & 7.46 & 28.20 & 7.29 & 0.001 & 0.148 & 56 & 27 & 62 & 66 \\\hline
O & 36.46 & 6.00 & 36.03 & 5.97 & 36.46 & 5.98 & 0.483 & 0.035 & 51 & 36 & 54 & 53 \\\hline
A & 29.93 & 6.74 & 27.25 & 7.49 & 29.68 & 6.86 & $<0.001$ & 0.170 & 57 & 28 & 61 & 57 \\\hline
C & 27.36 & 6.88 & 26.27 & 7.38 & 27.26 & 6.91 & 0.145 & 0.069 & 53 & 26 & 54 & 55 \\\hline
Imp & 4.59 & 2.04 & 5.08 & 2.01 & 4.62 & 2.05 & 0.020 & 0.113 & 54 & 4 & 55 & 51 \\\hline
SS & 7.14 & 2.15 & 7.29 & 2.36 & 7.15 & 2.18 & 0.525 & 0.031 & 51 & 7 & 55 & 50 \\\hline
\multicolumn{13}{|c|}{Month-based user definition}\\\hline
N  & 23.50 & 9.73 & 30.04 & 8.76 & 24.32 & 9.80 & $<0.001$ & 0.308 & 62 & 27 & 68 & 66 \\\hline
E & 29.09 & 7.53 & 24.58 & 8.65 & 28.43 & 7.89 & 0.001 & 0.233 & 59 & 27 & 64 & 65 \\\hline
O & 36.66 & 6.12 & 35.40 & 6.82 & 36.55 & 6.14 & 0.217 & 0.089 & 54 & 36 & 55 & 54 \\\hline
A & 29.64 & 6.82 & 25.83 & 7.42 & 29.27 & 6.81 & 0.001 & 0.242 & 60 & 28 & 59 & 57 \\\hline
C & 27.54 & 7.20 & 24.81 & 7.60 & 27.19 & 7.31 & 0.020 & 0.159 & 56 & 26 & 58 & 59 \\\hline
Imp & 4.53 & 2.07 & 5.30 & 1.89 & 4.63 & 2.07 & 0.010 & 0.170 & 57 & 4 & 65 & 52 \\\hline
SS & 7.08 & 2.14 & 7.53 & 2.31 & 7.12 & 2.18 & 0.198 & 0.092 & 54 & 7 & 59 & 52 \\\hline
\multicolumn{13}{|c|}{Week-based user definition}\\\hline
N  & 24.18 & 9.66 & 31.83 & 6.76 & 26.12 & 9.67 & $<0.001$ & 0.348 & 64 & 28 & 67 & 68 \\\hline
E & 29.86 & 8.29 & 24.03 & 8.84 & 28.25 & 8.81 & 0.003 & 0.239 & 59 & 25 & 73 & 70 \\\hline
O & 37.81 & 6.06 & 35.34 & 7.62 & 37.12 & 6.55 & 0.122 & 0.125 & 55 & 36 & 55 & 59 \\\hline
A & 29.94 & 6.48 & 25.38 & 5.46 & 28.81 & 6.57 & 0.001 & 0.285 & 61 & 27 & 62 & 59 \\\hline
C & 27.33 & 7.07 & 24.59 & 7.49 & 26.57 & 7.22 & 0.091 & 0.135 & 55 & 25 & 59 & 56 \\\hline
Imp & 5.01 & 2.17 & 5.24 & 1.79 & 5.06 & 2.06 & 0.576 & 0.046 & 52 & 5 & 46 & 56 \\\hline
SS & 7.33 & 2.19 & 7.28 & 2.45 & 7.30 & 2.25 & 0.911 & 0.005 & 50 & 7 & 57 & 48 \\\hline
\end{tabular}
\end{table}

We have employed LDA for separation of ecstasy users from heroin users. We have applied the formula (\ref{FisherLD}) for the linear discriminant direction with covariance matrices  for ecstasy users and heroin users calculated for the month-based definition of users. The intercept $\Theta$ was calculated for the most balanced separation measured by TER and THR. 

\begin{table}\centering
\caption{Coefficients of the linear discriminant for  separation of ecstasy users from heroin users for the month-based definition of users (7 attributes). TER=70.5\% THR=73.0\% (the sample of ecstasy AND heroin users); TER=69.6\% THR=62.2\%  (LOOCV).}
\begin{tabular}{|c|c|c|c|c|c|c|c|c|}\hline
&$\Theta$&N&E&O&A&C&Imp&SS\\\hline
Coefficients&29.896&-0.541&0.476&-0.050&0.469&-0.136&-0.441&-0.214\\\hline
SD of coefficients&1.331&0.012&0.010&0.013&0.008&0.011&0.015&0.018\\\hline
\end{tabular}
\end{table}

\begin{table}\centering
\caption{Coefficients of the linear discriminant for  separation of ecstasy users from heroin users for the month-based definition of users (10 attributes). TER=75.0\% THR=73.0\% (the sample of ecstasy AND heroin users); TER=71.6\% THR=64.9\%  (LOOCV).}
\begin{tabular}{|c|c|c|c|c|c|c|c|c|c|c|c|}\hline
&$\Theta$&Age&Edu&N&E&O&A&C&Imp&SS&Gndr\\\hline
Coefficients&0.915&0.011&0.534&-0.401&0.379&-0.039&0.411&-0.176&-0.417&-0.092&0.166\\\hline
SD of coefficients&0.023&0.015&0.010&0.011&0.010&0.010&0.008&0.009&0.012&0.014&0.011\\\hline
\end{tabular}
\end{table}

The ranking of attributes for the month-based ecstasy users / heroin users separation, based on the size of the coefficients of the linear discriminant, is:
$$\mbox{N, E, A, Imp, SS, C, O (7 attributes)}.$$
It is important to note that the values of the coefficients of N, E, A, and Imp do not differ much and for SS, C, and O the coefficients are much smaller.

For 10 attributes the ranking by the linear discriminant coefficients is:
$$\mbox{Edu, Imp, A, N, E, C, Gndr, SS, O, Age}.$$
Again, for the leading group of attributes, Edu, Imp, A, N, and E,   the coefficients decay slowly, and for the other attributes are much smaller. These observations, together with analysis of attribute means (Fig. \ref{EHGraphs}), convinces us that the main and statistically significant differences
 between ecstasy users and heroin users are in Edu, Imp, A, N, and E.

\section{Significant difference between benzodiazepines, ecstasy, and heroin \label{Sec:Benz/Ecs/Her}} 

\begin{figure}
\centering
{\includegraphics[width=0.9\textwidth]{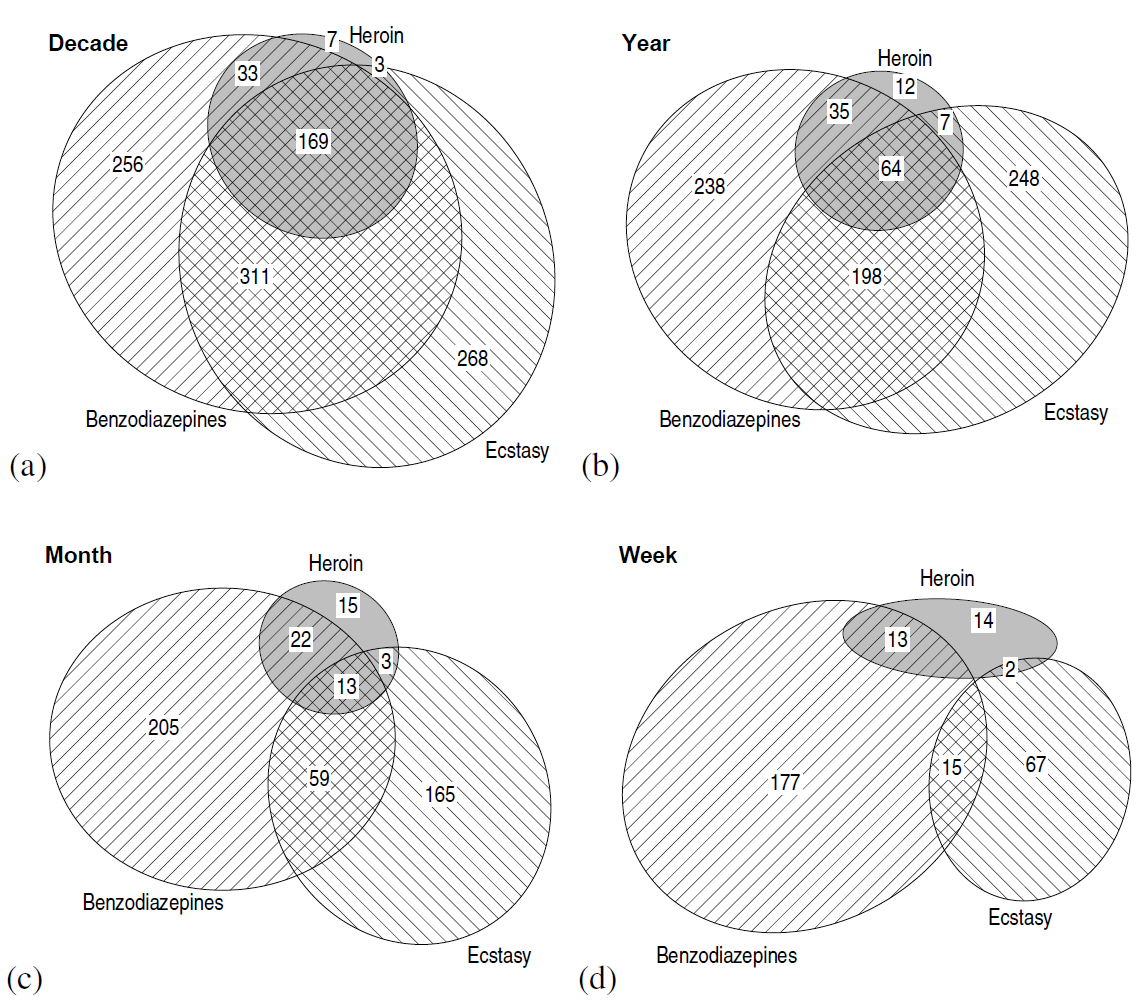}}
\caption {Venn diagrams of relations between benzodiazepines, ecstasy and heroin use for (a) decade-, (b) year-, (c) month-, and (d) week-based definitions of users.}
 \label{Fig:BEHVenn}
\end{figure}

Analysis of differences between various groups of drug users can be extended to all pairs of drugs. Let us add for investigation the centre of third pleiad, benzodiazepines. The relations between groups of users of benzodiazepines, ecstasy and heroin is shown in Fig.~\ref{Fig:BEHVenn}.

\begin{figure}
\centering
\includegraphics[width=0.84\textwidth]{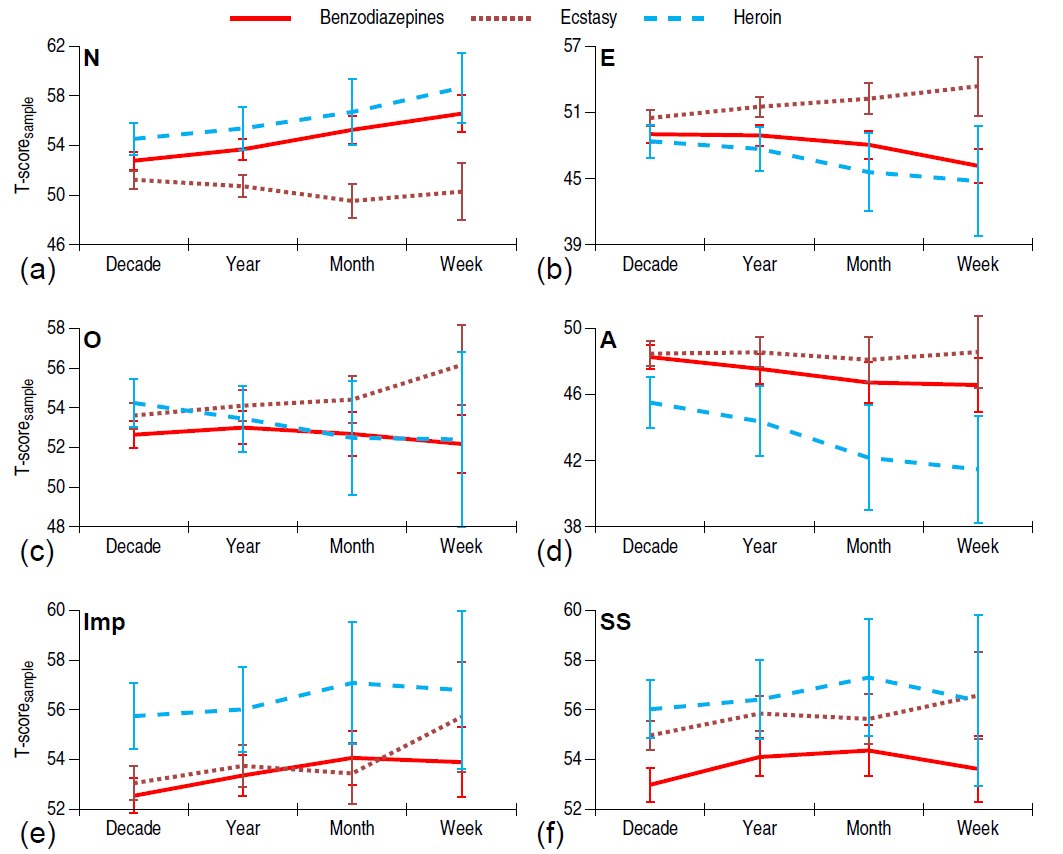}
\caption {Mean values with their 95\% confidence intervals for significantly different psychological traits  of  benzodiazepines, ecstasy and heroin users: (a) N, (b) E, (c) O, (d) A, (e) Imp, and (f) SS.}
\label{BEHGraphs}
\end{figure}

In Fig.~\ref{BEHGraphs} the mean values of personality traits N, E, O, A, Imp, SS are given with 95\% CI  for groups of  benzodiazepines, ecstasy, and heroin users, and different recency of use. (C is not given because the difference in C between different groups of drug users is insignificant.)

\begin{table}
\centering
\caption{$p$-values of differences of mean values for benzodiazepines (benzos), ecstasy and heroin for all definitions of users.}
\label{table:BEHsign}
\begin{tabular}{|c|c|c|c|c|c|c|c|c|c|c|c|c|c|c|}
\cline{1-13}
Factors & \multicolumn{4}{c|}{Benzos and ecstasy}& \multicolumn{4}{c|}{Benzos and heroin} & \multicolumn{4}{c|}{Ecstasy and heroin} \\ \cline{2-13}
	&Decade & Year &Month &Week &Decade & Year &Month &Week &Decade & Year &Month &Week \\ \hline
N& \textbf{0.003} & \textbf{\textless0.001} & \textbf{\textless0.001} & \textbf{\textless0.001} & \textbf{0.021} & 0.088 & 0.322 & 0.188 & \textbf{\textless0.001} & \textbf{\textless0.001} & \textbf{\textless0.001} & \textbf{\textless0.001} \\\hline
E& \textbf{0.007} & \textbf{\textless0.001} & \textbf{\textless0.001} & \textbf{\textless0.001} & 0.443 & 0.275 & 0.189 & 0.590 & \textbf{0.012} & \textbf{0.001} & \textbf{0.001} & \textbf{0.003} \\\hline
O& \textbf{0.045} & 0.058 & \textbf{0.035} & \textbf{0.002} & \textbf{0.024} & 0.629 & 0.898 & 0.921 & 0.366 & 0.483 & 0.217 & 0.122 \\\hline
A& 0.697 & 0.128 & 0.141 & 0.150 & \textbf{0.002} & \textbf{0.007} & \textbf{0.010} & \textbf{0.006} & \textbf{0.001} & \textbf{\textless0.001} & \textbf{0.001} & \textbf{0.001} \\\hline
C& 0.439 & 0.943 & 0.412 & 0.557 & \textbf{0.021} & 0.135 & 0.052 & 0.146 & 0.071 & 0.145 & \textbf{0.020} & 0.091 \\\hline
Imp& 0.315 & 0.520 & 0.457 & 0.172 & \textbf{\textless0.001} & \textbf{0.007} & \textbf{0.028} & 0.099 & \textbf{\textless0.001} & \textbf{0.020} & \textbf{0.010} & 0.576 \\\hline
SS& \textbf{\textless0.001} & \textbf{0.001} & 0.081 & \textbf{0.009} & \textbf{\textless0.001} & \textbf{0.011} & \textbf{0.025} & 0.139 & 0.117 & 0.525 & 0.198 & 0.911 \\\hline
\end{tabular}
\end{table}

We can see from Table~\ref{table:BEHsign} that mean values in the groups of benzodiazepines and ecstasy users are statistically significantly different with confidence level 95\% for N (higher for benzodiazepines) and E (higher for ecstasy) for all definition of users, for O (higher for ecstasy) for all definitions of users excluding the year-based, and for SS (higher for ecstasy) for all definitions excluding the month-based. Benzodiazepines and heroin are statistically significantly different with confidence level 95\% for A for all definition of users, for Imp (higher for heroin) and SS (higher for heroin) for all definitions of users excluding the week-based. 
Heroin and ecstasy are statistically significantly different with confidence level 95\% for N (higher for heroin), E (higher for ecstasy) and A (higher for ecstasy) for all definition of users, and for Imp (higher for heroin) for all definitions excluding the week-based. 

 \section{A tree of linear discriminants: no essential improvements}

\begin{figure}
 \centering
\includegraphics[width=0.95\textwidth]{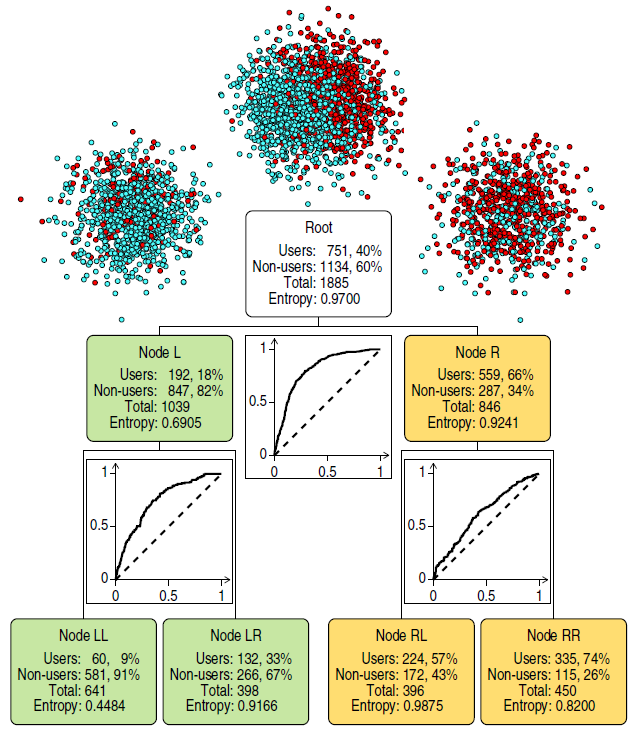}
\caption {A two-level classification tree for ecstasy users and non-users (decade-based definition of users) with linear discriminant classifiers at the nodes. A data cloud is visualised by projection  onto the plane of the two first principal components for the root and the nodes of the first levels (above nodes). Users are represented by blue (light) circles, non-users by red (dark) circles.  The ROC curves for the linear discriminants at each branching node are below the nodes}
 \label{TreeEcstasy}
\end{figure}

Performance of simple LDA is not much worse than the quality of the best selected classifiers (see Table \ref{tab:12}). Moreover, straightforward attempts to improve performance by creation of  a simple  decision tree, or simple kNN classifiers, fail. In Fig. \ref{TreeEcstasy} a two-level decision tree with a hierarchy of linear discriminants is presented. The PCA visualisation demonstrates that after application of the first linear discriminant, there remains a typical `flies-and-mosquito' mixture without any apparent user -- non-user separation in the groups. (Principal components were recalculated for each node.) Use of kNN classifiers does not give any essential improvement over LDA either (Table \ref{kNNecstasy}).

\begin{table}
\centering
\caption{The numbers of false positive (FP) and false negative (FN) errors for ecstasy user/non-user decision tree classifiers (decade-based definition of users) with linear discriminant at each node, and with four different criteria of threshold selection: Accuracy, Sp+Sn, Balance (Sn=Sp), and Information Gain (IG).}
\label{DTlevels}
\begin{tabular}{|c|c|c|c|c|c|c|c|c|}
\cline{1-9}
  &\multicolumn{2}{c|}{Accuracy}	&\multicolumn{2}{c|}{Sp+Sn}& \multicolumn{2}{c|}{Balanced}&	\multicolumn{2}{c|}{IG}\\ \cline{2-9}
Level	&FP	&FN&		FP	&FN	&	FP	&FN	&	FP&	FN	\\ \hline
1&	232&	224&	330&	155&		287	&192&		401&	130\\ \hline
2&	232&	224&	330&	155&	287&	192&		401&	130\\ \hline
3&	232&	224&	210&	250&		189&	287&		251&	220\\ \hline
4&	238&	203&	268&	190&		268&	179&		251&	220\\ \hline
5&	228&	169&	218&	206&		189&	240&		210&	253\\ \hline
6&	173&	177&	203&	160&		177&	171&		242&	201\\ \hline
7&	156&	185&	121&	186&		168&	173&		193&	198\\ \hline
\end{tabular}
\end{table}

\begin{table}
\centering
\caption{Performance of kNN user/non-user classifiers for ecstasy (decade-based definition of users) for different k and for the standard Euclidean distance.}
\label{kNNecstasy}
\begin{tabular}{|c|c|c|c|c|c|c|c|c|c|c|c|c|c|c|c|}
\cline{1-16}
$k$ &5& 6& 7& 8& 9& 10& 11& 12& 13& 14& 15& 16& 17& 18& 19\\ \hline
Sp (\%)& 59.7& 70.5&65.5& 72.0& 68.3& 65.2 &70.5 &66.9& 71.2& 68.9& 66.4& 70.2& 67.5& 70.4& 68.6\\ \hline
Sn (\%) &81.9&72.8& 80.2& 72.2& 78.3& 81.5& 75.8& 78.8& 74.7& 77.8& 80.2& 77.5& 80.0& 76.0& 79.5\\ \hline
\end{tabular}
\end{table}

\section{Visualisation on non-linear PCA canvas \label{Sec:VisNLinPCA}}

Principal components provide us with a canvas for convenient visualisation of the data distribution (see Fig. \ref{TreeEcstasy}). This works well if the data are distributed near a low-dimensional plane.
Manifold learning methods, using non-linear PCA, generalise this idea to data approximation by smooth non-linear manifolds of small dimension  \cite{Gorban08}.  
These methods allow us to approximate data more successfully, and to find in non-linear two-dimensional visual data maps effects which can only be captured in higher-dimensional linear principal components \cite{Gorban10}. 

In a series of works, the metaphor of elastic membrane and plate was used to construct one-, two- and three-dimensional
principal manifold approximations for various data topologies \cite{Gorban10}. The error of mean-squared distance approximation, combined with the elastic energy of the membrane, serves as a functional to be optimised.

\begin{figure}
 \centering{
 \includegraphics[width=0.8\textwidth]{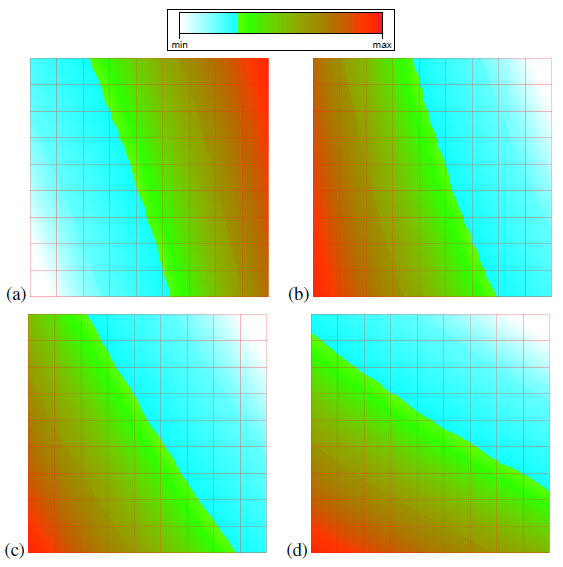}}
\caption{Visualisation of various functions on the elastic map (internal coordinates): (a) Age, (b) Imp, (c) SS, (d) O (Age and attributes apparently correlated with Age on the maps)
 \label{VisAgeImpSSA}}
\end{figure}

\begin{figure}
 \centering{
\includegraphics[width=0.8\textwidth]{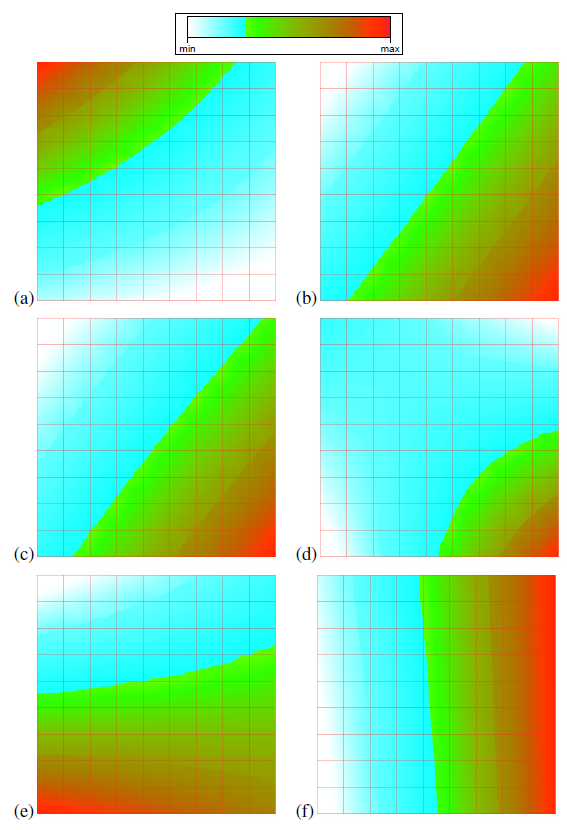}}
\caption{Visualisation of various functions on the elastic map (internal coordinates): (a) N, (b) A, (c) C, (d) Edu,  (e) E, (f) Gndr
 \label{VisNACEdu}}
\end{figure}

We have generated an elastic map for the whole dataset in the 10-dimensional space of quantified attributes (Fig. \ref{VisAgeImpSSA}). Neither three-dimensional PCA view and the elastic map view reveal any significantly non-linear, non-ellipsoidal peculiarities in data distribution. In Figs. \ref{VisAgeImpSSA} -- \ref{VisNACEdu} the attributes are visualised.

\begin{figure}
 \centering{
{\large(a)}\includegraphics[width=0.95\textwidth]{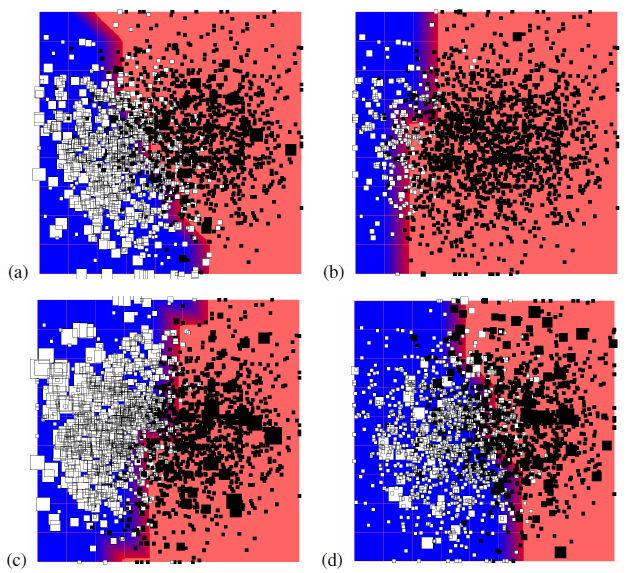}}
\caption {Visualisation of linear discriminant classifiers on the non-linear PCA elastic map canvas:( a) Ecstasy, (b) Heroin, (c) Benzodiazepines, (d) The group of `Illicit drugs'. White squares -- users, black squares -- non-users. Light (red) background -- LDA predicts users, dark (blue) background -- LDA predicts non-users}
 \label{LDAVis}
\end{figure}

We can see that the attributes Imp, SS, and O generate similar colorings, which are opposite to the coloring for Age. It should be mentioned that this similarity suggests strong correlations of attributes on the two-dimensional map but does not imply strong correlations in the higher-dimensional data space. Analogously, attributes N,  A,  C, and Edu have similar coloring on the map with N opposite to other three attribute colorings (Fig. \ref{VisNACEdu}a-d). The colorings for E and Gndr differ from all others, and are independent on the map (Fig. \ref{VisNACEdu}e,f). 
Linear discriminant separation of drug users from non-users is visualised on the elastic map in Fig. \ref{LDAVis}. Of course, on the non-linear canvas  the linear discriminant is represented by a curve, not a straight line. The quality of the linear discriminant separation is visibly high.

\section{Risk maps}

\label{Risk evaluation}

For every set of attributes, we can evaluate the conditional probability density to be a drug user for each value of the attributes. Visualisation of the conditional density of drug use can be thought of as a risk map \cite{Mirkes14a,Mirkes14b}. Let us recall that the probability of being a drug user in the data base is higher than in the populations. In application of the risk maps to real cases, risk evaluation should be renormalised to the population a priori probability of drug use. These maps can be used without such a renormalisation if we consider, not the absolute risk, but the relative risk for comparison of different values of attributes.

\begin{figure}
 \centering{
  \includegraphics[width=0.8\textwidth]{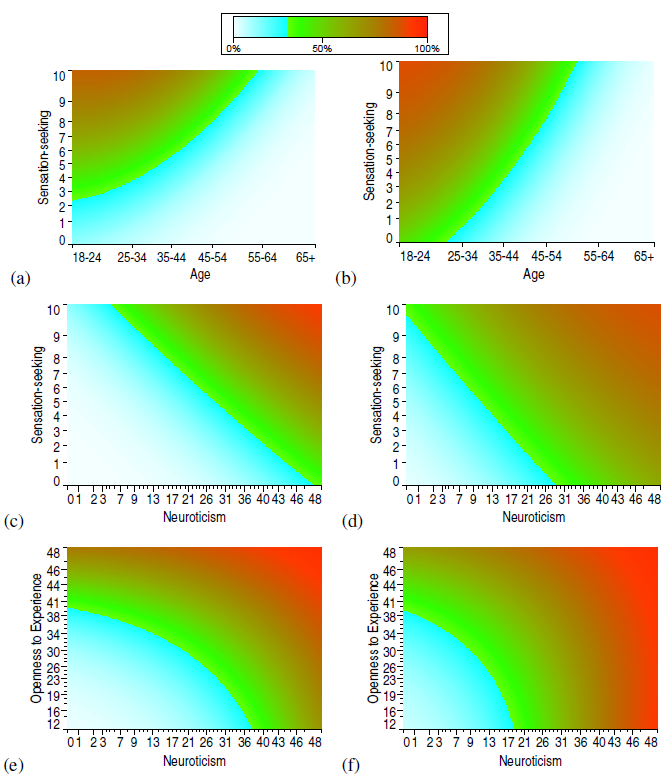} }
\caption {Simplest examples of risk maps of: ecstasy consumption for female (a) and male (b), heroin consumption for female (c) and male (d), and benzodiazepines consumption for female (e) and male (f)}
 \label{RiskMapsfig:8}
\end{figure}

In Fig. \ref{RiskMapsfig:8} we have presented the simplest risk maps produced by the bi-Gaussian approximation of the probability density: the densities of users and of non-users of a specific drug were approximated by two-dimensional Gaussian distributions, and then the conditional probability density $\rho_u$ of drug use has been evaluated:
$$\rho_u=\frac{n_u N(\mu_u, S_u)}{n_u N(\mu_u, S_u)+n_{n-u} N(\mu_{n-u}, S_{n-u})},$$
where $n_u$ and $n_{n-u}$ are the number of users and non-users of the drug respectively, $S_u$ and $S_{n-u}$ are the empiric covariance matrices of the users and non-users respectively, and $N(\mu, \Sigma)$ is the normal distribution with mean $\mu$ and covariance matrix $\Sigma$.  For the user/non-user separation this approximation corresponds to so-called quadratic discriminant analysis \cite{MarksDunn1974}.

\begin{figure}
 \centering{
 \includegraphics[width=0.8\textwidth]{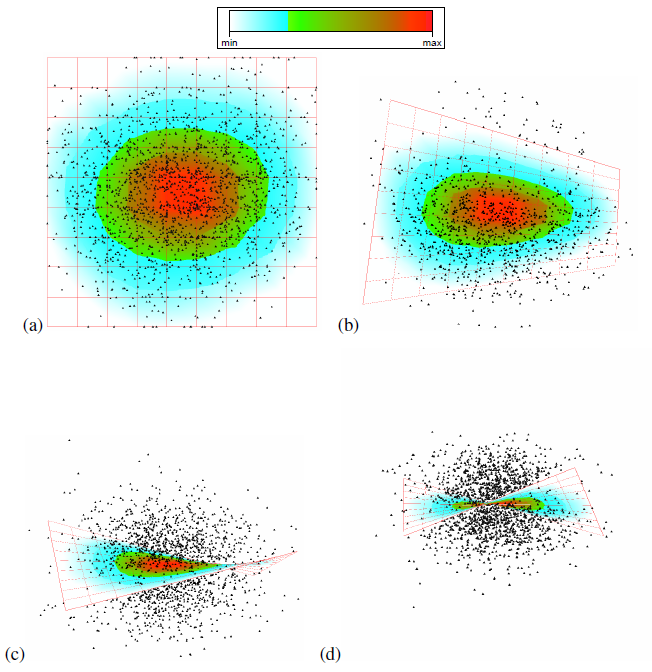}}
\caption{Elastic maps and density visualisation for the database of drug users: (a) Visualisation of the density of the data cloud on the elastic map presented in the internal coordinates, (b)-(d) Elastic map embedded in the 3-dimensional principal component space with various points of view. Data points are in black}
 \label{ElMapGeneral}
\end{figure}

%% file: summary.tex
%%%%%%%%%%%%%%%%%%%%% chapter.tex %%%%%%%%%%%%%%%%%%%%%%%%%%%%%%%%%
%FINAL
% sample chapter
%
% Use this file as a template for your own input.
%
%%%%%%%%%%%%%%%%%%%%%%%% Springer-Verlag %%%%%%%%%%%%%%%%%%%%%%%%%%
%\motto{Use the template \emph{chapter.tex} to style the various elements of your chapter content.}
%\chapter{Chapter Heading}
%\label{intro} % Always give a unique label
% us \chaptermark{}
% to alter or adjust the chapter heading in the running head

\chapter*{Summary}
\markboth{Summary}{Summary}
\addcontentsline{toc}{chapter}{Summary}

What have all of these data tables told us? We have asked whether or not a psychological predisposition to drug consumption exists. Now,  we can formulate the answer in brief:
\begin{itemize}
\item There is a significant difference in the psychological profiles of drug users and non-users.
\item The psychological predisposition to using different drugs may be different.
\item Machine learning algorithms can solve the user/non-user classification problem for many drugs with impressive sensitivity and specificity. Therefore, the question of predictability of the risk of drug consumption on the basis of personality traits and simple demographic attributes is answered in the affirmative.
\item Simple robust Fisher's linear discriminants are successfully employed for this classification.
\item We describe the groups of drugs which have correlated use (correlation pleiades) and we can collect users of these groups of drugs together for the purpose of analysis.
\item Mean profiles of users of different drugs are different. (This is explicitly proved for benzodiazepines, ecstasy, and heroin; this is obvious from the attached tables for many more pairs of substances.)
\end{itemize}

Our study demonstrates strong correlations between personality profiles and the risk of drug use. This result partially supports observations from the previous research \cite{Sutina13,Haider02,Vollrath02,Flory02,Terracciano08,Turiano12,Stewart00,Roncero14}. For example, individuals involved in the use of `heavy' drugs like heroin and methadone are more likely to have higher scores for N, and lower scores for A and C . In addition, they have significantly higher O, and so the typical profile is  N$\Uparrow$,  O$\Uparrow$, A$\Downarrow$, and C$\Downarrow$.
The profile is different for recent users (within the last year) of `party drugs' like ecstasy, LSD, and amyl nitrite. For them, N is not high, and the typical profile is O$\Uparrow$, A$\Downarrow$, and C$\Downarrow$.

We have analysed, in full detail, the average differences in the groups of drug users and non-users for 18 drugs (Tables~\ref{tab:5}, \ref{tab:5a}, \ref{tab:5b}, and \ref{tab:5c}). In addition to this analysis, we have achieved a much more detailed understanding of the relationship between personality traits, demographic data, and the use of individual drugs or drug clusters by an individual subject.

The database we analysed contains 1,885 participants and 12 features (input attributes). These features include the five personality traits (NEO-FFI-R); impulsivity (BIS-11), sensation-seeking (ImpSS), level of education, age, gender, country of residence, and ethnicity. The data set includes information on the consumption of 18 central nervous system psychoactive drugs: alcohol, amphetamines, amyl nitrite, benzodiazepines, cannabis, chocolate, cocaine, caffeine, crack, ecstasy, heroin, ketamine, legal highs, LSD, methadone, mushrooms, nicotine, as well as VSA (output attributes). 

In our analysis we used three different techniques for  ranking features. After input feature ranking we excluded ethnicity and country of residence since the dataset has not enough data for most of the ethnicities and countries to prove the value of this information. As a result, 10 input features remained: age, Edu., N, E, O, A, C, Imp, SS, and gender. Our aim was to predict the risk of drug consumption for an individual.

All input features are ordinal or nominal. To apply data mining methods which were developed for continuous input features we applied the CatPCA technique to quantify the data.

We used four different definitions of drug users based on the recency of the last consumption of the drug: the decade-based,  year-based, month-based and  week-based user/non-user separation (Fig.~\ref{Categoriesfig:1}).  

Our findings have allowed us to draw a number of important conclusions about the associations between personality traits and drug use. All five personality factors are relevant traits to be taken into account when assessing the risk of an individual consuming drugs.

The mean scores for the groups of users of all 18 drugs are high or neutral for N and O, and moderately low  for A and C. The only exception is for crack usage for the week-based classification problem, which has a moderately low $(-)$ O score (see Table~\ref{tab:5c}). Users of legal drugs (alcohol, chocolate, and caffeine) have neutral  A and C  scores, and nicotine users have moderately low $(-)$ C score. For LSD users in the year-based classification problem and for LSD and magic mushrooms users in the week-based classification problem, the A score  is neutral.

The impact of the E score was found to be drug specific. For example, for the decade-based user/non-user definition the E score is negatively correlated with the consumption of  crack, heroin, VSA, and methadone (the E score is $(-)$ for their users). It has no predictive value for other drugs for the decade-based classification (the E score for users was $(0)$),  whereas in the year-, month-, and week-based classification problems all three possible values of E score were observed (see Tables~\ref{tab:5}, \ref{tab:5a}, \ref{tab:5b} and \ref{tab:5c}).

We confirm the findings of previous researchers that higher scores for N and O, and lower scores for C and A, lead to increased risk of drug use \cite{Belcher2016}. The O score is marked by curiosity and open-mindedness (and correlated with intelligence), and it is therefore understandable why higher O may be sometimes be associated with drug use \cite{Wilmoth12}. Flory et al. \cite{Flory02} found marijuana use to be associated with lower A and C, and higher O. These findings have been partially confirmed by our study. Our results improve the understanding of the pathways leading to drug consumption. In particular, these pathways should form paths in the correlation pleiades of drugs.

Our study has demonstrated that different attributes are important for different drugs. The detailed profiles for users and non-users of all drugs, and groups of drugs, are collected in Tables \ref{table4}, \ref{table4b}
Using these tables we can compare, discuss, and verify almost all results and hypotheses concerning the psychological profiles of drug users. For example:	
\begin{enumerate}
\item In the paper \cite{Terracciano08} significant differences in the NEO-PI-R  mean profiles of current cocaine/heroin users and non-users was found: for users, N$\Uparrow $ and C$\Downarrow $. In our table, for the month-based user definition (Table \ref{table4b}), the mean profiles for cocaine users differ from non-users but the differences in A and O are also significant ($p\leq 0.001$): N$\Uparrow $, O$\Uparrow $, A$\Downarrow $, and C$\Downarrow $. For heroin, our table gives the result N$\Uparrow $, E$\Downarrow $, A$\Downarrow $, and C$\Downarrow$. The mean heroin user profile differs significantly from the profile for cocaine users: heroin users are significantly less extroverted than non-users, whereas the E-score for cocaine users is found to be slightly higher than for non-users ($p=0.129$). Joining heroin users and cocaine users in one category concealed the deviations in E-score. The mean O-score of heroin users is higher than for non-users but not significantly higher (the empirical value of the O-score for heroin users is approximately the same as for cocaine users but the significance is different because of the different sample sizes).
\item Again in \cite{Terracciano08} and in \cite{Fridberg2011} significant deviations of NEO-PI-R  mean profiles for current cannabis users, from the means for non-users, are given as O$\Uparrow $, A$\Downarrow $, and C$\Downarrow $. We additionally see in Table \ref{table4b} that we have N$\Uparrow $. The differences in the empirical value of N-scores between cannabis users and non-users are approximately the same in our tables as in  \cite{Terracciano08} ($\approx 1$), but the sample sizes are different and therefore, the statistical significance differs.
\item In \cite{Flory02} a strong difference in N between cannabis users and non-users is proposed, and no significant differences in E and O. This result strongly contradicts our observations, as well those of other earlier works \cite{Terracciano08,Grossman1974}, which report a high difference in O and no (or more modest) difference in N.
\item Combinations of high and low scores in three NEO-FFI personality dimensions,  neuroticism (N), extraversion (E), and conscientiousness (C), result in eight different personality types. Smoking, consumption of
alcohol and drugs, and risky sexual behaviour were studied in  a sample of 683 university students in \cite{Vollrath02}. Two types, E$\Uparrow $, C$\Downarrow $ (Impulsives, Hedonists) and N$\Uparrow $,  C$\Downarrow $ (Insecures) were particularly inclined to engage in multiple, risky health behaviours, whereas the type with  E$\Downarrow $, C$\Uparrow $  (Sceptics, Brooders) abstained from risky behaviour. Let us look at our data with the month-based user definition (Table \ref{table4b}).
    {\begin{itemize} \item Users of amphetamines, benzodiazepines, cannabis, cocaine, crack, heroin, legal highs, and nicotine belong to the type  N$\Uparrow $,  C$\Downarrow $ (Insecures) and do not belong to the type E$\Uparrow $, C$\Downarrow $  (Impulsives, Hedonists).
    \item Users of ecstasy and LSD belong to the type E$\Uparrow $, C$\Downarrow $  and do not belong to the type N$\Uparrow $ ,  C$\Downarrow $. 
    \item Users of methadone belong to both types (the intersection). 
     \item It is worth mentioning that users of VSA do not belong to these types but have significant deviations in  O$\Uparrow $ and in A$\Downarrow $ (insufficient significance of deviation in N and C may be caused by the small sample size with VSA). 
     \item Users of magic mushrooms have significant deviations in  O$\Uparrow $ and in C$\Downarrow $ but do not belong to both types (insignificance of deviation of E$\Uparrow $ may be caused by the small number of magic mushroom users). 
     \item Users of ketamine also have the profile O$\Uparrow $ and in C$\Downarrow $ (insignificance of deviation of N$\Uparrow $ may be caused by the sample size). 
     \item Users of amyl nitrite have profile A$\Downarrow $, C$\Downarrow $.
     \item The profile with C$\Uparrow $ does not exist among user profiles. 
     \end{itemize}}
     The hypothesis we make above about types of risky behaviour is partially supported by the data. Moreover, we suggest that the type  E$\Uparrow $, C$\Downarrow $  (Impulsives, Hedonists) is more typical among ecstasy and LSD consumers, whereas the type  N$\Uparrow $,  A$\Downarrow $ is more expected among heroin  users. Detailed comparison of   ecstasy and heroin users demonstrates that they are significantly different. Heroin users have higher N, and lower E and A. We also suggest that a high O-score is typical for many drug users (besides users of heroin, crack, and amyl nitrite) and therefore the O score cannot be excluded from the typology of risky behaviour. Moreover, very low  A$\Downarrow $ is typical for VSA users. This is especially interesting because low A is the main significant predictor of violence in FFM  and  is central to the dark behaviours \cite{Pailing2014}.  These comments may help in the further development of the typology of risky behaviour.
\end{enumerate}

We tested eight types of classifiers for each drug for the decade-based user definition. Leave one out cross-validation (LOOCV) was used to evaluate sensitivity and specificity.
 In this study we select as the best, the classification method which provides the maximal value of the least of sensitivity and specificity. If there is a tie on this basis, as there is in two cases, the method with maximal sum of the sensitivity and specificity is selected as the best.  There were classifiers with sensitivity and specificity greater than 70\% for the decade-based user/non-user separation for all drugs except magic mushrooms, alcohol, and cocaine  (Table~\ref{tab:12}). This accuracy was unexpectedly high for this type of problem. The poorest result was obtained for the prediction of alcohol consumption.

The best set of input features was defined for each drug (Table~\ref{tab:12}). An exhaustive search was performed to select the most effective subset of input features, and the best data mining methods to classify users and non-users for each drug.   There were 10 input features. Each of them is an important factor for risk evaluation for the use of some drugs. However, there was no single most effective classifier using all input features. The maximal number of  attributes used in the best classifiers is six (out of 10) and the minimal number is two.

Table~\ref{tab:12} shows the best sets of attributes  for user/non-user classification for different drugs and for the decade-based classification problem. This table, together with its analogues for pleiades of drugs and all decade-year-month-week classification problems  (Table~\ref{tab:12a}), are important outputs of the analysis.

The decision tree (DT) for crack consumption used only two features, E and C, and provided sensitivity of 80.63\%, and specificity of 78.57\%. The DT for VSA  used age, Edu., E, A, C, and SS, and provided sensitivity 83.48\% and specificity 77.64\% (Table~\ref{tab:12}).

Age was employed in the best classifiers for 14 drugs for the decade-based classification problem, and so was a very widely used feature. Gender was used in the best methods for 10 drugs. We found some unexpected outcomes. For example, the fraction of females which are alcohol users is greater than the fraction of males but a majority of males consume caffeine (coffee).

Most of the features which are not used in the best classifiers are redundant but not uninformative.
For example, the best classifier for ecstasy consumption used age, SS, and gender, and had sensitivity 76.17\% and specificity 77.1	6\%. There is another DT which utilizes age, Edu., O, C, and SS with sensitivity 77.23\% and specificity 75.22\%; a DT with inputs age, Edu., E, O, and A, with sensitivity 73.24\% and specificity 78.22\%, and an advanced $k$NN classifier with inputs age, Edu., N, E, O, C, Imp., SS, and gender, with sensitivity 75.63\% and specificity 75.75\%. This means that for evaluating the risk of ecstasy usage all input attributes are informative but the required information can be extracted from a subset of attributes.

We have demonstrated that there are three groups of drugs with strongly correlated consumption. That is, drug usage has a `modular structure'. The idea of merging correlated attributes into `modules' is popular in biology. These modules are called the `correlation pleiades' \cite{Terentjev31,Berg60,Mitteroecker07} (see Section `Pleiades of drugs').
The modular structure contains three modules: the heroin pleiad, ecstasy pleiad, and benzodiazepines pleiad:
\begin{itemize}
\item The  \emph{Heroin pleiad (heroinPl)} includes crack, cocaine, methadone, and heroin.
\item The \emph{Ecstasy pleiad (ecstasyPl)} includes amphetamines, cannabis, cocaine, ketamine, LSD, magic mushrooms, legal highs, and ecstasy.
\item The \emph{Benzodiazepines pleiad (benzoPl)} contains methadone, amphetamines, and cocaine.
\end{itemize}
The modular structure is well represented in the correlation graph Fig~\ref{Strongdrugusfig:5}.
We define groups of users and non-users for each pleiad. In most of the databases the classes of users and non-users for most of the individual drugs are imbalanced (see Table~\ref{tab:1a}), but merging the users of all drugs into one class `drug users' does not seem to be the best solution because of physiological, psychological, and cultural differences in the usage of different drugs.
We propose instead to use correlation pleiades for the analysis of drug usage as a solution to the class imbalance problem because for all three pleiades the classes of users and non-users are better balanced (Table~\ref{tab:13a}), and the consumption of different drugs from the same pleiad is correlated.

We have applied the eight methods described in the `Risk evaluation methods' Section and selected the best one for each problem for each of the pleiades. The results of the classifier selection are presented in Table~\ref{tab:12a} and the quality of the classification is high. The majority of the best classifiers for pleiades of drugs has a better accuracy than the classifiers for individual drug usage (see Tables \ref{tab:12} and~\ref{tab:12a}). The best classifiers for pleiades of drugs use more input features than the best classifiers for the corresponding individual drugs. The classification problems for pleiades of drugs are more balanced. Therefore, we expect that the classifiers for pleiades are more robust than the classifiers for individual drugs.

The user/non-user classifiers can also be used for the formation of risk maps. Risk maps are useful tools for the visualisation of data and for generating hypotheses about the problem under consideration.

The {\em limitations} of the proposed analysis are obvious:
\begin{itemize}
\item Collection of data through the Internet with elements of the snowball technology (existing participants recruit future participants from amongst their acquaintances) created a biased sample. 
\item Our definitions of `substance use' do not assume any form of dependence, and are much more inclusive and wider than the definition of substance dependence disorder \cite{DSM52013}. Therefore, for example, the majority of participants  were considered as users of alcohol and caffeine. For these substances, non-users form special categories of interest. Similar differences in definitions are common problems. For example, standard epidemiological surveys often use data from official records with unavoidable potential biases, whereas personal interviews give qualitatively different information. 
\end{itemize}

Now, after we finished this book, we see that our analysis is not exhaustive and we are ready to restart. Many questions should be answered. Some of them require just an extension of our analysis: for example, analyse differences between heroin and cocaine users or between cannabis and nicotine users. Some questions are qualitatively different. For example, we considered groups of substance users but did not consider groups of poly drug users separately. Nevertheless, these groups are not small (Figs.~\ref{Fig:VennHE}, \ref{Fig:BEHVenn}) and require special attention because of synergistic effects between different drugs. In a comparative analysis of ecstasy and heroin users we have to create classifiers for three groups: users of heroin but not ecstasy, users of ecstasy but not heroin, and users of ecstasy and heroin. For three drugs  (Fig.~\ref{Fig:BEHVenn}) we have to analyse seven groups. Such expansion of the data analysis will increase the research report enormously and will, perhaps, require a book which is three times longer.
%%%%%%%%%%%%%%%%%%%%%%%%%%%%%%%%%%%%%%%%%%%%%%%%%%

\chapter*{Discussion }
\markboth{Discussion }{Discussion }
\addcontentsline{toc}{chapter}{Discussion }

These results are important as they examine the question of the relationship between drug use and personality comprehensively and engage in the challenge of untangling correlated personality traits (the FFM, impulsivity, and sensation-seeking \cite{Whiteside2001}), and clusters of substance misuse (the correlation pleiades).  The work acknowledged the breadth of a common behaviour which may be transient and leave no impact, or may significantly harm an individual.  We examined drug use behaviour comprehensively in terms of the many kinds of substances that may be used (from the legal and anodyne, to the deeply harmful), as well as the possibility of behavioural over-claiming. Through inclusion of different timescales we were able to explore persistence of use, perhaps related to trends and fashions (e.g. the greater use of LSD in the 1960s and 1970s, the rise of ecstasy in the 1980s, some people being one-off experimenters with recreational drugs, and others using such substances on a daily basis).

We defined substance use in terms of behaviour rather than legality, as legislation in the field is variable.  Our data were gathered before  ‘ legal highs’ emerged as a health concern \cite{Gibbons2012} so we did not differentiate, for example, synthetic cannabinoids and cathinone-based stimulants; these substances have since been widely made illegal.  We were nevertheless able to accurately classify users of these substances (reciprocally, our data were gathered before cannabis decriminalisation in parts of North America, but again, we were able to accurately classify cannabis users).  We included control participants who had never used these substances, those who had used them in the distant past, up to and including persons who had used the drug in the past day, avoiding the Procrustean data-gathering and classifying methods which may prevent an accurate picture of drug use behaviour and risk \cite{Nutt2007}.  Such rich data and the complex methods used for analysis necessitated a large and substantial sample.

The study was atheoretical regarding the morality of the behaviour, and did not medicalise or pathologise participants, optimising engagement by persons with heterogeneous drug-use histories.  Our study used a rigorous range of data-mining methods beyond those typically used in studies examining the association of drug use and personality in the psychological and psychiatric literature, revealing that decision tree methods were most commonly effective for classifying drug users.  We found that high N, low A, and low C are the most common personality correlates of drug use, these traits being sometimes seen, in combination, as an indication of higher-order stability and behavioural conformity, and, inverted, are associated with externalisation of distress \cite{Digman1997,DeYoung2002,DeYoung2008}. Deviation from this rule (N$\Uparrow$,  A$\Downarrow$, C$\Downarrow$) for some drugs is also interesting. LSD use correlates with high O and low C (and does not correlate significantly with high N, at least, for recent LSD users). High O is correlated with the use of many drugs. Nevertheless, there exist significant differences between profiles of different drugs. We analysed this difference in detail for benzodiazepines, ecstasy, and heroin. This may be important for risk assessment and treatment planning.

 Low stability is also a marker of negative urgency \cite{Settles12} whereby people act rashly when distressed.  Our research points to the importance of individuals acquiring emotional self-management skills preceding distress as a means to reduce self-medicating drug-using behaviour, and the risk to health that injudicious or chronic drug use may cause.

In this book we have hopefully demonstrated how to use data analytic methods for investigating a fascinating data set, and found some really interesting patterns in users of different drugs. Of course we have not discovered all of the information that is hiding in the data. We invite other interested parties, be they undergraduate, post graduate, or full time researcher, to join in with us in exploring this data to uncover more interesting relationships, which may help various agencies to manage the challenges associated with drug use.

%% file: appendix.tex
%%%%%%%%%%%%%%%%%%%%% appendix.tex %%%%%%%%%%%%%%%%%%%%%%%%%%%%%%%%%%Final
% sample appendix
%FINAL
% Use this file as a template for your own input.
%
%%%%%%%%%%%%%%%%%%%%%%%% Springer-Verlag %%%%%%%%%%%%%%%%%%%%%%%%%%

\appendix
\motto{}
\chapter{Main tables}
\label{Appendix 1} % Always give a unique label
%\chaptermark{}
% to alter or adjust the chapter heading in the running head
\abstract*{Psychological profiles of drug users and non-users, correlation coefficients, coefficients of linear discriminants between users and nonusers, and other tables are presented.}
\section{Psychological profiles of drug users and non-users}

{ Mean for groups of users and non-users.} In this Appendix, we present mean T-$score_{sample}$ (\ref{eq:2}) for groups of users and non-users for decade, year, month, and week based user definitions respectively. Column $p$-value assesses the significance of differences of  mean scores for groups of users and non-users: it is the probability of observing by chance the same or greater differences for mean scores if both groups have the same mean. Rows `\#' contain number of users and non-users for the drugs. These tables include all information about five factor personality profiles for various definitions of users of 18 drugs, and four groups of drugs:
\begin{itemize}
\item The heroin pleiad: crack, cocaine, methadone, and heroin;
\item The ecstasy pleiad:  amphetamines, cannabis, cocaine, ketamine, LSD, magic mushrooms, legal highs, and ecstasy;
\item The benzodiazepines pleiad includes methadone, amphetamines, cocaine, and benzodiazepines.
\item  The union of these three pleiades and Volatile Substance Abuse (VSA): amphetamines, amyl nitrite, benzodiazepines, cannabis,  cocaine,  crack, ecstasy, heroin, ketamine, legal highs, LSD, methadone, magic mushrooms (MMushrooms),  and VSA, which we call for short `illicit drugs' (with some abuse of language).
\end{itemize}

\def\mysp{\hskip.4em\relax}

\footnotesize
\begin{center}
\begin{longtable}{|c||>{\mysp}c<{\mysp}>{\mysp}c<{\mysp}>{\mysp}c<{\mysp}>{\mysp}c<{\mysp}>{\mysp}r@{\:}||>{\mysp}c<{\mysp}>{\mysp}c<{\mysp}>{\mysp}c<{\mysp}>{\mysp}c<{\mysp}>{\mysp}r@{\:}|}
\caption{Mean T-$score_{sample}$ (MT) and 95\% CI for it for groups of users and non-users with decade and year based definitions of users}
\label{table4}
\\\hline
Factor&\multicolumn{5}{c||}{Decade based}&\multicolumn{5}{c|}{Year based}\\\cline{2-11}
&\multicolumn{2}{c|}{Users}&\multicolumn{2}{c|}{Non-users}& \multicolumn{1}{c||}{$p$-value}&\multicolumn{2}{c|}{Users}&\multicolumn{2}{c|}{Non-users}& \multicolumn{1}{c|}{$p$-value}\\\cline{2-5}\cline{7-10}
&\multicolumn{1}{c|}{MT}&\multicolumn{1}{c|}{95\% CI}&\multicolumn{1}{c|}{MT}&\multicolumn{1}{c|}{95\% CI}& &\multicolumn{1}{c|}{MT}&\multicolumn{1}{c|}{95\% CI}&\multicolumn{1}{c|}{MT}&\multicolumn{1}{c|}{95\% CI}&\\\hline
\endfirsthead
\multicolumn{11}{c} {\tablename\ \thetable.\textit{ Continued}} \\\hline
Factor&\multicolumn{5}{c||}{Decade based}&\multicolumn{5}{c|}{Year based}\\\cline{2-11}
&\multicolumn{2}{c|}{Users}&\multicolumn{2}{c|}{Non-users}& \multicolumn{1}{c||}{$p$-value}&\multicolumn{2}{c|}{Users}&\multicolumn{2}{c|}{Non-users}& \multicolumn{1}{c|}{$p$-value}\\\cline{2-5}\cline{7-10}
&\multicolumn{1}{c|}{MT}&\multicolumn{1}{c|}{95\% CI}&\multicolumn{1}{c|}{MT}&\multicolumn{1}{c|}{95\% CI}& &\multicolumn{1}{c|}{MT}&\multicolumn{1}{c|}{95\% CI}&\multicolumn{1}{c|}{MT}&\multicolumn{1}{c|}{95\% CI}&\\\hline
\endhead
\hline
\multicolumn{11}{r}{\textit{Continued on the next page}} \\
\endfoot
\hline
\endlastfoot
&\multicolumn{10}{c|}{Alcohol}\\*\nobreakhline
\# &\multicolumn{2}{c}{1817}&\multicolumn{2}{|c|}{68}&&\multicolumn{2}{c}{1749}&\multicolumn{2}{|c|}{136}&\\*\nobreakhline
N  &50.07&49.61, 50.53&48.12&45.70, 50.54& $0.119$&49.96&49.49, 50.43&50.55&48.87, 52.23& $0.501$\\*
E  &50.00&49.54, 50.46&50.00&47.63, 52.36& $0.997$&50.12&49.64, 50.59&48.52&46.89, 50.14& $0.064$\\*
O  &50.04&49.58, 50.50&48.86&46.51, 51.22& $0.331$&50.07&49.60, 50.54&49.11&47.46, 50.76& $0.271$\\*
A  &49.91&49.45, 50.37&52.49&50.24, 54.74& $0.028$&49.95&49.48, 50.42&50.70&49.03, 52.37& $0.392$\\*
C  &49.88&49.42, 50.34&53.28&51.01, 55.54& $0.004$&49.94&49.47, 50.41&50.73&49.03, 52.44& $0.378$\\*
Imp&50.14&49.68, 50.60&46.23&44.09, 48.37& $0.001$&50.08&49.61, 50.55&48.96&47.31, 50.62& $0.203$\\
\hline
\end{longtable}
\end{center}

\begin{center}
\begin{longtable}{|c||>{\mysp}c<{\mysp}>{\mysp}c<{\mysp}>{\mysp}c<{\mysp}>{\mysp}c<{\mysp}>{\mysp}r@{\:}||>{\mysp}c<{\mysp}>{\mysp}c<{\mysp}>{\mysp}c<{\mysp}>{\mysp}c<{\mysp}>{\mysp}r@{\:}|}
\caption{Mean T-$score_{sample}$ (MT) and 95\% CI for it for groups of users and non-users with month and week based definitions of users}
\label{table4b}
\\\hline
Factor&\multicolumn{5}{c||}{Month based}&\multicolumn{5}{c|}{Week based}\\\hline
&\multicolumn{2}{c|}{Users}&\multicolumn{2}{c|}{Non-users}& \multicolumn{1}{c||}{$p$-value}&\multicolumn{2}{c|}{Users}&\multicolumn{2}{c|}{Non-users}& \multicolumn{1}{c|}{$p$-value}\\\cline{2-5}\cline{7-10}
&\multicolumn{1}{c|}{MT}&\multicolumn{1}{c|}{95\% CI}&\multicolumn{1}{c|}{MT}&\multicolumn{1}{c|}{95\% CI}& &\multicolumn{1}{c|}{MT}&\multicolumn{1}{c|}{95\% CI}&\multicolumn{1}{c|}{MT}&\multicolumn{1}{c|}{95\% CI}&\\\hline
\endfirsthead
\multicolumn{11}{c} {\tablename\ \thetable.\textit{ Continued}} \\\hline
Factor&\multicolumn{5}{c||}{Month based}&\multicolumn{5}{c|}{Week based} \\\hline
&\multicolumn{2}{c|}{Users}&\multicolumn{2}{c|}{Non-users}& \multicolumn{1}{c||}{$p$-value}&\multicolumn{2}{c|}{Users}&\multicolumn{2}{c|}{Non-users}& \multicolumn{1}{c|}{$p$-value}\\\cline{2-5}\cline{7-10}
&\multicolumn{1}{c|}{MT}&\multicolumn{1}{c|}{95\% CI}&\multicolumn{1}{c|}{MT}&\multicolumn{1}{c|}{95\% CI}& &\multicolumn{1}{c|}{MT}&\multicolumn{1}{c|}{95\% CI}&\multicolumn{1}{c|}{MT}&\multicolumn{1}{c|}{95\% CI}&\\\hline
\endhead
\hline
\multicolumn{11}{r}{\textit{Continued on the next page}} \\
\endfoot
\hline
\endlastfoot

&\multicolumn{10}{c|}{ {Alcohol}}\\*\nobreakhline
\# &\multicolumn{2}{c}{1551}&\multicolumn{2}{|c|}{334}&&\multicolumn{2}{c}{1264}&\multicolumn{2}{|c|}{621}&\\*\nobreakhline
N  &49.85&49.35, 50.35&50.69&49.61, 51.78& $0.167$&49.82&49.28, 50.36&50.37&49.56, 51.19& $0.267$\\*
E  &50.62&50.12, 51.11&47.14&46.11, 48.17&$<0.001$&50.90&50.35, 51.44&48.17&47.38, 48.97&$<0.001$\\*
O  &50.22&49.73, 50.72&48.96&47.87, 50.06& $0.040$&50.08&49.54, 50.63&49.83&49.03, 50.62& $0.602$\\*
A  &50.10&49.60, 50.60&49.55&48.50, 50.60& $0.357$&50.05&49.50, 50.60&49.89&49.09, 50.69& $0.745$\\*
C  &50.15&49.66, 50.65&49.28&48.22, 50.35& $0.146$&50.19&49.64, 50.74&49.61&48.81, 50.42& $0.244$\\*
Imp&50.06&49.57, 50.56&49.71&48.61, 50.82& $0.571$&50.19&49.64, 50.73&49.62&48.81, 50.43& $0.253$\\
\hline
\end{longtable}
\end{center}
\normalsize

\begin{landscape}
\section{Correlation between consumption of different drugs}
\label{Appendix 2}
 % Always give a unique label
%\chaptermark{}
% to alter or adjust the chapter heading in the running head

In this section we show Pearson's correlation coefficients (PCCs) between drug consumptions for  decade and year based user/non-user separation.

\def\mysp{\hskip.35em\relax}
\begin{table}
\footnotesize{
\caption {PCCs between drug consumptions with decade based user/non-user separation. Benz. stays for for Benzodiazepines.}
\label{tab:14}
\begin{tabular}{|l|r<{\mysp}r<{\mysp}r<{\mysp}r<{\mysp}r<{\mysp}r<{\mysp}r<{\mysp}r<{\mysp}rr<{\mysp}r<{\mysp}r<{\mysp}rr<{\mysp}r<{\mysp}rr|}
\hline
\multicolumn{1}{|c}{\rotatebox{90}{Drug}}&\multicolumn{1}{|c}{\rotatebox{90}{Amphetamines}}&\multicolumn{1}{|c}{\rotatebox{90}{Amyl nitrite}}&\multicolumn{1}{|c}{\rotatebox{90}{Benzodiazepines}}&\multicolumn{1}{|c}{\rotatebox{90}{Cannabis}}&\multicolumn{1}{|c}{\rotatebox{90}{Chocolate}}&\multicolumn{1}{|c}{\rotatebox{90}{Cocaine}}&\multicolumn{1}{|c}{\rotatebox{90}{	Caffeine}}&\multicolumn{1}{|c}{\rotatebox{90}{	Crack}}&\multicolumn{1}{|c}{\rotatebox{90}{Ecstasy}}&	\multicolumn{1}{|c}{\rotatebox{90}{Heroin}}&\multicolumn{1}{|c}{\rotatebox{90}{Ketamine}}&\multicolumn{1}{|c}{\rotatebox{90}{Legal highs}}&\multicolumn{1}{|c}{\rotatebox{90}{LSD}}&\multicolumn{1}{|c}{\rotatebox{90}{	Methadone}}&\multicolumn{1}{|c}{\rotatebox{90}{MMushrooms}}&\multicolumn{1}{|c}{\rotatebox{90}{Nicotine}}&\multicolumn{1}{|c|}{\rotatebox{90}{VSA}} \\\hline
Alcohol     &$0.074^1$&$0.074^1$&$0.051^2$&$0.119^1$&$0.099^1$&$0.111^1$&$0.157^1$&$0.027^4$&$0.105^1$& $0.033^4$&$0.078^2$&$0.061^2$&$0.069^2$&$-0.007^4$&$0.071^1$&$0.113^1$& $0.046^3$\\\hline
Amphetamines&         &$0.372^1$&$0.463^1$&$0.469^1$&$0.013^4$&$0.580^1$&$0.106^1$&$0.323^1$&$0.597^1$& $0.359^1$&$0.412^1$&$0.481^1$&$0.490^1$& $0.415^1$&$0.481^1$&$0.343^1$& $0.304^1$\\\hline
Amyl nitrite&         &         &$0.226^1$&$0.292^1$&$0.028^4$&$0.381^1$&$0.060^2$&$0.144^1$&$0.392^1$& $0.137^1$&$0.345^1$&$0.268^1$&$0.213^1$& $0.084^1$&$0.271^1$&$0.196^1$& $0.130^1$\\\hline
Benz.       &         &         &         &$0.354^1$&$0.006^4$&$0.428^1$&$0.055^2$&$0.326^1$&$0.383^1$& $0.395^1$&$0.303^1$&$0.348^1$&$0.352^1$& $0.468^1$&$0.366^1$&$0.260^1$& $0.294^1$\\\hline
Cannabis    &         &         &         &         &$0.046^3$&$0.453^1$&$0.113^1$&$0.216^1$&$0.521^1$& $0.217^1$&$0.302^1$&$0.526^1$&$0.421^1$& $0.299^1$&$0.497^1$&$0.533^1$& $0.237^1$\\\hline
Chocolate	&         &         &         &         &         &$0.006^4$&$0.122^1$&$0.032^4$&$0.040^4$&$-0.026^4$&$0.035^4$&$0.017^4$&$0.029^4$& $0.007^4$&$0.024^4$&$0.037^4$&$-0.021^4$\\\hline
Cocaine	    &         &         &         &         &         &         &$0.099^1$&$0.396^1$&$0.633^1$& $0.414^1$&$0.454^1$&$0.445^1$&$0.442^1$& $0.354^1$&$0.480^1$&$0.362^1$& $0.277^1$\\\hline
Caffeine    &         &         &         &         &         &         &         &$0.035^4$&$0.107^1$& $0.026^4$&$0.058^3$&$0.085^1$&$0.075^1$& $0.039^4$&$0.100^1$&$0.145^3$& $0.053^3$\\\hline
Crack       &         &         &         &         &         &         &         &         &$0.280^1$& $0.509^1$&$0.255^1$&$0.203^1$&$0.268^1$& $0.367^1$&$0.276^1$&$0.191^1$& $0.278^1$\\\hline
Ecstasy     &         &         &         &         &         &         &         &         &         & $0.301^1$&$0.511^1$&$0.586^1$&$0.599^1$& $0.315^1$&$0.599^1$&$0.370^1$& $0.289^1$\\\hline
Heroin      &         &         &         &         &         &         &         &         &         &          &$0.274^1$&$0.237^1$&$0.347^1$& $0.494^1$&$0.306^1$&$0.185^1$& $0.293^1$\\\hline
Ketamine    &         &         &         &         &         &         &         &         &         &          &         &$0.393^1$&$0.462^1$& $0.246^1$&$0.436^1$&$0.243^1$& $0.192^1$\\\hline
Legal highs &         &         &         &         &         &         &         &         &         &          &         &         &$0.519^1$& $0.334^1$&$0.575^1$&$0.364^1$& $0.314^1$\\\hline
LSD         &         &         &         &         &         &         &         &         &         &          &         &         &         & $0.343^1$&$0.680^1$&$0.289^1$& $0.299^1$\\\hline
Methadone   &         &         &         &         &         &         &         &         &         &          &         &         &         &          &$0.343^1$&$0.234^1$& $0.277^1$\\\hline
MMushrooms  &         &         &         &         &         &         &         &         &         &          &         &         &         &          &         &$0.324^1$& $0.253^1$\\\hline
Nicotine    &         &         &         &         &         &         &         &         &         &          &         &         &         &          &         &         & $0.221^1$\\\hline
\end{tabular}\\
Note: $^1p <0.001$, $^2p <0.01$, $^3p <0.05$, $^4p >0.05$. \emph{p}-value is the probability to observe by chance the same or greater correlation for uncorrelated variables.}
\end{table}
\end{landscape}

\begin{landscape}
\def\mysp{\hskip.26em\relax}

\begin{table}
\footnotesize{
\caption {PCCs between drug consumptions with year based user/non-user separation. Benz. stays for for Benzodiazepines.}
\label{tab:14a}
\begin{tabular}{|l|r<{\mysp}r<{\mysp}r<{\mysp}rr<{\mysp}r<{\mysp}rr<{\mysp}rr<{\mysp}r<{\mysp}r<{\mysp}rr<{\mysp}r<{\mysp}rr|}
%                   1       2           3     45        6        78        90        1        2        34             
\hline
\multicolumn{1}{|c}{\rotatebox{90}{Drug}}&\multicolumn{1}{|c}{\rotatebox{90}{Amphetamines}}&\multicolumn{1}{|c}{\rotatebox{90}{Amyl nitrite}}&\multicolumn{1}{|c}{\rotatebox{90}{Benzodiazepines}}&\multicolumn{1}{|c}{\rotatebox{90}{Cannabis}}&\multicolumn{1}{|c}{\rotatebox{90}{Chocolate}}&\multicolumn{1}{|c}{\rotatebox{90}{Cocaine}}&\multicolumn{1}{|c}{\rotatebox{90}{	Caffeine}}&\multicolumn{1}{|c}{\rotatebox{90}{	Crack}}&\multicolumn{1}{|c}{\rotatebox{90}{Ecstasy}}&	\multicolumn{1}{|c}{\rotatebox{90}{Heroin}}&\multicolumn{1}{|c}{\rotatebox{90}{Ketamine}}&\multicolumn{1}{|c}{\rotatebox{90}{Legal highs}}&\multicolumn{1}{|c}{\rotatebox{90}{LSD}}&\multicolumn{1}{|c}{\rotatebox{90}{	Methadone}}&\multicolumn{1}{|c}{\rotatebox{90}{MMushrooms}}&\multicolumn{1}{|c}{\rotatebox{90}{Nicotine}}&\multicolumn{1}{|c|}{\rotatebox{90}{VSA}} \\\hline
Alcohol     &$0.046^4$&$0.061^3$&$0.048^2$&$0.078^3$& $0.077^1$&$0.124^1$&$0.111^1$& $0.058^1$&$0.107^1$& $0.030^1$&$0.072^4$&$0.093^1$&$0.084^1$&$-0.005^1$&$0.075^4$&$0.081^1$&$0.055^3$\\\hline
Amphetamines&         &$0.222^1$&$0.436^1$&$0.421^1$& $0.003^4$&$0.453^1$&$0.072^2$& $0.193^1$&$0.461^1$& $0.305^1$&$0.325^1$&$0.471^1$&$0.392^1$& $0.382^1$&$0.375^1$&$0.311^1$&$0.173^1$\\\hline
Amyl nitrite&         &         &$0.199^1$&$0.185^1$& $0.016^4$&$0.262^1$&$0.050^3$& $0.077^1$&$0.276^1$& $0.100^1$&$0.280^1$&$0.277^1$&$0.120^1$& $0.091^1$&$0.159^1$&$0.139^1$&$0.107^1$\\\hline
Benz.       &         &         &         &$0.334^1$&$-0.009^4$&$0.365^1$&$0.062^2$& $0.232^1$&$0.304^1$& $0.318^1$&$0.263^1$&$0.318^1$&$0.212^1$& $0.464^1$&$0.271^1$&$0.261^1$&$0.183^1$\\\hline
Cannabis    &         &         &         &         & $0.020^4$&$0.392^1$&$0.074^2$& $0.165^1$&$0.484^1$& $0.199^1$&$0.277^1$&$0.516^1$&$0.433^1$& $0.301^1$&$0.470^1$&$0.517^1$&$0.164^1$\\\hline
Chocolate   &         &         &         &         &          &$0.008^4$&$0.089^1$&$-0.037^4$&$0.057^3$&$-0.003^4$&$0.000^4$&$0.026^4$&$0.044^4$& $0.006^4$&$0.019^4$&$0.009^4$&$-0.012^4$\\\hline
Cocaine     &         &         &         &         &          &         &$0.069^2$& $0.322^1$&$0.535^1$& $0.358^1$&$0.379^1$&$0.394^1$&$0.302^1$& $0.314^1$&$0.346^1$&$0.329^1$&$0.187^1$\\\hline
Caffeine    &         &         &         &         &          &         &         & $0.023^4$&$0.059^3$& $0.023^4$&$0.026^4$&$0.067^3$&$0.032^4$& $0.027^4$&$0.050^3$&$0.105^1$&$0.042^4$\\\hline
Crack       &         &         &         &         &          &         &         &          &$0.156^1$& $0.350^1$&$0.180^1$&$0.147^1$&$0.139^1$& $0.265^1$&$0.181^1$&$0.126^1$&$0.145^1$\\\hline
Ecstasy     &         &         &         &         &          &         &         &          &         & $0.190^1$&$0.455^1$&$0.502^1$&$0.509^1$& $0.245^1$&$0.480^1$&$0.343^1$&$0.174^1$\\\hline
Heroin      &         &         &         &         &          &         &         &          &         &          &$0.217^1$&$0.185^1$&$0.170^1$& $0.385^1$&$0.171^1$&$0.149^1$&$0.121^1$\\\hline
Ketamine    &         &         &         &         &          &         &         &          &         &          &         &$0.373^1$&$0.351^1$& $0.202^1$&$0.362^1$&$0.222^1$&$0.151^1$\\\hline
Legalhighs  &         &         &         &         &          &         &         &          &         &          &         &         &$0.434^1$& $0.309^1$&$0.485^1$&$0.348^1$&$0.220^1$\\\hline
LSD         &         &         &         &         &          &         &         &          &         &          &         &         &         & $0.234^1$&$0.627^1$&$0.267^1$&$0.174^1$\\\hline
Methadone   &         &         &         &         &          &         &         &          &         &          &         &         &         &          &$0.253^1$&$0.211^1$&$0.167^1$\\\hline
MMushrooms  &         &         &         &         &          &         &         &          &         &          &         &         &         &          &         &$0.282^1$&$0.174^1$\\\hline
Nicotine    &         &         &         &         &          &         &         &          &         &          &         &         &         &          &         &         &$0.145^1$\\\hline

\end{tabular}\\
Note: $^1p <0.001$, $^2p <0.01$, $^3p <0.05$, $^4p >0.05$. \emph{p}-value is the probability to observe by chance the same or greater correlation for uncorrelated variables.}
\end{table}
\end{landscape}

\section{Linear discriminants for user/non-user separation}
\normalsize{

Linear discriminants separate users from non-users  by linear inequalities:}
$$\Theta+\sum c_i z_i > 0$$ for users and $\leq 0$ for non-users, where $\Theta$ are the threscholds, $z_i$ are the attributes, and $c_i$ are the coefficients. Tables \ref{TabDiscr1}-\ref{TabDiscr4} contain the coefficients $c_i$ of linear discriminants for user/nonuser separation in 10-dimensional space (7 psychological attributes, age, education, gender). The attributes in these tables are quantified and transformed to $z$-scores with zero mean and unite variance (positive values of the Gndr $z$-score corresponds to female). The last rows of the tables include the standard deviaton of the coefficients in LOOCV. For 7-dimensional space of psychological attributes taken separately (T-scores), the linear discriminants are presented in tables \ref{TabDiscr5}-\ref{TabDiscr8}.

Performance of linear discriminants in user/non-user separation is evaluated by several methods (tables \ref{Tab:DiscrPerf1}--\ref{Tab:DiscrPerf4} for 10-dimensional data space and tables \ref{Tab:DiscrPerf5}--\ref{Tab:DiscrPerf8} for 7-dimensional space of T-scores of psychological attributes). First of all, we calculated the linear discriminant using the whole sample (see tables \ref{TabDiscr1}--\ref{TabDiscr4}) and find all their errors. For each solution of the classification problem we have several numbers, $P$ (positive), the number of samples recognised as positive, and  $N$ - negative, the number of samples recognised as negative. $P+N$ is the total number of samples. $P$=TP+FP (True Positive plus False Positive) and $N$=TN+FN (True Negative plus False Negative). Sensitivity is Sn=TP/(TP+FN)$\times$100\% and Specificity is Sp=TN/(TN+FP)$\times$100\%. Accuracy is Acc=(TP+TN)/($P$+$N$)$\times$100\%. We calculate these performance indicators for the total sample and for the LOOCV procedure. In LOOCV the linear discriminant is calculated for the set of all samples excluding the example left out for testing. The test was performed for all samples with the corresponding redefining of Sn, Sp, and Acc. In LOOCV the linear discriminants are calculated for each testing example. Each of these discriminants is a separate classification model. Stability of classification can be measured by the number of examples which change their class at least once. We took the basis model for the total sample and find how many true positive examples of this model became  FN examples of a LOOCV model at least once. This number measured in \% of  TP+FP of the basic model is TP$\to$FN. Analogously, we defined FP$\to$TN,  TN$\to$FP, and FN$\to$TP. The last two numbers are measured in \% of  TN+FN of the basic model.

\footnotesize{
\begin{table}\centering
\caption{Coefficients of linear discriminant for user/non-user separation  and decade-based definition of users (10 attributes) \label{TabDiscr1}}
\begin{tabular}{|l|r|r|r|r|r|r|r|r|r|r|r|}\hline
\multicolumn{1}{|c|}{Drug}&\multicolumn{1}{c|}{$\Theta$}&\multicolumn{1}{c|}{Age}&\multicolumn{1}{c|}{Edu}&\multicolumn{1}{c|}{N}&\multicolumn{1}{c|}{E}& \multicolumn{1}{c|}{O}&\multicolumn{1}{c|}{A}&\multicolumn{1}{c|}{C}&\multicolumn{1}{c|}{Imp}&\multicolumn{1}{c|}{SS}&\multicolumn{1}{c|}{Gndr}\\\hline
Alcohol&0.358&-0.463&0.266&-0.042&0.030&-0.103&-0.117&-0.361&-0.031&0.710&0.230\\\hline
Amphetamines&-0.236&-0.569&-0.236&0.081&-0.146&0.307&-0.017&-0.202&0.119&0.559&-0.362\\\hline
Amyl nitrite&-0.261&-0.549&0.251&0.110&0.180&-0.142&-0.031&-0.229&-0.063&0.556&-0.451\\\hline
Benz.&-0.098&-0.067&-0.243&0.566&-0.152&0.595&-0.164&-0.128&0.088&0.378&-0.214\\\hline
Cannabis&0.434&-0.639&-0.241&-0.043&-0.213&0.518&-0.127&-0.193&0.043&0.352&-0.196\\\hline
Chocolate&0.210&-0.169&-0.387&-0.149&-0.303&0.673&0.135&-0.168&-0.408&0.109&0.170\\\hline
Cocaine&-0.198&-0.441&-0.121&0.220&0.068&0.276&-0.287&-0.235&0.090&0.682&-0.226\\
\hline
\end{tabular}
\end{table}

\begin{table}\centering
\caption{Coefficients of linear discriminant for user/non-user separation  and year-based definition of users (10 attributes) \label{TabDiscr2}}
\begin{tabular}{|l|r|r|r|r|r|r|r|r|r|r|r|}\hline
\multicolumn{1}{|c|}{Drug}&\multicolumn{1}{c|}{$\Theta$}&\multicolumn{1}{c|}{Age}&\multicolumn{1}{c|}{Edu}&\multicolumn{1}{c|}{N}&\multicolumn{1}{c|}{E}& \multicolumn{1}{c|}{O}&\multicolumn{1}{c|}{A}&\multicolumn{1}{c|}{C}&\multicolumn{1}{c|}{Imp}&\multicolumn{1}{c|}{SS}&\multicolumn{1}{c|}{Gndr}\\\hline
Alcohol&0.218&-0.585&0.342&-0.228&0.101&-0.133&-0.031&-0.165&-0.211&0.575&0.241\\\hline
Amphetamines&-0.543&-0.643&-0.249&0.063&-0.176&0.347&-0.103&-0.201&0.241&0.418&-0.293\\\hline
Amyl nitrite&-0.527&-0.647&0.085&0.124&0.075&-0.036&-0.249&-0.142&-0.191&0.420&-0.507\\\hline
Benz.&-0.296&-0.272&-0.263&0.594&-0.110&0.437&-0.187&-0.007&0.102&0.419&-0.282\\\hline
Cannabis&0.130&-0.569&-0.310&-0.069&-0.190&0.521&-0.020&-0.194&-0.034&0.411&-0.245\\\hline
Chocolate&0.169&-0.259&-0.217&0.036&-0.168&0.575&0.064&-0.236&-0.571&0.072&0.367\\\hline
Cocaine&-0.494&-0.685&-0.007&0.211&0.177&0.077&-0.247&-0.175&0.086&0.521&-0.280\\
\hline
\end{tabular}
\end{table}

\begin{table}\centering
\caption{Coefficients of linear discriminant for user/non-user separation  and month-based definition of users (10 attributes) \label{TabDiscr3}}
\begin{tabular}{|l|r|r|r|r|r|r|r|r|r|r|r|}\hline
\multicolumn{1}{|c|}{Drug}&\multicolumn{1}{c|}{$\Theta$}&\multicolumn{1}{c|}{Age}&\multicolumn{1}{c|}{Edu}&\multicolumn{1}{c|}{N}&\multicolumn{1}{c|}{E}& \multicolumn{1}{c|}{O}&\multicolumn{1}{c|}{A}&\multicolumn{1}{c|}{C}&\multicolumn{1}{c|}{Imp}&\multicolumn{1}{c|}{SS}&\multicolumn{1}{c|}{Gndr}\\\hline
Alcohol&0.130&-0.263&0.590&0.096&0.588&-0.111&0.078&-0.083&-0.193&0.402&0.058\\\hline
Amphetamines&-0.543&-0.643&-0.249&0.063&-0.176&0.347&-0.103&-0.201&0.241&0.418&-0.293\\\hline
Amyl nitrite&-0.821&-0.361&-0.229&-0.223&-0.114&-0.178&-0.144&-0.018&-0.088&0.749&-0.365\\\hline
Benz.&-0.416&-0.115&-0.243&0.711&-0.128&0.284&-0.180&0.072&0.167&0.418&-0.292\\\hline
Cannabis&-0.122&-0.542&-0.394&-0.132&-0.166&0.547&-0.037&-0.132&0.015&0.355&-0.250\\\hline
Chocolate&0.132&0.138&-0.284&-0.161&0.107&0.379&-0.004&-0.440&-0.488&0.193&0.501\\\hline
Cocaine&-0.597&-0.624&0.029&0.345&0.212&-0.007&-0.305&0.054&0.062&0.523&-0.270\\
\hline
\end{tabular}
\end{table}

\begin{table}\centering
\caption{Coefficients of linear discriminant for user/non-user separation  and week-based definition of users (10 attributes) \label{TabDiscr4}}
\begin{tabular}{|l|r|r|r|r|r|r|r|r|r|r|r|}\hline
\multicolumn{1}{|c|}{Drug}&\multicolumn{1}{c|}{$\Theta$}&\multicolumn{1}{c|}{Age}&\multicolumn{1}{c|}{Edu}&\multicolumn{1}{c|}{N}&\multicolumn{1}{c|}{E}& \multicolumn{1}{c|}{O}&\multicolumn{1}{c|}{A}&\multicolumn{1}{c|}{C}&\multicolumn{1}{c|}{Imp}&\multicolumn{1}{c|}{SS}&\multicolumn{1}{c|}{Gndr}\\\hline
Alcohol&0.063&0.053&0.640&0.157&0.522&-0.283&0.069&-0.125&-0.105&0.412&-0.099\\\hline
Amphetamines&-0.546&-0.441&-0.070&-0.071&-0.295&0.348&-0.207&-0.359&0.584&0.120&-0.248\\\hline
Amyl nitrite&-0.747&0.023&0.173&-0.389&-0.351&-0.203&0.006&-0.213&0.214&0.695&-0.281\\\hline
Benz.&-0.546&0.027&-0.331&0.680&-0.371&0.273&-0.230&0.091&0.231&0.287&-0.137\\\hline
Cannabis&-0.240&-0.502&-0.402&-0.120&-0.172&0.594&-0.036&-0.108&-0.023&0.335&-0.253\\\hline
Chocolate&0.109&0.390&0.195&0.117&0.444&-0.034&0.114&-0.172&0.283&-0.461&0.513\\\hline
Cocaine&-0.806&-0.543&0.015&0.358&0.274&-0.118&-0.299&-0.048&-0.042&0.534&-0.331\\
\hline
\end{tabular}
\end{table}

\begin{table}\centering
\caption{Performance and stability of linear discriminant for decade-based definition of users  (10 attributes).  \label{Tab:DiscrPerf1}}
\begin{tabular}{|l|>{\quad}c<{\quad}|>{\quad}c<{\quad}|>{\quad}c<{\quad}|>{\quad}c<{\quad}|>{\quad}c<{\quad}|>{\quad}c<{\quad}|c|c|c|c|}\hline
&\multicolumn{3}{c|}{Total sample}&\multicolumn{3}{c|}{LOOCV}&\multicolumn{4}{c|}{Stability indicators}\\\cline{2-11}
 Drug  &Sn &Sp&Acc&Sn &Sp& Acc &TP$\to$FN&FN$\to$TP&FP$\to$TN&TN$\to$FP\\\hline
Alcohol&65.1&67.6&65.2&64.9&57.4&64.6&5.8&8.5&11.8&7.4\\\hline
Amphetamines&72.0&71.9&71.9&71.1&71.5&71.4&1.2&0.9&0.7&0.5\\\hline
Amyl nitrite&63.8&63.6&63.6&62.2&63.0&62.9&2.4&1.6&1.3&1.8\\\hline
Benz.&67.4&67.5&67.4&66.8&67.1&67.0&0.4&0.4&0.6&1.3\\\hline
Cannabis&78.1&78.2&78.1&77.9&77.6&77.8&0.4&0.6&0.8&1.0\\\hline
Chocolate&57.9&57.1&57.9&57.5&37.1&57.1&11.2&14.3&14.1&28.6\\\hline
Cocaine&67.7&67.6&67.6&67.0&67.3&67.2&0.9&0.6&0.5&1.0\\\hline
Caffeine&68.7&70.3&68.8&68.6&62.2&68.5&5.5&8.8&8.1&13.5\\\hline
Crack&69.6&69.1&69.1&65.4&68.9&68.5&5.2&2.1&1.7&1.7\\\hline
Ecstasy&74.4&74.7&74.6&74.2&74.3&74.2&0.3&0.7&0.5&0.4\\\hline
Heroin&70.8&69.9&70.0&68.4&69.8&69.7&2.8&6.1&1.3&2.2\\\hline
Ketamine&68.0&67.8&67.8&66.6&67.5&67.3&1.4&1.4&0.5&1.4\\\hline
Legal highs&79.0&79.2&79.1&78.9&78.9&78.9&0.1&0.8&0.3&0.4\\\hline
LSD&76.8&76.7&76.8&76.5&76.7&76.6&0.5&0.7&0.5&0.7\\\hline
Methadone&70.5&70.8&70.7&69.3&70.2&70.0&1.2&0.7&1.4&0.5\\\hline
MMushrooms&75.6&75.7&75.6&75.1&75.3&75.2&0.9&0.1&0.7&0.5\\\hline
Nicotine&70.1&70.4&70.2&69.7&69.1&69.5&0.8&0.9&1.6&1.1\\\hline
VSA&74.8&74.8&74.8&74.3&74.7&74.5&2.2&2.2&0.7&1.6\\\hline
Heroin pleiad&69.6&70.1&69.9&69.1&69.4&69.3&0.5&0.7&0.9&0.4\\\hline
Ecstasy pleiad&78.6&78.7&78.6&78.6&78.2&78.5&0.5&0.8&0.4&0.4\\\hline
Benz. pleiad&70.7&70.7&70.7&70.2&69.8&70.0&0.9&0.8&0.9&0.8\\ \hline
Illicit drugs&77.9&78.2&78.1&77.5&77.8&77.7&0.4&0.6&0.7&0.9\\\hline
\end{tabular}
\end{table}

\begin{table}\centering
\caption{Performance and stability of linear discriminant for  year-based definition of users  (10 attributes). \label{Tab:DiscrPerf2}}
\begin{tabular}{|l|>{\quad}c<{\quad}|>{\quad}c<{\quad}|>{\quad}c<{\quad}|>{\quad}c<{\quad}|>{\quad}c<{\quad}|>{\quad}c<{\quad}|c|c|c|c|}\hline
&\multicolumn{3}{c|}{Total sample}&\multicolumn{3}{c|}{LOOCV}&\multicolumn{4}{c|}{Stability indicators}\\\cline{2-11}
Drug  &Sn &Sp&Acc&Sn &Sp& Acc &TP$\to$FN&FN$\to$TP&FP$\to$TN&TN$\to$FP\\\hline
Alcohol&63.1&63.2&63.1&62.7&58.1&62.4&3.3&4.6&5.9&4.4\\\hline
Amphetamines&72.5&72.7&72.7&71.6&72.4&72.2&0.9&0.5&0.9&0.9\\\hline
Amyl nitrite&69.2&68.6&68.6&62.4&68.3&67.9&4.5&2.3&2.7&2.6\\\hline
Benz.&69.7&69.4&69.5&68.2&69.3&69.0&1.9&0.6&0.5&0.7\\\hline
Cannabis&79.3&79.3&79.3&78.9&79.0&78.9&0.7&0.8&0.3&0.2\\\hline
Chocolate&58.1&60.0&58.1&57.4&37.8&57.0&13.6&16.1&17.8&13.3\\\hline
Cocaine&71.0&71.5&71.4&69.8&71.1&70.8&1.0&0.2&1.0&1.2\\\hline
Caffeine&68.0&67.2&68.0&67.7&54.1&67.2&8.9&7.3&11.5&6.6\\\hline
Crack&73.4&73.9&73.9&69.6&73.6&73.4&1.3&2.5&3.4&3.7\\\hline
Ecstasy&75.4&75.7&75.6&75.0&75.6&75.4&0.2&0.6&0.8&0.3\\\hline
Heroin&73.7&73.7&73.7&70.3&73.6&73.4&2.5&2.5&2.1&2.3\\\hline
Ketamine&71.6&71.6&71.6&68.8&71.4&71.1&2.9&1.9&2.1&1.9\\\hline
Legal highs&77.7&77.7&77.7&77.0&77.4&77.3&0.9&0.4&0.5&0.5\\\hline
LSD&80.0&80.0&80.0&79.2&79.9&79.7&0.5&0.5&0.3&0.9\\\hline
Methadone&70.6&70.6&70.6&69.7&70.4&70.3&0.9&0.9&0.7&1.7\\\hline
MMushrooms&77.9&77.9&77.9&77.4&77.7&77.7&0.5&0.2&0.2&0.6\\\hline
Nicotine&70.4&70.4&70.4&70.1&70.2&70.1&0.3&0.8&0.4&0.4\\\hline
VSA&72.6&72.7&72.7&71.6&72.5&72.4&2.1&2.1&1.7&3.9\\\hline
Heroin pleiad&71.5&72.0&71.8&71.3&71.7&71.6&0.3&0.7&0.8&1.0\\\hline
Ecstasy pleiad&80.2&80.3&80.2&80.0&80.0&80.0&0.1&0.6&0.3&0.5\\\hline
Benz. pleiad&73.0&72.7&72.8&72.4&72.3&72.4&0.6&0.4&0.5&0.4\\\hline
Illicit drugs&80.2&79.6&79.8&79.5&79.4&79.4&1.0&0.4&0.4&0.3\\\hline
\end{tabular}
\end{table}

\begin{table}\centering
\caption{Performance and stability of linear discriminant for month-based definition of users  (10 attributes).  \label{Tab:DiscrPerf3}}
\begin{tabular}{|l|>{\quad}c<{\quad}|>{\quad}c<{\quad}|>{\quad}c<{\quad}|>{\quad}c<{\quad}|>{\quad}c<{\quad}|>{\quad}c<{\quad}|c|c|c|c|}\hline
&\multicolumn{3}{c|}{Total sample}&\multicolumn{3}{c|}{LOOCV}&\multicolumn{4}{c|}{Stability indicators}\\\cline{2-11}
Drug  &Sn &Sp&Acc&Sn &Sp& Acc &TP$\to$FN&FN$\to$TP&FP$\to$TN&TN$\to$FP\\\hline
Alcohol&60.9&61.1&60.9&60.4&58.4&60.1&2.3&2.6&3.0&2.4\\\hline
Amphetamines&68.9&68.1&68.2&65.5&68.0&67.7&2.9&3.4&1.2&2.0\\\hline
Amyl nitrite&70.7&75.3&75.2&63.4&75.1&74.9&2.4&2.4&8.5&5.2\\\hline
Benz.&68.9&68.7&68.8&68.2&68.5&68.5&1.3&1.3&1.1&2.0\\\hline
Cannabis&78.6&78.3&78.4&77.7&77.9&77.8&1.1&0.0&0.6&0.9\\\hline
Chocolate&57.9&56.6&57.8&57.2&48.5&56.7&9.7&8.0&9.1&8.1\\\hline
Cocaine&69.8&70.0&70.0&67.3&69.9&69.7&3.1&2.5&2.3&2.7\\\hline
Caffeine&60.8&60.3&60.7&60.5&54.5&60.2&7.3&5.2&6.6&5.8\\\hline
Crack&80.0&75.3&75.4&65.0&75.2&75.1&15.0&15.0&4.8&11.6\\\hline
Ecstasy&72.5&72.4&72.4&71.3&72.3&72.1&1.7&1.3&1.2&1.2\\\hline
Heroin&79.2&77.6&77.7&69.8&77.5&77.2&9.4&7.5&3.2&3.8\\\hline
Ketamine&72.2&73.3&73.2&64.6&73.1&72.8&3.8&1.3&3.4&2.7\\\hline
Legal highs&72.6&72.5&72.5&71.4&72.4&72.3&1.2&1.7&0.9&1.2\\\hline
LSD&76.5&76.4&76.4&74.7&76.3&76.2&1.2&1.2&1.0&1.9\\\hline
Methadone&68.4&68.1&68.2&66.7&67.7&67.6&1.8&3.5&1.6&2.6\\\hline
MMushrooms&75.5&74.3&74.4&73.6&74.3&74.2&2.5&2.5&0.9&2.0\\\hline
Nicotine&66.4&66.2&66.3&65.8&66.0&65.9&1.1&0.7&0.3&0.7\\\hline
VSA&76.5&75.6&75.6&58.8&75.6&75.3&5.9&5.9&2.9&6.1\\\hline
Heroin pleiad&69.3&69.4&69.4&68.0&69.2&69.0&1.6&1.3&1.5&0.9\\\hline
Ecstasy pleiad&79.4&79.6&79.5&79.2&79.1&79.2&0.5&0.5&0.7&0.7\\\hline
Benz. pleiad&70.5&70.3&70.3&70.1&69.9&70.0&0.4&0.8&0.7&1.0\\\hline
Illicit drugs&78.9&78.8&78.8&78.4&78.5&78.5&0.4&0.3&0.5&0.4\\\hline
\end{tabular}
\end{table}

\begin{table}\centering
\caption{Performance and stability of linear discriminant for week-based definition of users  (10 attributes).  \label{Tab:DiscrPerf4}}
\begin{tabular}{|l|>{\quad}c<{\quad}|>{\quad}c<{\quad}|>{\quad}c<{\quad}|>{\quad}c<{\quad}|>{\quad}c<{\quad}|>{\quad}c<{\quad}|c|c|c|c|}\hline
&\multicolumn{3}{c|}{Total sample}&\multicolumn{3}{c|}{LOOCV}&\multicolumn{4}{c|}{Stability indicators}\\\cline{2-11}
Drug  &Sn &Sp&Acc&Sn &Sp& Acc &TP$\to$FN&FN$\to$TP&FP$\to$TN&TN$\to$FP\\\hline
Alcohol&60.4&59.9&60.3&59.3&58.9&59.2&2.0&0.9&1.0&1.3\\\hline
Amphetamines&66.9&66.7&66.7&62.0&66.5&66.1&4.3&2.5&3.0&2.8\\\hline
Amyl nitrite&70.6&77.6&77.5&52.9&77.4&77.1&0.0&5.9&14.7&11.8\\\hline
Benz.&70.4&70.1&70.1&67.0&69.8&69.5&3.4&1.7&2.1&2.7\\\hline
Cannabis&75.9&75.9&75.9&75.3&75.5&75.4&0.9&0.0&0.5&0.5\\\hline
Chocolate&57.8&57.7&57.8&57.2&54.4&56.6&3.6&2.5&3.3&1.3\\\hline
Cocaine&75.0&75.1&75.1&61.7&75.1&74.6&8.3&6.7&4.1&4.4\\\hline
Caffeine&59.5&58.1&59.4&58.5&52.0&57.7&5.8&2.8&6.2&5.3\\\hline
Crack&90.9&85.9&85.9&54.5&85.8&85.6&27.3&9.1&4.1&7.7\\\hline
Ecstasy&72.6&71.5&71.5&67.9&71.4&71.2&3.6&8.3&2.4&5.3\\\hline
Heroin&79.3&80.1&80.1&65.5&80.0&79.8&6.9&6.9&4.6&4.8\\\hline
Ketamine&73.0&71.9&71.9&62.2&71.6&71.5&8.1&8.1&3.5&6.5\\\hline
Legal highs&70.2&71.0&71.0&66.4&70.9&70.6&3.8&5.3&2.5&1.9\\\hline
LSD&81.2&80.6&80.6&76.8&80.3&80.2&4.3&2.9&1.5&2.2\\\hline
Methadone&70.2&69.4&69.5&65.3&69.3&69.1&3.3&4.1&2.6&3.2\\\hline
MMushrooms&70.5&68.7&68.7&61.4&68.4&68.2&6.8&9.1&4.4&7.9\\\hline
Nicotine&63.9&63.8&63.8&63.5&63.5&63.5&0.5&0.8&0.3&0.6\\\hline
VSA&81.0&78.7&78.7&61.9&78.6&78.4&4.8&0.0&5.2&7.7\\\hline
Heroin pleiad&70.7&70.3&70.3&65.8&69.8&69.4&4.9&1.6&2.1&1.6\\\hline
Ecstasy pleiad&77.3&77.3&77.3&76.6&77.0&76.9&0.6&0.3&0.6&0.8\\\hline
Benz. pleiad&68.3&68.5&68.4&66.9&68.3&68.0&1.1&1.1&1.2&0.9\\\hline
Illicit drugs&77.2&77.1&77.2&76.8&76.7&76.8&0.7&0.8&0.7&0.5\\\hline
\end{tabular}
\end{table}

\begin{table}\centering
\caption{Coefficients of linear discriminant for user/non-user separation  and decade-based definition of users (7 attributes) \label{TabDiscr5}}
\begin{tabular}{|l|r|r|r|r|r|r|r|r|}\hline
\multicolumn{1}{|c|}{Drug}&\multicolumn{1}{c|}{$\Theta$}&\multicolumn{1}{c|}{N}&\multicolumn{1}{c|}{E}& \multicolumn{1}{c|}{O}&\multicolumn{1}{c|}{A}&\multicolumn{1}{c|}{C}&\multicolumn{1}{c|}{Imp}&\multicolumn{1}{c|}{SS}\\\hline
Alcohol&-22.282&0.124&0.149&-0.068&-0.102&-0.460&0.056&0.856\\\hline
Amphetamines&-36.596&0.040&-0.197&0.363&-0.063&-0.343&0.040&0.840\\\hline
Amyl nitrite&-29.255&0.083&0.177&-0.023&-0.104&-0.308&-0.195&0.904\\\hline
Benz.&-56.705&0.532&-0.211&0.597&-0.205&-0.186&0.105&0.478\\\hline
Cannabis&-22.278&-0.004&-0.242&0.571&-0.143&-0.357&-0.014&0.684\\\hline
Chocolate&6.158&-0.133&-0.369&0.603&0.271&-0.308&-0.467&0.309\\\hline
Cocaine&-43.257&0.203&0.037&0.303&-0.288&-0.293&0.024&0.834\\\hline
Caffeine&-9.371&-0.121&0.479&0.313&0.026&-0.682&-0.139&0.416\\\hline
Crack&-48.494&0.145&-0.293&0.326&-0.202&-0.243&0.415&0.718\\\hline
Ecstasy&-35.896&0.045&-0.017&0.407&-0.085&-0.342&-0.156&0.827\\\hline
Heroin&-55.851&0.381&-0.283&0.526&-0.322&-0.124&0.249&0.563\\\hline
Ketamine&-34.711&-0.003&-0.094&0.478&-0.197&-0.333&-0.024&0.782\\\hline
Legal highs&-30.803&-0.017&-0.227&0.532&-0.040&-0.303&-0.118&0.747\\\hline
LSD&-34.972&-0.108&-0.258&0.632&-0.103&-0.162&-0.068&0.694\\\hline
Methadone&-27.409&0.208&-0.416&0.620&-0.262&-0.253&0.058&0.513\\\hline
MMushrooms&-29.576&-0.147&-0.239&0.631&-0.106&-0.265&0.002&0.665\\\hline
Nicotine&-38.644&0.171&-0.092&0.397&-0.093&-0.393&0.005&0.801\\\hline
VSA&-34.395&0.071&-0.209&0.326&-0.060&-0.392&0.012&0.829\\\hline
Heroin pleiad&-30.383&0.113&-0.149&0.441&-0.270&-0.331&0.007&0.767\\\hline
Ecstasy pleiad&-17.510&-0.002&-0.222&0.542&-0.180&-0.385&-0.028&0.689\\\hline
Benz. pleiad&-35.808&0.232&-0.136&0.492&-0.201&-0.347&-0.042&0.723\\\hline
Illicit drugs&-18.501&0.096&-0.272&0.537&-0.169&-0.409&0.014&0.657\\\hline \hline
SD&$\leq 0.970$&$\leq 0.008$&$\leq 0.008$&$\leq 0.007$&$\leq 0.007$&$\leq 0.008$&$\leq 0.008$&$\leq 0.009$\\\hline
\end{tabular}
\end{table}

\begin{table}\centering
\caption{Coefficients of linear discriminant for user/non-user separation  and year-based definition of users (7 attributes) \label{TabDiscr6}}
\begin{tabular}{|l|r|r|r|r|r|r|r|r|}\hline
\multicolumn{1}{|c|}{Drug}&\multicolumn{1}{c|}{$\Theta$}&\multicolumn{1}{c|}{N}&\multicolumn{1}{c|}{E}& \multicolumn{1}{c|}{O}&\multicolumn{1}{c|}{A}&\multicolumn{1}{c|}{C}&\multicolumn{1}{c|}{Imp}&\multicolumn{1}{c|}{SS}\\\hline
Alcohol&-16.181&-0.136&0.270&-0.082&-0.024&-0.236&-0.314&0.865\\\hline
Amphetamines&-39.624&0.061&-0.213&0.424&-0.136&-0.365&0.166&0.769\\\hline
Amyl nitrite&-5.472&0.060&0.023&0.157&-0.388&-0.292&-0.319&0.796\\\hline
Benz.&-64.199&0.566&-0.171&0.456&-0.220&-0.112&0.089&0.611\\\hline
Cannabis&-24.008&-0.064&-0.230&0.555&-0.040&-0.354&-0.093&0.706\\\hline
Chocolate&0.709&0.114&-0.159&0.582&0.264&-0.380&-0.606&0.203\\\hline
Cocaine&-46.414&0.224&0.152&0.187&-0.299&-0.292&0.030&0.846\\\hline
Caffeine&-8.114&-0.222&0.503&0.076&-0.063&-0.636&0.052&0.530\\\hline
Crack&-52.425&0.244&-0.153&0.140&-0.219&-0.282&0.399&0.782\\\hline
Ecstasy&-42.579&-0.019&0.078&0.373&-0.045&-0.356&-0.096&0.846\\\hline
Heroin&-44.733&0.361&-0.327&0.368&-0.397&-0.108&0.218&0.641\\\hline
Ketamine&-42.342&0.056&-0.003&0.417&-0.143&-0.363&-0.047&0.818\\\hline
Legal highs&-30.027&-0.062&-0.181&0.460&-0.091&-0.300&-0.098&0.803\\\hline
LSD&-40.183&-0.141&-0.220&0.650&-0.039&-0.118&-0.115&0.693\\\hline
Methadone&-28.093&0.234&-0.417&0.575&-0.296&-0.206&0.024&0.556\\\hline
MMushrooms&-36.864&-0.142&-0.182&0.669&-0.093&-0.215&-0.053&0.664\\\hline
Nicotine&-35.594&0.122&-0.126&0.376&-0.036&-0.441&0.019&0.795\\\hline
VSA&-55.631&0.382&0.009&0.213&-0.038&-0.379&-0.014&0.815\\\hline
Heroin pleiad&-32.917&0.154&-0.147&0.385&-0.310&-0.312&0.044&0.782\\\hline
Ecstasy pleiad&-20.439&-0.059&-0.264&0.548&-0.102&-0.354&-0.045&0.699\\\hline
Benz. pleiad&-41.126&0.256&-0.164&0.438&-0.230&-0.287&0.018&0.761\\\hline
Illicit drugs&-22.701&0.033&-0.295&0.549&-0.125&-0.351&-0.012&0.686\\\hline \hline
SD&$\leq 1.250$&$\leq 0.011$&$\leq 0.011$&$\leq 0.009$&$\leq 0.009$&$\leq 0.009$&$\leq 0.008$&$\leq 0.012$\\\hline
\end{tabular}
\end{table}

\begin{table}\centering
\caption{Coefficients of linear discriminant for user/non-user separation  and month-based definition of users (7 attributes) \label{TabDiscr7}}
\begin{tabular}{|l|r|r|r|r|r|r|r|r|}\hline
\multicolumn{1}{|c|}{Drug}&\multicolumn{1}{c|}{$\Theta$}&\multicolumn{1}{c|}{N}&\multicolumn{1}{c|}{E}& \multicolumn{1}{c|}{O}&\multicolumn{1}{c|}{A}&\multicolumn{1}{c|}{C}&\multicolumn{1}{c|}{Imp}&\multicolumn{1}{c|}{SS}\\\hline
Alcohol&-63.256&0.190&0.823&0.002&0.087&-0.005&-0.268&0.456\\\hline
Amphetamines&-24.230&0.011&-0.223&0.361&-0.254&-0.499&0.375&0.605\\\hline
Amyl nitrite&-6.284&-0.243&-0.183&-0.082&-0.187&-0.092&-0.117&0.918\\\hline
Benz.&-69.623&0.702&-0.189&0.286&-0.215&0.006&0.148&0.568\\\hline
Cannabis&-25.637&-0.145&-0.215&0.606&-0.069&-0.322&-0.051&0.674\\\hline
Chocolate&19.518&-0.121&0.254&0.308&0.168&-0.595&-0.621&0.240\\\hline
Cocaine&-57.737&0.366&0.148&0.082&-0.370&-0.014&-0.001&0.837\\\hline
Caffeine&-91.137&0.351&0.687&0.026&0.052&-0.080&0.567&0.271\\\hline
Crack&-46.531&0.459&-0.220&-0.077&-0.357&0.096&0.066&0.771\\\hline
Ecstasy&-27.572&-0.169&0.101&0.462&-0.191&-0.354&-0.139&0.752\\\hline
Heroin&-30.020&0.302&-0.409&0.232&-0.499&-0.107&0.345&0.555\\\hline
Ketamine&-6.182&-0.169&-0.210&0.643&-0.232&-0.464&-0.044&0.493\\\hline
Legal highs&-17.306&-0.067&-0.240&0.474&-0.201&-0.388&-0.053&0.721\\\hline
LSD&-47.031&-0.012&-0.097&0.762&-0.023&-0.225&-0.155&0.578\\\hline
Methadone&-17.825&0.260&-0.501&0.436&-0.286&-0.238&-0.004&0.594\\\hline
MMushrooms&-29.897&-0.229&-0.273&0.762&-0.037&-0.217&-0.019&0.494\\\hline
Nicotine&-34.684&0.072&-0.032&0.342&-0.022&-0.590&0.204&0.698\\\hline
VSA&-70.559&0.137&0.068&0.230&-0.208&-0.073&0.182&0.918\\\hline
Heroin pleiad&-41.444&0.277&-0.272&0.284&-0.346&-0.105&0.113&0.791\\\hline
Ecstasy pleiad&-19.318&-0.141&-0.257&0.573&-0.125&-0.359&0.014&0.664\\\hline
Benz. pleiad&-47.877&0.416&-0.199&0.300&-0.258&-0.246&0.128&0.745\\\hline
Illicit drugs&-21.767&-0.051&-0.295&0.567&-0.129&-0.344&0.009&0.674\\\hline \hline
SD&$\leq 1.041$&$\leq 0.009$&$\leq 0.009$&$\leq 0.009$&$\leq 0.008$&$\leq 0.007$&$\leq 0.007$&$\leq 0.010$\\\hline
\end{tabular}
\end{table}

\begin{table}\centering
\caption{Coefficients of linear discriminant for user/non-user separation  and week-based definition of users (7 attributes) \label{TabDiscr8}}
\begin{tabular}{|l|r|r|r|r|r|r|r|r|}\hline
\multicolumn{1}{|c|}{Drug}&\multicolumn{1}{c|}{$\Theta$}&\multicolumn{1}{c|}{N}&\multicolumn{1}{c|}{E}& \multicolumn{1}{c|}{O}&\multicolumn{1}{c|}{A}&\multicolumn{1}{c|}{C}&\multicolumn{1}{c|}{Imp}&\multicolumn{1}{c|}{SS}\\\hline
Alcohol&-52.758&0.202&0.791&-0.273&0.041&-0.002&-0.176&0.476\\\hline
Amphetamines&-17.930&-0.054&-0.311&0.402&-0.237&-0.472&0.577&0.356\\\hline
Amyl nitrite&5.116&-0.466&-0.348&-0.188&-0.084&-0.186&0.249&0.723\\\hline
Benz.&-51.825&0.707&-0.438&0.240&-0.229&0.052&0.224&0.382\\\hline
Cannabis&-28.488&-0.136&-0.215&0.658&-0.074&-0.282&-0.079&0.641\\\hline
Chocolate&-22.538&0.221&0.565&-0.149&0.195&0.020&0.302&-0.693\\\hline
Cocaine&-42.900&0.347&0.198&-0.011&-0.407&-0.090&-0.103&0.810\\\hline
Caffeine&-53.485&0.422&0.821&-0.118&-0.135&-0.146&0.301&-0.061\\\hline
Crack&-22.095&0.304&-0.279&-0.282&-0.463&0.148&0.119&0.707\\\hline
Ecstasy&-60.127&-0.095&0.168&0.695&-0.128&-0.355&0.242&0.528\\\hline
Heroin&-37.099&0.548&-0.325&0.265&-0.547&-0.069&0.389&0.261\\\hline
Ketamine&1.992&-0.284&-0.395&0.574&-0.307&-0.228&-0.030&0.535\\\hline
Legal highs&0.347&-0.015&-0.352&0.406&-0.315&-0.454&-0.022&0.636\\\hline
LSD&-35.647&0.001&0.054&0.716&0.008&-0.371&-0.309&0.501\\\hline
Methadone&-14.932&0.317&-0.636&0.331&-0.310&-0.106&0.084&0.521\\\hline
MMushrooms&-66.891&-0.143&-0.071&0.715&0.073&-0.114&0.059&0.665\\\hline
Nicotine&-42.383&0.175&-0.013&0.222&0.012&-0.574&0.284&0.714\\\hline
VSA&-75.762&0.043&-0.068&0.419&-0.191&0.250&0.056&0.846\\\hline
Heroin pleiad&-25.458&0.341&-0.397&0.242&-0.379&-0.205&0.116&0.684\\\hline
Ecstasy pleiad&-19.877&-0.146&-0.272&0.603&-0.124&-0.338&-0.004&0.642\\\hline
Benz. pleiad&-43.258&0.498&-0.349&0.260&-0.250&-0.274&0.343&0.554\\\hline
Illicit drugs&-22.772&-0.055&-0.329&0.575&-0.113&-0.322&0.004&0.665\\\hline \hline
SD&$\leq 0.707$&$\leq 0.006$&$\leq 0.006$&$\leq 0.006$&$\leq 0.007$&$\leq 0.007$&$\leq 0.009$&$\leq 0.007$\\\hline
\end{tabular}
\end{table}

\begin{table}\centering
\caption{Performance and stability of linear discriminant for  decade-based definition of users  (7 attributes).  \label{Tab:DiscrPerf5}}
\begin{tabular}{|l|>{\quad}c<{\quad}|>{\quad}c<{\quad}|>{\quad}c<{\quad}|>{\quad}c<{\quad}|>{\quad}c<{\quad}|>{\quad}c<{\quad}|c|c|c|c|}\hline
&\multicolumn{3}{c|}{Total sample}&\multicolumn{3}{c|}{LOOCV}&\multicolumn{4}{c|}{Stability indicators}\\\cline{2-11}
 Drug  &Sn &Sp&Acc&Sn &Sp& Acc &TP$\to$FN&FN$\to$TP&FP$\to$TN&TN$\to$FP\\\hline
Alcohol&67.1&64.7&67.0&66.9&55.9&66.5&7.6&4.7&11.8&4.4\\\hline
Amphetamines&69.1&69.3&69.2&68.5&68.7&68.6&0.7&1.0&1.1&1.2\\\hline
Amyl nitrite&61.9&61.8&61.9&61.1&61.7&61.5&1.6&2.4&0.9&2.0\\\hline
Benz.&67.6&67.3&67.4&67.0&66.8&66.9&0.8&0.7&0.8&0.9\\\hline
Cannabis&75.0&75.2&75.1&74.9&74.4&74.7&0.5&0.5&0.8&0.6\\\hline
Chocolate&56.7&54.3&56.7&56.4&42.9&56.2&11.4&13.1&5.7&20.0\\\hline
Cocaine&66.4&66.5&66.5&66.1&66.2&66.2&0.4&0.6&0.7&0.7\\\hline
Caffeine&69.9&70.3&69.9&69.7&56.8&69.4&6.5&7.5&10.8&8.1\\\hline
Crack&66.0&65.9&65.9&62.8&65.6&65.3&4.7&1.6&2.1&2.6\\\hline
Ecstasy&71.6&71.6&71.6&71.4&71.4&71.4&0.5&0.8&0.4&0.7\\\hline
Heroin&70.3&70.8&70.7&68.4&70.6&70.3&1.9&1.4&2.1&1.6\\\hline
Ketamine&66.9&66.8&66.8&65.1&66.6&66.4&2.3&2.0&1.2&1.1\\\hline
Legal highs&73.4&73.3&73.3&73.0&72.9&72.9&0.4&0.1&0.5&0.6\\\hline
LSD&72.4&72.4&72.4&71.6&72.2&72.0&0.7&0.4&0.6&0.9\\\hline
Methadone&69.3&69.5&69.4&68.6&69.2&69.1&1.2&1.4&1.0&1.1\\\hline
MMushrooms&71.0&70.9&71.0&70.0&70.6&70.4&1.2&0.9&1.1&0.9\\\hline
Nicotine&66.3&66.0&66.2&66.1&65.2&65.8&0.9&0.7&1.0&1.1\\\hline
VSA&70.4&70.5&70.5&67.8&70.4&70.1&3.0&1.7&1.9&0.8\\\hline
Heroin pleiad&68.1&68.3&68.2&67.5&67.7&67.6&0.6&0.4&0.7&0.1\\\hline
Ecstasy pleiad&76.2&75.5&76.0&75.8&75.2&75.6&1.1&0.3&0.2&0.5\\\hline
Benz. pleiad&68.7&68.8&68.8&68.6&68.5&68.5&0.2&0.8&0.4&0.6\\\hline
Illicit drugs&74.9&75.2&75.1&74.5&74.9&74.8&0.2&0.9&0.6&0.7\\\hline
\end{tabular}
\end{table}

\begin{table}\centering
\caption{Performance and stability of linear discriminant for year-based definition of users  (7 attributes).
 \label{Tab:DiscrPerf6}}
\begin{tabular}{|l|>{\quad}c<{\quad}|>{\quad}c<{\quad}|>{\quad}c<{\quad}|>{\quad}c<{\quad}|>{\quad}c<{\quad}|>{\quad}c<{\quad}|c|c|c|c|}\hline
&\multicolumn{3}{c|}{Total sample}&\multicolumn{3}{c|}{LOOCV}&\multicolumn{4}{c|}{Stability indicators}\\\cline{2-11}
Drug  &Sn &Sp&Acc&Sn &Sp& Acc &TP$\to$FN&FN$\to$TP&FP$\to$TN&TN$\to$FP\\\hline
Alcohol&58.0&58.1&58.0&57.7&54.4&57.5&4.7&5.3&3.7&5.9\\\hline
Amphetamines&70.6&70.4&70.5&69.5&70.2&70.0&1.8&0.9&0.8&0.9\\\hline
Amyl nitrite&65.4&65.2&65.3&63.2&65.0&64.8&2.3&3.0&2.7&3.7\\\hline
Benz.&68.0&68.2&68.2&67.3&67.9&67.7&1.3&1.1&0.8&0.9\\\hline
Cannabis&75.0&74.8&74.9&74.9&74.6&74.7&0.4&0.5&0.2&0.3\\\hline
Chocolate&58.0&60.0&58.0&57.9&40.0&57.5&15.5&15.3&20.0&6.7\\\hline
Cocaine&68.1&68.3&68.3&67.4&68.1&68.0&0.2&0.7&1.3&1.1\\\hline
Caffeine&65.2&65.6&65.3&65.1&63.9&65.1&5.4&7.0&3.3&6.6\\\hline
Crack&69.6&68.4&68.5&62.0&68.3&68.0&5.1&2.5&2.9&5.3\\\hline
Ecstasy&70.6&70.1&70.2&69.8&70.0&69.9&1.0&0.4&0.4&0.8\\\hline
Heroin&71.2&70.9&70.9&64.4&70.7&70.3&6.8&3.4&2.4&2.4\\\hline
Ketamine&68.3&68.8&68.8&67.3&68.6&68.5&1.4&1.0&2.0&2.0\\\hline
Legal highs&73.0&73.3&73.2&72.5&72.7&72.7&0.7&0.7&1.1&0.7\\\hline
LSD&72.1&72.0&72.0&71.1&71.8&71.6&1.1&1.1&0.9&1.0\\\hline
Methadone&69.1&69.3&69.3&68.1&69.1&68.9&0.9&2.2&1.0&1.4\\\hline
MMushrooms&71.7&71.8&71.8&71.0&71.7&71.5&0.9&0.9&0.8&0.9\\\hline
Nicotine&67.5&67.8&67.6&67.3&67.3&67.3&0.3&0.5&0.8&0.4\\\hline
VSA&68.4&68.5&68.5&63.2&68.3&68.1&5.3&5.3&2.8&3.5\\\hline
Heroin pleiad&68.5&68.6&68.6&67.9&68.5&68.3&1.0&0.3&0.8&0.8\\\hline
Ecstasy pleiad&75.9&76.1&76.0&75.7&75.6&75.6&0.4&0.6&0.9&0.4\\\hline
Benz. pleiad&70.1&69.8&69.9&69.3&69.2&69.2&1.1&0.6&0.7&0.8\\\hline
Illicit drugs&76.5&76.4&76.4&76.2&76.3&76.2&0.6&0.3&0.3&0.3\\\hline
\end{tabular}
\end{table}

\begin{table}\centering
\caption{Performance and stability of linear discriminant for month-based definition of users  (7 attributes).  \label{Tab:DiscrPerf7}}
\begin{tabular}{|l|>{\quad}c<{\quad}|>{\quad}c<{\quad}|>{\quad}c<{\quad}|>{\quad}c<{\quad}|>{\quad}c<{\quad}|>{\quad}c<{\quad}|c|c|c|c|}\hline
&\multicolumn{3}{c|}{Total sample}&\multicolumn{3}{c|}{LOOCV}&\multicolumn{4}{c|}{Stability indicators}\\\cline{2-11}
Drug  &Sn &Sp&Acc&Sn &Sp& Acc &TP$\to$FN&FN$\to$TP&FP$\to$TN&TN$\to$FP\\\hline
Alcohol&58.5&57.8&58.4&57.8&56.0&57.5&2.4&1.7&3.3&2.4\\\hline
Amphetamines&68.1&67.7&67.7&64.7&67.5&67.2&4.6&0.8&1.8&1.4\\\hline
Amyl nitrite&70.7&70.8&70.8&61.0&70.6&70.4&2.4&2.4&3.9&6.7\\\hline
Benz.&68.2&68.0&68.0&67.6&68.1&68.0&1.3&1.3&1.2&2.4\\\hline
Cannabis&73.7&73.7&73.7&73.2&73.6&73.4&0.6&0.1&0.3&0.5\\\hline
Chocolate&56.7&56.6&56.7&56.0&46.5&55.5&11.1&11.5&11.1&11.1\\\hline
Cocaine&67.3&68.0&68.0&65.4&67.8&67.6&3.8&3.1&2.3&1.9\\\hline
Caffeine&58.7&59.5&58.8&59.3&53.7&58.9&4.0&6.8&5.8&4.1\\\hline
Crack&80.0&81.4&81.4&45.0&81.4&81.0&5.0&5.0&7.9&4.1\\\hline
Ecstasy&66.7&66.6&66.6&65.4&66.5&66.4&1.7&1.7&1.6&1.6\\\hline
Heroin&75.5&75.8&75.8&67.9&75.6&75.4&9.4&3.8&4.3&3.5\\\hline
Ketamine&67.1&66.7&66.7&59.5&66.4&66.1&8.9&2.5&4.8&6.3\\\hline
Legal highs&68.5&68.7&68.7&66.4&68.4&68.1&2.5&2.1&1.2&1.0\\\hline
LSD&69.9&70.0&70.0&69.3&69.8&69.8&0.6&2.4&1.8&3.2\\\hline
Methadone&66.7&66.9&66.9&63.7&66.6&66.4&2.9&2.3&2.1&1.6\\\hline
MMushrooms&71.7&71.4&71.4&67.9&71.1&70.9&4.4&3.1&1.3&1.4\\\hline
Nicotine&64.1&64.2&64.1&63.5&63.7&63.6&1.0&0.7&0.8&0.5\\\hline
VSA&70.6&70.3&70.3&55.9&70.2&69.9&11.8&5.9&4.2&7.8\\\hline
Heroin pleiad&67.3&67.0&67.1&64.7&66.6&66.3&3.2&1.9&1.5&1.3\\\hline
Ecstasy pleiad&74.4&74.4&74.4&74.0&74.2&74.1&0.3&0.3&0.4&0.8\\\hline
Benz. pleiad&68.9&69.0&69.0&68.6&68.8&68.7&0.2&0.6&1.0&1.8\\\hline
Illicit drugs&74.6&74.4&74.5&74.2&74.2&74.2&0.6&0.4&0.2&0.6\\\hline
\end{tabular}
\end{table}

\begin{table}\centering
\caption{Performance and stability of linear discriminant for week-based definition of users  (7 attributes). \label{Tab:DiscrPerf8}}
\begin{tabular}{|l|>{\quad}c<{\quad}|>{\quad}c<{\quad}|>{\quad}c<{\quad}|>{\quad}c<{\quad}|>{\quad}c<{\quad}|>{\quad}c<{\quad}|c|c|c|c|}\hline
&\multicolumn{3}{c|}{Total sample}&\multicolumn{3}{c|}{LOOCV}&\multicolumn{4}{c|}{Stability indicators}\\\cline{2-11}
Drug  &Sn &Sp&Acc&Sn &Sp& Acc &TP$\to$FN&FN$\to$TP&FP$\to$TN&TN$\to$FP\\\hline
Alcohol&56.0&55.6&55.9&55.2&54.1&54.9&1.4&1.3&1.6&2.4\\\hline
Amphetamines&66.3&66.3&66.3&63.2&65.9&65.6&4.9&3.1&2.3&2.5\\\hline
Amyl nitrite&70.6&85.0&84.8&47.1&84.9&84.5&0.0&5.9&13.1&4.6\\\hline
Benz.&69.8&69.5&69.5&67.0&69.3&69.1&3.9&1.7&2.0&2.6\\\hline
Cannabis&71.1&71.2&71.2&70.8&70.7&70.8&0.8&0.3&0.5&0.5\\\hline
Chocolate&54.7&54.9&54.7&54.3&51.4&53.7&3.0&3.4&3.8&3.0\\\hline
Cocaine&71.7&69.0&69.1&65.0&68.8&68.7&6.7&3.3&3.2&6.5\\\hline
Caffeine&54.3&53.7&54.2&53.8&50.2&53.4&4.6&5.4&2.6&5.7\\\hline
Crack&81.8&82.3&82.3&45.5&82.2&82.0&9.1&9.1&7.7&6.7\\\hline
Ecstasy&70.2&71.4&71.4&65.5&71.2&70.9&7.1&6.0&3.8&3.3\\\hline
Heroin&79.3&80.7&80.6&65.5&80.6&80.4&6.9&0.0&5.3&3.0\\\hline
Ketamine&64.9&66.7&66.6&56.8&66.5&66.3&5.4&10.8&8.0&6.7\\\hline
Legal highs&69.5&68.6&68.7&63.4&68.5&68.1&4.6&2.3&2.3&3.1\\\hline
LSD&69.6&70.6&70.6&65.2&70.5&70.3&4.3&2.9&4.4&3.3\\\hline
Methadone&68.6&68.0&68.1&64.5&67.9&67.7&5.8&2.5&2.7&4.0\\\hline
MMushrooms&72.7&72.9&72.9&63.6&72.8&72.6&6.8&2.3&5.5&4.9\\\hline
Nicotine&62.5&62.7&62.6&62.1&62.3&62.2&0.4&1.2&0.9&0.9\\\hline
VSA&71.4&74.3&74.3&61.9&74.2&74.1&9.5&4.8&8.3&5.5\\\hline
Heroin pleiad&68.5&68.4&68.4&65.2&68.1&67.9&3.3&2.7&1.8&1.8\\\hline
Ecstasy pleiad&72.3&72.5&72.4&72.0&72.1&72.0&0.5&0.1&0.5&0.4\\\hline
Benz. pleiad&68.0&68.3&68.3&66.9&68.1&67.9&0.8&0.3&1.0&1.2\\\hline
Illicit drugs&73.4&72.9&73.2&72.8&72.6&72.7&0.9&0.2&0.2&0.7\\\hline
\end{tabular}
\end{table}
}

%% file: AuthorsBio.tex
\chapter*{Authors}

\noindent{\bf Elaine Fehrman}, BSc (Hons), BA DipSW, MSc, is employed as an HCPC-registered Advanced Practitioner at Rampton Hospital, which is one of three high secure hospitals in England and Wales. She is the substance misuse treatment clinical lead for the National High Secure Healthcare Service for Women at Rampton Hospital, and has several years clinical practice experience working with offenders with mental health problems in low, medium, and high security, along with the community and prison. She has a keen interest in improving the recovery and treatment outcomes for patients. Elaine is currently completing her Doctorate in Forensic Psychology, and is undertaking research investigating the psychological sequelae of trauma in high risk mentally disordered offenders with substance misuse issues, and to determine if gender differences exist. The research will have important implications for informing treatment interventions at Rampton Hospital with this cohort with interrelated mental health issues.
\vspace{5mm}

\noindent{\bf Vincent Egan} is an associate professor of forensic psychology practice in the Department of Psychiatry and Applied Psychology at the University of Nottingham, and a HCPC-accredited forensic and clinical psychologist. He obtained a BSc (Honours) degree in Psychology from the University of London in 1984, A PhD in Psychology from the University of Edinburgh in 1991, and a Doctorate in Clinical Psychology from the University of Leicester in 1996. After post-doctoral work in the Department of Psychiatry at the University of Edinburgh, he was employed as a Clinical Psychologist by the Central Nottinghamshire Health Service NHS Trust at the East Midlands Centre for Forensic Mental Health, then became, successively, director of the MSc in Forensic Psychology at Glasgow Caledonian University, then director of the accredited course in Forensic Psychology at the University of Leicester.  \href{https://scholar.google.com/citations?user=9vw4Oa4AAAAJ&hl=en}{Vinsent Egan} on GOOGLE Scholar.
 \vspace{5mm}

\noindent{\bf Alexander N. Gorban} (PhD, ScD, Professor) holds a Personal Chair in applied 
mathematics at the University of Leicester since 2004. He worked for Russian Academy of
Sciences, Siberian Branch (Krasnoyarsk, Russia), and ETH Z\"urich (Switzerland), was a
visiting Professor and Research Scholar at Clay Mathematics Institute (Cambridge, MA),
IHES (Bures-sur-Yvette, \^{I}le de France), Cou\-rant Institute of Mathematical Sciences
(New York), and Isaac Newton Institute for Mathematical Sciences (Cambridge, UK). His
main research interests are machine learning, data mining and model reduction problems,  dynamics of systems of physical, chemical and biological kinetics, and biomathematics. He has been accorded the title of Pioneer of Russian Neuroinformatics (2017) for his extraordinary contribution into theory and applications of artificial neural networks, received Lifetime Achievement Award (MaCKIE-2015) in recognition of outstanding contributions to the research field of (bio)chemical kinetics, and was awarded by Prigogine medal  (2003) for achievements in non-equilibrium themodynamics and physical kinetics. \href{https://scholar.google.com/citations?hl=en&user=D8XkcCIAAAAJ}{Alexander N. Gorban} on GOOGLE Scholar.
\vspace{5mm}

\noindent {\bf Jeremy Levesley } (PhD, FIMA) is a Professor in the Department of Mathematics at the University of Leicester. His research area is kernel based approximation methods in high dimensions, in Euclidean space and on manifolds. He is interested in developing research at the interface of mathematics and medicine, and sees interpretation of medical data sets as a key future challenge for mathematics. \href{https://scholar.google.com/citations?user=I3QvO2MAAAAJ&hl=en}{Jeremy Levesley} on GOOGLE Scholar.
\vspace{5mm}

\noindent {\bf Evgeny M. Mirkes} (PhD, ScD) is a Research Fellow at the University of Leicester.
He worked for Russian Academy of Sciences, Siberian Branch, and Siberian Federal
University (Krasnoyarsk, Russia). His main research interests are biomathematics, data
mining and software engineering, neural networks, and artificial intelligence. He led and
supervised many medium-sized research and industrial projects in data analysis and development of
decision-support systems for pattern recognition, computational diagnosis and treatment planning.
\href{https://scholar.google.com/citations?user=yKkB_D0AAAAJ&hl=en}{Evgeny M. Mirkes}  on GOOGLE Scholar.
\vspace{5mm}

\noindent{\bf Awaz K. Muhammad} is a PhD student in applied mathematics at the University of Leicester and a lecturer (on leave) of  University of Salahaddin, Erbil, Kurdistan Region, Iraq, sertified SAS programmer. She  was awarded BSc degree in Mathematics in 2004, and   MSc degree in Mathematics in 2010 from the University of Salahaddin, Erbil, Iraq. College of Sciences Education. Her  PhD viva was held in autumn 2017 in Leicester. Her main research interest is statistical programming, statistical analysis of big data sets, and reliable and robust statistical inference.